\documentclass[11 pt]{article}

\usepackage{authblk} % author package
\usepackage{tabularx} % extra features for tabular environment
\usepackage{amsmath,amssymb}  % improve math presentation
\usepackage{mathtools} % includes \coloneqq

\usepackage{graphicx} % takes care of graphic including machinery
\usepackage{subcaption} % package for subfigure
\usepackage{float}
\usepackage[margin=1in,letterpaper]{geometry} % decreases margins
\usepackage{cite} % takes care of citations
\usepackage[final]{hyperref} % adds hyperlinks inside the generated pdf file
\hypersetup{
  colorlinks   = true, %Colours links instead of ugly boxes
  urlcolor     = black, %Colour for external hyperlinks
  linkcolor    = blue, %Colour of internal links
  citecolor   = blue %Colour of citations
}
\usepackage[nobiblatex]{xurl}
\usepackage{lscape} % package for landscape
\usepackage{booktabs} % package for tex professional table
\usepackage{bm} % bold math 

\usepackage{setspace} % set space for text, figures, and tables 
\usepackage[flushleft]{threeparttable} %package for tablenotes
\usepackage{natbib}
\usepackage{bibentry}
\usepackage[nottoc,numbib]{tocbibind} % add Reference to table of contents
\usepackage{sectsty} % to use \sectionfont
\usepackage{titlesec}
\titleformat{\paragraph}{\normalfont\normalsize\bfseries}{\theparagraph}{1em}{}
\titlespacing*{\paragraph}{0pt}{3.25ex plus 1ex minus .2ex}{1.5ex plus .2ex}

\usepackage{tikz} % graph in latex
\usetikzlibrary{positioning,arrows.meta} %tikz diagram
\usepackage{pgfplots}
\pgfplotsset{compat=newest} % This setting allows a node to go to the correct place
\usepgfplotslibrary{fillbetween}
\usepackage{siunitx} % to use mm in tikz
\usetikzlibrary{{shapes.misc}}
\usepackage{pdflscape} %pdf landscape page

\newenvironment{chapquote}[2][2em]
  {\setlength{\@tempdima}{#1}%
   \def\chapquote@author{#2}%
   \parshape 1 \@tempdima \dimexpr\textwidth-2\@tempdima\relax%
   \itshape}
  {\par\normalfont\hfill--\ \chapquote@author\hspace*{\@tempdima}\par\bigskip}
\makeatother

\newcommand{\comment}[1]{}

\newcommand{\lnb}[1]{%
  \ln\left(#1\right)%
}

\usepackage{setspace} % set space for text, figures, and tables 
\begin{document}
%The legibility of text can be enhanced by separating paragraphs with an amount of white space that will vary according to some design aesthetic.
\setlength{\parskip}{3pt plus1pt minus1pt}

\setlength{\abovedisplayskip}{3pt} % the space above equations 
\setlength{\belowdisplayskip}{3pt} % the space below equations 

%--------------------------------------------------------
\begin{titlepage}
\newgeometry{top=0.6in, bottom=0.8in, left=0.7in, right=0.7in}  % Adjust as needed
\vspace{-30mm}
\title{A Unified Credit Expansion Theory on Housing Cycle: \\
Causal Evidence for Within- and Cross-Metro Patterns in the \\
Prior, Boom, Bust, and Recovery Periods }
%\author[1]{}
\author[1]{\vspace{-2mm} \textbf{Bo Li}  \thanks{ \protect\linespread{1}\protect\selectfont {\scriptsize Bo Li is with the Department of Finance, Georgia Institute of Technology (bli96@gatech.edu). I am extremely grateful for the advice and constant encouragement from my ASU dissertation committee, Laura Lindsey (chair), Oliver Boguth, and Rawley Heimer. This paper benefits from invaluable discussions with Greg Bochak, Anthoney Zhang, Ruiyuan Chen, Emil Verner, Paul Willen, Anthony DeFusco, Sudheer Chava, Rohan Ganduri, Joseph Hall, Rik Sen, Ruchi Singh, Gen Li (discussant), Heejin Yoon (discussant), Leili Rostami (discussant), Huiming Zhang (discussant), Christopher Hansman, Christoph Schiller, David Schreindorfer, Yuri Tserlukevich, Ilona Babenka, Mark Seasholes, Maziar Kazemi, Sreedhar Bharath, Michael Barnett, Hendrik Bessembinder, Shasta Shakya, Kimberly Winson, my ASU classmates, and participants at the seminars of Arizona State University, Georgia Institute of Technology, 2024 Eastern Finance Annual Conference, 2024 Economics Graduate Student Conference (WashU), 2025 Financial Management Association Annual Conference, 2025 Southern Finance Association Annual Conference, and 2026 Southwestern Finance Association Annual Conference.  All errors are my own. The author declares that he has no financial relationships or other potential conflicts of interest relevant to this research.} } }

\date{\vspace{-7mm} This Version: May 16th, 2026}

\maketitle

\vspace{-8mm}

\begin{abstract}
\fontsize{11}{11}\selectfont
\noindent During the 1999-2019 U.S. housing cycle, three empirical facts present a puzzle: in the boom period, the correlation between income growth and mortgage growth is (1) negative across ZIP codes within a metropolitan area, but (2) positive across metropolitan areas, and (3) the metropolitan areas that experience the worst bust also show the strongest recovery. I develop a unified credit expansion theory that explains both within- and cross-metro patterns in the prior, boom, bust, and recovery periods (including the three facts above) and generates new testable implications of ``double differences" (cross ZIP codes and cross metros) for the four periods. Following the idea of ``Economic Base Theory", I construct local economic exposure to net export growth as the driving force of local economy and credit expansion. For the identification strategy, I use a new instrumental variable approach from the International trade literature for the following empirical results. First, I show that high-net-export-growth metros experience a stronger boom-bust-recovery housing cycle due to credit expansion in private-label (non-jumbo) mortgages (PLNJMs), rather than in government-sponsored enterprise mortgages (GSEMs), because only the former can legally respond to local economic conditions. Second, for the ``double differences", I define a low-minus-high (LMH) factor as the private-label (non-jumbo) mortgage (and house price) growth in low-income ZIP codes minus that in high-income ZIP codes within the same metropolitan area. I show that this low-minus-high factor (as a measure of credit expansion) in the high-net-export-growth metros is more positive during the boom period, more negative during the bust period, and slightly more positive in the recovery period than in the low-net-export-growth metros. Lastly, I employ five tests to demonstrate that ``speculation" is unlikely to play a dominant role in this housing cycle. 

\vspace{8mm}

\noindent \textbf{Keywords}: housing cycle, credit supply, private-label mortgages \\
\noindent  \textbf{JEL Classification}: G01, G21, R31 \\

\end{abstract}

\thispagestyle{empty}
\end{titlepage}
%------------------------------------------------------------------
\restoregeometry

%--------------------------------------------------------
\begin{titlepage}
\newgeometry{top=1.4in, bottom=0.8in, left=0.7in, right=0.7in}  % Adjust as needed

\centering

{\Large A Unified Credit Expansion Theory on Housing Cycle: \\
\vspace{2mm}
Causal Evidence for Within- and Cross-Metro Patterns in the \\
\vspace{2mm}
Prior, Boom, Bust, and Recovery Periods }

%\vspace{20mm}

%\title{\textbf{Credit Expansion and Housing Cycle}}
%\author[1]{\textbf{Bo Li}  \thanks{ \protect\linespread{1}\protect\selectfont {\footnotesize Bo Li is with the Department of Finance, Arizona State University (boli15@asu.edu). I am extremely grateful for the advising and constant encouragement from Laura Lindsey (chair), Oliver Boguth, and Rawley Heimer. This paper also benefits from invaluable discussion with Christoph Schiller, David Schreindorfer, Yuri Tserlukevich, Ilona Babenka, Shasta Shakya, Mark Seasholes, and Sreedhar Bharath.} } }
%\date{Nov 16th, 2023}
%\maketitle

\vspace{16ex}

\begin{abstract}
\fontsize{11}{11}\selectfont
\noindent During the 1999-2019 U.S. housing cycle, three empirical facts present a puzzle: in the boom period, the correlation between income growth and mortgage growth is (1) negative across ZIP codes within a metropolitan area, but (2) positive across metropolitan areas, and (3) the metropolitan areas that experience the worst bust also show the strongest recovery. I develop a unified credit expansion theory that explains both within- and cross-metro patterns in the prior, boom, bust, and recovery periods (including the three facts above) and generates new testable implications of ``double differences" (cross ZIP codes and cross metros) for the four periods. Following the idea of ``Economic Base Theory", I construct local economic exposure to net export growth as the driving force of local economy and credit expansion. For the identification strategy, I use a new instrumental variable approach from the International trade literature for the following empirical results. First, I show that high-net-export-growth metros experience a stronger boom-bust-recovery housing cycle due to credit expansion in private-label (non-jumbo) mortgages (PLNJMs), rather than in government-sponsored enterprise mortgages (GSEMs), because only the former can legally respond to local economic conditions. Second, for the ``double differences", I define a low-minus-high (LMH) factor as the private-label (non-jumbo) mortgage (and house price) growth in low-income ZIP codes minus that in high-income ZIP codes within the same metropolitan area. I show that this low-minus-high factor (as a measure of credit expansion) in the high-net-export-growth metros is more positive during the boom period, more negative during the bust period, and slightly more positive in the recovery period than in the low-net-export-growth metros. Lastly, I employ five tests to demonstrate that ``speculation" is unlikely to play a dominant role in this housing cycle.

\vspace{8mm}

\noindent \textbf{Keywords}: housing cycle, credit supply, private-label mortgages \\
\noindent  \textbf{JEL Classification}: G01, G21, R31 \\

\end{abstract}

\thispagestyle{empty}
\end{titlepage}
%------------------------------------------------------------------
\restoregeometry

\clearpage 
\thispagestyle{empty}
\renewcommand{\baselinestretch}{1}\selectfont
\tableofcontents
\renewcommand{\baselinestretch}{1.4}
\thispagestyle{empty}
%\thispagestyle{empty}

%\clearpage
%\thispagestyle{empty}

%----------------------------------------------------------------------
% section 1: Introduction

\clearpage 
\pagenumbering{arabic}
\setcounter{page}{1}

%--------------------------------------------------------------
%--------------------------------------------------------------
% This is the beginning of the entire Section of Introduction
%--------------------------------------------------------------
%--------------------------------------------------------------
\clearpage

\begin{chapquote}{Griffin, Kruger, and Maturana, \textit{JFE, 2021}}
\noindent ``Ten years after the financial crisis, the central question of what explains the rise and fall in house prices remains unresolved.”
\end{chapquote}

\vspace{-3mm}
\section{Introduction}

The U.S. housing price boom and bust in the 2000s are unprecedented. During the bust period, U.S. households experienced the deepest recession since the Great Depression, with widespread mortgage defaults \citep{mayer2009rise, keys2010did}, a large drop in consumption \citep{mian2013household, kaplan2020non}, and a substantial rise in unemployment \citep{hoynes2012suffers, mian2014explains}. Understanding the causes of such a large housing boom and bust is crucial for understanding the economic connections among credit, the housing cycle, and economic activity. It is also helpful for the design of housing policy and the regulation of mortgage credit.

In the literature on 1999-2011 U.S. housing cycle, three empirical facts present a puzzle: in the boom period, the correlation between income growth and mortgage growth is (1) negative across ZIP codes within a metropolitan area \citep{mian2009consequences}, but (2) positive across metropolitan areas \citep{adelino2016loan}, and (3) the metropolitan areas that experience the worst bust also show the strongest recovery \citep{chodorow20242000s}. The first empirical fact is consistent with the ``credit expansion" view, in which lenders' credit expansion plays the dominant role, while the second empirical fact is claimed to be consistent with the ``speculation/demand" view, in which households' speculation or demand is the dominant factor. For the third empirical fact, \cite{chodorow20242000s} argue that a fundamental factor drives long-term local economic growth, and they use a model of fundamental economic growth and extrapolated beliefs to rationalize this fact.

To preview, I develop a unified credit expansion theory that explains empirical patterns both across ZIP codes within a metro and cross metros in the prior, boom, bust, and recovery periods (including the three facts above) and generates new testable implications of ``double differences" (cross ZIP codes and cross metros) for the four periods. Following the idea of ``Economic Base Theory", I construct local economic exposure to net export growth as the driving force of local economy and credit expansion. For the identification strategy, I employ a new instrumental variable approach from the International trade literature for the empirical results consistent with my model. My core finding is that credit expansion, rather than speculation, plays the dominant role in causing the 1999-2019 U.S. housing cycle and can unify all empirical facts. Therefore, the policy implies that the new regulatory design to prevent a recurrence of the same housing cycle should focus more on lenders. In the remainder of the introduction, I will describe the model's intuition, identification strategy, main findings, and my contribution.

\begin{figure}[H]

\begin{center}

%\resizebox{4.5in}{3.2in}{%
\resizebox{\textwidth}{!}{%
\includegraphics[]{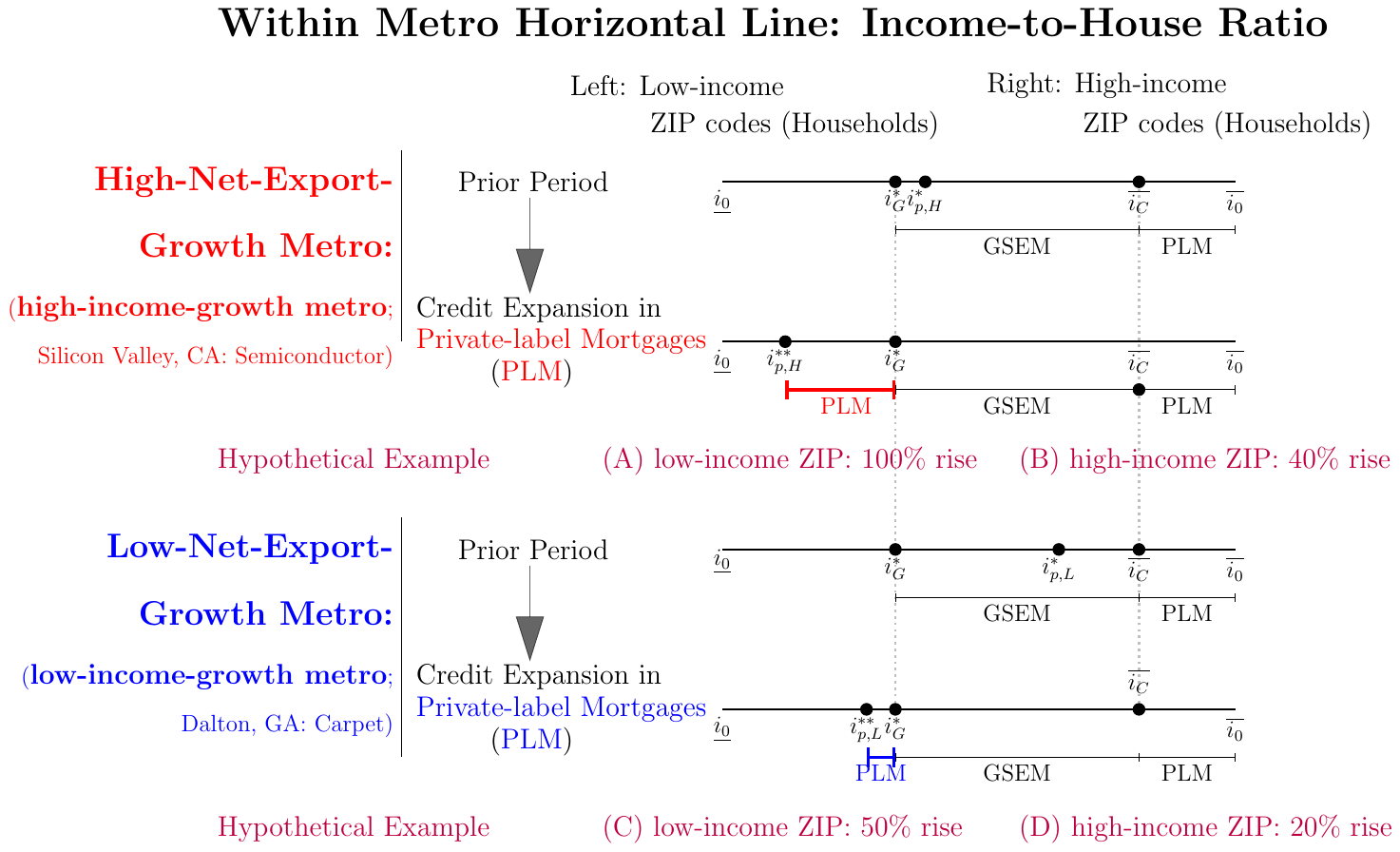}
} %end of resizebox

\end{center}

%-----------------------------------------------
% Figure setting: caption and label
%-----------------------------------------------
\caption{Model Intuition: Before and After Credit Expansion}
\label{fig_ModelIntuition}

\end{figure}

%----------------------------------------------------
%\subsection{Model}
%----------------------------------------------------
\noindent \textbf{Intuition of the Model} I develop a unified credit expansion theory to explain both within- and cross-metro patterns in the prior, boom, bust, and recovery periods. As Figure (\ref{fig_ModelIntuition}) shows, within each metropolitan area, ZIP codes (and households) are sorted by income-to-house ratio, meaning low-income ZIP codes (and households) are in the left tail. Before credit expansion, government-sponsored enterprise mortgages (GSEMs) dominate the middle-class households due to lower rates resulting from implicit government guarantees and large scale, while private-label mortgages (PLMs) occupy the market above the conforming loan limits on the right. When mortgage credit expansion occurs, low-income ZIP codes experience greater mortgage growth (and house price growth) than high-income ZIP codes within a metropolitan area.

%----------------------------------------------------------------------
%----------------------------------------------------------------------
% Beginning of Figure_ MortMkt_Bef&Aft_Axis
%----------------------------------------------------------------------
%----------------------------------------------------------------------

%----------------------------------------------------------------------
% end of Figure_ MortMkt_Bef&Aft_Axis
% begins in Line 
%----------------------------------------------------------------------

Then, I decompose metropolitan areas into high- and low-net-export-growth metros, where I proxy for local economic conditions using local economic exposure to net export growth (hereafter, net export growth). Here, I compare private-label (non-jumbo) mortgages (PLNJMs) and government-sponsored enterprise mortgages (GSEMs) because both fall below the conforming loan limits. Since only the former (PLNJMs) can legally respond to local economic conditions, while the latter (GSEMs) cannot, the model predicts that credit expansion in PLNJMs is stronger in the high-net-export-growth metros. Since net export growth also drives local income growth\footnote{For empirical causal evidence, please see my verification tests in Appendix Section (\ref{subsec:NEG_LocalEconConditions})}, my model reconciles the following two empirical facts: in the boom period, the correlation between income growth and mortgage growth is (1) negative across ZIP codes within a metropolitan area \citep{mian2009consequences}, but (2) positive across metropolitan areas \citep{adelino2016loan}, where (2) is equivalent to the prediction that credit expansion in PLNJMs is stronger in the high-net-export-growth metros. A hypothetical numeric example can illustrate how my model reconciles these two empirical facts. In Figure (\ref{fig_ModelIntuition}), I assume the (A) low-income ZIP codes experience a 100\% increase in PLMs and (B) high-income ZIP codes experience a 40\% increase in the high-net-export-growth metro, while (C) low-income ZIP codes experience a 50\% increase in PLMs and (B) high-income ZIP codes experience a 20\% increase in the low-net-export-growth metro. Within the same metro, (A)-(B) comparison or (C)-(D) comparison shows a negative correlation between income growth and mortgage growth. However, across metros, the correlation becomes positive as (A) is higher than (C) and (B) is higher than (D). Given that net export growth determines the local economic conditions in the long term\footnote{For evidence in the literature and my empirical causal evidence, please see my verification tests in Appendix Section (\ref{subsec:NEG_LocalEconConditions})}, my model predicts that high-net-export-growth metros experience a stronger boom, a worse bust, and a stronger recovery than the low-net-export-growth metros. This prediction explains the (3) empirical fact above. 

Further, as for the ``double differences", I define a low-minus-high (LMH) factor as the growth of private-label (non-jumbo) mortgage (and house price) in low-income ZIP codes minus that in high-income ZIP codes within the same metropolitan area. Credit expansion means that the LMH factor is positive during the boom period and negative during the bust period in metropolitan areas. Because credit expansion in PLNJMs is stronger in the high-net-export-growth metros, my model predicts that this low-minus-high factor in the high-net-export-growth metros is more positive during the boom period, more negative during the bust period, and slightly more positive in the recovery period than in the low-net-export-growth metros. I summarize twelve predictions of my model in the following table:

%----------------------------------------------------------
\begin{table}[H]
\renewcommand{\arraystretch}{1.1}

\centering

\resizebox{0.9\columnwidth}{!}{%
%----------------------------------------------------------
\begin{tabular}{lllll}
\toprule
Three Dimensions                                                                                                                                  & \begin{tabular}[c]{@{}l@{}}{\small Prior Period}\\{\small  (91-99) }\end{tabular} & \begin{tabular}[c]{@{}l@{}}{\small Boom Period}\\{\small  (99-06)} \end{tabular} & \begin{tabular}[c]{@{}l@{}}{\small Bust Period}\\ {\small (07-11)} \end{tabular} & \begin{tabular}[c]{@{}l@{}}{\small Recovery Period}\\ {\small (11-19) }\end{tabular} \\
\hline
\begin{tabular}[c]{@{}l@{}}First Difference Cross Metro: \\ {\small Growth (high-net-export-growth metro) - }\\ {\small Growth (low-net-export-growth metro)}\end{tabular}                              & H1: Zero                                                   & H2: Positive                                              & H3: Negative                                              & H4: Positive                                                  \\
\hline
\begin{tabular}[c]{@{}l@{}}First Difference Within Metro (LMH): \\ {\small Growth  (low-income ZIP) -} \\ {\small Growth (high-income ZIP)}\end{tabular}                  & H5: Zero                                                   & H6: Positive                                              & H7: Negative                                              & H8: Slightly Positive                                                     \\
\hline
\begin{tabular}[c]{@{}l@{}}Double Difference Across Metro\\ {\small LMH (high-net-export-growth metro) -}\\ {\small LMH (low-net-export-growth metro)}\end{tabular} & H9: Zero                                                   & H10: Positive                                              & H11: Negative                                              & H12: Slightly Positive     \\       \bottomrule                                          
\end{tabular}
%----------------------------------------------------------
} % end of resize box
\end{table}
%----------------------------------------------------------

\noindent \textbf{Intuition of Identification Strategy} I operationalize the idea of the ``economic base theory" \citep{tiebout1962community} and construct a variable that captures the long-term incentive of mortgage credit expansion: metropolitan exposure to net export growth of manufacturing industries (hereafter ``net export growth"). local economics argues that the tradable sector is the key driver of long-term local economic growth and house prices, as it brings wealth from outside into the local economy, whereas the nontradable sector only recycles money within the local economy. Therefore, credit expansion would be stronger in areas with stronger growth in the tradable sector. Ideal measurement requires census-style data on the accounting records of all firms in the tradable sector, which are unavailable. Instead, I use manufacturing employment at the industry-by-metropolitan level as shares, and time-series changes in net export growth as shifts. This shift-share design provides a useful measure of the local tradable sector. I use the instrumental variable approach developed by \cite{feenstra2019us} as my identification strategy. \cite{feenstra2019us} develop their IVs from a general equilibrium setting that can isolate the exogenous part of U.S. imports and exports. Intuitively, their IVs capture the exogenous parts of net export growth due to (1) rising world demand reflected in U.S. export growth, (2) rising world supply reflected in U.S. import growth, and (3) tariff changes. They also use high-dimensional fixed effects to remove the potentially endogenous parts and predetermined parts: (1) U.S. industry-by-year supply shocks in exports, (2) U.S. industry-by-year demand shocks in imports, and (3) predetermined bilateral distance between the U.S. and partner countries. They construct IVs for exports and imports separately, and I combine them into a single IV for net export growth. 

This identification strategy has two advantages. First, its flexibility allows me to construct an instrumental variable for the boom period from 1999 to 2005, which aligns with the exogenous timing of securitization innovation in the Copula model in 1999 \citep{li2000copula,salmon2009recipe} and incorporates all events related to credit expansion. In addition to the Copula model, these events include the "global saving glut" \citep{bernanke2005global, bernanke2007global}, the deregulation of the mortgage market in 2004 and 2005 \citep{di2017credit,lewis2023creditor}, and the political campaign of the mortgage industry \citep{mian2013political}. Therefore, this IV strategy enables tests that capture exogenous credit expansion arising from these events and assess its overall effect. Conducting empirical tests on a single event can facilitate the design of a clean identification strategy but limit the assessment of the overall impact of credit expansion. Second, this identification strategy separates government-sponsored enterprise mortgages (GSEMs) and private-label mortgages (PLMs), the former of which are legally prohibited from considering local economic conditions. This separation is crucial for empirical tests that show credit expansion occurs only in PLMs, not in GSEMs. This separation also facilitates the design of the placebo test that supports the exclusion restriction and against the ``speculation/demand" view.

%----------------------------------------------------
%\subsection{Findings}
%----------------------------------------------------
\noindent \textbf{Findings} I illustrate my empirical causal evidence in three parts: (1) cross-metro evidence supporting credit expansion, (2) ``double difference' evidence supporting credit expansion, and (3) cross-metro evidence against speculation. The within-metro predictions of my model have been well documented in the literature, including \citep{mian2009consequences} and related papers.

%------------------------------------
\begin{figure}[H]

\begin{center}

\resizebox{5in}{!}{%
%\resizebox{\textwidth}{!}{%
\includegraphics[width=16cm, height=11cm]{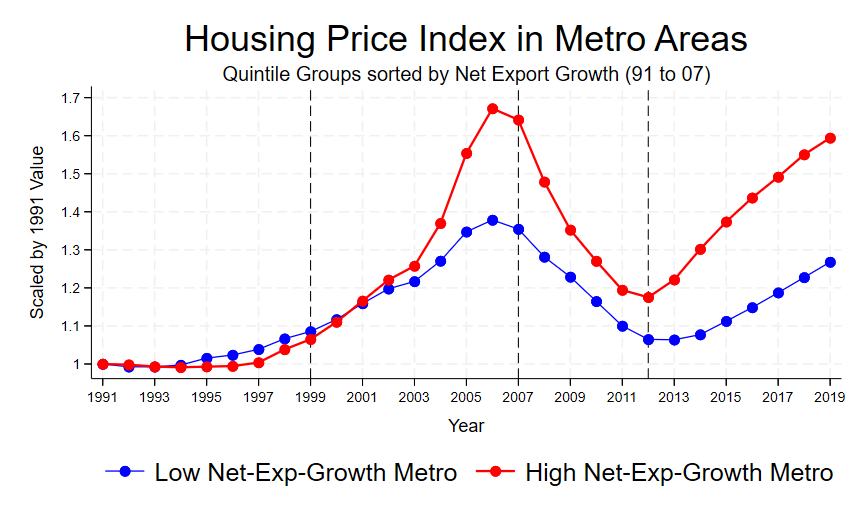}
} %end of resizebox

\end{center}
\vspace{-4mm}
%---------------
% Figure setting: caption and label
%---------------
\caption{Housing Price Growth (1991-2019) across Metropolitan Areas}
\label{fig_HPI_91to19_NEG_Intro}

\end{figure}
%------------------------------------

First, I show that the empirical causal evidence across metropolitan areas supports credit expansion theory. In particular, with the IV strategy above, I show that high-net-export-growth metros experience a stronger boom-bust-recovery housing cycle due to credit expansion in private-label (non-jumbo) mortgages (PLNJMs), rather than in government-sponsored enterprise mortgages (GSEMs), because only the former can legally respond to local economic conditions. This cross-metro pattern in house prices is clearly evident in Figure (\ref{fig_HPI_91to19_NEG_Intro}). In terms of economic significance, a one-standard-deviation increase in PLNJM growth (1999-2005) accounts for $23.34\%$ (equivalent to $109.60\%$ of a one-standard-deviation increase) in house price growth (1999-2005) and $14.24\%$ (equivalent to $115.22\%$ of a one-standard-deviation drop) in house price bust (2007-2009).

The exclusion restriction of this IV strategy is that net export growth (1999-2005) affects house price growth (1999-2005) only through its effect on the growth of private-label (non-jumbo) mortgages (PLNJMs). I use four tests to support this exclusion restriction. First, if net export growth (1999-2005) significantly increases middle-class household speculation or demand for mortgages, the increase will first appear in government-sponsored enterprise mortgages (GSEMs) because GSEMs offer lower rates and are easier for lenders to originate. However, net export growth (1999-2005) does not increase growth in GESMs (1999-2005), contrary to the household speculation or demand view. Second, when I use the difference in growth rates between PLNJMs and GSEMs as the mortgage growth rate that captures credit expansion after fully controlling household demand, the 2SLS estimates imply a similar magnitude of economic impact on housing price growth. Third, I show that, in a prior period (1991-1999) without credit expansion but with strong differential net export growth across metropolitan areas, net export growth does not increase house price growth or mortgage growth. This placebo result provides evidence directly against the demand channel. Fourth, I use the square of the IV as a second but redundant IV and then employ the over-identification test to provide suggestive evidence supporting the exclusion restriction.

In the second part, I show causal evidence of ``double differences" in mortgage and house price growth that are consistent with my model. I define a low-minus-high (LMH) factor as mortgage (and house price) growth in low-income ZIP codes minus that in high-income ZIP codes within the same metropolitan area. Consistent with the within-metro fact \citep{mian2009consequences}, this low-minus-high factor for private-label (non-jumbo) mortgages and house prices is positive in the boom period and negative in the bust period. Using the identification strategy discussed above, I show the evidence of ``double differences": this low-minus-high factor for the private-label (non-jumbo) mortgages and house prices in the high-net-export-growth metros is not more positive in the prior period, more positive during the boom period, more negative during the bust period, and slightly more positive in the recovery period than that in the low-net-export-growth metros. In the numeric example above and in the Figure (\ref{fig_ModelIntuition}), the ``double differences" can be illustrated as the $(A-B) - (C-D)=30\%$, which is positive in the boom period.

In the third part, I employ three tests to show that credit expansion, rather than speculation, plays the dominant role in causing the cross-metro house price cycle above. In these tests, I use a specific measure of pure speculation: ``non-owner-occupied" private-label mortgages for home purchases \citep{gao2020economic}. Inspired by this measure, I use ``owner-occupied" private-label mortgages for home purchase as a measure of pure credit expansion. First, I show that pure credit expansion induced by net export growth can explain more than half of the pure cross-metropolitan variation in speculation. The residuals give us a measure of credit-independent speculation. My second test shows that, compared with credit-independent speculation, pure credit expansion is the dominant factor in explaining the house price boom. Specifically, pure credit expansion accounts for approximately four times the house price growth that speculation accounts for. My last test focuses on the prior period (91-99) with substantial net export growth but no credit expansion. I show that, without credit expansion, pure speculation does not increase as a response to the net export growth. Taken together, the results of my three tests are more consistent with the notion that credit expansion rather than speculation plays the dominant role in the house price boom and bust. My argument against demand does not mean demand does not play any role. Instead, I mean only that credit expansion plays the dominant role relative to speculation.

I conduct multiple robustness tests to support my main conclusion that private-label (non-jumbo) mortgages growth during the boom period causes the house price boom and bust across metros. First, I consider the state-level anti-predatory lending law and the 2004 preemption of national banks by the Office of the Comptroller of the Currency \citep{di2017credit}. Although preempted states experience a stronger housing boom and bust cycle caused by credit expansion, my main conclusion still holds in all states. Second, I test state-level differences in recourse law \citep{ghent2011recourse}. Even though non-resource states experience a stronger housing bust caused by credit expansion, my main conclusion is robust for all states. Third, I investigate state-level differences in the judicial requirement of foreclosure laws \citep{mian2015foreclosures}. Though non-judicial states experience a stronger house price bust caused by credit expansion, my main conclusion still holds in all states. Fourth, my main conclusion still holds after controlling the sand-state dummy. Fifth, my main conclusion is robust to state capital gain tax.

Please note that ``causal evidence" in this paper only means that I find an incentive (net export growth) that induces credit expansion in private-label (non-jumbo) mortgages to be much stronger in the high-net-export-growth metropolitan areas in the cross-section. I do not find the incentives that caused the aggregate credit expansion in the early 2000s. For the causes of the aggregate credit expansion in the boom period (1999-2005), the literature has documented evidence from financial innovation in securitization (notably the Copula formula by \cite{li2000copula} \citep{salmon2012formula}), international capital flow (``global saving glut") by \cite{bernanke2005global, bernanke2007global}),  mortgage market deregulations \citep{di2017credit,lewis2023creditor, mian2013political}, and political campaign \citep{mian2013political}. I define 1999-2005 as the boom period to capture all major events above and align with the mortgage boom period commonly used in the literature (see \cite{griffin2021drove} for a review). 

While my results support the ``credit expansion" view, my paper does not exclude the ``speculation/demand" phenomenon, which is significant in dollar amount. However, when I decompose the mortgage and house price trends across metros, within metros across ZIP codes, and in double differences, the evidence is consistent: "credit expansion" plays the dominant role. 

During 1999–2007, the expansion likely went beyond an efficient reallocation of credit (due to technology innovation in securitization, ``global saving glut", and deregulation), given the magnitude of the subsequent bust. My paper is not designed to identify the micro mechanisms behind the aggregate mortgage credit expansion. In particular, my paper cannot separately quantify the following three micro mechanisms: (i) “technology innovation going wrong” (e.g., underpricing of risks due to the Copula model) and resulting loosening of lending criteria; (ii) reduced screening incentives and efforts in lenders; and (iii) fraud due to misaligned incentives along the originate-to-distribute chain.

%--------------------------------------------------
%\subsection{Contribution to the Literature}
%--------------------------------------------------

\noindent \textbf{Related Literature} This paper contributes to several related literatures. First, I contribute to the theoretical literature on the U.S. housing cycle during the 2000s. Two views have been proposed to explain this housing cycle. The first view is ``credit expansion", which argues that the excessive credit expansion by lenders causes this housing cycle. The second view is the ``speculation/demand" view, which claims that household speculation/demand driven by expectation/belief of future house price growth plays the dominant role. Many studies use general equilibrium models to study the housing boom and bust \citep{justiniano2019credit,kaplan2020housing,greenwald2025credit}. Different from their approach, my framework views each metropolitan area is isolated from others and the local economic growth only impacts local credit expansion in private-label mortgages. This framework is suitable to uncover the mortgage and house price patterns in three dimensions: within-metro across ZIP codes, across metropolitan areas, and across ZIP codes across metro (``double differences"). In total, my model yields twelve predictions across three dimensions and four periods, all of which are consistent with the empirical tests. To the best of my knowledge, my model explains the most comprehensive set of empirical facts in the literature.

This paper contributes to the empirical literature on the U.S. housing cycle in four distinct dimensions. First, in the empirical literature on the 1999-2011 U.S. housing cycle, many papers focus on specific events that contributed to credit expansion in a particular dimension. \cite{bernanke2005global, bernanke2007global} focus on international capital flows (the "global saving glut"), \cite{di2017credit} studies the 2004 preemption of national banks from the state-level anti-predatory lending law, \cite{lewis2023creditor} studies the 2005 Bankruptcy Abuse Prevention and Consumer Protection Act, and \cite{mian2013political} focuses on the mortgage industry campaign contributions. In addition, \cite{favara2015credit} exploit the 1994-2005 state-level branching deregulation, \cite{adelino2012credit} use exogenous changes in the conforming loan limit, and \cite{ghent2011recourse} study the impact of state-level non-recourse/recourse laws. Instead of focusing on a single event, I define 1999–2005 as the boom period to capture the combined influence of the significant forces listed above. This choice allows me to assess the aggregate impact of credit expansion on the housing cycle and compare it with arguments from the speculation/demand view.

Second, I distinguish between government-sponsored enterprise mortgages (GSEMs) and private-label mortgages (PLMs) based on an institutional difference: the former are legally prohibited from considering local economic conditions. Accordingly, I design an identification strategy and introduce a new method from International economics. This identification strategy enables me to demonstrate that credit expansion in PLMs, rather than in GSEMs, causes the housing boom and bust. The institutional difference described above is first documented by \cite{hurst2016regional}, who show that government-sponsored enterprise mortgages do not account for local risk in mortgage rate setting due to a legal constraint, resulting in a local welfare transfer. \cite{mian2022credit} exploits lenders' reliance on non-core deposit funding and provides evidence that credit expansion through private-label mortgages can explain various forms of speculation. But their analysis focuses on ZIP codes within counties and metropolitan areas. \cite{keys2010did,keys2012lender} find the screening differences of the lenders in securitized non-agency vs. agency mortgages at the cutoff point of the FICO score of 620.

Third, many empirical studies on a particular sector of mortgages find evidence consistent with the ``speculation/demand" view. \cite{adelino2016loan} documents the robust positive relationship between mortgage growth and house price growth across metropolitan areas (or counties) as evidence consistent with the ``speculation/demand" view. By both a model illustration and empirical causal evidence, I show that this fact is also consistent with the ``credit expansion" view, where the high-net-export-growth metro experience higher income growth and pronounced credit expansion in mortgages, which is also evident in the stronger low-minus-high factor in these metropolitan areas (``double differences") . An extensive literature examines speculative component in mortgages or home purchases in the housing cycle. Using credit-report data, \cite{haughwout2011real} documents the role of real estate investors (buy-and-hold or buy-and-flip) during the boom-bust cycle, particularly in the sand states. Using deed-level data, \cite{chinco2016misinformed} show that speculation by out-of-town second-house buyers played an important role in the housing boom and bust in twenty-one cities. \cite{bhutta2015ins} highlights the fastest growth of investors (borrowers with more than one mortgaged property) among mortgage inflow during the boom period. \cite{gao2020economic} examines speculation measured by non-owner-occupied mortgages and its impact on economic consequences, using the state capital gain tax as an instrument. \cite{defusco2022speculative} study the important role of short-term real estate investors in shaping house prices and volume dynamics. Parallel to the residential market, \cite{nathanson2018arrested} study the land market speculation by home-builders and illustrate how disagreement can cause strongest house price boom and bust in cities with intermediate elastic housing supply. There is also a distinct and parallel strand of literature that focuses on the impact and implications of household expectations regarding house prices \citep{piazzesi2009momentum,case2012have,case2003there}. In response to this literature, I measure ``speculation" by non-owner-occupied private-label (non-jumbo) mortgages and measure ``pure credit expansion" by owner-occupied private-label (non-jumbo) mortgages. Then, I employ three tests to show that credit expansion, rather than speculation, plays the dominant role in causing the cross-metro house price cycle. In fact, much of speculation is fueled by credit expansion trend, a view consistent with evidence from \cite{mian2022credit}.

Fourth, guided by my model, I document new empirical facts of ``double differences", where the positive low-minus-high factors for private-label (non-jumbo) mortgages and house prices (as evidence of credit expansion) are more pronounced in high-net-export-growth metropolitan areas. Such empirical facts are new and crucial to the unified credit expansion theory, which explains the empirical facts in three dimensions: within-metro across ZIP codes, cross-metro, and cross ZIP codes across metropolitan areas (``double difference"). Under this theory, the positive correlation between mortgage growth and house price growth across metropolitan areas is a natural effect of credit expansion, with one modification: credit expansion is more pronounced in high-net-export-growth metropolitan areas, whose income growth is higher.

Parallel to my paper, many empirical studies examine the micro mechanisms by which credit expansion causes housing booms and busts. First, \cite{mian2009consequences} show that the sharp increase in subprime mortgages leads to the housing cycle. Second, \cite{anenberg2019measuring, mian2022credit} argue that easier securitization increases mortgage credit, thereby fueling the housing cycle. \cite{keys2010did, keys2013mortgage, purnanandam2011originate} provide evidence that securitization lowers lenders' incentive to screen and subsequently results in poor mortgage performance. \cite{piskorski2015asset, griffin2016did, griffin2016facilitated, mian2017fraudulent} find evidence that financial fraud induced by moral hazard plays a vital role in mortgage expansion and subsequent crisis. Complement to these studies, my paper provides an empirical framework that organizes evidence along three margins: within-metro across ZIP codes, cross-metros, and ``double differences".

The remainder of this article is organized as follows: Section (\ref{sec:model}) illustrates the model, Section (\ref{sec:ResearchDesign}) presents the research design, and Section (\ref{sec:data}) describes the data. For empirical evidence supporting credit expansion, Section (\ref{sec:Evidence_Metro_CreditExpansion}) presents evidence across metropolitan areas and Section (\ref{sec:Empirical_DoubleDifference}) presents evidence of ``double differences". Then, Section (\ref{sec:Empirical_Metro_AgainstSpeculation}) provides evidence against speculation across metropolitan areas and Section (\ref{sec:Conclusion}) concludes.

\section{The Model}\label{sec:model}

In this section, I build a simple theoretical model to illustrate how credit expansion alone can reconcile the two seemingly opposing empirical facts: the correlation between income growth and mortgage growth is negative across ZIP codes within metropolitan areas but positive across metropolitan areas. Thus, this section addresses the potential concern that two seemingly opposing empirical facts may result in internal conflicts within the ``credit expansion'' view. I can summarize the model intuition as follows. First, when the funding cost of private-label mortgages declines within metropolitan areas, lenders can extend cheap credit to low-income borrowers traditionally excluded from the mortgage markets due to usury law. Since households with sharp rising wealth tend to self-relocate to rich ZIP codes and households with sharp declining wealth tend to self-relocate to poor ZIP codes, income growth in poor ZIP codes is usually lower than in rich ZIP codes, even income growth distribution is the same across all occupations, industries and initial wealth level.\footnote{This within-metropolitan cross-ZIP codes self reallocation is in the spirit of ``voting by foot" by \cite{tiebout1956pure}} Thus, the negative correlation between income growth and mortgage growth within metropolitan areas means that poor ZIP codes see higher mortgage growth. Second, given the above within-metropolitan fact, poor ZIP codes in the high-net-export-growth metropolitan areas experience relatively higher private-label mortgage growth than those in the low net-export-growth areas due to the differences in net export growth. Thus, the correlation between income growth and mortgage growth is positive across metropolitan areas. In summary, credit expansion induced by net export growth alone can reconcile empirical facts within and across metropolitan areas.

\comment{
Two key contributing factors are: (1) Government-Sponsored Enterprise Mortgages (GSEM) cannot consider differences in local economic conditions, which is net export growth in my model (we refer (1) as legal constraint)\footnote{In my research design in section \ref{sec:ResearchDesign}, I illustrate why net export growth captures the growth of economic base, which is the driving force of local economy.}; (2) a decline in funding cost in the private-label mortgages (PLMs), as documented by \cite{justiniano2022mortgage} (I refer (2) as mispriced credit expansion). I build this model because the majority of the literature does not recognize the fundamental economic incentives shaping the direction of credit expansion and housing cycle, as shown in my new empirical fact. Given the credit expansion documented by \cite{justiniano2022mortgage}, the legal constraint determines that only PLMs respond to net export growth and cause a stronger housing cycle in the HNEG area. 
}

\subsection{Mortgage First \& Secondary Market Structure}

First, let us describe the market structure in the mortgage first and secondary markets in the United States. In the mortgage first market, each household enters into a mortgage contract with a lender separately if the mortgage application is approved. Mortgage lenders include deposit institutions (commercial banks, thrifts, and credit unions) and non-deposit institutions (independent mortgage companies).

\begin{figure}[H]

\begin{center}

\resizebox{6in}{!}{%
%\resizebox{\textwidth}{!}{%
\includegraphics[]{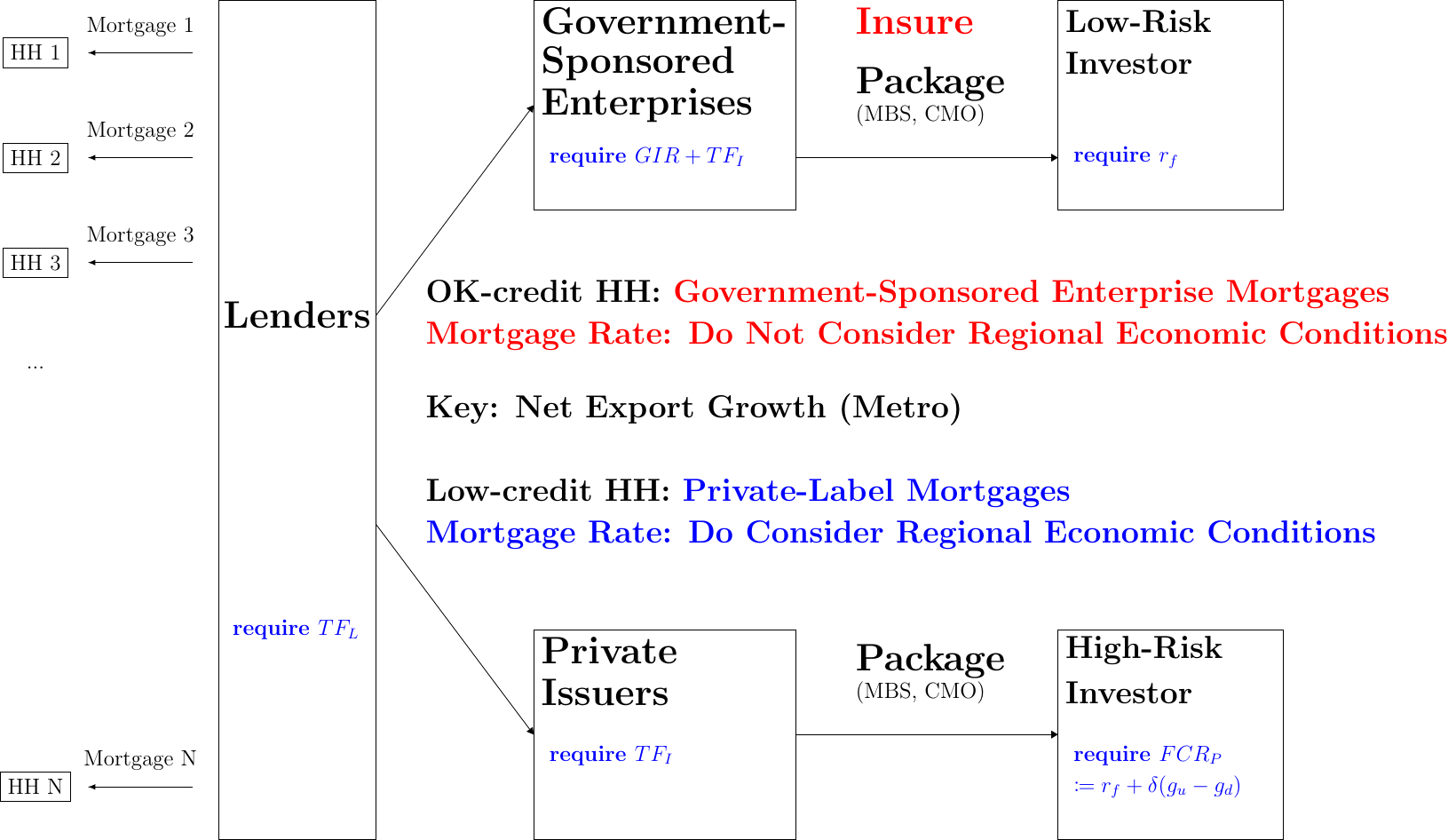}
} %end of resizebox

\end{center}

%-----------------------------------------------
% Figure setting: caption and label
%-----------------------------------------------
\caption{Mortgage First and Secondary Markets}
\label{fig_MortMkt_RequiredReturn}

\end{figure}

After origination, most mortgages in the United States are sold to a secondary market rather than staying on lenders' balance sheets. For example, about 80 percent of mortgages were sold and securitized between 2004 to 2006 \citep{keys2013mortgage}. In the secondary mortgage market, when the mortgages are qualified for purchase by government-sponsored enterprises (GSEs include Fannie Mae, Freddie Mac, and Ginnie Mae)\footnote{Ginnie Mae is a government agency. For simplicity, I call these three agencies GSEs.}, lenders sell these mortgages to GSEs. These mortgages are called government-sponsored enterprise mortgages (GSEM). GSEs insure GSEMs against loss of principal and interest via charging insurance fees.\footnote{This is a simplified assumption just for modeling. In reality, Ginnie Mae does not buy loans or issue MBS while approved issuers create the MBS. Ginnie Mae only provides insurance for timely payment of principal and interest to investors, with prepayment risk uncovered.} GSEs then package these mortgages into mortgage-backed securities (MBS), collateralized mortgage obligations (CMO), and other types of securities. Eventually, these securities are sold to investors with relatively low-risk tolerance. Typical investors include mutual funds, pension funds, and insurance companies. Because of lower funding costs due to implicit guarantees by the government and the large scale of the GSEM market \citep{hurst2016regional}, lenders will first choose to sell mortgages to GSEs if these mortgages are qualified. It is important to note that GSEs are not allowed by law to consider local economic conditions in setting mortgage rates.

However, if mortgages are not qualified as GSEMs, lenders will sell mortgages to private issuers. Typical private issuers include large commercial banks, large mortgage companies, investment banks, and real estate investment trusts (REITs). These mortgages are called private-label mortgages (PLMs). Contrary to GSEs, private issuers consider local economic conditions when setting mortgage rates and do not provide insurance against principals or interests. Like GSEs, private issuers package these mortgages into MBS, CMO, and other types of securities (usually by creating special purpose vehicles (SPV)). Ultimately, these securities are sold to investors with relatively high risk tolerance who bear the default risk of the underlying mortgages. Typical investors include hedge funds, investment banks, and REITs.

Now, let us describe each player's required returns and fees in the mortgage first and secondary market. Figure \ref{fig_MortMkt_RequiredReturn} depicts the entire market. Since the lenders now play the role of originate-to-distribute, they only charge transaction fees (underwriting services, etc.) $TF_{L}$ (as a percentage of house value). In the GSEM market, since GSEs provide insurance and bear the default risk, they charge an insurance fee of $GIR$ (as a percentage of house value). They also charge a transaction fee for securitization $TF_{I}$ (as a percentage of house value). Low-risk investors only charge a (close to) risk-free rate $r_f$ since they are fully insured against any potential loss. I define the funding cost of GSEM as $FCR_G \coloneqq r_f + GIR $. In sum, to fund a mortgage in the GSEM market, the total required return is $ 1+ FCR_G + TF$, where $TF \coloneqq TF_{L} + TF_{I} $. 

In the PLM market, however, private issuers only charge transaction fees for securitization  $TF_{I}$ without providing insurance. As default risk is eventually borne by high-risk investors, they charge risk-adjusted return (defined as funding cost for PLM market) $FCR_{P} \coloneqq  r_f + \delta (g_{u} - g_{d})$. $\delta$ is the risk-averse parameter, $g_{u}$ and  $g_{d}$ are household income growth rates for up-state and down-state, respectively. Thus, $g_{u} - g_{d}$ approximates the uncertainty in the household income growth, and $\delta (g_{u} - g_{d})$ approximates the risk premium in mortgage rate.  In sum, to fund a mortgage in the PLM market, the total required return is $ 1+ FCR^{*}_{P} + TF$, where again $TF \coloneqq TF_{L} + TF_{I} $.

\subsection{Basic Setup}
Now let us set up the environment of the model. There are two areas in the economy: an area with high net export growth (HNEG area) and an area with low economic growth (LNEG area). In the model, I assume net export growth can increase local employment growth, household income growth, and population growth (including migration).\footnote{For the economic empirical literature of trade impact on these local economic conditions and my verification tests, please refer to Section \ref{subsec:NEG_LocalEconConditions}} Thus, the HNEG area has an additional positive employment growth rate ($\textcolor{red}{+ \mathit{\Delta}e}$), an additional positive household income growth rate ($\textcolor{red}{+ \mathit{\Delta}g}$),  and an additional positive population growth rate ($\textcolor{red}{+ \mathit{\Delta}p}$)  while the LNEG area has an additional negative employment growth rate ($\textcolor{blue}{- \mathit{\Delta}e}$), an additional negative household income growth rate ($\textcolor{blue}{- \mathit{\Delta}g}$), and an additional negative population growth rate ($\textcolor{blue}{- \mathit{\Delta}p}$). Each area has both secondary mortgage markets described above, namely GSEM and PLM markets. 

In each area, there are an infinite number of homogeneous lenders who play the role of originate-to-distribute. The market players and their required returns and fees are described in the previous section. In each area, there are an infinite number of heterogeneous households that differ only in the initial income-to-house ratio $i_0 \coloneqq \frac{I_0}{H_0}$, where $I_0$ and $H_0$ are the household income and house value at time 0. To make households from both areas (HNEG and LNEG metros) comparable, I assume the household income-to-house ratios in time 0 follow the same distribution in both the LNEG area and the HNEG area.\footnote{It is important to note that I do not assume the initial house prices are the same in the HNEG area and the LNEG area. The initial house price and income in the HNEG metro can be higher than those in the LNEG metro.} In addition, I assume the house price is the same within the HNEG area ($H_{H,0}$) and within the LNEG area ($H_{L,0}$) because my focus is on the cross-area comparison of the market share of GSEM and PLM. This simplified assumption effectively avoids the unnecessary complexity of income-house matching within each area. With the simplified assumptions above, I can achieve a closed-form solution that has an intuitive economic interpretation and a graph demonstration.

There are two periods ($t = 0 \ \& \ 1$) in the model. In the first period ($t=0$), each household starts with no initial wealth and earns income $I_0$, which is observable to the lenders. For the simplicity of modeling, I assume the household consumes all income $I_0$ in time 0. In the first period ($t=0$), each household enters into a mortgage contract with a lender by pledging the house as collateral, if their income-to-house ratio $i_0$ is qualified for the mortgage. I assume utility-maximizing households get higher utility from living in a house than renting an apartment. Thus households will get mortgages as long as they can. Otherwise, they rent an apartment and pay rent at the end of time 1. 

In the second period ($t=1$), there is uncertainty at the individual household level: the income growth could have two states with equal probability ($\frac{1}{2}$)\footnote{Equal probability of two states are meant to be simple for illustration purpose. For a general probability $p \in [\frac{1}{2}, 1]$ for up-state, my conclusions still hold. See Appendix \ref{sec:App_Model_P} for proof.}: an up-state with income growth rate $g_{u}$ and a down-state with income growth rate $g_{d}$, where $g_{u} > g_{d}$. In the up-state, the household income growth rate is $g_{u} \textcolor{red}{+ \mathit{\Delta} g}$ in the HNEG area and $g_{u} \textcolor{blue}{- \mathit{\Delta} g}$ in the LNEG area. Similarly, in the down-state, the household income growth rate is $g_{d} \textcolor{red}{+ \mathit{\Delta} g}$ in HNEG area and $g_{d} \textcolor{blue}{- \mathit{\Delta} g}$ in LNEG area. Since uncertainty exists only at the individual household level, house prices only vary with area-specific economic growth rate, meaning $H_0(1\textcolor{red}{+\mathit{\Delta} g +\mathit{\Delta} e +\mathit{\Delta} p })$ in HNEG area and $H_0(1\textcolor{blue}{-\mathit{\Delta} g -\mathit{\Delta} e -\mathit{\Delta} p })$ in LNEG area.  I further assume that in time 1, the household income is sufficient to pay for the mortgage amount in the up-state, while not sufficient in the down-state. This two-state assumption is meant to be simple so that I can get a closed-form solution of mortgage rate with easy economic interpretation. With this assumption, to make the mortgage rate linearly decrease with income-to-house ratio $i_0$ (which is realistic in practice), I further assume that households would forgo both the house (as collateral) and disposable income (as a penalty) to the lender in case of default. The loss of disposable income in the down-state is motivated by three reasons. First, the foreclosure sale of the house usually cannot cover the entire loss of the mortgage, especially during the housing crisis. Second, mortgages in thirty-nine U.S. states are recourse loans, meaning lenders may be able to collect on a debt of households by obtaining a deficiency judgment when proceedings from a foreclosure sale are not enough to cover the loss of a mortgage (see \cite{ghent2011recourse}). Third, the standard mortgages in the United States have a time to maturity of 30 years such that some portion of payments is paid before default.

Next, I will focus on the break-even conditions in the following three situations: (1) GSEM in both HNEG area and LNEG area, (2) PLM in HNEG area, and (3) PLM in LNEG area. Since lenders only play the role of originate-to-distribute, the break-even condition includes required returns and fees by all parties in the market depicted in Figure (\ref{fig_MortMkt_RequiredReturn}).

%-------------------------------------------
% Beginning of GSEM
%-------------------------------------------
\subsubsection{GSEM in Both HNEG area and LNEG area}

According to the analysis by \cite{hurst2016regional}, the mortgage evaluation and pricing system by GSEs does not consider differences in local economic conditions. In other words, the following two decisions are made without local economic conditions: (1) whether a mortgage can be bought by GSE as an agency mortgage and (2) what the mortgage rate is set on the loan. I model these features as that GSEs ignore differences in employment growth rate ($\mathit{\Delta}e$), household income growth rate ($\mathit{\Delta}g$), and population growth rate ($\mathit{\Delta}p$) in both HNEG area and LNEG area. In addition, GSEs require an insurance premium $GIR$ in equilibrium so that GSEs charge $GIR + R_{G}$ in the non-default case while insuring against any loss in the default case.

Due to legal constraints that prevent considering differences in local economic conditions, lenders could only perceive the household disposable income in time 1 would be $I_0 (1+g_{u} - c)$ in up-state with probability $\frac{1}{2}$ and $I_0 (1+g_{d} - c)$ in down-state with probability $\frac{1}{2}$, where $c$ is the living cost that is the same for all households.\footnote{In this section, for simplicity, I omit the subscript of HNEG area and LNEG area in $I_{0}$ and $H_{0}$ since ultimately only $i_0$ matters and $i_0$ has same distribution among households in both areas.} The same $c$ is meant to simplify the situation and is not crucial to the final conclusion. The full mortgage amount is $H_0$ and mortgage payment at time 1 is $H_0(1+FCR_G + R_G + TF)$. $FCR_G$ is the funding cost rate, which includes the insurance rate $GIR$ required by GSEs and the risk-free rate $r_f$ required by low-risk investors. $R_G$ is the return paid to GSEs in up-state (when there is no default) to compensate for the potential loss in down-state due to insurance.  $TF$ is the sum of transaction fees required by lenders ($TF_{L}$) for mortgage processing and by GSEs ($TF_{I}$) for packaging mortgages. 

To make the model simple but still contains uncertainty in mortgage decision-making, I assume (1) all households can pay the full mortgage amount in up-state and (2) all households default in down-state. The first assumption means that even a household with the lowest income can pay the full mortgage amount in up-state in time 1: $\underline{I_0}(1+g_{u} - c) \geq H_0 ( 1 + FCR_G + R_G + TF) $. This inequality implies 
\begin{equation*}
\frac{\underline{I_0}}{H_0} \geq \frac{1+FCR_G + R_G + TF}{1+g_{u} - c} \coloneqq \underline{i_G}
\end{equation*}
The second assumption means that even a household with the highest income could default in down-state in time 1: $\overline{I_0}(1+g_{d} - c) < H_0 ( 1 + FCR_G + R_G + TF) $. This inequality implies 
\begin{equation*}
\frac{\overline{I_0}}{H_0} < \frac{1+FCR_G + R_G + TF}{1+g_{d} - c} \coloneqq \overline{i_G}
\end{equation*}
Therefore, I only assume that households have income-to-house ratio in time 0: $i_0 \in [\underline{i_G}, \overline{i_G}  )$. This assumption matches the practice. For the households that would default with probability 1 ($i_0<\underline{i_G}$ in my setting), lenders would not grant mortgages at all. For the households that would not default at all ($i_0>\overline{i_G}$ in my setting), they could always get mortgages by paying the lowest rate (insurance rate $GIR$ plus risk-free rate $r_f$), thus having no variation in mortgage rates. In either case, there is no economic trade-off in mortgage lending that is key to my model demonstration. I purposefully ignore these two extreme groups of households to make my model simple and focus my attention on the key economic mechanism of interest.

Now let us focus on the key break-even condition for the GSEM market in both HNEG area and LNEG area: the expected payoff from households shall be just enough to compensate the required return of ultimate low-risk investors ($r_f*H_0$), the insurance required by GSEs ($GIR*H_0$), and the transaction fees ($TF*H_0$) in a competitive market: 
\begin{equation}\label{eq:BE_G}
\resizebox{0.9\textwidth}{!}{$
    \frac{1}{2}H_0 \underbrace{ ( 1 + FCR_G + R_G + TF)}_{MR^{*}_G} + \frac{1}{2} [\underbrace{RR* H_0}_{\text{Recovery Value}} + \underbrace{I_0 (1+g_{d} -c)}_{\text{Disposable Income} }  ] = H_0(1+FCR_G + TF)
$} %end of \resizebox
\end{equation}
The first term on the left-hand-side of the equality is the product of probability in up-state ($\frac{1}{2}$) and the payment of the mortgage full amount $H_0 ( 1 + FCR_G + R_G + TF)$. The second term is the product of probability in down-state ($\frac{1}{2}$) and the sum of the following two terms: (1) the recovery value of the mortgage from foreclosure sale $RR* H_0$ and (2) the disposable income $I_0(1+g_{d} -c)$. A perfectly competitive market means that the mortgage rate $ MR^{*}_G $ is set to make the equality hold for each $i_0$ ex-ante : 
\begin{equation}\label{eq:MR_G}
\begin{split}
    MR^{*}_G  & \coloneqq 1 + FCR_G + R_{G} + TF \\
    & = [ 2(1 + FCR_G + TF) - RR] - (1+g_{d} -c)*i_0 \\
    & = C^{*}_G - K^{*}_G*i_0  
\end{split}
\end{equation}

The above mortgage rate has very clear economic interpretations. The mortgage rate in GSEM (in a non-default state) is positively related to funding cost ($FCR_G = r_f + GIR$, where $ r_f$ is the risk-free rate required by low-risk investors and $GIR$ is the insurance rate required by GSEs in equilibrium) and the transaction fee ($TF$) required by lenders for mortgage processing ($TF_{L}$) and GSEs for packaging ($TF_{I}$). In addition, the mortgage rate is negatively related to the recovery value of the mortgage from foreclosure sale ($RR$) and the disposable income in the default case ($(1+g_{d} -c)*i_0$), since these two proceeds are used to cover loss given default.

%-------------------------------------------
% End of GSEM
%-------------------------------------------

%-------------------------------------------
% Beginning of PLM in HNEG area
%-------------------------------------------
\subsubsection{PLM in HNEG area}
In the GSEM market, since GSEs provide insurance against principal and interest, low-risk investors only require a risk-free rate as compensation. However, in the PLM market, lenders would consider local economic conditions and high-risk investors would require a higher return to compensate for default risk due to a lack of insurance. I model these features in the following ways. First, PLM market (lenders, private issuers, and high-risk investors) would consider differences in local employment growth rate, household income growth rate, and population growth rate in both up-state and down-state. Second, high-risk investors would require a higher return $FCR^{*}_{P} = r_f + \delta*(g_{u} - g_{d})$, where $\delta$ is a risk-averse parameter and $(g_{u} - g_{d})$ (difference in household income growth rate for two states) is a measure of risk. 

By observing household income $I_{H,0}$ at time 0, lenders know the household disposable income in time 1 would be $I_{H,0} (1+g_{u} \textcolor{red}{+ \mathit{\Delta}g}- c)$ in up-state with probability $\frac{1}{2}$ and $I_{H,0} (1+g_{d} \textcolor{red}{+ \mathit{\Delta}g}- c)$ in down-state with probability $\frac{1}{2}$. The full mortgage amount is $H_{H,0}(1+FCR^{*}_{P} + R_{P,H} + TF)$. $FCR^{*}_{P}$ is the funding cost required by high-risk investors in equilibrium because of default risk due to a lack of insurance. $R_{P,H}$ is the premium that goes to high-risk investors in up-state to compensate for the potential loss in down-state.  $TF$ is the transaction fee rate required by lenders for mortgage processing and by private issuers for packaging (securitization). 

To make the model simple but still contain uncertainty in mortgage decision-making, I assume (1) all households can pay the full mortgage amount in up-state and (2) all households default in down-state.  I get the following two boundary conditions for income-to-house ratios\footnote{The derivation is in Appendix Section \ref{subsec:app_model}} :
\begin{equation*}
\begin{split}
\frac{\underline{I_{H,0}}}{H_{H,0}} & \geq \frac{ 1 + FCR^{*}_{P} + R_{P,H} + TF}{1+g_{u} \textcolor{red}{+ \mathit{\Delta}g} - c} \coloneqq \underline{i_{P,H}} \\
\frac{\overline{I_{H,0}}}{H_{H,0}} & < \frac{ 1 + FCR^{*}_{P} + R_{P,H} + TF}{1+g_{d} \textcolor{red}{+ \mathit{\Delta}g} - c} \coloneqq \overline{i_{P,H}}
\end{split}
\end{equation*}
Again, I only assume that households have income-to-house ratio in time 0: $i_0 \in [\underline{i_{P,H}}, \overline{i_{P,H}} )$. 

Now let us focus on the key break-even condition for the PLM market in HNEG area: the expected payoff from households shall be just enough to compensate the required return of ultimate high-risk investors ($FCR^{*}_{P} = r_f + \delta*(g_{u} - g_{d})$) and the transaction fees $TF$ in a competitive market: 
\begin{equation}\label{eq:BE_P_H}
\resizebox{0.9\textwidth}{!}{$
    \frac{1}{2}H_{H,0} \underbrace{( 1 + FCR^{*}_{P} + R_{P,H} + TF)}_{ MR^{*}_{P,H}} + \frac{1}{2} [\underbrace{RR* H_{H,0}(1\textcolor{red}{+ \mathit{\Delta}e + \mathit{\Delta}g  + \mathit{\Delta}p })}_{\text{Recovery Value}} + \underbrace{I_{H,0}(1+g_{d}\textcolor{red}{+ \mathit{\Delta}g} -c)}_{\text{Disposable Income}}] = H_{H,0}(1+FCR^{*}_{P} + TF)
$} %end of resizebox
\end{equation}

The first term on the left-hand-side of the equality is the product of probability in up-state ($\frac{1}{2}$) and the payment of the mortgage full amount $H_{H,0} ( 1 + FCR^{*}_{P} + R_{P,H} + TF)$. The second term is the product of probability in down-state ($\frac{1}{2}$) and the sum of the following two terms: (1) the recovery value of the mortgage from foreclosure sale $RR* H_{H,0}(1\textcolor{red}{+ \mathit{\Delta}e + \mathit{\Delta}g + \mathit{\Delta}p })$ and (2) the disposable income $I_{H,0}(1+g_{d}\textcolor{red}{+ \mathit{\Delta}g} -c)$. The additional house price growth happens due to the additional positive growth in employment, household income, and population caused by higher net export growth. In the empirical literature, there is evidence that employment growth, household income growth, and population growth (including migration) can push up housing demand and then housing price, especially in the long run \citep{olsen1987demand}.  Thus, we can have
\begin{equation}\label{eq:MR_P_H}
\begin{split}
    MR^{*}_{P,H}  & \coloneqq 1 + FCR^{*}_{P} + R_{P,H} + TF \\
    & = [ 2(1 + FCR^{*}_{P} + TF) - RR(1\textcolor{red}{+ \mathit{\Delta}e + \mathit{\Delta}g + \mathit{\Delta}p })] - (1+g_{d} \textcolor{red}{+ \mathit{\Delta}g}-c)*i_0 \\
    & = C^{*}_{P,H} - K^{*}_{P,H}*i_0  
\end{split}
\end{equation}

The above mortgage rate has very clear economic interpretations. The mortgage rate in PLM in HNEG area (in a non-default state) is positively related to funding cost ($FCR^{*}_{P} = r_f + \delta*(g_{u} - g_{d})$ ) and the transaction fee ($TF$) charged by lenders for mortgage processing and private issuers for packaging (securitization). In addition, the mortgage rate is negatively related to the recovery value of the mortgage from foreclosure sale ($RR(1\textcolor{red}{+ \mathit{\Delta}e + \mathit{\Delta}g + \mathit{\Delta}p })$) and the disposable income in the default case ($(1+g_{d} \textcolor{red}{+ \mathit{\Delta}g}-c)*i_0$), since these proceeds are used to cover the loss given default. 

%-------------------------------------------
% end of PLM in HNEG area
%-------------------------------------------

%-------------------------------------------
% Beginning of PLM in LNEG area
%-------------------------------------------
\subsubsection{PLM in LNEG area}
Similar to PLM in HNEG-are, by observing household income $I_{L,0}$ at time 0, lenders consider the household disposable income in time 1 would be $I_{L,0} (1+g_{u} \textcolor{blue}{- \mathit{\Delta}g}- c)$ in up-state with probability $\frac{1}{2}$ and $I_{L,0} (1+g_{d} \textcolor{blue}{- \mathit{\Delta}g}- c)$ in down-state with probability $\frac{1}{2}$. The full mortgage amount is $H_{L,0}(1+FCR^{*}_{P} + R_{P,L} + TF)$. $FCR^{*}_{P}= r_f + \delta*(g_{u} - g_{d})$ is the funding cost required by high-risk investors in equilibrium because of default risk due to a lack of insurance. $R_{P,L}$ is the premium that goes to high-risk investors in up-state to compensate for the potential loss in down-state.  $TF$ is the transaction fee rate charged by lenders for mortgage processing and by private issuers for packaging (securitization).

Assuming (1) all households can pay the full mortgage amount in up-state and (2) all households default in down-state, I get the following two boundary conditions for income-to-house ratios\footnote{The derivation is in Appendix Section \ref{subsec:app_model}} :
\begin{equation*}
\begin{split}
\frac{\underline{I_{L,0}}}{H_{L,0}} & \geq \frac{ 1 + FCR^{*}_{P} + R_{P,L} + TF}{1+g_{u} \textcolor{blue}{- \mathit{\Delta}g} - c} \coloneqq \underline{i_{P,L}} \\
\frac{\overline{I_{L,0}}}{H_{L,0}} & < \frac{ 1 + FCR^{*}_{P} + R_{P,L} + TF}{1+g_{d} \textcolor{blue}{- \mathit{\Delta}g} - c} \coloneqq \overline{i_{P,L}}
\end{split}
\end{equation*}

Therefore, I only assume that households have income-to-house ratio in time 0: $i_0 \in [\underline{i_{P,L}}, \overline{i_{P,L}}  )$. Again, I purposefully ignore these two extreme groups of households to make my model simple and focus my attention on the key economic mechanism of interest.

Now let us focus on the key break-even condition for the PLM market in the LNEG area: the expected payoff from households shall be just enough to compensate the required return of the ultimate high-risk investors ($FCR^{*}_{P} = r_f + \delta*(g_{u} - g_{d})$) and the transaction fees $TF$ in a competitive market: 
\begin{equation}\label{eq:BE_P_L}
\resizebox{0.9\textwidth}{!}{$
    \frac{1}{2}H_{L,0} \underbrace{( 1 + FCR^{*}_{P} + R_{P,L} + TF)}_{MR^{*}_{P,L}} + \frac{1}{2} [\underbrace{RR* H_{L,0}(1\textcolor{blue}{- \mathit{\Delta}e - \mathit{\Delta}g  - \mathit{\Delta}p })}_{\text{Recovery Value}} + \underbrace{I_{L,0}(1+g_{d}\textcolor{blue}{- \mathit{\Delta}g} -c)}_{\text{Disposable Income}}] = H_{L,0}(1+FCR^{*}_{P} + TF)
$} %end of \resizebox
\end{equation}

The first term on the left-hand-side of the equality is the product of the probability in up-state ($\frac{1}{2}$) and the payment of the mortgage full amount $H_{L,0} ( 1 + FCR^{*}_{P} + R_{P,L} + TF)$. The second term is the product of probability in down-state ($\frac{1}{2}$) and the sum of the following two terms: (1) the recovery value of collateral from foreclosure sale $RR* H_{L,0}(1\textcolor{blue}{- \mathit{\Delta}e - \mathit{\Delta}g - \mathit{\Delta}p  })$ and (2) the disposable income $I_{L,0}(1+g_{d}\textcolor{blue}{- \mathit{\Delta}g} -c)$. The house price growth happens due to the additional negative growth in employment, household income, and population caused by lower net export growth (essentially higher net import growth). Thus, we can have
\begin{equation}\label{eq:MR_P_L}
\begin{split}
    MR^{*}_{P,L}  & \coloneqq  1 + FCR^{*}_{P} + R_{P,L} + TF \\
    & = [2(1 + FCR^{*}_{P} + TF) - RR(1\textcolor{blue}{- \mathit{\Delta}e - \mathit{\Delta}g - \mathit{\Delta}p })] - (1+g_{d} \textcolor{blue}{- \mathit{\Delta}g}-c)*i_0 \\
    & = C^{*}_{P,L} - K^{*}_{P,L}*i_0  
\end{split}
\end{equation}

The above mortgage rate has very clear economic interpretations. The mortgage rate in PLM in LNEG area (in a non-default state) is positively related to funding cost ($FCR^{*}_{P} = r_f + \delta*(g_{u} - g_{d})$ ) required by high-risk investors and the transaction fees ($TF$) charged by lenders for mortgage processing and by private issuers for packaging (securitization). In addition, the mortgage rate is negatively related to the recovery value of the mortgage from foreclosure sale ($RR(1\textcolor{blue}{- \mathit{\Delta}e - \mathit{\Delta}g - \mathit{\Delta}p  })$) and the disposable income ($(1+g_{d} \textcolor{blue}{- \mathit{\Delta}g}-c)*i_0$), since these proceeds are used to cover loss given default.

%-------------------------------------------
% end of PLM in LNEG area
%-------------------------------------------

%-------------------------------------------
% Combine
%-------------------------------------------
\subsection{The Entire Mortgage Market: Combining GSEM and PLM}

I now combine the above three cases to get a full picture of the entire mortgage market. First, I define the domain of income-to-house ratio $ i_0 \in [\underline{i_0},\overline{i_0})$ where $ \underline{i_0} \coloneqq max \{ \underline{i_G} ,  \underline{i_{P,H}} , \underline{i_{P,L}} \} $ , \ and $\overline{i_0} \coloneqq min\{\overline{i_C}, \ \overline{i_{P,H}}, \ \overline{i_{P,L}} \} $. For the reasons stated above, my focus of $ i_0 \in [\underline{i_0},\overline{i_0})$ can help us illustrate the key economic trade-offs in a simple model.

%% add two ceiling $\overline{MR_G} $ and $\overline{MR_P} $

Second, let us add lower and higher bounds of income-to-house ratios required by government-sponsored enterprise mortgages. The US government essentially subsidizes homeownership of residences by providing insurance via GSEs at a very low cost. However, such a subsidy has a lower bound criteria for household income, credit scores, income-to-mortgage ratios, and other credit factors. To incorporate these criteria into my model, I set up a corresponding lower bound of income-to-house ratio $i^{*}_G$ (in both the HNEG and LNEG areas). In addition, the Federal Housing Finance Agency (FHFA) sets conforming loan limits (CLLs) for GSEM for each county each year. Mortgages above the limits are called "jumbo" mortgages and cannot be purchased by GSEs. The setup of CLLs essentially prevents the government from providing subsidies to high-income residents buying luxury houses. To incorporate this feature into my model, I set up a conforming income-to-house ratio $\overline{i_C} ( < \overline{i_0} )$ above which GSEs cannot purchase mortgages.

Third, let us add one ceiling for the total mortgage rates $\overline{MR_P} $ for the PLM market. The upper bound $\overline{MR_P}$ for PLM can be motivated by usury law, which is used to protect consumers from predatory lending and protect the less sophisticated consumers from irrational behavior due to poor financial knowledge.\footnote{In some states, the usury law limits are not binding due to (1) exemptions by other state laws, (2) override by contract laws, and (3) opt-out authority from federal usury laws. In these cases, I argue that the mortgage rate limits shall be interpreted broadly, including credit rationing motive.} Alternatively, the mortgage rate ceiling can be motivated by credit rationing \citep{hodgman1960credit, stiglitz1981credit}. Mortgage lenders have the incentive to limit PLM rates since higher rates could only attract risky borrowers (1) whose information is too limited for banks to evaluate credit risks (like many subprime mortgage borrowers) and (2) who have strong moral hazard motives. Effectively, $\overline{MR_P}$ and break-even PLM rate functions can give us a lower bound of the income-to-house ratio.

%------------------------------------------
\subsubsection{Mortgage Market in HNEG area in normal time}
%------------------------------------------

Now I consider the mortgage market in HNEG area that contains both the GSEMs and PLMs in normal times when $FCR^{*}_{P}$ is much higher than $FCR_G$. Based on the break-even mortgage rate functions in Equation (\ref{eq:MR_G}) and (\ref{eq:MR_P_H}), I know the absolute value of slope $K^{*}_{P,H}$ is higher than the one of slope $K^{*}_{G}$ because the PLM market considers the differential higher household income growth rate $\mathit{\Delta} g$ in HNEG area. Compared to GSEM, for each unit increase of income-to-house ratio $i_0$ in time 0, the PLM will be perceived as being compensated more by disposable income in default (in down-state), thereby resulting in a lower mortgage rate in equilibrium. 

Now let us compare the constant part $C^{*}_{P,H}$ and $C^{*}_{G}$. For simplicity, I assume the transaction fees are the same in GSEM and PLM markets since these fees cover the similar mortgage processing and packaging (securitization) process. I assume $2FCR^{*}_{P} - RR*( \mathit{\Delta}e + \mathit{\Delta}g + \mathit{\Delta}p  ) > 2FCR_G$ so that the constant part $C^{*}_{P,H}$ is higher than $C^{*}_{G}$. This assumption matches the reality in the sense that the funding cost of PLM ($FCR^{*}_{P}$) is much higher than the one of GSEM ($FCR_G$) due to a lack of insurance and potential default risk. In essence, mortgage insurance provided by GSEs is a government subsidy for home ownership so that it is charged at a very low rate $GIR$.\footnote{This insurance rate is lower than the economic break-even insurance rate because, in the Great Recession, the US government had to pay for defaults of a large amount of agency-backed MBS since GSEs failed to do so.} Therefore, the product of recovery rate and small growth rate differences in employment, household income, and population between the two areas are smaller than the twice difference between two funding costs: $2FCR^{*}_{P} - RR*( \mathit{\Delta}e + \mathit{\Delta}g + \mathit{\Delta}p  ) > 2FCR_G$. Conversely, if I assume $2FCR^{*}_{P} - RR*(\mathit{\Delta}e  + \mathit{\Delta}g + \mathit{\Delta}p  ) < 2FCR_G$, together with the fact that $ |K^{*}_{P,H} > K^{*}_{G} | $, I could get the conclusion that mortgage rate in PLM for conforming loans is even lower than mortgage rate in GSEM. This conclusion apparently contradicts reality.

%--------------------------------------------------------------
% Beginning of figure_ 
% Mortgage Lending in High-NEG Areas
% ends in Line 
%--------------------------------------------------------------

\begin{figure}[H]

\begin{center}

%\resizebox{4.5in}{3.2in}{%
\resizebox{0.7\textwidth}{!}{%
\includegraphics[]{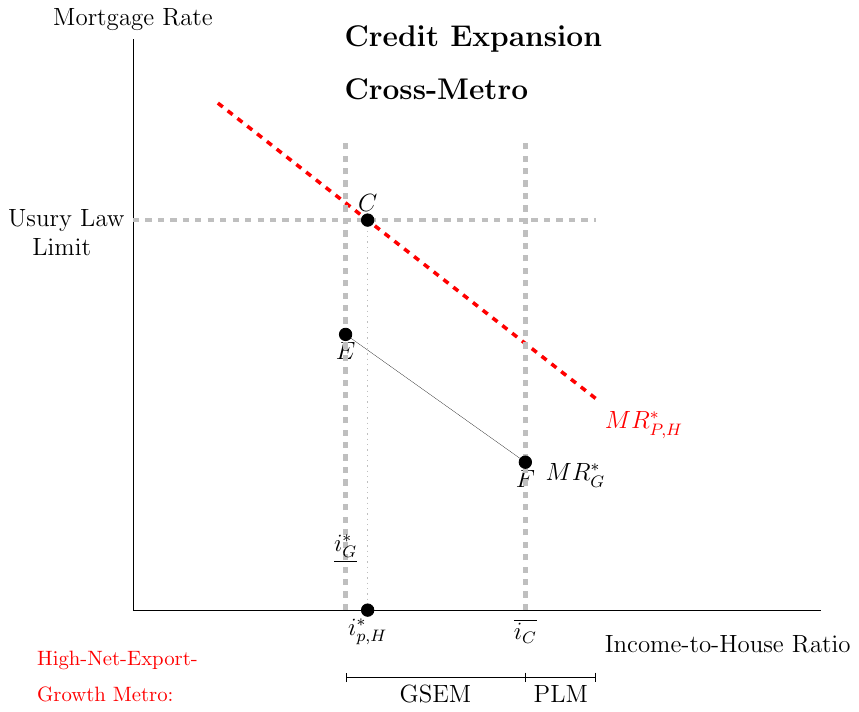}
} %end of resizebox

\end{center}

%-----------------------------------------------
% Figure setting: caption and label
%-----------------------------------------------
\caption{Mortgage Lending in High-NEG Areas}
\label{fig_HNEG_area_Bef}

\end{figure}
%----------------------------------------------------------------------
% End of figure_ 
% Mortgage Lending in High-NEG Areas
% starts in 
%----------------------------------------------------------------------

%\vspace{-0.5cm}

The above comparison of two lines of break-even mortgage rates can be shown in Figure (\ref{fig_HNEG_area_Bef}). After imposing the ceiling mortgage rate $\overline{MR_G}$ for the GSEM and $\overline{MR_P}$ for the PLM, I can solve for the corresponding two lower bounds for income-to-house ratio: 
\begin{equation}
    i^{*}_G = \frac{ 2(1 + FCR_G + TF) - RR - \overline{MR_G}}{1+g_{d} - c}
\end{equation}
\begin{equation}
    i^{*}_{P,H} = \frac{ 2(1 + FCR^{*}_{P} + TF) - RR(1 \textcolor{red}{ + \mathit{\Delta}e + \mathit{\Delta}g + \mathit{\Delta}p  } ) - \overline{MR_P}}{1+g_{d} \textcolor{red}{+ \mathit{\Delta}g} - c}
\end{equation}
For an easy demonstration, I further assume that $i^{*}_{P,H} > i^{*}_G$.\footnote{This additional assumption is made for easy illustration of the key economic prediction: the differential higher relative increase in PLMs in the HNEG area than the LNEG area after the decline of the funding cost of PLM. By assuming $i^{*}_{P,H} > i^{*}_G$, I effectively let PLM share only cover the ``jumbo" mortgages where $i_0 >= \overline{i_0}$. Of course, this assumption may not be realistic, so empirical tests of key predictions by the model are necessary.} Therefore, in the mortgage lending market in HNEG area, lenders with income-to-house ratio $i_0 \in [\overline{i_C}, \overline{i_0} )$ can only get mortgages from the PLM market since the government does not sponsor home ownership for the very rich group of households buying luxury houses. Lenders with income-to-house ratio $i_0 \in [i^{*}_G, \overline{i_C} )$ would choose the GSEM market because (1) the lower rate in the GSEM market than the one in PLM market when $i_0 \in [i^{*}_{P,H}, \overline{i_C} )$ and (2) GSEM is the only one available when $i_0 \in [i^{*}_{G}, i^{*}_{P,H} )$. Lastly, lenders with income-to-house ratio $i_0 \in [\underline{i_0}, i^{*}_G)$ cannot get mortgages because of the legal ceiling limit of PLM rates.

\subsubsection{Mortgage Market in LNEG area in normal time}

Now I consider the mortgage market in LNEG area that contains both the GSEMs and PLMs in normal time, when $FCR^{*}_{P}$ is much higher than $FCR_G$. Based on the break-even mortgage rate functions in Equation (\ref{eq:MR_G}) and (\ref{eq:MR_P_L}), I know the absolute value of slope $K^{*}_{P,L}$ is lower than the one of slope $K^{*}_{G}$ because the PLM market considers the differential lower household income growth rate $\mathit{\Delta} g$ in LNEG area. For each unit increase of income-to-house ratio $i_0$ in time 0, the PLM will be perceived as being compensated less by disposable income in default (in down-state), thereby charging a higher mortgage rate when there is no default (in up-state). I can also know that the constant part $C^{*}_{P,L}$ is higher than $C^{*}_{G}$ because $2FCR^{*}_{P} + RR*(\mathit{\Delta}e + \mathit{\Delta g} + \mathit{\Delta}p  ) > 2FCR_G$ given that the funding cost in PLM is higher than the one in GSEM ($FCR^{*}_{P} > FCR_G$).

%--------------------------------------------------------------
% Beginning of figure_ 
% Mortgage Lending in Low-NEG Areas in normal time
% ends in Line
%--------------------------------------------------------------

\begin{figure}[H]

\begin{center}

%\resizebox{4.5in}{3.2in}{%
\resizebox{0.7\textwidth}{!}{%
\includegraphics[]{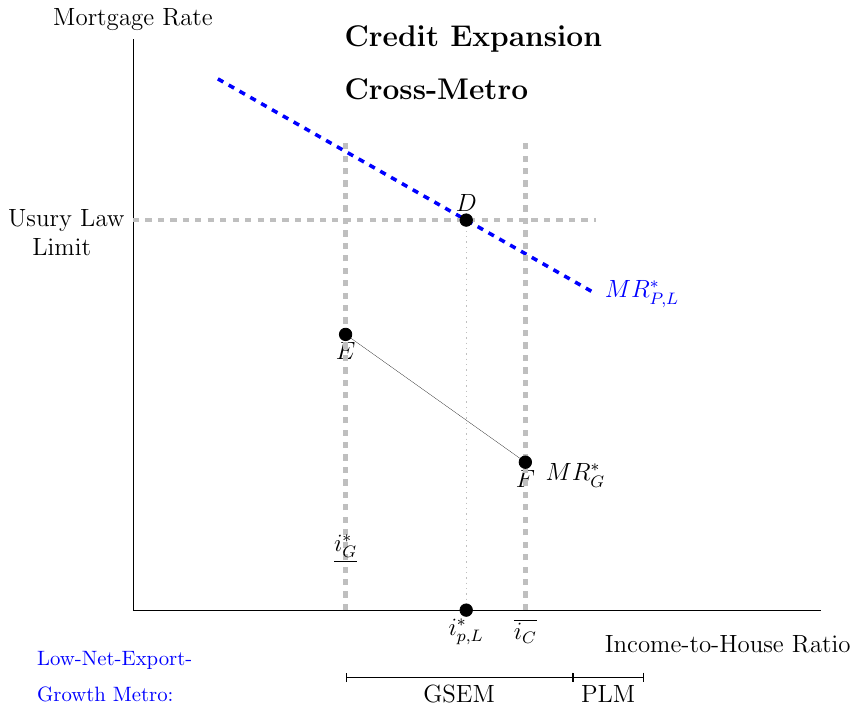}
} %end of resizebox

\end{center}

%-----------------------------------------------
% Figure setting: caption and label
%-----------------------------------------------
\caption{Mortgage Lending in Low-NEG Areas in Normal Time}
\label{fig_LNEG_area_Bef}

\end{figure}
%----------------------------------------------------------------------
% End of figure_ 
% Mortgage Lending in Low-NEG Areas in normal time
% starts in Line 
%----------------------------------------------------------------------

The above comparison of two lines of break-even mortgage rates can be shown in Figure (\ref{fig_LNEG_area_Bef}). After imposing the ceiling mortgage rate $\overline{MR_G}$ for the GSEM and $\overline{MR_P}$ for the  PLM, I can additionally solve for the corresponding lower bounds for income-to-house ratio in the PLM market in LNEG area: 

\begin{equation}
    i^{*}_{P,L} = \frac{ 2(1 + FCR^{*}_{P} + TF) - RR(1 \textcolor{blue}{ - \mathit{\Delta}e -\mathit{\Delta}g - \mathit{\Delta}p }   ) - \overline{MR_P}}{1+g_{d} \textcolor{blue}{- \mathit{\Delta}g} - c}
\end{equation}

By comparing numerator and denominator, I can conclude $i^{*}_{P,L} > i^{*}_{P,H}$. From previous assumption that $i^{*}_{P,H} > i^{*}_G$, I can  
easily know $i^{*}_{P,L} > i^{*}_G$. Therefore, in the mortgage lending market in LNEG area, lenders with income-to-house ratio $i_0 \in [\overline{i_C}, \overline{i_0} )$ can only get mortgages from the PLM market since the government does not sponsor home ownership for the very rich group of households buying luxury houses. Lenders with income-to-house ratio $i_0 \in [i^{*}_G, \overline{i_C} )$ would choose the GSEM market because (1) the lower rate in the GSEM market than the one in PLM market when $i_0 \in [i^{*}_{P,L}, \overline{i_C} )$ and (2) GSEM is the only one available when $i_0 \in [i^{*}_{G}, i^{*}_{P,L} )$. Lastly, lenders with income-to-house ratio $i_0 \in [\underline{i_0}, i^{*}_G)$ cannot get mortgages because of the legal ceiling limit of PLM rates.

\subsubsection{Compare Markets in HNEG area and LNEG area in normal time}

%--------------------------------------------------------------
% Beginning of figure_ Mortgage Markets Before the Decline of Funding Cost in PLM Market. 
%--------------------------------------------------------------

\begin{figure}[H]

\begin{center}

%\resizebox{4.5in}{3.2in}{%
\resizebox{0.7\textwidth}{!}{%
\includegraphics[]{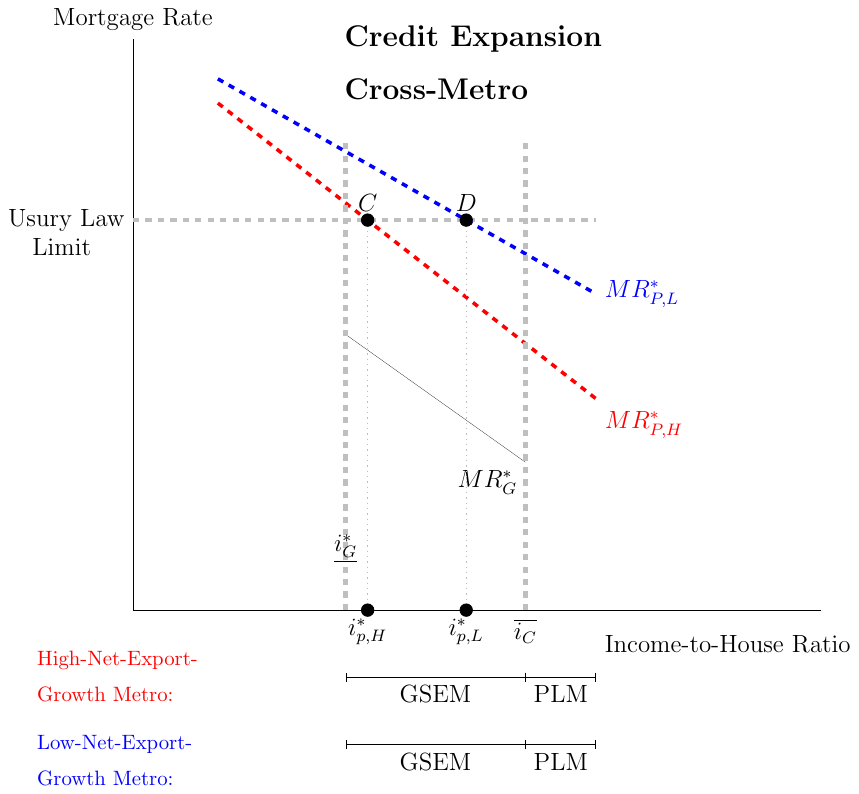}
} %end of resizebox

\end{center}

%-----------------------------------------------
% Figure setting: caption and label
%-----------------------------------------------
\caption{Mortgage Markets Before the Decline of Funding Cost in PLM Market}
\label{fig_MortMkt_Bef}

\end{figure}
%----------------------------------------------------------------------
% end of figure_ Mortgage Markets Before the Decline of Funding Cost in PLM Market
%----------------------------------------------------------------------

Now let us compare mortgage markets in HNEG area and LNEG area in normal times. Here I can compare two markets in one figure since I have assumed that income-to-house ratio $i_0$ follows the same distribution in both areas. First, since the GSEM market does not consider the differences in local economic conditions, the GSEM rate is the same in both HNEG area and LNEG area. Of course, $MR^{*}_{G}$ is the lowest for any $i_0$ since GSEs provide mortgage insurance at a rate lower than the economic break-even cost, as a subsidy to home ownership by the government. Second, when $i_0=\underline{i_0}$, I have $MR^{*}_{P,L} > MR^{*}_{P, H}$ since the PLM ultimate investors in LNEG area get lower recovery value and lower disposable income in case of default. To achieve the break-even condition, the mortgage rate must be higher for PLM in LNEG area than the one in HNEG area for the same income-to-house ratio. Third, the absolute value of the slope $K^{*}_{P,L} < K^{*}_{P,H} $, because, for each unit increase of income-to-house ratio $i_0$, the increase in disposable income in case of default in LNEG area is lower than the one in HNEG area. Thus, PLM rate declines with $i_0$ less steeply in LNEG area than in HNEG area. These patterns can be seen in Figure (\ref{fig_MortMkt_Bef}).

%--------------------------------------------------------------
% This is start of new subsubsection
%--------------------------------------------------------------

\subsection{When Funding Cost of PLM Declines}
Now I study the changes when the funding cost of PLM declines relative to the one of GSEM. For the empirical motivation and evidence, please refer to \cite{justiniano2022mortgage}, which finds the sharp decline of PLM rates in the summer of 2003. I can model this situation simply by letting $FCR^{*}_{P}$ decline to $FCR_{P}^{**}$ while holding $FCR_G$ unchanged. The timing of credit expansion I chose in the empirical setting is the end of 1999, based on the timing of the invention of the Copula formula in securitization technology. It took some time for the finance industry and investors to accept complex financial products such as Collateralized Debt Obligations. 

%--------------------------------------------------------------
% This is the total graph that contains all elements of the following graphs 
% 
%--------------------------------------------------------------

\begin{figure}[H]

\begin{center}

%\resizebox{4.5in}{3.2in}{%
\resizebox{0.7\textwidth}{!}{%
\includegraphics[]{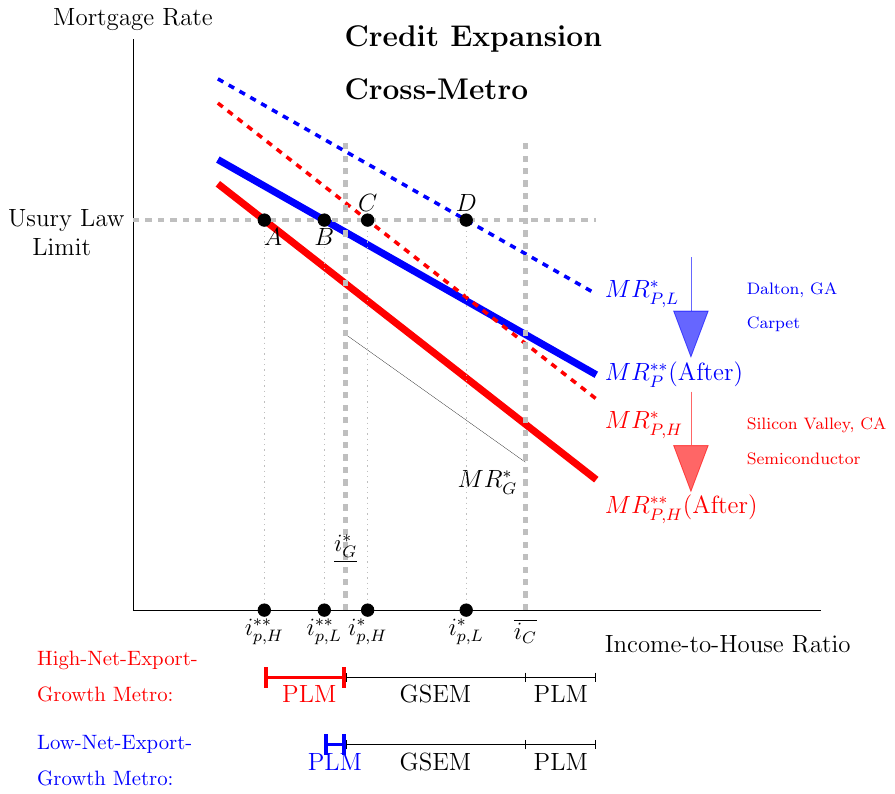}
} %end of resizebox

\end{center}

%-----------------------------------------------
% Figure setting: caption and label
%-----------------------------------------------
\caption{Mortgage Market Before and After A Decline of Funding Cost in PLM}
\label{fig_MortMkt_Bef&Aft}

\end{figure}
%----------------------------------------------------------------------
% end of figure 
%----------------------------------------------------------------------

The above decline will result in a downward shift of PLM rates in both areas. Such a shift can be seen in Figure (\ref{fig_MortMkt_Bef&Aft}). The after-shift lines are plotted in bold lines. Mortgage rate $MR^{**}_{P,H}$ and $MR^{**}_{P,L}$ interacts with legal upper limit $\overline{MR_P}$ at point A with $i_0 = i^{**}_{P,H}$ and point B with $i_0 = i^{**}_{P,L}$, respectively. 

By replacing $FCR^{*}_{P}$ with $FCR_{P}^{**}$, I can get: 

\begin{equation}\label{eq:i_newPH}
    i^{**}_{P,H} = \frac{ 2(1 + FCR^{**}_P + TF) - RR(1\textcolor{red}{ + \mathit{\Delta}e + \mathit{\Delta}g + \mathit{\Delta}p  } ) - \overline{MR_P}}{1+g_{d} \textcolor{red}{+ \mathit{\Delta}g} - c}
\end{equation}

\begin{equation}\label{eq:i_newPL}
    i^{**}_{P,L} = \frac{ 2(1 + FCR^{**}_P + TF) - RR(1 \textcolor{blue}{- \mathit{\Delta}e - \mathit{\Delta}g  - \mathit{\Delta}p  } ) - \overline{MR_P}}{1+g_{d} \textcolor{blue}{- \mathit{\Delta}g} - c}
\end{equation}

By comparing both numerators and denominators, I can easily know $i^{**}_{P,H}< i^{**}_{P,L}$ so point A lies to the left of point B. I further assume $i^{**}_{P,L} < i^{*}_G$, which matches the reality that PLM does increase in LNEG area after its funding cost declines.\footnote{Please see the empirical results in this paper.} I also assume that $ i^{**}_{P,H} > \underline{i_0} $ since, in practice, not every household in HNEG area can be approved for PLM.

Now, in the mortgage market in HNEG area, households with income-to-house ratio $i_0 \in [\overline{i_C}, \overline{i_0}  ) \cup [i^{**}_{P,H}, i^{*}_G )  $ can only get mortgages from the PLM market. Households with income-to-house ratio $i_0 \in [i^{*}_G, \overline{i_C} )$ would choose the GSEM market due to the lower rate. Lastly, households with income-to-house ratio $ i_0 \in [\underline{i_0}, i^{**}_{P,H}) $ cannot get mortgages because of the legal ceiling limit of PLM rates.

Similarly, now, in the mortgage lending market in LNEG area, households with income-to-house ratio $i_0 \in [\overline{i_C}, \overline{i_0} ) \cup [i^{**}_{P,L}, i^{*}_G ) $ can get mortgages from the PLM market. Households with income-to-house ratio $i_0 \in [i^{*}_G, \overline{i_C} )$ would choose the GSEM market due to the lower rate. Lastly, households with income-to-house ratio $i_0 \in [\underline{i_0}, i^{**}_{P,L} )$ cannot get mortgages because of the legal ceiling limit of PLM rates.

%--------------------------------------------------------------
% This is start of new subsection
%--------------------------------------------------------------

\subsection{Summary: How the Model Reconciles Two Empirical Facts}

%----------------------------------------------------------------------
%----------------------------------------------------------------------
% Beginning of figure_ MortMkt_Bef&Aft_Axis
%----------------------------------------------------------------------
%----------------------------------------------------------------------

\begin{figure}[H]

\begin{center}

%\resizebox{4.5in}{3.2in}{%
\resizebox{0.8\textwidth}{!}{%
\includegraphics[]{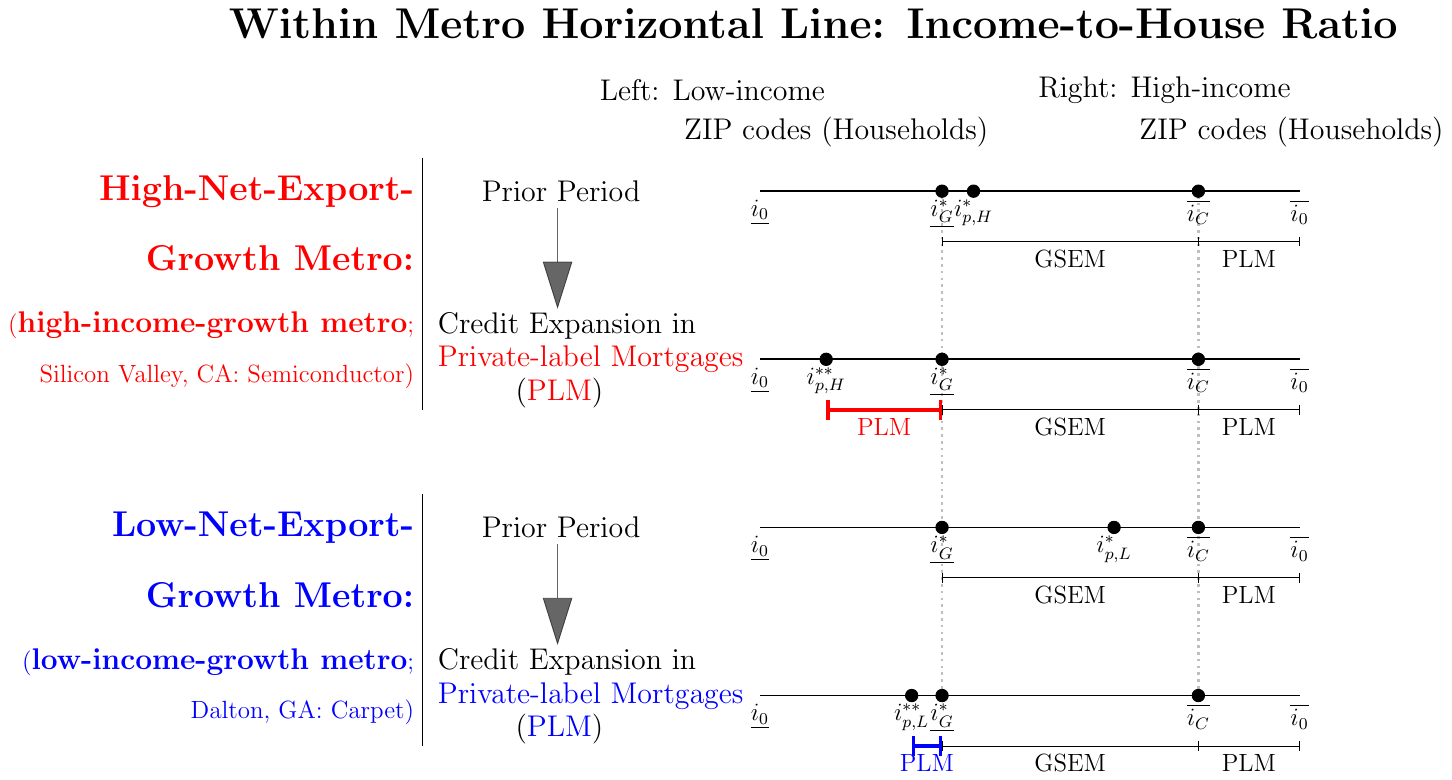}
} %end of resizebox

\end{center}

%-----------------------------------------------
% Figure setting: caption and label
%-----------------------------------------------
\caption{Compare: Normal Time vs $FCR^{*}_{P}$ Declines}
\label{fig_MortMkt_Bef&Aft_Axis}

\end{figure}
%----------------------------------------------------------------------
% end of figure_ MortMkt_Bef&Aft_Axis
% begins in Line 
%----------------------------------------------------------------------

Figure (\ref{fig_MortMkt_Bef&Aft_Axis}) summarizes the mortgage market shares by GSEM and PLM both in normal times (when the funding cost of PLM ($FCR^{*}_{P}$) is much higher than the one of GSEM ($FCR_G$)) and when $FCR^{*}_{P}$ declines. In the latter case, my model predicts that credit expansion in PLM is stronger in the HNEG area than in the LNEG area. This stronger increase in PLM share in the HNEG area shows up in the relationship $i^{**}_{P,H}< i^{**}_{P,L}$. Comparing Equation (\ref{eq:i_newPH}) and (\ref{eq:i_newPL}), a sufficient condition is that $\mathit{\Delta}g > 0$, $\mathit{\Delta}e > 0$, and $\mathit{\Delta}p > 0$. In economic terms, this sufficient condition means there are higher positive growth rates in employment, household income, and population in the high-net-export-growth metropolitan areas in the future.\footnote{As argued in Section \ref{sec:Empirical_Metro_AgainstSpeculation}, realized effects at the household level could be considered by government-sponsored enterprise mortgages, which seems not big enough to show up in cross-metro patterns of GSEMs. In contrast, unrealized and predicted effects at the metropolitan level can only be considered by private-label mortgages, which is large enough to show up in tests in Section \ref{sec:Evidence_Metro_CreditExpansion}.}  It is through these three economic mechanisms (unrealized prediction) that net export growth drives the stronger credit supply in PLM in the HNEG area between 1999-2005. For empirical tests verifying that higher net export growth causes higher employment growth, higher household income growth, and higher population growth, see Appendix Section \ref{subsec:NEG_LocalEconConditions}.  

The prediction that credit expansion in PLM is stronger in the HNEG area implies that the positive correlation between mortgage growth and income growth across metropolitan areas is a natural result of credit expansion in response to net export growth. This phenomenon happens because (1) within each metro, ZIP codes with low income-to-house ratio experience the majority of PLM growth, and (2) ZIP codes with low income-to-house ratio in the HNEG areas experience much higher growth in PLMs than the ones in the LNEG areas. This empirical fact alone is not sufficient to support the ``speculation" view as \cite{adelino2016loan}. Therefore, the empirical evidence provided in support of this model argues against the ``speculation" interpretation by \cite{adelino2016loan}. 

The model also predicts that a natural result of credit expansion is that within-metropolitan areas (and within-county), credit expansion is stronger in low income-to-house-ratio areas. This prediction is consistent with the empirical evidence by \cite{mian2009consequences} that subprime zip codes experience stronger mortgage growth within counties.

Thus, the simple model reconciles two seemly opposing empirical facts: the correlation between income growth and mortgage growth is positive across metropolitan areas \citep{adelino2016loan} but negative areas across ZIP codes within metropolitan areas \citep{mian2009consequences}. These two empirical facts can coexist within a credit expansion framework.

%--------------------------------------------------------------
%--------------------------------------------------------------
% This is the end of the entire section of Model 
%--------------------------------------------------------------
%--------------------------------------------------------------

%----------------------------------------------------------------------

%----------------------------------------------------------------------
% section 3: Research Design

%--------------------------------------------------------------------------------------
% new section
%--------------------------------------------------------------------------------------

%\clearpage

\section{Research Design}\label{sec:ResearchDesign}
In this section, I illustrate empirical research design. First, I measure the fundamental incentive for mortgage credit by operationalizing the central ideal of the ``Economic Base theory" by \cite{tiebout1962community}. I use metropolitan exposure to net export growth of manufacturing industries (hereafter ``net export growth") as a proxy for the growth of the local tradable sector, which, by theory, is the fundamental force of long-term house price growth. To overcome the endogeneity issue of OLS, I use an instrumental variable approach by \cite{feenstra2019us} as an identification strategy.

\subsection{Operationalize the ``Economic Base Theory"}
In this subsection, I operationalize the central idea of ``economic base theory" by using metropolitan exposure to net export growth of manufacturing industries (hereafter ``net export growth") as a proxy for the growth of the local tradable sector. This net export growth measure can fulfill the requirement of the treatment variable in the introduction: enough area coverage (U.S. mainland), enough time coverage (91-09), enough geographic variation, and a feature of persistence.

To capture the incentive to credit expansion in mortgages, I revisit the ``economic base theory" in regional economics. Regional economics defines the ``economic base'' (tradable sector) as the economic activities that a local area provides for the world beyond its boundaries \citep{tiebout1962community, nijkamp1987regional}. Thus, the tradable sector brings wealth into the local area, and most of the money will be reused locally via a multiplier effect through the nontradable sector. Based on this theory, the tradable sector is the most important driving force for local economic growth, land use, and house prices in the long term \citep{nijkamp1987regional,thrall2002business,ling2013real}. Therefore, the tradable sector growth can predict demand-side factors that shape the long-term growth of housing prices, such as employment growth, household income growth, and population growth.\footnote{\cite{olsen1987demand} surveys on the demand factors of housing, including the above three factors.} From this perspective, in a credit expansion, credit growth would be stronger in areas with stronger tradable sector growth due to at least two reasons: (1) higher foreclosure price of house and (2) higher income growth that can be recoursed by lenders given default.

Ideal measurement of local tradable sector growth requires census-style data covering the accounting data of all firms in the tradable sector, which does not exist. Instead, I use the local employment composite (share) as a proxy for the composite of the local tradable sector. Then, I use the large time-series change in U.S. net export growth in manufacturing industries as a proxy for the relative growth (shift) at the industry level. Aggregating the above two proxies (share and shift) results in a good measure of the relative growth of the local treatable sector across metropolitan areas.\footnote{We acknowledge that the use of manufacturing does not include several other factors in the tradable sector: college town, retirement community, other tradable goods industries (e.g., natural resource), and tradable services (e.g., information technology and medical sector). Instead, I will use control variables to account for most of the above factors in Section \ref{subsec:causal_house_price_boom_bust}.} To account for the commuting within local areas, I aggregate net export growth at the metropolitan level. 

I implement the above approach in two steps in data. First, net export measure in industry $g$ year $t$ is defined as $ \text{NetExp}_{g,t}= \frac{Export_{g,t} - Import_{g,t}}{Y_{g,91}} $, where $Export_{g,t}$, $Import_{g,t}$, and $Y_{g,91}$ are US export, import and domestic production in industry $g$ in year $t$ (or 1991), respectively. All terms are measured in 2007 U.S. dollars. The scaling factor, U.S. domestic production in 1991, is chosen to avoid the potential response of domestic production to trade in later years in my study.\footnote{The same choice of domestic production in 1991 as the denominator is taken by other papers like \cite{barrot2022import}.} Second, similar to \cite{acemoglu2016import}, I use local employment data (share) to aggregate net export growth at the metropolitan level across the period: 
\begin{equation}\label{eq:NEG_m}
    \triangle_{t_{1},t_{2}}\text{NetExp}_{m} = \sum_{g} \big[  (L_{m,g,t_{0}}/L_{m.t_{0}}) * (\text{NetExp}_{g,t_{2}} - \text{NetExp}_{g,t_{1}}) \big]
\end{equation}
where $L_{m,g,t_{0}}$ and $L_{m.t_{0}}$ are the employment of industry $g$ and total employment in metropolitan area $m$ in year $t_{0}$, respectively. I choose $t_{0} = t_{1}-1$ to make sure the employment share is pre-existing to the trade measure so that all changes in net export growth ($\triangle_{t_{1},t_{2}}\text{NetExp}_{m}$) is completely driven by changes in trade measures (shift) rather than employment composite (share). Year $t_{1}$ to $t_{2}$ is the period of interest.

The net export growth above satisfies the requirement of the treatment variable. First, it has enough area coverage (U.S. mainland) and enough time coverage (91-09). Second, its geographic variation is large because the local tradable sector tends to cluster within several related industries due to economies of scale. Internal economies of scale increase the size of local tradable firms by at least three mechanisms: input purchase at a volume discount, fixed cost of plant operating, and learning in operation \citep{worldbank2009ch4}. \cite{kwon2022100} recently study the corporate concentration (e.g., asset share or sales share of top-size business) by digitizing Statistics of Income (SOI) and the associated Corporate Source Book from the Internal Revenue Service (IRS) from 1918 to 2018. They find a persistent increase in the shares accounted for top-size corporations throughout the last century. External economies of scale attract firms in the same and related industries by at least three channels: specialization of suppliers, labor market pooling, and knowledge spillovers \citep{krugman2018internationalCh7}. Third, two other trends also strengthen the local industry clustering. The first trend is the long-term reduction in transportation costs due to the invention of new technology, construction of infrastructure, adoption of information technology, and reorganization of logistic methods \citep{redding2015transportation}. The second trend is the increased labor mobility across areas within the United States due to increased education attainment, easy availability of housing, encouraging public policies, and long-standing cultural \citep{molloy2011internal,greenwood1997internal}. Famous examples of local industry clustering include the semiconductor industry concentrated in Silicon Valley, the automobile industry concentrated in Detroit \citep{holmes2004spatial}, and the carpet industry concentrated in Dalton, Georgia \citep{krugman1991increasing}.

Fourth, many global events between the 1980s and 1990s promoting international trade also added to the local industry clustering at the global level. In the 1980s, many developing countries rapidly abolished earlier import substitution strategies and sometimes undertook radical trade liberalization. In addition, China's economic transition accelerated in the 1990s \citep{amiti2010anatomy}, and India's reforms were initiated in 1991 \citep{rodrik2004hindu}. Besides, Global negotiation among the General Agreement on Tariffs and Trade (GATT) members resulted in 33 percent between 1980-1987, by 38 percent between 1995-1999 \citep[p. 207]{worldtradereport2007}. Moreover, the North American Free Trade Agreement (NAFTA) was established in 1994 \citep{krugman2008trade} and the World Trade Organization (WTO) in created in 1995. Further, the Dissolution of the Soviet Union in 1991 triggered many countries in Central and Eastern Europe and the former Soviet Union to apply to accede to the GATT and then the WTO \citep{broadman2006disintegration}.

Fifth, the persistence feature of net export growth at the metropolitan level arises from three dimensions. At the industry level, both comparative advantages (due to technology level or natural endowment) and horizontal specialization (due to economies of scale) across nations make trade patterns persistent over time. At the local area level, clusters of manufacturing industries have been established over decades, involving extensive plants built and the migration by workers. Such industry cluster formation with huge costs is unlikely to change rapidly in a short period of time. At the individual level, human capital accumulation takes both time and investment, making job reallocation across different industries very difficult both locally and remotely.

\subsection{Gravity Model-based Instrumental Variable}\label{subsec:GIV_exports}

In this subsection, I start with the potential OLS bias and then illustrate how the instrumental variable approach can achieve causal estimates. I start with OLS specification that regresses the house price growth on the growth of private-label (non-jumbo) mortgages (PLNJM) at the county level:
\begin{equation}\label{eq:OLS_HP_PLNJM}
\begin{split}
     \triangle_{99,05} Ln(HP_{c}) = \beta * \triangle_{99,05} Ln(\text{Private-Mort}_{c}) + \gamma* Controls_{c}  + \epsilon_{c} 
\end{split}
\end{equation}

\noindent Here, the dependent variable $\triangle_{99,05} Ln(H.P._{c})$ is the house price growth (normalized to 2007 USD) at county $c$ 99-05. The independent variable $\triangle_{99,05} Ln(PLNJM_{c})$ is the growth of dollar amount (normalized to 2007 USD) of private-label (non-jumbo) mortgages (PLNJM) at county $c$ 99-05.

Potential omitted variables could bias down $\beta$. For example, the rapid net export growth and, hence, mortgage growth in 1999-2005 have been anticipated by employees in Silicon Valley so that house prices would grow before 1999 to reflect such expectations. In this case, $\beta$ could be biased downward because some of the effect of the mortgage on house price shows up in house price growth in early periods, reducing the house price growth 1999-2005.

To overcome the potential endogenous concern regarding OLS specification, I use gravity model-based instrumental variable introduced by \cite{feenstra2019us} for net export growth.\footnote{I am very grateful for help from Robert Feenstra, Hong Ma, and Yuan Xu in replicating their instrumental variables.} They construct IV for exports and imports separately, I take a further step to combine both as an IV for net exports. To illustrate the key idea, I show the export model while keeping the import model in Appendix Section \ref{subsec:GIV_imports}. 

As for the instrumental variable for U.S. exports at the industry-year level, \cite{feenstra2019us} builds on the idea that eight other high-income countries' exports can instrument for the U.S. exports since they both capture the world's rising demand. In addition, they also incorporate tariff changes, which are exogenous to foreign firms. Lastly, this method corrects for the supply shocks in the home country by using a fixed effect to remove them.

To predict U.S. export, the gravity model-based IV begins with a simple symmetric constant-elasticity equation in \cite{romalis2007nafta} for export: 
\begin{equation}{\label{eq:gravity_export}}
    \frac{X^{US,j}_{s,v,t}}{X^{i,j}_{s,v,t}} = \Bigg( \frac{w^{US}_{s,t}*d^{US,j}*\tau^{US,j}_{s,t}}{w^{i}_{s,t}*d^{i,j} * \tau^{i,j}_{s,t}} \Bigg) ^{1-\sigma}
\end{equation}
In the above formula, $X^{U.S.,j}_{s,v,t}$ is U.S. export to country $j$ in product variant $v$ in industry $s$ in year $t$. $X^{i,j}_{s,v,t}$ is the similar term but representing export from country $i$ to $j$. $w^{US}_{s,t}$ and $w^{i}_{s,t}$ are the relative marginal cost of production in industry $s$ in the US and country $i$. $\tau^{U.S.,j}_{s,t}$ and $\tau^{i,j}_{s,t}$ are the \textit{ad valorem} total import tariff imposed by country $j$ on exports from the U.S. and country $i$, respectively; $d^{U.S.,j}$ and $d^{i,j}$ are the bilateral distance and other fixed trade costs from U.S. to country $j$ and from country $i$ to country $j$. Lastly, $\sigma$ denotes the constant elasticity of substitution ($\sigma > 1$). 

The intuition of this gravity-style model is quite straightforward. Competing with country $i$, U.S. exports to the country $j$ are decreasing with the ratio of relative marginal cost, the ratio of bilateral distance, and the ratio of \textit{ad valorem} total import tariff. 

Suppose that there are $N^{i}_{s,t}$ identical product varieties exported by country $i$ to the country $j$ in the industry $s$ and year $t$, \cite{feenstra2019us} re-arranges the above equation, multiply both sides with $N^{i}_{s,t}$, and sum over countries $ i \neq US$:
\begin{equation*}
    X^{US,j}_{s,v,t}*\sum_{i\neq US} \big[ N^{i}_{s,t}(w^{i}_{s,t} d^{i,j})^{1-\sigma} \big] = (w^{US}_{s,t}d^{US,j}\tau^{US,j}_{s,t})^{1-\sigma} * \sum_{i\neq US} \big[N^{i}_{s,t}X^{i,j}_{s,v,t}  (\tau^{i,j}_{s,t})^{\sigma-1}\big]
\end{equation*}

Since above equation holds for any countries $i \neq US$, one can choose the set of countries that have similar economic conditions (so that they are closely competing with U.S. exports) to make my prediction more accurate. Following \cite{autor2013china}, \cite{feenstra2019us} use the eight high-income countries.

Again, I can multiple $N^{U.S.}_{s,t}$ (number of variants of products exported by U.S.) on both sides and denote the sectoral exports $X^{U.S.,j}_{s,t} \equiv X^{U.S.,j}_{s,v,t}N^{U.S.}_{s,t}$ and $X^{i,j}_{s,t} \equiv X^{i,j}_{s,v,t}N^{i}_{s,t}$, then I have
\begin{equation*}
    X^{US,j}_{s,t}*\sum_{i\neq US} \big[ N^{i}_{s,t}*(w^{i}_{s,t} d^{i,j})^{1-\sigma} \big] = N^{US}_{s,t}*(w^{US}_{s,t}d^{US,j}\tau^{US,j}_{s,t})^{1-\sigma} * \sum_{i\neq US} \big[X^{i,j}_{s,t} * (\tau^{i,j}_{s,t})^{\sigma-1}\big]
\end{equation*}

After a few re-arrangements, one can get the formula for $ X^{U.S.,j}_{s,t}$:
\begin{equation}
    X^{US,j}_{s,t} =   \frac{N^{i}_{s,t}*(w^{US}_{s,t}d^{US,j}\tau^{US,j}_{s,t})^{1-\sigma}}{\sum_{i\neq US} \big[ N^{i}_{s,t}*(w^{i}_{s,t} d^{i,j})^{1-\sigma} \big]}  * \bigg( \sum_{k\neq US} X^{k,j}_{s,t} \bigg)  * \sum_{i\neq US} \bigg[  \frac{ X^{i,j}_{s,t} }{\sum_{k\neq US} X^{k,j}_{s,t}}  * (\tau^{i,j}_{s,t})^{\sigma-1} \bigg]
\end{equation}
 
Note in the above formula, I multiply and divide by $\sum_{k\neq U.S.} X^{k,j}_{s,t}$ to prepare for the regression setup in the next step. Now I can take logs of the above equation and get the regression-style formula:
\vspace{-1mm}
\begin{equation} \label{eq:exp_gravityRegression}
\resizebox{0.92\textwidth}{!}{%
\begin{math}
\begin{aligned}
\lnb{X^{US,j}_{s,t}} & = \underbrace{\lnb{\sum_{k\neq US}X^{k,j}_{s,t}} }_{\text{Term 0}} +  \underbrace{\lnb{N^{US}_{s,t}(w^{US}_{s,t})^{1-\sigma}}}_{\text{Ind-Year FE: } \alpha^{US}_{s,t}} + \underbrace{(1-\sigma)\lnb{d^{US,j}}}_{\text{Importing-country FE: } \delta^{US,j}}  \\
& + \underbrace{ (1 - \sigma) \lnb{\tau^{US,j}_{s,t}} }_{\text{Term 1}} + \underbrace{ (\sigma-1) \lnb{ \Bigg\{  \sum_{i\neq US} \bigg[ \frac{ X^{i,j}_{s,t} }{\sum_{k\neq US} X^{k,j}_{s,t}} (\tau^{i,j}_{s,t})^{\sigma -1} \bigg] \Bigg\} ^{\frac{1}{\sigma-1}}  }}_{\text{Term 2:} (\sigma -1)\lnb{T^{j}_{s,t}} } + \epsilon^{j}_{s,t} \\ 
\end{aligned} 
\end{math}
} %end of \scalemath \resizebos
\end{equation}
We can see that U.S. exports to the country $j$ in the industry $s$ year $t$ can be decomposed into six terms. ``Term 0'' is the exports from eight other high-income countries (whose economic development is similar to the U.S.) to the destination country $j$, which reflects the world demand. The second term $\alpha^{U.S.}_{s,t}$, which reflects the U.S. supply shocks by industry and year, is potentially endogenous. For instance, the U.S. computer chip industry exports a lot, partially because of its superior investment in human capital and innovation. \cite{feenstra2019us} remove this term by the U.S. industry-year fixed effects. The third term $\delta^{U.S.,j}$ reflects the distance from the U.S. to the destination market $j$ and all other industry- and year-invariant trade costs. For example, the US exports more to Mexico than the mid-east, partially because of the lower transportation cost associated with shorter distances. Since this term is predetermined and endogenous to firms' exporting decisions, \cite{feenstra2019us} remove it by importing-country fixed effects.  ``Term 1" represents the tariffs on U.S. exports imposed by destination country $j$, which is out of control by U.S. firms. This term is retained to capture the shocks from tariffs. ``Term 2" is the weighted average tariffs on non-US exports imposed by destination country $j$, which is again out of control of U.S. firms. Intuitively, when this weighted average tariff on non-US exports rises, destination country $j$ will import more from the U.S. as substitutions. This term is retained to reflect such substitution effect. The last term $\epsilon^{j}_{s,t} = - \lnb{ \sum_{k\neq US} [ N^{i}_{s,t}(w^{i}_{s,t}d^{i,j})^{1-\sigma} ] } $ is unobserved and remains in the regression error term. 

After the above regression, I can construct predicted U.S. exports that are isolated from supply shocks and the predetermined factor by the above ``Term 0", ``Term 1", and ``Term 2":
\begin{equation} \label{eq:gravityPreUSExp}
\lnb{\widehat{X^{US,j}_{s,t}}} = \lnb{\sum_{k\neq US}X^{k,j}_{s,t}} + \hat{\beta_1} *\lnb{\tau^{US,j}_{s,t}} + \hat{\beta_2}* \lnb{T^{j}_{s,t}}
\end{equation}

%\subsection{Data Implementation}\label{subsec:Data_Implementation}

\noindent \textbf{Data Implementation} There are four detailed steps implementing the above procedures in data to get net export and its GIV at the metropolitan level across periods. First, I estimate Eq (\ref{eq:exp_gravityRegression}) at the 6-digit H.S. industry level (5673 industries) and get predicted U.S. export (isolated from supply shocks) by Eq (\ref{eq:gravityPreUSExp}). Second, I aggregate predicted U.S. exports across importing countries and crosswalk 6-digit H.S. code (5673 industries) to 4-digit revised SIC system (392 manufacturing industries) via the crosswalk with weights by \cite{acemoglu2016import}. I end up with predicted U.S. exports to the world at the industry $g$ year $t$ level. We get this aggregation as $ \widehat{X^{US}_{g,t}} = \sum_{s\in g}\sum_{j} \widehat{ X^{US,j}_{s,t} }$. I do a similar work to get predicted U.S. imports from the world $ \widehat{M^{U.S.}_{g,t}}$. Third, I derive the gravity model-based instrumental variable for net export at the industry-by-year level 
\begin{equation}\label{eq:givNEP_gt}
\text{givNetExp}_{g,t}^{US}= \frac{\widehat{X^{US}_{g,t}} - \widehat{M^{US}_{g,t}} }{ Y_{g,91} }
\end{equation}
where $Y_{g,91}$ represents US domestic production in year 1991. Fourth, similar to my construction in net export measure, I use employment data from County Business Pattern to aggregate $givNEP$ at the metropolitan area across periods. 
\begin{equation}\label{eq:delta_givNEP_m}
    \triangle_{t_{1},t_{2}}\text{givNetExp}_{m} = \sum_{g} \big[  (L_{m,g,t^{\prime}_{0}}/L_{m.t^{\prime}_{0}}) * (\text{givNetExp}_{g,t_{2}} - \text{givNetExp}_{g,t_{1}}) \big]
\end{equation}
where $L_{m,g,t^{\prime}_{0}}$ and $L_{m.t^{\prime}_{0}}$ are the employment of industry $g$ and total employment, respectively, in metropolitan area $m$ in year $t^{\prime}_{0}$. Following \cite{acemoglu2016import}, I choose $t^{\prime}_{0} = t_{1}-3$ to avoid potential data error covariance between dependent variable and independent variable.

\noindent \textbf{Relevance Condition} Intuitively, the gravity model-based IV captures the exogenous part of net export growth due to (1) rising world demand reflected in net export growth by other eight high-income counties and (2) tariff changes, after controlling for U.S. industry-by-year supply shocks. This relevance condition of this GIV is satisfied since it is derived from the general equilibrium model by \cite{romalis2007nafta} and specific decomposition by \cite{feenstra2019us}. I will test this condition in data by robust F-statistics \citep{kleibergen2006generalized} and efficient F-statistics \cite{olea2013robust}. 

\noindent \textbf{Exclusion Restriction} There are exclusion restrictions at two levels. The first level refers to the gravity model-based instruments by \cite{feenstra2019us}. They have removed supply-side shocks by industry fixed-effect in predicted U.S. exports and demand-side shocks by industry fixed-effect in predicted U.S. imports. Thus, exclusion restriction holds for the gravity model-based IV. The second level refers to my regression specification Eq (\ref{eq:OLS_HP_PLNJM}), in which gravity model-based IV serves as the IV for the growth in private-label (non-jumbo) mortgages. Exclusion restriction here requires that net export growth impacts house prices only through its role in contributing to the growth of private-label (non-jumbo) mortgages. Empirically, the IV estimation of the impact of private-label (non-jumbo) mortgages on house prices does not separate demand and supply channels brought about by exogenous changes in net exports. Instead, I rely on additional tests to address the concern of demand-side (speculation) channels in \ref{sec:Empirical_Metro_AgainstSpeculation}.
 
As with all instrumental variable estimates, my 2SLS estimates represent the local average treatment effects on compliers \citep{imbens1994identification}. In my setting, compliers are metropolitan-by-year observations that experience more industry-by-year U.S. net export to the world following increases in gravity model-predicted U.S. net export.
 
%-------------------------------------------------------
%-------------------------------------------------------

%--------------------------------------------------------------------------------------
% end of the entire section 
%--------------------------------------------------------------------------------------

%----------------------------------------------------------------------

%----------------------------------------------------------------------
% section 4: Data Source

%--------------------------------------------------------------------------------------------------------------------------------------------------------------------------

%\clearpage
\section{Data Sources}\label{sec:data}
%--------------------------------------------------------------------------------------------------------------------------------------------------------------------------

I combine several data sources to study how credit expansion causes a stronger housing boom and bust in the high-net-export-growth metropolitan areas (HNEG areas). The granular international trade and tariff data are new to the literature on housing, allowing the use of the gravity model-based instrumental variable from international economics \citep{feenstra2019us} as my identification strategy. 

%--------------------------------------------------------------------------------------------------------------------------------------------------------------------------
\subsection{Data for Net Export Growth}
%---------------------------------------------------------------------------------------------------------------------------

%---------------------------------------------------------------------------------------
\noindent \textbf{Trade Flow Data} International trade flow data for 1991–2011 are from the UN Comtrade Database.\footnote{UN Comtrade Database is available here: \url{https://comtrade.un.org/data/}.} This database records bilateral imports and exports data for detailed products recorded under the six-digit Harmonized Commodity Description and Coding System (HS code). For the gravity model-based IV, I use trade data for ten countries: USA, China, and the eight other high-income countries (Australia, Denmark, Finland, Germany, Japan, New Zealand, Spain, and Switzerland) used by \cite{david2013china}. To deflate trade value to 2007 USD dollar, I use the Personal Consumption Expenditures Chain-type Price Index from Federal Reserve in St. Louis.\footnote{Personal Consumption Expenditures Chain-type Price Index from in St. Louis Federal Reserve is available at \url{https://fred.stlouisfed.org/series/PCEPI}}. To crosswalk these trade data from a six-digit HS code to a four-digit SIC code, I use the crosswalk file and revised SIC system (392 manufacturing industries) in \cite{acemoglu2016import}.\footnote{This crosswalk file is also available from Prof. David Dorn's website: \url{https://www.ddorn.net/data.htm}. The file name is "[D4] HS 1996/2002 6-digit to sic87dd 4-digit". The further refined SIC system (392 manufacturing industries) and crosswalk file are available from \cite{acemoglu2016import}.}  \footnote{To make sure my calculation is correct, I calculate China's exports to the US and eight other high-income countries at the industry-year level from 1991 to 2007 and compare them to data provided by David Dorn's website. The trade data is within the section [D] Industry Trade Exposure via his data page \url{https://www.ddorn.net/data.htm}. Correlations between my calculation and his corresponding data are 0.9983 for China's export to the U.S. and 0.9973 for China's export to eight other high-income countries.}

%---------------------------------------------------------------------------------------
\noindent \textbf{Tariff Data} I obtain Bilateral tariff schedule data at five-digit SITC products between 1984 to 2011 from \cite{feenstra2014international}.\footnote{The original data are collected from the TRAINS and IDB databases accessed via the World Bank's WITS website and various other resources, with multiple steps of cleaning and replacing missing values by other resources. The World Bank's WITS website is \url{https://wits.worldbank.org/}. The complete procedure of data work is described in Appendix C in \cite{feenstra2014international}.} To crosswalk tariff data from a five-digit SITC system to a six-digit HS system, I follow the methods in \cite{feenstra2019us}. Specifically, I first convert the HS 2007 version to the HS 2002 version using the crosswalk files from the Trade Statistics Branch (TSB) of the United Nations Statistics Division.\footnote{The crosswalk files between different HS versions are available from the UN Comtrade database: \url{https://unstats.un.org/unsd/trade/classifications/correspondence\%2Dtables.asp}.} Then I employ a crosswalk from \cite{feenstra2005world} to match each six-digit HS code to one 5-digit SITC2. In the cases that one six-digit HS code is matched to multiple SITC2 codes, I follow \cite{feenstra2019us} and employ the one with the highest value share. 

%---------------------------------------------------------------------------------------
\noindent \textbf{Manufacturing Production Data} I obtain the U.S. 4-digit SIC manufacturing industry total domestic production (vship) in 1991 from the NBER-Center for Economics Studies (NBER-CES) Manufacturing Industry Database. I use the industry total domestic production in the first year 1991 in my sample as the denominator to scale the trade value so the trade change afterward cannot respond to it.\footnote{This choice of scaling is also used by \cite{barrot2022import}.} 

%---------------------------------------------------------------------------------------
\noindent \textbf{Manufacturing Employment Data} I obtain Employment data in detailed manufacturing industries at the county-by-year level from County Business Patterns Database at U.S. Census.\footnote{County Business Patterns Database is available here: \url{https://www.census.gov/programs-surveys/cbp/data/datasets.html}.} Following \cite{acemoglu2016import}, I use manufacturing employment data to aggregate net export growth and its IV at the metropolitan areas across periods. For details, refer to Section \ref{subsec:GIV_exports}.

%--------------------------------------------------------------------------------------------------------------------------------------------------------------------------
\subsection{Data for Mortgages and House Prices}\label{subsec:Data_Mortgage_HousePrice}
%--------------------------------------------------------------------------------------------------------------------------------------------------------------------------

%---------------------------------------------------------------------------------------
\noindent \textbf{Mortgage Data} Detailed loan-level mortgage data are from the Home Mortgage Disclosure Act (HMDA) database.\footnote{The Federal Financial Institutions Examination Council (FFIEC) releases 2017-2021 HMDA data at \url{https://ffiec.cfpb.gov/data\%2Dpublication/2021}. The Consumer Financial Protection Bureau (CFPB) releases 2007-2017 HMDA data at \url{https://www.consumerfinance.gov/data\%2Dresearch/hmda/historic\%2Ddata/}. CFPB provides links to 1990-2006 HMDA contained in National Archives at \url{https://github.com/cfpb/HMDA_Data_Science_Kit/blob/master/hmda_data_links.md}. Alternatively, the University of Michigan provides 1990-2006 HMDA data at \url{https://www.openicpsr.org/openicpsr/project/151921/version/V1/view}.} In 1975, Congress enacted HMDA to improve public access to information of mortgage loans. Any financial institution must report HMDA data to its designated regulator if it meets certain criteria, such as a threshold for asset size and if the institution has a home office or branch in a Metropolitan Statistical Area (MSA).\footnote{For a depository institution that has a home office or a branch in an MSA, it must file HMDA data if it has made a home purchase loan or has refinanced a home purchase loan and has assets above a threshold set up annually by the CFPB. For a non-depository institution (e.g., a mortgage company that raises funds instead from banks or capital markets), it must file HMDA data if at least 10\% of the loan portfolio is comprised of home purchase loans and if its asset is above \$ 10 million.} This annual database contains information on loan applications (regardless of whether they are approved or denied), borrower demographics, lender identifiers, and loan specifics such as purpose, amount, and location. The HMDA database provides near-universal coverage of the mortgage market. \cite{avery20102008} conclude that commercial banks filing HMDA in 2008 carried 93\% of the total mortgage dollars outstanding on commercial bank portfolios.\footnote{Even though lenders with offices only in non-metropolitan areas are exempt from filing HMDA, 83.2\% of the population in 2006 lived in metropolitan areas \citep{dell2012credit}.}

I use the following filtering criteria for the raw HMDA data. First, I keep originated loans and drop applications that are denied, withdrawn, or not accepted for other reasons. Second, for loan types, I keep conventional and Federal Housing Administration-insured (FHA-insured) loans and drop mortgages insured by the Veterans Administration, Farm Service Agency, or Rural Housing Service. Third, for loan purposes, I use home purchase mortgages for most empirical tests and include refinancing mortgages for some robustness tests. Fourth, for occupancy types, we keep both owner-occupied and non-owner-occupied mortgages and treat ``not applicable" occupancy as owner-occupied.\footnote{According to the HMDA manual (\url{https://www.ffiec.gov/hmda/pdf/1998guide.pdf}), this ``not applicable" occupancy very likely refers to a multifamily dwelling the borrower lives in. In terms of magnitude, this ``not applicable" occupancy is only around 0.59\% of the number of owner-occupied loans and 3.5\% of the number of non-owner-occupied loans as of 2007.}

I use the HMDA database to construct loan volume (number and dollar amount) at the county-by-year level for private-label mortgages (PLMs) and government-sponsored enterprise mortgages (GSEMs). According to the HMDA examination procedures, an institution must report the type of entities that purchase the mortgages that are originated (or purchased) and then sold in the same calendar year.\footnote{The ``Home Mortgage Disclosure Act
Examination Procedures" is available at \url{https://www.federalreserve.gov/boarddocs/caletters/2009/0910/09\%2D10_attachment.pdf}. These procedures imply that the HMDA can potentially under-estimate the mortgages that are sold as GSEMs and PLNJMs since the mortgages originated near the end of a calendar year need some time for sale. Nonetheless, this potential underestimation can only bias my results to zero.} I use the classification of PLMs and GSEMs by \cite{mian2022credit}.\footnote{\cite{mian2022credit} group five categories as PLMs when a mortgage is sold: (1) into private securitization, (2) to a commercial bank, savings bank, or savings affiliation affiliate, (3) to a life insurance company, credit union, mortgage bank, or finance company, (4) to an affiliate institution, and (5) to other types of purchaser.}

%---------------------------------------------------------------------------------------
\noindent \textbf{Conforming Loan Limits Data} I obtain conforming loan limits by year and county from Federal Housing Finance Agency.\footnote{Before and in 2007, FHFA sets conforming loan limits only at the national level: \url{https://www.fhfa.gov/AboutUs/Policies/Documents/Conforming\%2DLoan\%2DLimits/loanlimitshistory07.pdf}. From 2008 onward, FHFA sets conforming loan limits by year and by county: \url{https://www.fhfa.gov/DataTools/Downloads/Pages/Conforming\%2DLoan\%2DLimit.aspx}. } Conforming loan limits are, in general, different for dwellings with different units in each year (and county after 2007). Since the HMDA data does not contain information on the number of units in a home between 1991 and 2009, I use the 1-unit conforming loan limit for all mortgages. Thus my measure of non-jumbo mortgages is potentially an underestimation of the true size, avoiding an upward bias of my results.

%---------------------------------------------------------------------------------------
\noindent \textbf{Consistent Counties Covered by HMDA} Although HMDA only mandates mortgage reports in metropolitan areas, many institutions still report all mortgages regardless of location or requirement on institutions. Thus, from the HMDA dataset, I cannot infer the required counties based on non-missing data. Instead, I need to find counties that are consistently covered by HMDA based on the scope of metropolitan areas defined and updated over time by the U.S. Office of Management and Budget. \cite{avery2007opportunities} point out that metropolitan boundaries were reasonably stable from 1996 to 2003 but changed a lot in 2004 HMDA due to the updated 2003 version of metropolitan areas. Thus, I define ``HMDA consistent counties after 1996" as counties that are covered in the following two versions of the metropolitan definition: (1) metropolitan areas definition (MSA and CMSA code) of 1999 version\footnote{Metropolitan areas definition (MSA and CMSA code) of 1999 can be found here: \url{https://www2.census.gov/programs\%2Dsurveys/metro\%2Dmicro/geographies/reference\%2Dfiles/1999/historical\%2Ddelineation\%2Dfiles/99mfips.txt}. At the moment of data work, the above link does not work. I use crosswalk from the county to the MSA in 2000 from Missouri Census Data Center at \url{https://mcdc.missouri.edu/applications/geocorr2000html}.} and (2) metropolitan areas definition (CBSA code) of 2003 version\footnote{Metropolitan areas definition (CBSA code) of 2003 can be found here: \url{https://www2.census.gov/programs\%2Dsurveys/metro\%2Dmicro/geographies/reference\%2Dfiles/2003/historical\%2Ddelineation\%2Dfiles/030606omb\%2Dnecta\%2Dcnecta.xls}.}. In addition, I require number of loans is positive in the initial year (1996) for three mortgage categories: retained in the balance sheet, GSEM and PLNJM. Consequently, ``HMDA consistent counties after 1996" contains 800 counties. Likewise, for the sample period that starts from 1991 or 1992, I define ``HMDA consistent counties after 1990" as counties that are covered in the following two sets of metropolitan areas: (1) metropolitan areas definition (MSA and CMSA code) of 1990 version\footnote{I use crosswalk from the county to the MSA in 1990 from Missouri Census Data Center at \url{https://mcdc.missouri.edu/applications/geocorr1990.html}.} and (2) the above ``HMDA consistent counties after 1996". Again, I require the number of loans in 1991 is positive for three mortgage categories: retained in the balance sheet, GSEM and PLNJM. Consequently, ``HMDA consistent counties after 1990" contains 712 counties.

%---------------------------------------------------------------------------------------
\noindent \textbf{U.S. House Price Data} The U.S. house price index data based on repeated sales at the county and ZIP levels are from the Federal Housing Finance Agency.\footnote{The data are available at \url{https://www.fhfa.gov/DataTools/Downloads}.FHFA working paper \cite{bogin2019missing} describes the construction of the index and tests its accuracy via various methods. } This database has reasonably good coverage of the US counties in metropolitan areas in and after 1991 (870 out of 1084 counties in metropolitan areas in 1991).

%---------------------------------------------------------------------------------------
\noindent \textbf{Merge House Price and Mortgage Data} For both figure and regression analysis regarding house prices, I require that the counties are covered by both house price data and mortgage data. The merged data set contains fewer counties compared to mortgage data since house price data covers fewer counties. Second, I require that each county contains both data in the period 1991-2011 (1999-2011) for the figure (regression) analysis. For more data details regarding mortgage and house prices at the ZIP level, refer to Section \ref{subsec:App_Data}.

%--------------------------------------------------------------------------------------------------------------------------------------------------------------------------
\subsection{Local Economic Conditions}
%--------------------------------------------------------------------------------------------------------------------------------------------------------------------------

%---------------------------------------------------------------------------------------
\noindent \textbf{BEA Employment Data} To test the impact of net export growth on total employment, I obtain annual employment data at the county level from the U.S. Bureau of Economic Analysis (BEA) because such data include both (1) wage and salary employment and (2) proprietor employment (self-employment).\footnote{The BEA employment data is available at \url{https://apps.bea.gov/regional/downloadzip.cfm}. Under the category ``Personal Income (State and Local)", "CAEMP25S" contains data from 1969 to 2000, and "CAEMP25N" contains data in and after 2001.} This coverage is better than County Business Pattern employment data, which does not include self-employment outside of establishments. 

%---------------------------------------------------------------------------------------
\noindent \textbf{IRS Household Income Data} To test the impact of net export growth on household income growth, I use the annual data at the county level from the Statistics of Income Division in the U.S. Internal Revenue Service (IRS).\footnote{For 1989 to 2018, the data is available at \url{https://www.irs.gov/statistics/soi\%2Dtax\%2Dstats\%2Dcounty\%2Ddata}.} This income data is based on the addresses reported on individual income tax returns filed with the IRS. I calculate average household income at the county level as the adjusted gross income divided by the number of returns (households). Then the income is adjusted to the 2007 USD by the Personal Consumption Expenditures Chain-type Price Index (PCEPI) from the Federal Reserve Bank of St. Louis.

%---------------------------------------------------------------------------------------
\noindent \textbf{Working-Age Population Data} In testing the impact of net export growth on the local population growth, I obtain the annual data on population at the county level from the U.S. Census.\footnote{The data is available here: https://www.census.gov/programs-surveys/popest.html}

%---------------------------------------------------------------------------------------
\noindent \textbf{Local Control Variables} Control variables at county and ZIP code levels are from U.S. Decennial Census Summary Files. For the year 1989, control variables at the county level are from 1990 (March) Census Summary File 1C and 3C, while control variables at the ZIP level are from Summary File 3B.\footnote{The 1990 U.S. Decennial Census File 1 database is available at \url{https://www.census.gov/data/datasets/1990/dec/summary-file-1.html} and File 3 database is available at \url{https://www.census.gov/data/datasets/1990/dec/summary-file-3.html}. } For the year 1999, control variables at the county level and the ZIP level are from both 2000 (March) Census Summary File 1 and 3.\footnote{The 2000 U.S. Decennial Census Files are available here: \url{https://www.census.gov/programs-surveys/decennial-census/guidance/2000.html}}

%--------------------------------------------------------------------------------------------------------------------------------------------------------------------------
\subsection{Twelve Counties Severely Affected by 2005 Hurricanes}\label{subsec:2005Hurricanes}
\noindent \textbf{Twelve Counties Severely Affected by 2005 Hurricanes}  I delete twelve ``deeply affected counties by 2005 Hurricanes'' since they experienced unusual growth in mortgages due to hurricane damage and subsequent government subsidies.\footnote{I try my best to present the most robust results. Since outliers only largely affect results in regression but not the illustration in figures, I include these twelve counties in the figures but remove them from regressions and summary statistics.} In 2005, three Category 5 hurricanes (Katrina, Rita, and Wilma) caused enormous fatalities and damage in Alabama, Florida, Louisiana, Mississippi, and Texas. Among them, Katrina was the most devastating natural disaster in United States history, causing 1392 fatalities, 200,000 homes destroyed \citep{deryugina2018economic}, and an estimated \$125 billion in damage in 2005 US dollars. The deeply affected counties have had unusual growth in mortgages in 2006. First, many households may choose to stay away permanently from the disaster areas. Second, lots of households may delay mortgage applications due to financial constraints caused by such hurricanes. Third, government assistance programs may replace much of the role played by PLM (and even GSEM). For example,  St. Bernard Parish in Louisiana (county FIPS code 22087) experienced a negative 66.6\% change in PLNJM between 2002 to 2006, mainly due to the Katrina hurricane. For the above reasons, I delete twelve ``deeply affected counties by 2005 Hurricanes''. I define “deeply affected” counties when more than 10 percent of occupied units were identified as major or severe damage by the Federal Emergency Management Agency as of February 2006.\footnote{ The Federal Emergency Management Agency reports the estimates of housing unit damage on Pages 15-19 of the following report at \url{ https://www.huduser.gov/publications/pdf/gulfcoast_hsngdmgest.pdf}.}. I use the percentage of major and severe damage since these damages are more likely to be reported by households and inspected by federal agencies, whereas minor damage might be under-reported. Eventually, I drop a list of the twelve most-affected counties.\footnote{These most-affected counties include Monroe County (FL, 12087), Cameron Parish (LA, 22023), Jefferson Parish (LA, 22051), Orleans Parish (LA,	22071), Plaquemines Parish (LA, 22075), St. Bernard Parish (LA,	22087), St. Tammany Parish (LA, 22103), Vermilion Parish (LA, 22113), Hancock County (MS, 28045), Harrison County (MS, 28047), Jackson County (MS, 28059), Stone County (MS, 28131).}

\subsection{Summary Statistics and Figures}
Summary statistics of key variables are reported in Table (\ref{table_SumStat}), separated into different periods. In the prior period (1991-1999), starting from 712 ``HMDA consistent counties after 1990”, I delete seven counties due to severe impact of the 2005 hurricanes described in section (\ref{subsec:Data_Mortgage_HousePrice}), resulting in 705 counties. Due to data availability, there are fewer observations for house price growth and housing supply elasticity. In the boom period (1999-2005) and bust period (2007-2009), starting from 800 ``HMDA consistent counties after 1996”, I delete eight counties due to severe impact of the 2005 hurricanes, resulting in 792 counties.  

First, we can see that the house price experienced a clear and pronounced cycle. While the mean of annualized growth rate across counties is only 1.30\% in the prior period (1991-1999), it is 3.82\% in the boom period (1999-2005) and -5.4\% in the bust period (2007-2009). 

Second, the government-sponsored enterprise mortgages (GSEM) play an less important role in the boom period: its mean annualized growth rate across counties is 17.07\% in the prior period but only 4.38\% in the boom period. 

Third, for the private-label (non-jumbo) mortgages (PLNJM), it is interesting that the mean of growth rates across counties is similar between the prior period (17.07\%) and boom period (16.98\%). The similar mean values result from the use of different base values.\footnote{To be specific, the 1991 PLNJM dollar amount is used for the growth rate between 1991 and 1999 whereas the 1999 PLNJM dollar amount is used for the growth rate between 1999 and 2005.} The similar mean growth rates mask the much larger increase in absolute dollar amount in the boom period in private-label mortgages. Figure (\ref{fig_GSEMvsPLNJM_91t19_combine}) shows the differential increase in dollar amount in PLNJM in the boom period, because time series values are scaled by 1991 dollar value only.  Figure (\ref{fig_GSEMvsPLNJM_91t19_combine}) and (\ref{fig_HPI_91to19_NEG_Intro}) hint that the impact of private-label (non-jumbo) mortgages will be much larger than the government-sponsored enterprise mortgages in the boom period since the former ones do show a differentially larger increase in the high-net-export metropolitan areas.

Fourth, the net export growth rates are a little bit more negative in the boom period (with a mean of -0.22\%  and standard deviation of 0.20\%) and in the prior period (with a mean of -0.18\%  and standard deviation of 0.17\%). This data pattern is also present in the gravity model-based net export growth rates. 

We include control variables at county $c$ only in the starting year in each period, which prevents any impact from net export growth during the period of interest. Control variables are used to neutralize factors that may affect credit expansion for reasons unrelated to the main hypothesis. Basic controls include the number of households, average household income, and the fraction of the labor force at the county $c$. Housing controls include the number of housing units, housing supply elasticity \citep{saiz2010geographic}, Wharton residential land use regulatory index \citep{gyourko2008new}, the house vacancy rate, and the fraction of renters in occupied housing units. Demographic controls include the fraction of the population (1) holding a Bachelor's degree or above, (2) being white in race, (3) of immigrants entering the U.S. between 1990 and 2000, and (4) of age 65 or above. 
%----------------------------------------------------------------------

%----------------------------------------------------------------------
% section 5: Empirical Results

%------------------------------------------------------------
%------------------------------------------------------------
%\clearpage
%------------------------------------------------------------
%------------------------------------------------------------
\section{Cross-Metro Empirical Causal Evidence: Supporting Credit Expansion}\label{sec:Evidence_Metro_CreditExpansion}
My main empirical tests in this section provide direct evidence that credit expansion in private-label mortgages (PLMs) rather than government-sponsored enterprise mortgages (GSEMs) causes the 1999-2011 housing price boom and bust. To compare these two types of mortgages under the same criteria, I use the conforming loan limits to get the non-jumbo category of private-label mortgages (PLNJMs) and I focus on the comparison between private-label (non-jumbo) mortgages (PLNJMs) and the government-sponsored enterprise mortgages (GSEMs).\footnote{Jumbo loans of private-label mortgages are smaller in number and dollar amount when compared to non-jumbo loans. Including jumbo ones only strengthens my results.} 1999 is the year when the Copula formula was invented \citep{li1999default} for securitization, which eventually led to the boom in private-label mortgage-backed securities \citep{salmon2009recipe,donnelly2010devil}.\footnote{Later, industry practitioners term the Copula model as ``the model kills Wall Street"\citep{salmon2009recipe}.} 

\comment{
My hypotheses in three parts are based on (1) the legal constraint that government-sponsored enterprise mortgages cannot consider local economic conditions (net export growth in my study) and (2) documented credit expansion in private-label mortgages (PLMs) by \cite{justiniano2022mortgage}. They use loan-level data and a regression model to construct the conditional spread between private-label mortgages (PLMs) and 10-year treasury yield after controlling the characteristics of the mortgages. They identify a sharp and persistent decrease in this conditional spread in the summer of 2003, which shows a sudden and persistent increase in credit supply in private-label mortgages. Based on this result by \cite{justiniano2022mortgage}, I interpret the private-label mortgage growth as credit expansion. I predict that when the funding cost declines, private-label (non-jumbo) mortgages expand more in the high-net-export-growth metropolitan areas in response to higher net export growth. 
}

Based on the intuition of the model described in the introduction, I conduct three tests for my set of results. My first test shows that, induced by net export growth, credit expansion in (non-jumbo) private-label mortgages (PLNJMs) cause the 1999-2009 housing price boom and bust across metropolitan areas in the USA. My second test shows that government-sponsored enterprise mortgages (GSEMs) do not respond to net export growth, and thus are unrelated to differential house price booms and busts across metropolitan areas. The second test can also help argue against the demand (or speculation) view that mortgage growth can be driven by demand (or speculation) induced by net export growth. Using the fact that there is no aggregate credit expansion during 1991-1999, my third test provides evidence supporting the exclusion restriction of my IV strategy: net export growth impacts house price growth only through private-label mortgages.

\comment{
We define the boom period as 1999-2005 (defined as the end of 1999 to the end of 2005) for two reasons. First, it includes the major housing market rise between 2002-2005, which is commonly used in the literature (see \cite{griffin2021drove}) for a review). Second, 1999 is one year before the formal publication of copula approach \citep{li2000copula}, a major financial innovation in the securitization of mortgages and debts.\footnote{\cite{salmon2009recipe} reports in the Wired Magazine and called the Copula by \cite{li2000copula} as the ``the formula that killed Wall Street". \cite{jones2009formula} reports in the Financial Times and called it ``The formula that felled Wall St".}
}

%------------------------------------------------------------
%------------------------------------------------------------
%------------------------------------------------------------
\subsection{House Price Boom (99-05) and Bust (07-09)}\label{subsec:causal_house_price_boom_bust}
%------------------------------------------------------------
%------------------------------------------------------------
%------------------------------------------------------------

Let us first focus on my main hypothesis: house prices as a result of credit expansion in response to net export growth across metropolitan areas. First, I expect private-label (non-jumbo) mortgages would experience stronger growth in the high-net-export-growth metropolitan areas during the boom period 1999-2005. Second, I expect that the drop in house prices will also be stronger in the high-net-export-growth metropolitan areas between 2007 and 2009.\footnote{The Great Recession period, 2007-2009, is defined by the US National Bureau of Economic Research at \url{https://www.nber.org/research/data/us-business-cycle-expansions-and-contractions}}

%\noindent \textbf{Hypothesis 1} In cross-section, growth in private-label (non-jumbo) mortgages (PLNJMs) in the boom period (1999-2005) causes the house price boom (1999-2005) and bust (2007-2009).

\noindent \textbf{House Price Growth in Boom (99-05)} To test the house price boom, I perform a regression with the following specifications. 
\begin{equation}\label{eq:HPI_reg_PLNJM}
    \triangle_{99,05} Ln(HPI_{c}) = \beta * \triangle_{99,05} Ln(PLNJM_{c}) + \gamma* \bm{Controls_{c}} + \epsilon_{c}
\end{equation}
The dependent variable $\triangle_{99,05} Ln(HPI_{c})$ is the growth rate of the house price index in county $c$ 99-05, and the key variable of interest $\triangle_{99,05} Ln(PLNJM_{c})$ is the growth rate of the dollar amount of the private-label (non-jumbo) mortgages (PLNJM) in county $c$ 99-05. $Controls_{c}$ indicates control variables at county $c$ in 1999. I use the gravity model-based instrument ($\triangle_{99,05}\text{givNetExp}_{m}$) as IV for $\triangle_{99,05} Ln(PLNJM_{c})$.\footnote{Compared to the HMDA sample starting from 1991, I have an increased sample size here due to the larger coverage of metropolitan definition after 1996. Another slight difference in the sample size arises from the data availability in house prices.}

Each regression is weighted by the natural logarithm of the number of house units in 1999. Logarithm rather than the absolute number of house units is chosen to prevent results from being dominated by a few super-populous counties. To take into account that households might commute to work across counties within a metropolitan area, I measure net export growth in the metropolitan area.\footnote{According to US Census, ``the general concept of a metropolitan statistical area is that of a core area containing a substantial population nucleus, together with adjacent communities having a high degree of economic and social integration with that core." (\url{https://www.census.gov/programs-surveys/metro-micro/about.html})} Furthermore, I cluster standard errors at the metropolitan area (MA) level.

Table (\ref{table_HPI.D99t05.PLNJM.4Reg}) reports OLS, reduced-form, second-stage, and first-stage results in panel A, B, C, and D, respectively.  First, panel A column (1) shows the positive and significant impact of PLNJM growth (99-05, annualized) on house price index growth (99-05, annualized) without any control. Through columns (2) to (4), the estimates show that the OLS coefficients of PLNJM are of the similar magnitude as more controls are added . Panel B reports the reduced-form estimates that share a similar pattern. The first-stage estimates in panel D are positive and significant.

Panel C reports 2SLS estimates for equation (\ref{eq:HPI_reg_PLNJM}).  Like reduced-form estimates, the 2SLS estimates are statistically significant at a one percent level and quite stable across various specifications. As for the weak IV concern, the clustered Kleibergen-Paap F-statistic \citep{kleibergen2006generalized} and the Montiel Olea-Pflueger Efficient F-Statistic \citep{olea2013robust} are both 13.06, which are larger than 10. Thus, my estimates are highly unlikely to be biased by a weak instrumental variable. As emphasized by \cite{jiang2017have}, it is important to compare the 2SLS estimates with OLS estimates to alleviate the ``blown-up" concern that statistical significance may pick up a weak IV that biases the estimate toward a large magnitude. Besides two large F-statistics, I also confirm that 2SLS estimates are only about three to four times the OLS estimates in my baseline specification in column (4). Thus, my estimates are very unlikely to be subject to the ``blow-up" concern of a weak IV.

Let us focus on economic meaning in 2SLS with controls in Table (\ref{table_HPI.D99t05.PLNJM.2SLS}). According to column (5) with all controls, one standard deviation in cross-sectional difference in annualized PLNJM growth results in $8.12\% \times 0.479 = 3.89\%$ difference in annualized house price index (HPI) growth across metropolitan areas, translating into $23.34\%$ difference from 1999 to 2005. One standard deviation in cross-section difference in annualized HPI growth is $3.55\%$, translating into $21.30\%$ from 1999 to 2005. The two results mean that, from 1999 to 2005, one standard deviation in PLNJM growth can explain $23.34\% / 21.30\% = 109.60\%$ of one standard deviation in HPI growth. 

Similarly, one standard deviation in cross-section housing supply elasticity (HSE) can explain $1.25 * 0.002 = 0.25\%$ difference in annualized HPI growth, translating into $1.49\%$ difference from 1999 to 2005. This result implies that, from 1999 to 2005, one standard deviation in HSE can explain $1.49\% / 21.30\% = 7.02\%$ of one standard deviation in HPI growth. Therefore, growth in private-label (non-jumbo) mortgages exhibits much higher explanatory power than housing supply elasticity \citep{saiz2010geographic}.

%------------------------------------------------------------
%------------------------------------------------------------
%------------------------------------------------------------
%------------------------------------------------------------
\textbf{House Price Boom and Bust}
We then compare the house price boom and bust in a single regression test. My stacked 2SLS regression is 
\begin{equation}\label{eq:HPIBoomBustonPLNJM}
\resizebox{0.92\textwidth}{!}{$
\begin{aligned}
\triangle_{99,05} \& \triangle_{07,09} Ln(HPI_{c}) & = \beta_{99,05} * \triangle_{99,05} Ln(PLNJM_{c}) \times Dum_{99,05} + \beta_{07,09} * \triangle_{99,05} Ln(PLNJM_{c}) \times Dum_{07,09} \\
& + \gamma_{99,05}* \bm{Controls_{c}} \times Dum_{99,05} + \gamma_{07,09}* \bm{Controls_{c}} \times Dum_{07,09} + \epsilon_{period, c}
\end{aligned}
$} %end of \resizebox
\end{equation}
Controls, weight, and standard errors are the same as Eq(\ref{eq:HPI_reg_PLNJM}). 

Table (\ref{table_HPI.D99t05vsD07t09.PLNJM.4Reg}) reports OLS, reduced-form, second stage, and the first stage of the stacked regression of house price index in the boom period (99-05) and the bust period (07-09) in panel A, B, C, and D.\footnote{Please note that I require counties to be consistently covered by HMDA and HPI databases since 1999. Because of the larger inclusion of metropolitan areas in 1999 than in 1991, my sample size during 1999-2005 period is larger than the one in 1991-1999 period. In addition, due to the availability of housing supply elasticity, columns (3) and (4) have smaller sample sizes.} First, panel A shows that the OLS coefficients of PLNJM growth (99-05) are positive in the boom period (99-05) but negative in the bust period (07-09). The same trend applies to reduced-form estimates in panel B and 2SLS estimates in panel C. Importantly, the coefficient magnitudes in 2SLS are between three and four times those of OLS results in baseline specification in column (3), much lower than the average of nine in the top three finance journals. Same as the Table (\ref{table_HPI.D99t05.PLNJM.4Reg}), first-stage estimates in panel D show the stable and strong positive correlation between the PLNJM and gravity model-based IV for Net Export Growth (GIV-NEG), with a large enough first-stage F-statistic. Therefore, the 2SLS do not have weak IV concerns. Let us turn our attention to the coefficient equality test of PLNJM growth in the boom and bust periods in 2SLS in the Table (\ref{table_HPI.D99t05vsD07t09.PLNJM.2SLS.wide}). For all specifications through columns (1)-(4), the chi-square statistics are large, and p-values are below 0.01, meaning the two coefficients are statistically different. 

In terms of economic meaning, one standard deviation in cross-sectional difference in annualized PLNJM growth (99-05) results in $8.01\% \times -0.886 = -7.12\%$ difference in the annualized house price drop (07-09). Since the lengths of the boom and bust periods are different, I need to consider the time horizon: a longer period of credit expansion results in a shorter period of bust period. One standard deviation in cross-sectional difference in six-year PLNJM growth (99-05) can result in $7.12\% \times 2 = 14.24\%$ cross-sectional difference in the two-year house price drop. For the annualized HPI drop, one standard deviation in cross-section difference is $6.18\%$, translating into $6.18\% \times 2 = 12.36\%$ from 2007 to 2009. The two results mean that one standard deviation in cross-sectional difference in six-year PLNJM growth (99-05) can explain $14.24\% / 12.36\% = 115.22\%$ of one standard deviation in two-year HPI drop (07-09). To sum up, six-year PLNJM growth can explain $109.60\%$ house price growth 1999-2005 and $115.22\%$ house price drop 2007-2009.

However, my specification in column (4) shows that the housing supply elasticity cannot impact house price drop after controlling all other factors. In terms of economic magnitude, one standard deviation in cross-section housing supply elasticity (HSE) can explain $1.246 * 0.001\%= 0.122\%$ difference in annualized HPI drop (07-09), translating into $0.243\%$ difference from 2007 to 2009. This figure implies that one standard deviation in HSE can explain $0.243\% / 12.36\% = 1.97\%$ of one standard deviation in house price drop from 2007 to 2009. This explanatory ratio of $1.97\%$ in the bust period 2007-2009 is much smaller than the ratio of $7.02\%$ in the boom period 1999-2005.

In summary, empirical tests in this subsection support the main conclusion that credit expansion in private-label mortgages induced by net export growth causes the 1999-2009 U.S. house price boom and bust across metropolitan areas. Further, in Section \ref{subsec:Empirical.Robustness}, I will show that this main conclusion is robust to state-level differences in anti-predatory lending laws \citep{di2017credit}, recourse laws \cite{ghent2011recourse}, judicial requirement in foreclosure \cite{mian2015foreclosures}, categorization of sand states \cite{choi2016sand}, and state capital gain tax \cite{gao2020economic}.

%------------------------------------------------------------
%------------------------------------------------------------
\subsection{Tests Supporting Exclusion Restriction} 
%------------------------------------------------------------
%------------------------------------------------------------

In this subsection, I design three tests to verify the exclusion restriction indirectly. 

First, the mortgage market setting has unique features that allow me to test it indirectly. In particular, net export growth can potentially cause an increase in mortgages only in two ways: (1) the credit expansion channel by lenders, and/or (2) the demand channel by borrowers. Since government-sponsored enterprise mortgages (GSEMs) are cheaper due to implicit government guarantees against credit risk and their large scale, middle-class household demand (particularly the first home for living) shall show up in GSEMs first. Lenders would also prefer to initiate a GSEM over a private-label (non-jumbo) mortgages as lenders can transfer credit risk completely and do not worry about resale or securitization.\footnote{In the secondary mortgage market, credit risk of mortgages cannot easily be completely transferred from lenders to investors of mortgage-backed securities. For example, \cite{acharya2013securitization} finds that many lenders as issuers of private-label mortgage-backed securities still provide guarantees against credit risk.} Therefore, if net export growth increases the demand of the middle class, the demand shall first show up in GSEMs. In my first test, I demonstrate that net export growth does not cause government-sponsored enterprise mortgage growth during the boom period (1999-2005). This test addresses the speculation (demand) view because demand does not appear in this cheaper mortgage type (GSEM). 

In my second test, I design a placebo test of the demand channel by using the fact that there is no significant credit expansion in private-label (non-jumbo) mortgages in the prior period (19991-1999). If net export growth were able to increase mortgage demand anyway in the boom period (1999-2005), then net export growth would increase mortgage demand and house prices in the prior period (1991-1999) as well. However, my test shows that, in a prior period (1991-1999) without substantial credit expansion, net export growth does not cause growth in private-label (non-jumbo) mortgages via a demand channel or house prices via cash purchase. Put differently, net export growth cannot directly cause house price growth via the demand channel without credit expansion in mortgages.

In my third test, I create a redundant IV (the square of net export growth) and take advantage of over-identification test to show that two IVs (net export growth and its square) are jointly correctly specified, meaning they are not correlated with the structural errors. I show that the square of net export growth is redundant in the sense that it does not provide additional information: the 2SLS estimates with two IVS are almost identical to the one with one IV (net export growth). Put together, this over-identification test provides suggestive evidence that my main specification with one IV is correctly specified.

%------------------------------------------------------------
%------------------------------------------------------------
\subsubsection{Private vs. Government Mortgages in Boom Period}\label{sec:PLNJM_vs_GSEM}
%------------------------------------------------------------
%------------------------------------------------------------

\comment{
\noindent \textbf{Hypothesis 1} In the boom period 1999-2005, credit expansion in private-label (non-jumbo) mortgages (PLNJM) rather than government-sponsored enterprise mortgages (GSEM) is stronger in counties that experience stronger net export growth. 
}

Motivated by the legal constraint that government-sponsored enterprise mortgages (GSEMs) cannot consider local economic conditions in setting up mortgage rates \citep{hurst2016regional}, I expect that, under my pure credit expansion theory, GSEMs do not respond significantly to net export growth and thus are not statistically related to the differential housing boom and bust across metropolitan areas. \comment{However, due to peer effect from neighbors who use credit expansion in private-label (non-jumbo) mortgages (PLNJM) for housing consumption, credit-qualified households may use government-sponsored enterprise mortgages (GSEMs) for their house consumption. In such cases, ignoring GSEMs can overestimate the impact of PLNJM. This section addresses such concern by showing the irrelevance of GSEMs to the housing cycle. Here, I distinguished two effects of net export growth: realized effects vs. unrealized but predicted effects in the future. Realized effects are realized at the household level, including income growth, credit score increase, etc. These realized effects at the household level can be considered by GSEMs. I argue such realized effects are still small in the sample periods, so they cannot be picked up in the regression analysis. In a similar spirit, my argument against demand does not mean that demand does not play any role. In fact, the realized effects above (caused by net export growth) can drive up housing demand in GSEMs, which seems not to show up in the regression. Overall, I only mean that credit expansion plays the dominant role. In contrast, unrealized but predicted effects in the future are forward-looking differences at the metropolitan level, including future average income growth, average credit score increase, average house price growth, etc. These unrealized but predicted effects associated with metropolitan areas can only be considered by PLMs, not GSEMs.}

Verifying the irrelevance of GSEMs can help argue against the demand view of mortgage growth. Since government-sponsored enterprise mortgages (GSEMs) are still cheaper than corresponding non-jumbo private-label mortgages (PLNJMs) for credit-qualified households, mortgage demand driven by net export growth would show up in both GSEMs and PLNJMs. That is to say, if the demand (speculation) channel exists, we would see higher mortgage growth in high-net-export-growth metros for both GSEMs and PLNJMs. However, in the empirical tests, we only see such a trend for PLNJMs rather than GSEMs. Therefore, such empirical evidence helps argue against the above demand (speculation) view.

%------------------------------------------------------------
%------------------------------------------------------------
\noindent \textbf{Private vs. Government Mortgage Growth in Boom Period (99-05)} Recall that I predict that GSEMs will not expand more in the boom period because GSEMs do not consider differences in local economic growth caused by exogenous net export growth. I use the following stacked regression specification: 
%\vspace{-2mm}
\begin{equation}\label{equ:reg_GSEM_vs_PLNJM_on_NetExp}
\resizebox{0.85\textwidth}{!}{$
    \begin{aligned}
        \triangle_{99,05} Ln(PLNJM_{c}) \quad  \text{stacked with} \quad  \triangle_{99,05} Ln(GSEM_{c}) & = \beta_{G} * \triangle_{99,05} \text{NetExp}_{m} \times Dum_{G} + \beta_{P} * \triangle_{99,05} \text{NetExp}_{m} \times Dum_{P} \\
        & + \gamma_{G}* \bm{Controls_{c}} \times Dum_{G} + \gamma_{P}* \bm{Controls_{c}} \times Dum_{P}
    \end{aligned}
$} %end of \resizebox    
\end{equation}

Table (\ref{table_GSEMvsPLNJM.D99t05.4Reg}) reports OLS, reduced-form, second stage, and the first stage of the stacked regression of GSEM and PLNJM in panels A, B, C, and D. Let us first focus on the results of private-label (non-jumbo) mortgages. First, panel A column (1) shows the positive and significant impact of net export growth (99-05, annualized) on the growth of private-label (non-jumbo) mortgages (PLNJM) (99-05, annualized), without any control. Through columns (2) to (4), the estimates show that the OLS coefficient of the next export growth is almost unaffected by the inclusion of the controls. Panel B shows that the reduced-form coefficients are significant and stable to the inclusion of various controls. Panel D shows that the first-stage coefficients are quite stable across various specifications. In column (4) with all control variables, the first-stage clustered Kleibergen-Paap F-statistic and the Montiel Olea-Pflueger Efficient F-statistic are 18.39. Thus, it is very unlikely that my estimates are biased by a weak instrument. Like reduced-form estimates, the 2SLS estimates in Panel C are statistically significant at a one percent level and quite stable across various specifications. Besides my two F-statistics, I also confirm that 2SLS is close to OLS estimates throughout various specifications, where my baseline results in column (4) indicate a ratio of $12.851/8.237=1.560$.

Let us focus on the economic meaning in 2SLS with controls in Table (\ref{table_GSEMvsPLNJM.D99t05.2SLS.wide}). According to column (4) with all controls, one standard deviation in cross-sectional difference in annualized net export growth results in $0.200\% \times 12.851 = 2.57\%$ difference in annualized private-label (non-jumbo) mortgages (PLNJM) growth, translating into $15.45\%$ difference at the county level from 1999 to 2005. One standard deviation in cross-section difference in annualized PLNJM growth is $8.14\%$, translating into $53.57\%$ from 1999 to 2005. A quick calculation means that, from 1999 to 2005, one standard deviation in net export growth can explain $15.45\% / 53.57\% = 28.84\%$ of one standard deviation in PLNJM growth. Likewise, one standard deviation in cross-section housing supply elasticity (HSE) can explain $1.245 * 0.004 = 0.50\%$ difference in annualized PLNJM growth, translating into $2.99\%$ difference at the county level from 1999 to 2005. A quick calculation implies that, from 1999 to 2005, one standard deviation in HSE can explain $2.99\% / 53.57\% = 5.58\%$ of one standard deviation in PLNJM growth. In comparison, the explanatory power of net export growth ($28.84\%$) is more than five times housing supply elasticity ($5.58\%$), a major factor documented in the literature \citep{saiz2010geographic, mian2011house}.

Next, let us focus on the results of government-sponsored enterprise mortgages (GSEMs). First, it is worth noting that the OLS coefficients of net export growth on GSEM growth are significant. In sharp contrast, their reduced-form counterparts are statistically insignificant across various specifications in panel B. This contrast highlights the necessity of instrumental variables since OLS estimates might pick up some non-exogenous part of net export growth or the expectation of the long-term trend. For example, due to the long-term trend of supply-side technology advances unique to US industries beyond eight other high-income countries, the booming local industries expand both exports and employment. Since such technological advances result in employment growth and income growth at the household level that can be considered by GSEMs, demand shows up in GSEMs' growth. In sharp contrast, my testing boom period is relatively short, and the exogenous change in net export may not precisely follow the above long-term trend of supply-side advances. Due to legal constraints \citep{hurst2016regional}, GSEMs cannot consider small differences across geographic areas caused by exogenous net export growth that are not reflected at the household level. My reduced-form estimates are consistent with GSEMs ignoring such differences.

Consistent with the above, the 2SLS estimates of the interaction term between net export growth and GSEM growth are mostly insignificant, especially when more controls are added. Let us turn our attention to the coefficient equality test of two interaction terms (net export growth interacted with Dum\_PLNJM and Dum\_GSEM) in 2SLS in Table (\ref{table_GSEMvsPLNJM.D99t05.2SLS.wide}). For all specifications, the chi-square statistics are large, and p-values are below 0.05, meaning the two coefficients are statistically different. In addition, the insignificant and close-to-zero coefficient of the interaction term housing supply elasticity $\times$ Dum\_GSEM confirms that GSEMs do not consider differences in land availability for housing across counties in metropolitan areas. 

In summary, my empirical evidence in this subsection confirms that exogenous net export growth does not cause growth in government-sponsored enterprise mortgages (GSEMs), which shall pick up the household demand (speculation) first due to lower rates. Put differently, the non-results of GSEMs argue against the demand (speculation) view of the housing cycle.

%------------------------------------------------------------
%------------------------------------------------------------
\subsubsection{Fully Controlling Demand via GSEM in Boom Period} 
%------------------------------------------------------------
%------------------------------------------------------------

Although the empirical test above shows that net export growth (1999-2005) has not significantly affected GSEMs, such an effect may be small and insignificant. After all, the net export growth can increase income, which ultimately increases the creditworthiness of households applying for GSEM. In this subsection, I define differential PLNJM-GSEM growth as the difference between PLNJM growth and GSEM growth during the boom period (1999-2005) and argue that it measures the differential credit expansion in PLNJM after controlling for GSEM demand from middle-class households who are eligible for GSEMs. Then I use the IV strategy to show that net export growth increases house price growth via the differential credit expansion in PLNJM. This test specification is as follows: 
\begin{equation}\label{eq:HPI_reg_PLNJM_GSEM}
    \triangle_{99,05} Ln(HPI_{c}) = \beta * (\triangle_{99,05} Ln(PLNJM_{c}) - \triangle_{99,05} Ln(GSEM_{c}) ) + \gamma* \bm{Controls_{c}} + \epsilon_{c}
\end{equation}

Table (\ref{table_HPI.D99t05.PLNJM_m_GSEM.4Reg}) reports OLS, reduced-form, second-stage, and first-stage results in panel A, B, C, and D, respectively.  First, panel A shows the OLS estimates of the differential growth rate (99-05, annualized) on house price growth (99-05, annualized). Through columns (2) to (4), the OLS coefficients are all postive and significant at 1\% level. A similar pattern appears in the reduced-form estimates in Panel B and 2SLS estimates in Panel C. The first-stage estimates in panel D are positive and significant. In column (4), the Cragg–Donald F-statistic is 10.41, and the clustered Kleibergen-Paap F-statistic and the Montiel Olea-Pflueger Efficient F-statistic are both 8.05. Thus, my estimates are highly unlikely to be biased by a weak instrumental variable. 

Let us focus on the economic meaning in the 2SLS estimate in column (5) with all controls in Table (\ref{table_HPI.D99t05.PLNJM_m_GSEM.2SLS}). One standard deviation in cross-sectional difference in annualized differential growth results in $9.774\% \times 0.712 = 6.96\%$ difference in annualized house price index (HPI) growth across metropolitan areas, translating into $41.76\%$ difference from 1999 to 2005. One standard deviation in cross-section difference in annualized HPI growth is $3.549\%$, translating into $21.30\%$ from 1999 to 2005. The two results mean that, from 1999 to 2005, one standard deviation in differential growth can explain $41.76\% / 21.30\% = 196.07\%$ of one standard deviation in HPI growth. Similarly, one standard deviation in cross-section housing supply elasticity (HSE) can explain $1.246 * 0.003 = 0.374\%$ difference in annualized HPI growth, translating into $2.24\%$ difference from 1999 to 2005. This result implies that, from 1999 to 2005, one standard deviation in HSE can explain $2.24\% / 21.30\% = 10.53\%$ of one standard deviation in HPI growth. Therefore, growth in private-label (non-jumbo) mortgages exhibits much higher explanatory power than housing supply elasticity \citep{saiz2010geographic}.

%------------------------------------------------------------
%------------------------------------------------------------
\subsubsection{Prior Period vs. Boom Period}\label{sec:ExclusionRestriction}
%------------------------------------------------------------
%------------------------------------------------------------

This subsection provides evidence supporting the exclusion restriction of our IV approach: net export growth impacts house prices only via its impact on PLNJMs. This subsection addresses the concern that net export growth directly drives stronger house price growth in high-net-export-growth metropolitan areas, or indirectly contributes to house price increases through a different channel. In the above two situations, the IV strategy can overestimate the impact of PLNJM growth. 

My test focuses on the prior period (1991-1999) when there is no credit expansion, and the funding cost of private-label (non-jumbo) mortgages (PLNJMs) is still high. In the prior period, two factors prevent PLNJMs from expanding more in the high-net-export-growth metropolitan areas: (1) securitization innovation, particularly the Copula formula \citep{li2000copula, salmon2012formula}, was developed in 1999, and (2) government-sponsored enterprise mortgages (GSEMs) dominate the mortgage markets because of low rates due to large scale and implicit government insurance \citep{sherlund2008jumbo}. Therefore, the prior period is ideal for a placebo test on the exclusion restriction of IV: net export growth impacts house prices only via its impact on PLNJMs. In other words, I can show that, in the prior period without credit expansion, net export growth did not cause higher growth in PLNJMs in the high-net-export-growth metropolitan areas. Consequently, it did not increase house prices in these areas. I choose a long period from 1991 to 1999 since I want to include two events: (1) the North American Free Trade Agreement enacted in 1994 and (2) the World Trade Organization established in 1995. These two events make the prior period an ideal setting for a placebo test: there is a substantial divergence in net export growth across metropolitan areas, but no credit expansion at the aggregate level. My argument against the demand (speculation) view does not mean demand does not play any role here. Theoretically, net export growth can have a positive effect on household income and credit scores, which drives up the demand for private-label (non-jumbo) mortgages and ultimately house prices. However, this demand channel does not appear to be significant in the empirical test. Overall, I mean that net export growth does not significantly cause house price growth. 

\comment{\noindent \textbf{Hypothesis 3} In cross-section, net export growth causes house price growth in the boom period (99-05) because it causes growth in private-label (non-jumbo) mortgages (PLNJM). Conversely, net export growth does not cause house price growth in the prior period (91-99) because it does not cause growth in private-label (non-jumbo) mortgages (PLNJM).}

In the first step, I test the relationship between net export growth and house price growth in both the prior (1991-1999) and boom (1999-2005) periods. Table (\ref{table_HPI.D91t99vsD99t05.4Reg}) reports OLS, reduced-form, second stage, and first stage of the stacked regression in panel A, B, C, and D, respectively. Regression specifications are similar to equation (\ref{eq:HPIBoomBustonPLNJM}) except that (1) the dependent variable stacks house price growth in two periods, and (2) the independent variables of interest are net export growth that interacts with dummy variables for two periods.\footnote{Because of the smaller size of metropolitan counties and the reduced availability of the housing price index in the early years, my sample size is smaller in the prior period (1991-1999). Due to data availability of housing supply elasticity, columns (3) and (4) have fewer observations.}\footnote{In the prior period, I winsorize the net export growth and its GIV at the 3\% and 97\%, since a few outliers (based on GIV net export growth) can make the coefficient of net export growth negative and significant. Even though the negative and significant coefficient does not violate my conclusion, I winterize the data to show the results for most observations. In the Appendix Section \ref{subsec:App_EmpCreditExpansion}, I will show more details.} First, panel A shows that the OLS coefficients of net export growth (99-05, annualized) are positive and statistically significant at a one percent level. However, the OLS estimates for net export growth (91-99, annualized) are insignificant. Similar patterns show up in reduced-form estimates in panel B and 2SLS estimates in panel C. First-stage estimates in panel D show the strong positive correlation between the net export growth and its gravity model-based instrumental variable is quite stable for both periods. In column (4) with all control variables, the first-stage clustered Kleibergen-Paap F-statistic and the Montiel Olea-Pflueger Efficient F-statistic are both 28.53 in the boom period (1999-2005) and 18.32 in the prior period (1991-1999). Thus, my estimates are very unlikely to be biased by a weak instrumental variable. Let us turn our attention to the coefficient equality test of net export growth in prior and boom periods in 2SLS in Table (\ref{table_HPI.D91t99vsD99t05.2SLS.wide}). In the prior period (91-99), as more controls are added, the coefficients for net export growth are insignificant in columns (3) and (4). The chi-square statistics are large, and p-values are below 0.01 for all specifications, meaning the two coefficients are statistically unequal. Taken together, the results of the above two tables verify that house price growth between 1991 and 1999 is not affected by net export growth directly (via cash purchase) or indirectly by other channels (including mortgages). 

As a second step, I test the relationship between net export growth and growth in private-label (non-jumbo) mortgages (PLNJM) in both the prior (1991-1999) and boom (1999-2005) periods. Table (\ref{table_PLNJM.D91t99vsD99t05.4Reg}) reports OLS, reduced-form, second stage, and first stage of the stacked regression of PLNJM in panel A, B, C, and D, respectively. \footnote{In the prior period, I drop four outliers, since these four outliers can make the coefficient of net export growth negative and significant. Even though the negative and significant coefficient does not violate my conclusion, I drop these outliers to to show the general results for most observations. In the Appendix Section \ref{subsec:App_EmpCreditExpansion}, I will show more details.} First, panel A shows that the OLS coefficients of net export growth (99-05, annualized) are positive and statistically significant at a one percent level and are stable across specifications. However, the OLS estimates for net export growth (91-99, annualized) are generally insignificant. Similar patterns apply to reduced-form estimates in panel B and 2SLS estimates in panel C. First-stage estimates in panel D show that the strong positive correlation between the net export growth and its gravity model-based instrumental variable is quite stable for both periods. In column (4) with all control variables, the first-stage clustered Kleibergen-Paap F-statistic and Montiel Olea-Pflueger Efficient F-statistic are 21.58 in the period period and 15.70 in the boom period. Thus, my estimates are very unlikely to be biased by a weak instrumental variable. Let us turn our attention to the coefficient equality test of net export growth in prior and boom periods in 2SLS in Table (\ref{table_PLNJM.D91t99vsD99t05.2SLS.wide}). For specifications in columns (1)-(4), the chi-square statistics are large, and p-values are below 0.01, meaning the two coefficients are statistically different. Taken together, results from the above two tables verify that PLNJM, between 1991 and 1999, is not affected by net export growth. It is also interesting that the specification in column (4) shows that PLNJM growth between 1991 and 1999 is unaffected by housing supply elasticity.

In summary, my empirical evidence in this subsection confirms that in the prior period (1991-1999), without credit expansion, net export growth does not cause growth in PLNJM or house prices via the demand channel, providing support for the exclusion restriction in my IV strategy. That is, net export growth can significantly affect house prices only through its effect on private-label (non-jumbo) mortgages.

%------------------------------------------------------------
%------------------------------------------------------------
\subsubsection{Over-identification Test} 
%------------------------------------------------------------
%------------------------------------------------------------

So far, I have not been able to test whether the regression specification in Equation (\ref{eq:HPI_reg_PLNJM}) is correct or not by an over-identification test (Hansen J test), primarily because I have only one instrument variable. In this subsection, I use the gravity model-based instrument ($\triangle_{99,05}\text{givNetExp}_{m}$) and its square as two separate instrumental variables for the growth rate of private-label mortgages in Equation (\ref{eq:HPI_reg_PLNJM}) so that I can conduct an over-identification test. This test can provide suggestive evidence of whether the specification in Equation (\ref{eq:HPI_reg_PLNJM}) is correctly specified in a statistical sense. 

Table (\ref{table_HPI.D99t05.PLNJM.4Reg.overID}) reports OLS, reduced-form, second-stage, and first-stage results in panels A, B, C, and D, respectively. In all four columns in panel D, the first-stage coefficient estimates are insignificant for the squared IV and F-statistics are not large enough, both of which indicate weak IV concern. However, 2SLS coefficient estimates in Panel C are significantly positive and have magnitudes (0.514) almost identical to the 2SLS estimates (0.479) in Table (\ref{table_HPI.D99t05.PLNJM.4Reg}) with only one IV. In addition, in column (1) of first-stage regression and column (5) of 2SLS regression, coefficients and standard errors of control variables in Table (\ref{table_HPI.D99t05.PLNJM.2SLS.overID}) with two IVs are almost identical to Table (\ref{table_HPI.D99t05.PLNJM.2SLS.overID}) with only one IV. The above results together suggest that the squared IV (square of gravity model-based instrument ($\triangle_{99,05}\text{givNetExp}_{m}^2$) ) does not provide additional information because the new 2SLS coefficient estimates are almost identical to the original ones. In other words, the square of gravity model-based instrument is redundant. I take advantage of this redundant IV to perform the over-identification test. Hansen J statistics are small enough, and p-values are large enough, especially in column (5) with full controls in Table (\ref{table_HPI.D99t05.PLNJM.2SLS.overID}) (0.545). These results suggest that my original specification with one instrumental variable in Equation (\ref{eq:HPI_reg_PLNJM}) is correctly specified in a statistical sense.

%------------------------------------------------------------
%------------------------------------------------------------
%------------------------------------------------------------
\subsection{The Recovery Period and the Long-Term Trend}\label{subsec:cross_metro_recovery_and_longterm_trend}
%------------------------------------------------------------
%------------------------------------------------------------
%------------------------------------------------------------

In this section, I show that, in the recovery period (2011-2019) and in the very long term (1991-2019), net export growth predicts cross-metro trends in mortgages and house prices. 

During the recovery period, local mortgage amounts and house prices gradually recover and converge to their long-term equilibrium levels, which are ultimately driven by local economic conditions. Guided by the ``Economic Base Theory" \citep{tiebout1956pure}, I proxy the local economic conditions by net export growth. First, in Figure (\ref{fig_HPI_91to19_NEG}), I show that house prices recover the most in the high-net-export-growth metropolitan areas between 2011 and 2019. These high-net-export-growth metros are exactly those that experience the strongest boom and deepest bust in mortgages and house prices. Similarly, Figure (\ref{fig_GSEMvsPLNJM_91t19_combine}) shows that both types of mortgages also recover between 2011 and 2019 with relatively weaker differences between the high- and low-net-export-growth metros. One possible reason is that private-label mortgages are subject to stringent post-crisis regulatory scrutiny. 

I want to show the cross-metro story in a consistent fashion through the boom-bust-recovery timeline. Thus, I would use net export growth (1991-2007) to predict the long-term house price growth (1991-2019). The rationale is that the recovery period (2012-2019) immediately after the Great Recession (2007-2011) cannot be treated as a regular period, and we cannot use contemporaneous change (e.g., NEG from 2012 to 2019) to infer causality. Instead, as a recovery period, the changes reflect two parts: (1) the long-term local economic growth due to net export growth, and (2) the short-term pronounced boom-bust-recovery cycle triggered by the excessive credit expansion during the boom period (1999-2006). During the boom, excessive credit expansion, driven by financial technology-enabled securitization (namely, the Copula formula) and misaligned incentives, overshoots in the high-net-export-growth metros. The overshoots trigger a stronger recession and pre-determine a potentially stronger recovery in the high-net-export-growth metropolitan areas. To be precise, a higher number of households in the high-net-export-growth metros are likely to avoid mortgage applications and home purchases due to the sharp decline in housing prices during the recession. Instead, they postpone their mortgage application and home purchase during the slow recovery period. At the same time, mortgage rates during the recession can be high due to banks’ limited funding and precautionary savings. Therefore, the recovery period changes are pre-determined by long-term economic conditions before the recession. Further, the recovery would eventually match the long-term equilibrium results, which reflect the long-term effect of net export growth. Therefore, I decide to use long-term net export growth just before the recession, which is from 1991 to 2007.

%\noindent \textbf{House Price Growth in the Long-Term (1991-2007)} 
To test the long-term growth trend in house prices, I perform a regression with the following specification. 
\begin{equation}\label{eq:HPID91t19_reg_NEG}
    \triangle_{91,19} Ln(HPI_{c}) = \beta * \triangle_{91,07} \text{NEG}_{m} + \gamma* \bm{Controls_{c}} + \epsilon_{c}
\end{equation}
where $\triangle_{91,07} \text{NEG}_{m})$ is instrumented by the ($\triangle_{91,07}\text{givNetExp}_{m}$).To reduce the impact of outliers, I winsorize at 2\% and 98\% for $\triangle_{91,07}\text{NetExp}_{m}$ and its IV $\triangle_{91,07}\text{givNetExp}_{m}$. 

Table (\ref{table_HPI.D91t19.NEG.D91t07.4Reg}) reports OLS, reduced-form, second-stage, and first-stage results in panel A, B, C, and D, respectively.  First, panel A shows the positive and significant impact of net export growth (91-07, annualized) on house price index growth (91-19, annualized), with all coefficients being significant at the 1\% level. Panel B reports the reduced-form estimates that share a similar pattern. The first-stage estimates in panel D are positive and significant. Panel C reports 2SLS estimates for equation (\ref{eq:HPID91t19_reg_NEG}). Like reduced-form estimates, the 2SLS estimates are statistically significant and quite stable across various specifications. As for the weak IV concern, the clustered Kleibergen-Paap F-statistic \citep{kleibergen2006generalized} and the Montiel Olea-Pflueger Efficient F-Statistic \citep{olea2013robust} are both 105.6 for the long period 1991-2007, which are larger than 10. Thus, my estimates are highly unlikely to be biased by a weak instrumental variable. I also confirm that, in column (4), the 2SLS estimate (0.971) is similar to the OLS estimate (1.374) in magnitude, thereby not subject to the ``blow-up" concern of a weak IV.  Table (\ref{table_HPI.D91t19.NEG.D91t07.2SLS}) shows all estimates of controls. The housing supply elasticity \citep{saiz2010geographic} is negatively significant at the 1\% level. Taken together, the next export growth (91-07) predicts the long-term housing price growth (91-19).

I also show that, even in the long term, net export growth (91-07) does not significantly increase the cross-metro growth rate (91-19) of government-sponsored enterprise mortgages (GSEMs) or private-label (non-jumbo) mortgages (PLNJMs). The results of OLS, reduced-form, second-stage, and first stage are reported in Table (\ref{table_GSEMvsPLNJM.D91t19.4Reg.NEG.D91t07}). These results also support the exclusion restriction of my IV strategy: net export growth is a slow-moving economic force that can only drive cross-sectional private-label (non-jumbo) mortgages when excessive credit expansion happens in the early 2000s. The literature has documented ample evidence that the credit boom in the early 2000s resulted from unprecedented events: financial innovation in securitization (notably the Copula formula), international capital flows (the "global saving glut"), mortgage market deregulation, and political campaigns. In the absence of those unprecedented events, even over the long term (1991-2019), the impact of net export growth is not significant for either type of mortgage, particularly the private-label (non-jumbo) mortgages, which can respond to local economic conditions.

%------------------------------------------------------------------------------
\noindent \textbf{Recovery Period: 2012-2019.} To show how the recovery trend is largely pre-determined by the long-term trend up to the peak of boom, I also show that the net export growth (1991-2007) can also predict the recovery trend in house price growth (2012-2019). Specifically, I perform a regression with the following specification. 
\begin{equation}\label{eq:HPID12t19_reg_NEG}
    \triangle_{12,19} Ln(HPI_{c}) = \beta * \triangle_{91,07} \text{NEG}_{m} + \gamma* \bm{Controls_{c}} + \epsilon_{c}
\end{equation}
where $\triangle_{91,07} \text{NEG}_{m})$ is instrumented by the ($\triangle_{91,07}\text{givNetExp}_{m}$). 

The recovery period is mixed with much impact from post-crisis tightened regulatory monitoring, weakened global demand for US exports, and trade war between US and China. These impact reflect in the data. To reduce the impact of outliers, I winsorize at 4\% and 96\%, a more aggressive level of winsorization than the previous analysis, for $\triangle_{91,07}\text{NetExp}_{m}$ and its IV $\triangle_{91,07}\text{givNetExp}_{m}$ in the recovery period sample (2012-2019). 

%-----------------------------------------------
%-----------------------------------------------
Table (\ref{table_HPI.D12t19.NEG.D91t07.4Reg}) reports OLS, reduced-form, second-stage, and first-stage results in panel A, B, C, and D, respectively. Even though recovery period trend is impacted by many factors above, the (4\%, 96\%) winsorization reduces noises from outliers and results are still strong. First, panel A shows the positive and significant impact of net export growth (91-07, annualized) on house price index growth (12-19, annualized), with all coefficients being significant. Panel B reports the reduced-form estimates that share a similar pattern. The first-stage estimates in panel D are positive and significant at 1\% level. Panel C reports 2SLS estimates for equation (\ref{eq:HPID12t19_reg_NEG}). Like reduced-form estimates, the 2SLS estimates are statistically significant and quite stable across various specifications. As for the weak IV concern, the clustered Kleibergen-Paap F-statistic \citep{kleibergen2006generalized} and the Montiel Olea-Pflueger Efficient F-Statistic \citep{olea2013robust} are both 191.5. Thus, my estimates are highly unlikely to be biased by a weak instrumental variable. I also confirm that, in column (4), the 2SLS estimate (3.028) is not very much higher than the OLS estimate (2.474) in magnitude, thereby not subject to the ``blow-up" concern of a weak IV.  Table (\ref{table_HPI.D12t19.NEG.D91t07.2SLS}) shows all estimates of controls. The housing supply elasticity \citep{saiz2010geographic} is not significant. Taken together, the next export growth (91-07) predicts the recovery-period housing price growth (12-19). 

%----------------------------------------------------------------------

%--------------------------------------------------------------
%--------------------------------------------------------------
% section 6: Empirical Results

%------------------------------------------------------------
%------------------------------------------------------------
%\clearpage
%------------------------------------------------------------
%------------------------------------------------------------
\section{``Double Differences" Empirical Causal Evidence: Supporting Credit Expansion}\label{sec:Empirical_DoubleDifference}
Beyond reconciling two seemingly opposing empirical facts within the ``credit expansion" view, this model provides new predictions of ``double differences" in mortgage growth and house price growth that further support the ``credit expansion" view. This section develops these new predictions and provides causal evidence and robustness checks.

%--------------------------------------------------------------
%------------------------------------------------------------
\subsection{Model-based New Predictions of ``Double Differences"}
%------------------------------------------------------------
%--------------------------------------------------------------

\noindent \textbf{Hypothesis} The 1999-2005 mortgage differential growth between the low-income ZIP codes and the high-income ZIP codes within the same metropolitan areas is positively caused by net export growth (1999-2005) across metropolitan areas. However, the corresponding 2005-2008 mortgage differential growth is negatively caused by net export growth (1999-2005) across metropolitan areas.

This hypothesis starts from the basic prediction that most of the credit expansion in mortgage growth is concentrated in low-income zip codes. In the argument by \cite{mian2009consequences}, an implicit assumption (and a fact in data) is that households choose ZIP codes within metropolitan areas primarily based on their income-to-house ratio. In my model, this assumption means that households with a low income-to-house ratio live primarily in low income ZIP codes, while households with a high income-to-house ratio live in high income ZIP codes. I adopt this assumption and create a differential mortgage growth between low-income ZIP codes and high-income ones within metropolitan areas: ``low-minus-high" differential mortgage growth. Due to credit expansion in response to net export growth, I predict that this ``low-minus-high" differential mortgage growth is positively correlated with net export growth (99-05) across metropolitan areas between 1999 and 2005 but negatively between 2005 and 2008.\footnote{This model-based new prediction is not a natural result by combining two empirical facts by \cite{mian2009consequences, adelino2016loan}. It could be the case that ``low-minus-high" differential mortgage growth is the same across metropolitan areas.}  This prediction can be seen intuitively from Figure \ref{fig_MortMkt_Bef&Aft_Axis}. Within metropolitan areas, credit expansion in private-label mortgages primarily concentrates on the low-income ZIP codes (left tail). Thus, the `low-minus-high" differential mortgage growth primarily captures the mortgage growth in the low-income ZIP codes. Due to the household advantages caused by net export growth, the `low-minus-high" differential mortgage growth is much higher in the high-net-export-growth metropolitan areas than in the low net-export-growth ones. In addition, since the credit expansion is unsustainable, the subsequent reversal is negative in magnitude.

\noindent \textbf{Hypothesis} The 2000-2006 house price differential growth between the low-income ZIP codes and the high-income ZIP codes within the same metropolitan areas is positively caused by ``low-minus-high" differential mortgage growth (1999-2005) across metropolitan areas. However, the corresponding 2007-2009 house price differential growth is negatively caused by ``low-minus-high" differential mortgage growth (1999-2005) across metropolitan areas.

The above hypothesis is derived by applying the same logic to differential house price growth. Although \cite{mian2009consequences} and \cite{adelino2016loan} also use counties, metropolitan areas are better units to study mortgage and housing markets in ZIP codes within an area of inter-connected economic conditions and commuting. First, counties are usually smaller than metropolitan areas. Some counties are in the central or richer places while others are on the outskirts or poor in metropolitan areas. Thus, the cross-ZIP code variation within metropolitan areas is usually larger than within counties. Second, due to the frequent commuting across counties within a metropolitan area, the net export growth measure is more appropriately aggregated at the metropolitan level than at the county level. Third, counties are administrative units and counties in western areas are often larger than ones in the eastern areas in USA.

%--------------------------------------------------------------
%--------------------------------------------------------------
\subsection{``Double Differences" in Mortgage Growth: Boom and Bust}\label{subsec:NewPrediction_Mortgage_Empirical}
%--------------------------------------------------------------
%--------------------------------------------------------------

This subsection provides descriptive evidence in figures and causal evidence in regressions for the model-based new prediction in differential mortgage growth (hypotheses 4).

First, Figure (\ref{fig_ZIP_HPIGrowth_cbQuint_zipHalf_Quart}) depicts the private-label (non-jumbo) mortgages growth (92-09) across ZIP code groups (high vs. low income) and across metropolitan areas (high vs. low net-export-growth). In Subfigure (a) with top and bottom income half ZIP codes, in the prior period (92-99), differential mortgage growth between high vs. low-income ZIP codes within metropolitan areas is very close to zero. In addition, differential mortgage growth between high vs. low net-export-growth metropolitan areas is also very close to zero. However, in the boom period (99-05), differential mortgage growth between high vs. low-income ZIP codes within metropolitan areas (hereafter ``low-minus-high" mortgage growth) is large. In addition, this ``low-minus-high" mortgage growth is larger in the high-net-export-growth metropolitan areas than the one in the low net-export-growth ones. Subfigure (b) shows similar trends with top and bottom quintile-income ZIP codes. However, since quintile groups compare two end tails, the differential ``low-minus-high" mortgage growth across metropolitan areas seems even more pronounced in Subfigure (b) than in Subfigure (a).  

To formally test the new prediction, I use the following regression specification:
\begin{equation}\label{eq:DD_PLNJMBoomBust_on_NEG}
\resizebox{0.92\textwidth}{!}{$
\begin{aligned}
\text{LMH}\triangle_{99,05} \&  \triangle_{05,08} Ln(PLNJM_{m}) & = \beta_{99,05} * \triangle_{99,05} \text{NetExp}_{m} \times Dum_{99,05} + \beta_{05,08} * \triangle_{99,05} \text{NetExp}_{m} \times Dum_{05,08} \\
& + \gamma_{99,05}* Dum_{99,05} + \gamma_{05,08}* Dum_{05,08} + \epsilon_{period, m}
\end{aligned}
$} %end of \resizebox
\end{equation}
where $\triangle_{99,05}\text{givNetExp}_{m}$ is the  IV for $\triangle_{99,05}\text{NetExp}_{m}$.

Table (\ref{table_ZIP.LMH.PLNJM.D99t05vsD05t08.4Reg}) reports the regression results for OLS, reduced-form, first-stage, and second-stage results. Four columns report the above four regressions for the sample of two, tertile, quartile, and quintile income groups of ZIP codes within metropolitan areas, respectively, based on 1999 IRS income data. Two clear patterns show up. First, when ZIP codes are divided into finer groups based on 1998 household income level, the results of ``double differences" are stronger in magnitude. The OLS, reduced-form, and 2SLS estimates in the boom period (99-05) are marginally significant in the samples of two income groups but significant in the samples of tertile, quartile, and quintile income groups. The coefficient equality tests reject equality in columns (1) at the 10\% level, but reject equality in columns (2), (3), and (4) at the 5\% level. Likewise, the adjusted R-squares in OLS and reduced-form regressions are gradually getting larger from column (1) to column (3), with column (4) being very close to column (3). Second, the results in the bust period (05-08) are more pronounced than in the boom period (99-05). All coefficients in the bust period are larger in magnitude than the corresponding ones in the boom period. The reduced-form and 2SLS estimates in the bust period (05-08) are not significant in the samples of the two income groups but are significant in the samples of the tertile, quartile, and quintile income groups, mostly at 1\% level.  OLS estimates are significant in the bust period in all columns at 5\% or 1\%, while OLS estimates are only significant in the boom period at 10\% or 5\%. In summary, the empirical tests show that net export growth (99-05) induces a much stronger ``low-minus-high" differential mortgage boom (99-05) and bust (05-08) cycle in the high-net-export-growth metropolitan areas than in the low net-export-growth ones.

\noindent \textbf{Exclusion Restriction}
One big concern regarding our identification strategy is that net export growth might also increase household mortgage demand, thereby contaminating our interpretation of credit expansion. I perform the following two tests to address such concerns. 

In the first test, I use government-sponsored enterprise mortgages (GSEMs) to substitute the private-label (non-jumbo) mortgages (PLNJM) in the above specification (\ref{eq:DD_PLNJMBoomBust_on_NEG}). If households mortgage demand increases, the increase will first show up in GSEMs for two reasons. On the one hand, households prefer GSEMs to PLNJM, where GSEMs have lower rates due to implicit government guarantees and the large scale \citep{sherlund2008jumbo}. On the other hand, lenders prefer initiating GSEMs, which guarantee resale and isolate default risk from lenders. Therefore, if net export growth also increases household mortgage demand and increases demand more in low-income ZIP codes within a metro, the GSEMs shall show a more substantial increase in the ``low-minus-high" GSEM growth in the high-net-export-growth metro than in the low-net-export-growth metro. Table (\ref{table_ZIP.LMH.GSEM.D99t05vsD05t08.4Reg}) shows results of OLS, reduced-form, first-stage, and second-stage regressions. All coefficient estimates in OLS, reduced-form, and 2SLS in the boom are insignificant, indicating that the impact of net export-growth on household demand (measured by government-sponsored enterprise mortgages) is negligible. Therefore, the exclusion restriction holds that net export growth increases private-label (non-jumbo) mortgages mainly through lender credit expansion.

In the second test, I stack the private-label (non-jumbo) mortgages (PLNJM) growth in the prior (92-99) and boom (99-05) periods, with specification similar to equation (\ref{eq:DD_PLNJMBoomBust_on_NEG}). Since the credit expansion technology, the Copula model for securitization, was published in 2000 \citep{li2000copula, salmon2009recipe}, the prior period (1992-1999) can serve as a perfect horizon for placebo test. If net export growth were to increase private-label (non-jumbo) mortgage growth via household demand without credit expansion (using the Copula model published in 2000), then this argument would also hold for the prior period (1992-1999). In table (\ref{table_ZIP.LMH.PLNJM.D92t99vsD99t05.4Reg}), the F-statistic in the prior period (1992-1999) is much larger, indicating the net export growth IV is much stronger. However, coefficient estimates in OLS, reduced-form, and 2SLS regressions in the prior period are insignificant. In addition, coefficients in the prior period are much smaller in magnitude compared to those in the boom period. Taken together, the prior period test indicates that, without credit expansion, the impact of net export growth on household demand is negligible in private-label (non-jumbo) mortgage growth. Therefore, the exclusion restriction holds that net export growth increases private-label (non-jumbo) mortgages mainly through lender credit expansion.

%--------------------------------------------------------------
%--------------------------------------------------------------
\subsection{``Double differences" in House Price Growth: Boom and Bust}\label{subsec:DoubleDifferences_HousePrice_Empirical}
%--------------------------------------------------------------
%--------------------------------------------------------------

Figure (\ref{table_ZIP.LMH.HPI.D00t06vsD07t09.PLNJM.4Reg}) plots the house price growth (92-09) across ZIP code groups (high vs. low income) across metropolitan areas (high vs. low net-export-growth). In Subfigure (a) with top and bottom half-income ZIP codes, in the prior period (92-99), differential house price growth between low- and high-income ZIP codes within metropolitan areas is very close to zero. In addition, differential house price growth between high- and low-net-export-growth metropolitan areas is very close to zero. However, in the boom period (00-06), differential house price growth between low- and high-income ZIP codes within metropolitan areas (hereafter ``low-minus-high" house price growth) is large.\footnote{Unlike analysis across metropolitan areas, I use 2000-2006 as the boom period for house price growth across ZIP codes within metropolitan areas because the ZIP-level double difference analysis requires more data variation, and house price growth shows a one-year lag in the data. This one-year lag (or even two-year lag) is evident in Figures (\ref{fig_ZIP_HPIGrowth_cbQuint_zipHalf_Quart}) and (\ref{table_ZIP.LMH.HPI.D00t06vsD07t09.PLNJM.4Reg}). The differential ``low-minus-high" mortgage growth across metropolitan areas is the largest ending in 2005. However, the differential ``low-minus-high" house price boom across metropolitan areas is the largest ending in 2006 (or even 2007).} In addition, this ``low-minus-high" house price growth is larger in the high-net-export-growth metropolitan areas than in the low net-export-growth ones. Subfigure (b) shows similar trends with top and bottom quintile-income ZIP codes. However, since quintile groups compare two end tails, the differential ``low-minus-high" house price boom and bust across metropolitan areas is even more pronounced in Subfigure (b) than in Subfigure (a).     

To formally test the new prediction, I use the following regression specification:
\begin{equation}\label{eq:DD_HPIBoomBust_on_DD_PLNJM}
\resizebox{0.92\textwidth}{!}{$
\begin{aligned}
\text{LMH}\triangle_{00,06} \&  \text{LMH}\triangle_{07,09} Ln(HPI_{m}) & = \beta_{00,06} * \text{LMH}\triangle_{99,05} Ln(PLNJM_{m}) \times Dum_{00,06} \\
& + \beta_{07,09} * \text{LMH}\triangle_{99,05} Ln(PLNJM_{m}) \times Dum_{07,09}  + \alpha_{00,06} + \alpha_{07,09} + \epsilon_{period, m}
\end{aligned}
$} %end of \resizebox
\end{equation}
where $\triangle_{99,05}\text{givNetExp}_{m}$ is the  IV for $\triangle_{99,05}\text{NetExp}_{m}$. Table (\ref{table_ZIP.LMH.HPI.D00t06vsD07t09.PLNJM.4Reg}) reports the results for OLS, reduced-form, first-stage, and second-stage regressions. Four columns report the above four regressions for the sample of two, tertile, quartile, and quintile income groups of ZIP code, respectively. Two patterns are apparent. First, when ZIP codes are further divided into finer groups based on 1998 household income, the empirical results of ``double differences" are more pronounced. The 2SLS estimates in the boom period (00-06) are insignificant in the samples of the two income groups but significant in the samples of the tertile, quartile, and quintile income groups. The coefficient equality tests reject equality in all columns. Second, the results in the bust period (05-08) are more pronounced than in the boom period (00-06). While some coefficients are insignificant or marginally significant in the boom period, their corresponding ones are all significant in the bust period. The coefficients of OLS, reduced-form, and 2SLS estimates in the bust period (07-09) are around four to seven times those in the boom period (00-06). In summary, the empirical tests show that induced by net export growth (99-05), the ``low-minus-high" differential mortgage growth causes a much stronger ``low-minus-high" differential house price boom (00-06) and bust (07-09) cycle in the high-net-export-growth metropolitan areas than in the low net-export-growth ones.

\noindent \textbf{Exclusion Restriction}
One major concern about our identification strategy is that net export growth might also directly raise house prices through cash purchases, rather than through mortgage credit expansion, thereby contaminating our interpretation of credit expansion. I use the prior period (92-99) to design a placebo test. If net export growth increases ``low-minus-high" house price growth through cash purchases, in addition to the credit expansion channel of private-label (non-jumbo) mortgages, then in the prior period (92-99) without credit expansion, net export growth would also increase ``low-minus-high" house price growth. The following regression specification stacks the prior (92-99) and boom (99-05) periods: 
\begin{equation}\label{eq:DD_HPIPriorBoom_on_NEG}
\resizebox{0.92\textwidth}{!}{$
\begin{aligned}
\text{LMH}\triangle_{92,99} \&  \text{LMH}\triangle_{00,06} Ln(HPI_{m}) & = \beta_{92,99} * \triangle_{92,99}\text{NetExp}_{m} \times Dum_{92,99} \\
& + \beta_{00,06} * \triangle_{99,05}\text{NetExp}_{m} \times Dum_{00,06}  + Dum_{92,99} + Dum_{00,06}  + \epsilon_{period, m}
\end{aligned}
$} %end of \resizebox
\end{equation}
where $\triangle_{92,99}\text{givNetExp}_{m}$ and $\triangle_{99,05}\text{givNetExp}_{m}$ are the  IV for $\triangle_{92,99}\text{NetExp}_{m}$ and $\triangle_{99,05}\text{NetExp}_{m}$. Table (\ref{table_ZIP.LMH.HPI.D92t99vsD00t06.NEG.4Reg}) reports the results for OLS, reduced-form, first-stage, and second-stage regressions. Four columns report the above four regressions for the sample of two, tertile, quartile, and quintile income groups of ZIP code, respectively. While the key coefficient estimates in the boom period (00-06) are significant, corresponding coefficient estimates in the prior period (92-99) are not significantly positive in the reduced-form and 2SLS specifications. Such a placebo test in the prior period (92-99) helps address the potential concern about the exclusion restriction: net export growth might drive house prices through channels other than private-label (non-jumbo) mortgages, such as cash purchases.

%--------------------------------------------------------------
%--------------------------------------------------------------
\subsection{``Double differences" in Recovery Period}\label{subsec:DoubleDifference_Recovery}
%--------------------------------------------------------------
%--------------------------------------------------------------

In this section, I will show that recovery-period trends in house prices and mortgages reflect the reversal of the crisis. To be specific, net export growth (91-07) weakly predicts a low-minus-high factor of house prices across metropolitan areas. However, net export growth (91-07) predicts both LMH PLNJM and LMH GSEM but does not predict the difference between them, which is the key feature of recovery periods (12-19) that is different from the boom period (00-06).

As emphasized previously, I aim to present the housing cycle consistently through the boom-bust-recovery timeline. Thus, I would use net export growth (1991-2007) to predict the ``double difference" of mortgage growth and house price growth in the recovery period  (2012-2019). The rationale is that the changes in recovery period reflect two parts: (1) the long-term local economic growth due to net export growth, and (2) the short-term pronounced boom-bust-recovery cycle triggered by the excessive credit expansion during the boom period (1999-2005). The excessive credit expansion during the boom period (1999-2005) triggers a strong recession and pre-determines a potentially strong recovery. To be precise, a higher number of households in the low-income ZIP codes are likely to avoid mortgage applications and home purchases due to the sharp decline in housing prices. Instead, they postpone their mortgage application and home purchase during the slow recovery period. And such a pattern is stronger in the high-net-export-growth metropolitan areas. At the same time, lenders may impose stricter mortgage constraints in low-income ZIP codes than in high-income ones, due to the greater decline in house prices. Again, such a lending pattern is also stronger in the high-net-export-growth metropolitan areas. Therefore, the recovery period changes are pre-determined by long-term economic conditions before the recession. Further, the recovery would eventually match the long-term equilibrium results, which reflect the long-term effect of net export growth. Therefore, I decide the independent variable to be long-term net export growth just before the recession, which is from 1991 to 2007. 

For ``double differences" across ZIP codes and metropolitan areas, it is infeasible to do a long-term predictive regression between 1991 and 2019 due to population growth and the resulting development of boundary ZIP codes in metropolitan areas. As the population grows rapidly between 1990 (with a population size of 248.7 million) and 2020 (with a population size of 331.4 million), metropolitan areas expand to include more outlying counties. Therefore, many low-income boundary ZIP codes of metropolitan areas in 1991 became middle-income ones, and many previously rural ZIP codes became new low-income ZIP codes in metropolitan areas.

\textbf{Hypothesis} The Low-minus-High Factor of PLNJM (or GSEM) growth (2012-2019) is defined as the mortgage growth in low-income ZIP codes minus the one in the high-income ZIP codes within the same metropolitan area. The Low-minus-High Factor for PLNJM (and GSEM) growth (2012-2019) is higher in metropolitan areas with high net export growth (1991-2007) before the Great Recession. However, the difference between LMH PLNJMU and LMH GSEM is not different between high- and low-net-export-growth metros, which is the distinct feature of the recovery period.  

Figure (\ref{fig_ZIP_D12t19_GSEMnPLNJM.Amt07_cbQuint_zipHalf_Quart}) shows the low-minus-high factor for GSEMs and PLNJM growth (2012-2019). The figure shows a slight positive trend of the low-minus-high factor for both mortgages. In addition, the low-minus-high factor appears more positive in the high-net-export-growth metropolitan areas. Furthermore, when within-metro ZIP Codes are divided into finer groups, the two patterns described above become stronger. To formally test the hypothesis above, I employ the following regression specification. 

\begin{equation}
    \text{LMH}\triangle_{12,19} Ln(\text{Mortgage}_{m}) = \beta * \triangle_{91,07} \text{NetExp}_{m} + \text{Constant} + \epsilon_{c}
\end{equation}
where gravity model-based instrumental variable ($\triangle_{91,07}\text{givNetExp}_{m}$) serves as the IV.

Table (\ref{table_ZIP.LMH.PLNJM.D12t19}) reports the results for OLS, reduced-form, first-stage, and second-stage regressions when the mortgage is the private-label (non-jumbo) mortgages. Four columns report the above four regressions for the sample of two, tertile, quartile, and quintile income groups of ZIP code, respectively. The OLS estimates of export growth (1991-2007) in Panel A are significant at the 1\% level for all four samples. When ZIP codes are further divided into finer groups based on the 2011 household income at the ZIP code level, the OLS estimates increase in magnitude. A similar pattern shows up in reduced-form estimates and 2SLS estimates. The first-stage estimates in Panel D are highly significant at the 1\% level. When comparing these results to the ones in Table (\ref{table_ZIP.LMH.PLNJM.D99t05vsD05t08.4Reg}), the 2SLS coefficients is smaller in the recovery period (2012-2019) than in the boom period (1999-2005). Such comparison is not surprising as the relatively weaker positive results of ``double differences" reflect the recovery from a pronounced bust in the low-income ZIP codes than in the high-income ones, particularly in the high-net-export-growth metros. After all, the recovery period does not experience the excessive credit expansion observed in the boom period. Instead, due to new and stringent regulations, particularly the Dodd–Frank Wall Street Reform and Consumer Protection Act, private-label mortgage credit is subject to close regulation. Thus, the positive ``double difference" is weaker and much smaller in magnitude compared to the boom period (1999-2005).

Table (\ref{table_ZIP.LMH.GSEM.D12t19}) reports the results for OLS, reduced-form, first-stage, and second-stage regressions for the government-sponsored enterprise mortgage (GSEM). The OLS estimates of export growth (1991-2007) in Panel A are insignificant for two groups or tertile groups. When ZIP codes are further divided into finer groups based on the 2011 household income, the OLS estimates become positively significant. A similar pattern shows up in reduced-form estimates and 2SLS estimates. The positive ``double difference" in GSEM is not surprising, as it reflects the recovery from a pronounced bust in the low-income ZIP codes than in the high-income ones, particularly in the high-net-export-growth metros. During the bust, it is natural for many middle-class families to wait until the recovery period to apply for mortgages and purchase homes. In addition, due to new and stringent regulations, particularly the Dodd–Frank Wall Street Reform and Consumer Protection Act, more households turn to GSEM after they have accumulated sufficient creditworthiness. Thus, the positive ``double difference" in GSEM during the recovery period reflects the reversal from the bust.

Then, I show that the difference between low-minus-high PLNJM growth and GSEM growth in the recovery period (2012-2019) is not more positive in the high-net-export-growth metropolitan areas. This feature distinguishes the recovery period (2012-2019) from the credit boom period (1999-2005), during which the difference is more positive in high-net-export-growth metros. Table (\ref{table_ZIP.LMH.GSEM.D12t19}) reports the results for OLS, reduced-form, first-stage, and second-stage regressions for the government-sponsored enterprise mortgage (GSEM). Table (\ref{table_ZIP.LMH.PLNJM_m_GSEM.D12t19}) reports the results for OLS, reduced-form, first-stage, and second-stage regressions for the difference between low-minus-high PLNJM growth and GSEM growth in the recovery period (2012-2019). Coefficient estimates of net export growth (91-07) are insignificant in OLS, reduced-form, and 2SLS specifications, while the first-stage estimates are highly significant at 1\% level. This non-positive result is unsurprising, as the stronger low-minus-high PLNJM growth and GSEM growth in the recovery period (2012-2019) in high-net-export-growth metropolitan areas both reflect a reversal from the bust (2007-2011), rather than credit expansion in one type of mortgage versus another.

\textbf{Hypothesis} The Low-minus-High Factor of house price growth (2012-2019)  is defined as the house price growth in low-income ZIP codes minus the one in the high-income ZIP codes within the same metropolitan area. This Low-minus-High Factor of house price growth (2012-2019) is higher in the metropolitan areas with high net export growth (1991-2007) before the Great Recession. 

Figure (\ref{fig_ZIP_D12t19_HPI_cbQuint_zipHalf_Quart}) shows the low-minus-high factor for house price growth (2012-2019). The figure shows a slight positive trend in the low-minus-high factor, and the low-minus-high factor appears more positive in the high-net-export-growth metropolitan areas. In addition, when within-metro ZIP Codes are divided into finer groups, the two patterns described above become stronger. To formally test the hypothesis above, I employ the following regression specification. 

\begin{equation}
    \text{LMH}\triangle_{12,19} Ln(HPI_{m}) = \beta * \triangle_{91,07} \text{NetExp}_{m} + \text{Constant} + \epsilon_{c}
\end{equation}
where gravity model-based instrumental variable ($\triangle_{91,07}\text{givNetExp}_{m}$) serves as the IV. 

Table (\ref{table_ZIP.LMH.HPI.D12t19}) reports the results for OLS, reduced-form, first-stage, and second-stage regressions.\footnote{To reduce the impact of outliers, I winsorize the ``double difference" of house price growth (12-19) at the (0.5\%, 99.5\%).} Four columns report the above four regressions for the sample of two, tertile, quartile, and quintile income groups of ZIP code, respectively. Two patterns are apparent. First, net export growth (1991-2007) is significant at the 1\% level for all four samples. When ZIP codes are further divided into finer groups based on 2011 household income at the ZIP code level, the OLS estimates increase. Although the first-stage regressions are highly significant at the 1\% level, the reduced-form estimates and 2SLS estimates are mostly weakly significant at the 10\% level. Such results are not surprising as this is the recovery period, where the weakly positive results of ``double differences" reflect the recovery from a pronounced bust in the low-income ZIP codes than in the high-income ones, particularly in the high-net-export-growth metros. After all, the recovery period does not experience excessive credit expansion. Instead, due to new and stringent regulations, particularly the Dodd–Frank Wall Street Reform and Consumer Protection Act, mortgage credit is subject to close regulation. Thus, the positive ``double difference" is weaker and much smaller in magnitude compared to the boom period (1999-2005).

%--------------------------------------------------------------
%--------------------------------------------------------------
\subsection{``Double differences": Robustness}
%--------------------------------------------------------------
%--------------------------------------------------------------

This subsection provides robustness tests of model-based predictions of ``double differences" using (1) the number of private-label (non-jumbo) mortgages, (2) the dollar amount of owner-occupied private-label (non-jumbo) mortgages, and (3) a complete set of controls.

First, I show that the model-based prediction of ``double differences" in mortgage growth applies to the number of private-label (non-jumbo) mortgages. The number of loans is also important because it represents the volume of transactions in the housing market. Figure (\ref{fig_ZIP_MortGrowth_PLNJMNL_cbQuint_zipHalf_Quart}) depicts the private-label (non-jumbo) mortgages number growth (92-09) across ZIP code groups (high vs. low income) and across metropolitan areas (high vs. low net-export-growth). Similar to the patterns of the dollar amount of private-label (non-jumbo) mortgages, Subfigure (a) and (b) show that in the prior period (92-99) ``low-minus-high" mortgage growth is very close to zero, and its difference between high vs. low net-export-growth metro areas is also very close to zero. However, in the boom period (99-05), ``low-minus-high" mortgage growth is large in absolute number and is larger in the high-net-export-growth metropolitan areas. Again, the above ``double difference" pattern seems to be stronger in the Subfigure (b) with finer-divided zip codes based on income level. 

For the regression tests, I use a specification similar to equation (\ref{eq:DD_PLNJMBoomBust_on_NEG}), except that the dependent variable is the ``low-minus-high" differential in mortgage number growth. Table (\ref{table_ZIP.LMH.nL.PLNJM.D99t05vsD05t08.4Reg}) reports the regression results for OLS, reduced-form, first-stage, and second-stage results. As before, the same two patterns are apparent. First, when ZIP codes are divided into finer groups based on 1998 household income, the results of ``double differences" are more pronounced. Second, the estimates in the bust period (05-08) are much stronger than the ones in the boom period (99-05). In general, the regressions show that net export growth (99-05) induces a much stronger ``low-minus-high" differential mortgage boom (99-05) and bust (05-08) cycle in the number of loans in the high-net-export-growth metropolitan areas than in the low-net-export-growth ones.

Second, I show that the model-based prediction of ``double differences" in differential mortgage growth applies to the dollar amount of owner-occupied private-label (non-jumbo) mortgages. which is used as a measure of pure credit expansion in Section \ref{sec:Empirical_Metro_AgainstSpeculation}. The goal of this robustness is to show that the ``double differences" pattern is independent of potential speculation. Figure (\ref{fig_ZIP_MortGrowth_PLNJMOwnAmt07_cbQuint_zipHalf_Quart}) depicts the owner-occupied private-label (non-jumbo) mortgages dollar growth (92-09) across ZIP code groups (high vs. low income) and across metropolitan areas (high vs. low net-export-growth). Similar to the results before, Subfigure (a) and (b) show that in the prior period (92-99) ``low-minus-high" owner-occupied mortgage growth is very close to zero, and its difference between high vs. low net-export-growth metro areas is also very close to zero. However, in the boom period (99-05), ``low-minus-high" owner-occupied mortgage growth is large in absolute magnitude and is larger in the high-net-export-growth metropolitan areas. Again, the above ``double difference" pattern appears to be stronger in the Subfigure (b) with finer-divided zip codes. 

For the regressions, I employ the specification in equation (\ref{eq:DD_PLNJMBoomBust_on_NEG}) except that the dependent variable is the ``low-minus-high" owner-occupied mortgage growth. Table (\ref{table_ZIP.LMH.PLNJM.Own.D99t05vsD05t08.4Reg}) reports the regression results for OLS, reduced-form, first-stage, and second-stage results. As before, the same two patterns are apparent. First, when ZIP codes are divided into more groups based on 1998 household income, the patterns of ``double differences" are more striking. Second, the estimates in the bust period (05-08) are much larger in magnitude than the ones in the boom period (99-05). In summary, the regressions show that net export growth (99-05) induces a much stronger "low-minus-high" owner-occupied mortgage boom (99-05) and bust (05-08) cycle in high-net-export-growth metropolitan areas than in the low-net-export-growth ones.

Third, I add a complete set of controls to the above specification (\ref{eq:DD_PLNJMBoomBust_on_NEG}), including basic, housing, and demographic controls. Table (\ref{table_ZIP.LMH.PLNJM.D99t05vsD05t08.FullContr.4Reg}) reports the results for OLS, reduced-form, first-stage, and second-stage regressions, while Table (\ref{table_ZIP.LMH.PLNJM.D99t05vsD05t08.FullContr.2SLS.wide}) shows coefficient estimates for all controls in all columns in 2SLS. Even with the full set of controls in all columns, most coefficient estimates remain significant. As before, when ZIP codes are divided into finer groups based on 1998 household income, the results of ``double differences" are more pronounced. In addition, the estimates in the bust period (05-08) are much stronger than the ones in the boom period (99-05). Further, compared to Table (\ref{table_ZIP.LMH.PLNJM.D99t05vsD05t08.4Reg}) without controls, coefficient estimates in Table (\ref{table_ZIP.LMH.PLNJM.D99t05vsD05t08.FullContr.4Reg}) do not change much in magnitude, further showing the robustness of my main results.

%--------------------------------------------------------------
%--------------------------------------------------------------
% This is the end of the entire section of Model 
%--------------------------------------------------------------
%--------------------------------------------------------------

%--------------------------------------------------------------
%--------------------------------------------------------------

%----------------------------------------------------------------------
% section 7: Empirical Results

%------------------------------------------------------------
%------------------------------------------------------------
%\clearpage
%------------------------------------------------------------
%------------------------------------------------------------
\section{Cross-Metro Empirical Causal Evidence: Against Speculation}\label{sec:Empirical_Metro_AgainstSpeculation}
%------------------------------------------------------------
%------------------------------------------------------------

In this section, I design three tests to address the potential concern from the ``speculation" view that, even though credit expansion occurs in the first place, speculation by borrowers may also dominate the mortgages by taking advantage of the credit expansion. This potential concern originated from the viewpoints of \cite{kindleberger1978manias} and \cite{minsky1986stabilizingan}. Since government-sponsored enterprise mortgages (GSEMs) do not respond to net export growth (see Table (\ref{table_GSEMvsPLNJM.D99t05.4Reg})), they cannot explain the cross-metro housing cycle. I only distinguish speculation and credit expansion within the private-label (non-jumbo) mortgages. I use the speculation measure by \cite{gao2020economic}, which is the ``non-owner-occupied'' home purchase mortgages. Following this logic, I use ``owner-occupied" home purchase mortgages to measure pure credit expansion. First, I demonstrate that the cross-metropolitan variation in speculation can be largely explained by pure credit expansion. I use the residuals from this first specification as my measure of the growth rate of credit-independent speculation, which is part of the growth rate of the non-owner-occupied private-label (non-jumbo) mortgages that cannot be explained by the growth rate of owner-occupied private-label (non-jumbo) mortgages. The second test shows that, compared to credit-independent speculation, credit expansion plays the dominant role in explaining the house price boom. Specifically, credit expansion explains four times the house price growth that speculation explains. The third test focuses on the prior period (91-99) where there is substantial differences in net export growth across metropolitan areas but no aggregate credit expansion. I show that, without credit expansion, speculation does not increase in response to local economic conditions caused by net export growth.

Before going into regression specification, it is helpful to see the general trend of credit expansion vs. speculation mortgage categories across time in Figure (\ref{fig_CreditExpansion_vs_Sepculation}). First, in the boom period (99-05), the credit expansion (owner-occupied home purchase) is much higher in absolute dollar amount than speculation (non-owner-occupied home purchase). Growth rates of both measures are stronger in the high-net-export-growth metropolitan areas. Such observations are consistent with my conclusion for tests 1 \& 2 discussed above. For example, in the peak year of 2005, for the top quintile metropolitan areas based on net export growth, pure credit expansion is $67.0$ Billion (07USD), whereas the speculation is only $13.9$ Billion (07USD). In addition, in the prior period, there is not much difference in speculation (non-owner-occupied home purchase) between the high and low net-export-growth metropolitan areas. This observation is consistent with the conclusion in my third test above. Please note that for ease of calculation, all dependent and independent variables are annualized in this section.

%------------------------------------------------------------
%------------------------------------------------------------
\subsection{Credit Expansion Causes Speculation}
%------------------------------------------------------------
%------------------------------------------------------------

In this subsection, I show that cross-metropolitan variation in speculation is caused by credit expansion. I use the following 2SLS specification: 
\begin{equation}\label{eq:PLNJMNonOwn_on_PLNJMOwn}
\resizebox{0.9\textwidth}{!}{$
\triangle_{99,05} Ln(PLNJM\_NonOwn_{c}) = \underbrace{\beta * \triangle_{99,05} Ln(PLNJM\_Own_{c})}_{\text{Credit-Induced Speculation}} + \underbrace{\gamma* \bm{Controls_{c}} + \epsilon_{c}}_{\text{Credit-Independent Speculation}}
$} %end of \resizebox
\end{equation} 
The left-hand-side dependent variable $\triangle_{99,05} Ln(PLNJM\_NonOwn_{c})$ is the growth rate of the dollar amount of non-owner-occupied private-label (non-jumbo) mortgages at county $c$ 99-05 and the key independent variable $\triangle_{99,05} Ln(PLNJM\_Own_{c})$ is the growth rate of the dollar amount of owner-occupied private-label (non-jumbo) mortgages at county $c$ 99-05. I use the gravity model-based instrumental variable ($\triangle_{99,05}\text{givNetExp}_{m}$) as IV for $\triangle_{99,05} Ln(PLNJM\_Own_{c})$. Controls, weights, and standard errors are the same as Eq (\ref{eq:HPI_reg_PLNJM}).

Table (\ref{table_PLNJMNonOwn.D99t05.PLNJMOwn.4Reg}) reports OLS, reduced-form, second stage, and the first stage of the results. First, panel A shows the positive and significant impact of owner-occupied private-label (non-jumbo) mortgages (PLNJM) (99-05, An) as credit expansion on the growth of non-owner-occupied PLNJM as speculation. This coefficient is quite stable to the inclusion of various controls. Panel B reports the reduced-form estimates and shows that gravity-model-based net export growth increases the speculation. First-stage estimates in panel D show that the strong positive correlation between the GIV and net export growth is quite stable across various specifications. In column (4) with all control variables, the first-stage clustered Kleibergen-Paap F-statistic and the Montiel Olea-Pflueger Efficient F-statistic are both 12.52. Thus, it is very unlikely that my estimates are biased by weak instruments. The 2SLS estimates in panel C are statistically significant at a one percent level and quite stable across various specifications. Besides my two F-statistics, I also confirm that 2SLS is close to OLS estimates throughout various specifications, where my baseline results in column (4) indicate a ratio of $1.395/1.140=1.224$, much lower than nine. In terms of economic meaning in 2SLS with full controls, one standard deviation of 6-year PLNJM (owner-occupied, credit expansion) growth can cause $1.395 * 0.467 = 65.20\%$ increase in 6-year PLNJM (non-owner-occupied, speculation) growth, which is $65.20\%/87.31\%  = 74.68\%$ of one standard deviation of PLNJM (non-owner-occupied) growth. Therefore, I can conclude that cross-metropolitan variation in speculation is largely driven by variation in credit expansion.

%-------------------------------------------------------
%-------------------------------------------------------
\subsection{Credit-Independent Speculation vs Pure Credit Expansion}
%------------------------------------------------------------
%------------------------------------------------------------
In this subsection, I use the decomposition from the first test to get credit-independent speculation. Then I show that, compared to the dominant role of pure credit expansion measured by owner-occupied private-label (non-jumbo) mortgages, credit-independent speculation can only explain a tiny portion of housing price boom.

To be specific, I use the following regression specification:
\begin{equation}\label{eq:HPIBoom_on_PureCredit_vs_Speculation}
\resizebox{0.9\textwidth}{!}{$
\triangle_{99,05} Ln(HPI_{c}) = \beta * \triangle_{99,05} Ln(PLNJM\_Own_{c}) + \gamma* \text{Credit-Independent Speculation}  +  \bm{Controls_{c}} + \epsilon_{c}
$} %end of \resizebox
\end{equation}
One key independent variable $\triangle_{99,05} Ln(PLNJM\_Own_{c})$ is the growth rate of the dollar amount (07USD) of owner-occupied private-label (non-jumbo) mortgages (PLNJM) as pure credit expansion at county $c$ 99-05. This variable is instrumented by the gravity model-based instrumental variable ($\triangle_{99,05}\text{givNetExp}_{m}$). The other key independent variable $\text{Credit-Independent Speculation}$ is derived from the regression Eq (\ref{eq:PLNJMNonOwn_on_PLNJMOwn}), which is the part of growth rate of the non-owner-occupied private-label (non-jumbo) mortgages that cannot be explained by the pure credit expansion.

Table (\ref{table_HPI.D99t05.PLNJMOwn_vs_NonOwnIndependent.4Reg}) reports OLS, reduced-form, second stage, and the first stage of the results. First, panel A shows the positive and significant impact of owner-occupied private-label (non-jumbo) mortgages (PLNJM) (99-05) as pure credit expansion on the house price growth. Panel B reports the reduced-form estimates and shows that GIV net export growth explains the housing price growth. First-stage estimates in panel D show that the strong positive correlation between the GIV net export growth and the growth rate of pure credit expansion. In column (4) with all control variables, the first-stage clustered Kleibergen-Paap F-statistic and the Montiel Olea-Pflueger Efficient F-statistics are both 13.14. Thus, it is very unlikely that my estimates are biased by weak instruments. The 2SLS estimates in panel C are statistically significant and quite stable across various specifications. In term of the economic meaning in 2SLS with full controls, one standard deviation of six-year owner-occupied PLNJM growth can cause  $0.509 * 0.467 = 23.79\%$ house price growth. In comparison, one standard deviation of six-year credit-independent non-owner-occupied PLNJM growth can cause $0.085 * 0.704  = 5.99\%$ house price growth. That is to say, the pure credit expansion can explain around four times ($23.79\%/5.99\%=3.98$) of the house price growth that is explained by credit-independent speculation.

Results together show that home consumption demand (``owner-occupied") induced by credit expansion plays a much more important role than speculation (``non-owner-occupied"). Therefore, I conclude that credit expansion can explain the majority of the house price growth, while credit-independent speculation can only explain a small portion of the house price growth.  This conclusion is slightly different from the view by \cite{minsky1986stabilizingan, kindleberger1978manias}.\footnote{This conclusion only refers to the first mortgage market. Though out of the scope of this paper, it might be the case that speculation by investors plays an important role in the secondary mortgage market.}

%-------------------------------------------------------
%-------------------------------------------------------
\subsection{Speculation Cannot Grow Without Credit Expansion}
%------------------------------------------------------------
%------------------------------------------------------------
In this subsection, I show that speculation cannot grow without credit expansion. Recall that in Eq (\ref{eq:PLNJMNonOwn_on_PLNJMOwn}), I have shown that net export growth causes the credit expansion that largely explains the cross-sectional variation of the speculation. To further show that credit expansion is a necessary condition for speculation, I focus on the prior period (91-99) when there is no aggregate credit expansion, and I show that net export growth alone cannot cause growth in speculation. 

Specifically, I use the following regression specification:
\begin{equation}\label{eq:SpeculationPrior_on_NEG}
\resizebox{0.9\textwidth}{!}{$
\begin{aligned}
\triangle_{91,99} \& \triangle_{99,05} Ln(PLNJM\_NonOwn_{c}) & =  \beta_{91,99} * \triangle_{91,99} \text{NetExp}_{m} \times Dum_{91,99} + \beta_{99,05} * \triangle_{99,05} \text{NetExp}_{m} \times Dum_{99,05} \\
& + \gamma_{91,99}* \bm{Controls_{c}} \times Dum_{91,99} + \gamma_{99,05}* \bm{Controls_{c}} \times Dum_{99,05}  + \epsilon_{period, c}
\end{aligned}
$} %end of \resizebox
\end{equation}
The left-hand-side dependent variable $\triangle_{91,99} \& \triangle_{99,05} Ln(PLNJM\_NonOwn_{c})$ is the stacked growth rate of the dollar amount (07USD) of non-owner-occupied private-label (non-jumbo) mortgages (PLNJM) at county $c$ 91-99 and 99-05, respectively. Again, the key independent variable, net export growth, is instrumented by the gravity model-based IV in two periods, respectively. 

Table (\ref{table_PLNJMNonOwn.D91t99vsD99t05.4Reg}) reports OLS, reduced-form, second stage, and the first stage of the results.\footnote{I drop two outliers, which makes the coefficient in the prior period (91-99) negatively significant. Even though this negatively significant coefficient in the prior period (91-99) does not contradict my conclusion, I drop these two outliers to reflect the estimate for most observations. These two outliers are Durham County (FIPS: 37063) and Boulder County (8013 ).} First, OLS regressions in panel A show that net export growth is positively correlated with the speculation in the boom period (99-05) but not in the prior period (91-99). The same pattern shows up in the reduced-form regressions. First-stage estimates in panel D and E show the strong positive correlation between the net export growth and its GIV in both periods. In column (4) with all control variables, the first-stage clustered Kleibergen-Paap F-statistic and the Montiel Olea-Pflueger Efficient F-statistics are both 25.75 in the prior (91-99) and 17.70 in the boom (99-05). The 2SLS estimates in panel C show that net export growth causes speculation in the boom period but bears no relation in the prior period (91-99). The above empirical results show that in the prior period (1991-1999), without credit expansion, net export growth alone cannot cause a rise in speculation. Put together, these three tests are more consistent with the view that credit expansion is a necessary condition for speculation. 

At this stage, I summarize the empirical evidence presented thus far that argues against the ``speculation” (or demand) view. First, in Section \ref{sec:PLNJM_vs_GSEM}, I use government-sponsored enterprise mortgages (GSEMs) as a control group to argue that the differentially higher growth rate in private-label mortgages (PLMs) in the high-net-export-growth metropolitan areas is not driven by demand. Since the mortgage rate of GSEM is always lower than PLM for the same borrower \citep{sherlund2008jumbo, justiniano2022mortgage}, the demand-driven mortgage by borrowers shall show up in the differentially higher growth in GSEMs in the high net-export growth areas. However, this demand-drive conjecture is not supported by empirical evidence. In fact, GSEMs experience no difference in growth rate between the high and low net-export-growth areas. Critical readers might think borrowers cannot know the differences between GSEM and PLM and only the lenders determine the choice of GSEM or PLM. Since government-sponsored enterprises (Fannie Mae and Freddie Mac) charge insurance fees and provide credit risk isolation, borrowers are more willing to originate GSEM than PLM for the same borrower whenever possible.

Second, in Section \ref{sec:ExclusionRestriction}, I use the prior period (1991-1999) to show that differentially higher growth in PLMs in the high-net-export-growth metropolitan areas is not likely driven by demand in response to net export growth. In the prior period (1991-1999), where there is no aggregate credit supply, the high-net-export-growth metropolitan areas did not experience a higher growth rate in PLMs or GSEMs. This prior period fact shows at least one scenario that, without credit expansion, net export growth does not drive demand for mortgages (GSEMs and PLMs). 

Third, Section \ref{sec:Empirical_Metro_AgainstSpeculation} is designed to address the potential concern from the “speculation” view that, even though credit expansion happens in the first place, speculation by borrowers may also play an important role by taking advantage of the credit expansion. Following \cite{gao2020economic}, I measure speculation by the “non-owner-occupied” private-label mortgages for home purchase. Following the same logic, I measure credit expansion by “owner-occupied” private-label mortgages for home purchases. My first test shows that the cross-metropolitan variation in speculation can be mostly explained by credit expansion. The second test shows that, compared to credit-independent speculation, credit expansion plays the dominant role in explaining the house price boom. Third, in the prior period (1991-1999) where there were substantial differences in net export growth across metropolitan areas but no credit expansion at the aggregate level, I show that speculation does not respond to the net export across metropolitan areas.

%----------------------------------------------------------------------

%----------------------------------------------------------------------
% section 8: Qualification

%\input{Discussion.CreditExpansion}
%----------------------------------------------------------------------

%----------------------------------------------------------------------
% section 9: Conclusion

%----------------------------------------------------------------------------
%\clearpage

\section{Conclusion}\label{sec:Conclusion}
The U.S. housing boom and bust in the 2000s were unprecedented, resulting in the deepest recession since the Great Depression. However, during this housing cycle, three empirical facts present a puzzle: (1) in the boom period, the correlation between income growth and mortgage growth is negative across ZIP codes within a metropolitan area, but (2) positive across metropolitan areas, and (3) the metropolitan areas that experience the worst bust also show the strongest recovery.

In this paper, I develop a unified credit expansion theory that explains both within- and cross-metro patterns in the prior, boom, bust, and recovery periods (including the three facts above) and generates new testable implications of `double differences' (cross ZIP codes and cross metro) for the four periods. For empirical identification, I operationalize the idea of `Economic Base Theory’ by constructing local economic exposure to net export growth, and use a new instrumental variable from the International trade literature.

First, I show that high-net-export-growth metros experience a stronger boom-bust-recovery housing cycle due to credit expansion in private-label (non-jumbo) mortgages (PLNJM), rather than in government-sponsored enterprise mortgages (GSEMs), because only the former can legally respond to local economic conditions. Second, for the `double differences', I define a low-minus-high (LMH) factor as the private-label (non-jumbo) mortgage (and house price) growth in low-income ZIP codes minus that in high-income ZIP codes within the same metropolitan area. I show that this low-minus-high factor in the high-net-export-growth metros is more positive during the boom period, more negative during the bust period, and weakly more positive in the recovery period than in the low-net-export-growth metros. Lastly, I employ five tests to demonstrate that `speculation’ is unlikely to play a dominant role in this housing cycle. Therefore, the policy implication is that the new regulatory design to prevent a recurrence of the same housing cycle should focus more on lenders.

Please note that “causal evidence” in this paper refers to a cross-sectional result, and this paper does not identify the forces behind the aggregate credit expansion during the boom period. Existing work points to several aggregate drivers, including securitization-related financial innovation, international capital inflows (``global saving glut”), mortgage-market deregulation, and political economy factors. While this paper supports the credit-expansion view and suggests that 1999–2007 credit growth likely overshot an efficient allocation, it does not disentangle three micro mechanisms: (i) “technology innovation going wrong” (e.g., underpricing of risks due to the Copula model) and resulting loosening of lending criteria; and (ii) reduced screening incentives and efforts in lenders; and (iii) fraud due to misaligned incentives along the originate-to-distribute chain.

%----------------------------------------------------------------------

%----------------------------------------------------------------------
% Section 8. bibliography
%----------------------------------------------------------------------

\clearpage

\begingroup
\setstretch{1.0}
%\bibliographystyle{jf}
%\printbibliography[title=References]
%\bibliography{Li_bibfile.bib}
\ifx\undefined\BySame
\newcommand{\BySame}{\leavevmode\rule[.5ex]{3em}{.5pt}\ }
\fi
\ifx\undefined\textsc
\newcommand{\textsc}[1]{{\sc #1}}
\newcommand{\emph}[1]{{\em #1\/}}
\let\tmpsmall\small
\renewcommand{\small}{\tmpsmall\sc}
\fi

\endgroup

%----------------------------------------------------------------------
% section 9: Figures and Tables

%------------------------------------------------------------
%------------------------------------------------------------
\pagebreak
%------------------------------------------------------------
%------------------------------------------------------------
\section{Figures and Tables}

%------------------------------------------------------------
% figure 0: fig_USMetroCty_NetExpGrowth

%------------------------------------------------------------
% fig_ZIP_MortGrowth_cbQuint_zipHalf&Quint

%------------------------------------
\begin{figure}[h!] 
    \centering
    \begin{subfigure}[t]{0.9\textwidth}
        \centering
        \includegraphics[height=7.5cm]{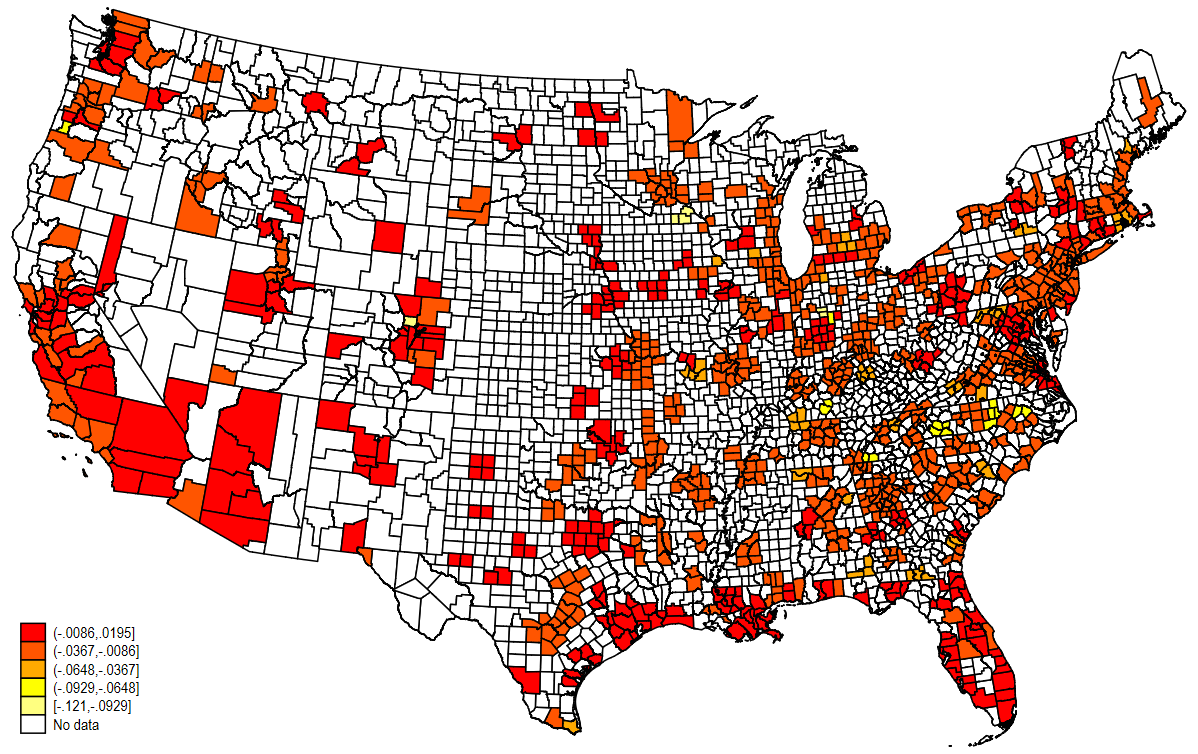}
        \caption{Net Export Growth (1991-1999)}
    \end{subfigure}%
    \hfill 
    \begin{subfigure}[t]{0.9\textwidth}
        \centering
        \includegraphics[height=7.5cm]{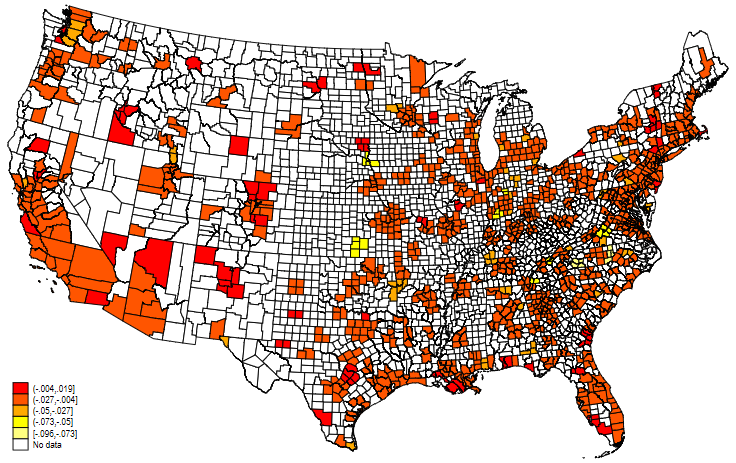}
        \caption{Net Export Growth (1999-2005)}
    \end{subfigure}
    \caption{\textbf{U.S. Mainland Metropolitan Heat Map of Net Export Growth: Two Periods}  \smallskip  \newline 
    {\footnotesize This figure displays the U.S. mainland metropolitan heat map of net export growth measure in two periods: subfigure (a) is for 1991-1999, and subfigure (b) is for 1999-2005. The net export growth measure at the metropolitan level for a period is defined in equation (\ref{eq:NEG_m}). In the above figures, each small area with a boundary is a county. Counties with white color are non-metropolitan areas in the 2003 CBSA version (1085 counties). Metropolitan counties are painted with colors ranging from yellow (for low net export growth) to red (for high net export growth) in five categories. 
        } %end of small font
    } % end of caption
    \label{fig_ZIP_MortGrowth_cbQuint_zipHalf&Quint}
    
\end{figure} 
%------------------------------------

%------------------------------------------------------------
% figure 1: fig_HPI_91t11

%------------------------------------------------------------
% fig_HPI_91to19_NEG

\begin{figure}[h!] 
    \centering
    \includegraphics[width=16cm, height=12cm]{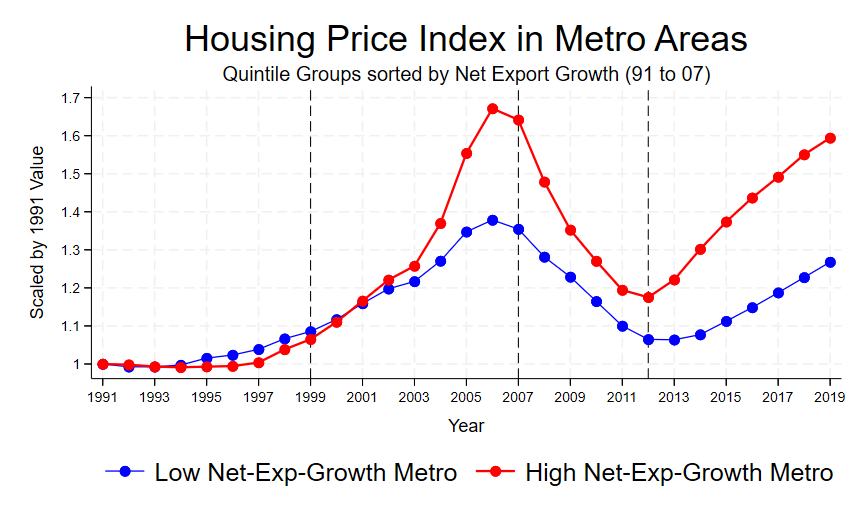}
    \caption{\textbf{Housing Price Index (1991-2019) in Metropolitan Areas, Sorted by High vs. Low Quintile of Net Export Growth (1991 to 2007)} \smallskip \newline 
    {\footnotesize This figure displays the time series of weighted average housing price indices (deflated to 2007 value) for high and low quintile groups of Metropolitan Areas (MSAs) from 1991 to 2011. For the entire period, the quintile groups are sorted by net export growth (1991-2007) at the metropolitan level (CBSA code, 2003 version). The sample comprises 301 metropolitan areas (712 counties) that have been consistently covered by the HMDA sample since 1990. 680 of the 712 counties consistently have non-missing house price indexes from 1991 to 2019. The low quintile group comprises 61 metros (84 counties), and the high quintile group comprises 60 metros (110 counties) throughout the entire period. The housing price index is weighted by the number of housing units in each group at the county level in each year. Throughout the period, the time series of the weighted-average housing price indices for each group are divided by their 1991 values, ensuring that both groups start at 1 in 1991. The red line represents the high-quintile group, whereas the blue line represents the low-quintile group.
    } %end of small font
    } % end of caption
    \label{fig_HPI_91to19_NEG}
    % note that \label is given after \caption.
\end{figure}

\pagebreak
%------------------------------------------------------------
%------------------------------------------------------------
% figure 2: fig_GSEMvsPLNJM_91t11_combine

%------------------------------------------------------------
% fig_GSEMvsPLNJM_91t19_combine
%------------------------------------------------------------

\begin{figure}[h!] 
    \centering
    \includegraphics[width=16cm, height=12cm]{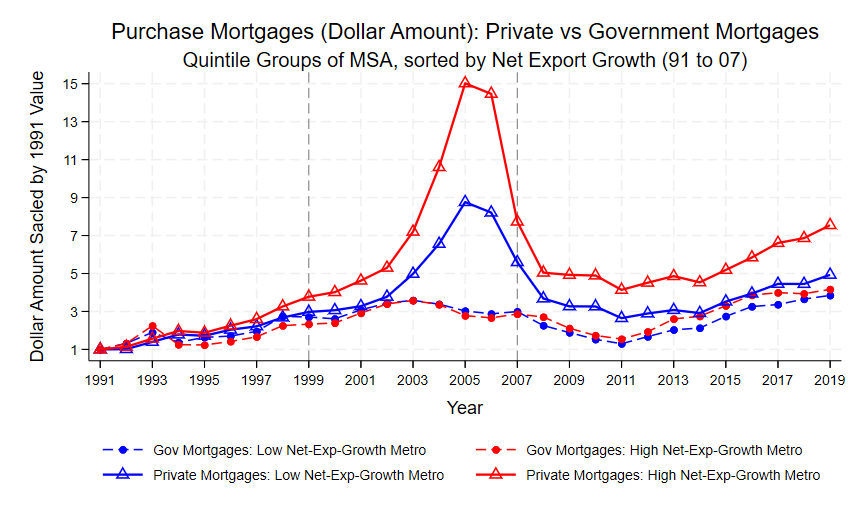}
    \caption{\textbf{Mortgage Growth (1991-2019) across Metropolitan Areas: GSEM vs. PLNJM, with High vs. Low Quintile Metropolitan Areas Sorted by Net Export Growth (1991-2007).}  \smallskip \newline 
    {\footnotesize This figure shows the time series of the weighted-average dollar amount of Government-Sponsored Enterprise Mortgages (GSEM) (in dashed lines with dots) and private-label (non-jumbo) mortgages (PLNJM) (in solid lines with triangles) for high and low quintile groups of metropolitan areas (MSA using CBSA03 code) from 1991 to 2019. Both types of mortgages consist solely of purchase loans and are deflated to 2007 U.S. dollars using the Personal Consumption Expenditures Chain-type Price Index (PCEPI) from the Federal Reserve Bank of St. Louis. For the entire period, the quintile groups are sorted by net export growth (1991-2007) at the MSA level. The whole sample includes 301 MSA (712 counties) that are consistently covered by the HMDA sample since 1990 due to the smaller coverage of metropolitan areas in the early years. The low quintile group comprises 61 MSAs (94 counties), and the high quintile group comprises 60 MSAs (112 counties) throughout the entire period. Throughout the entire period, the time series for the sum of each group's loan amounts is divided by its 1991 value, ensuring that both groups start at 1 in 1991. The red lines represent the high quintile group, while the blue lines represent the low quintile group.
        } %end of small font
    } % end of caption
    \label{fig_GSEMvsPLNJM_91t19_combine}
    % note that \label is given after \caption.
\end{figure}

\pagebreak
%------------------------------------------------------------
%------------------------------------------------------------
% figure 3: fig_CreditExpansion_vs_Sepculation

%------------------------------------------------------------
% figure 3: fig:CreditExpansion_vs_Sepculation

\begin{figure}[h!] 
    \centering
    \includegraphics[width=16cm, height=12cm]{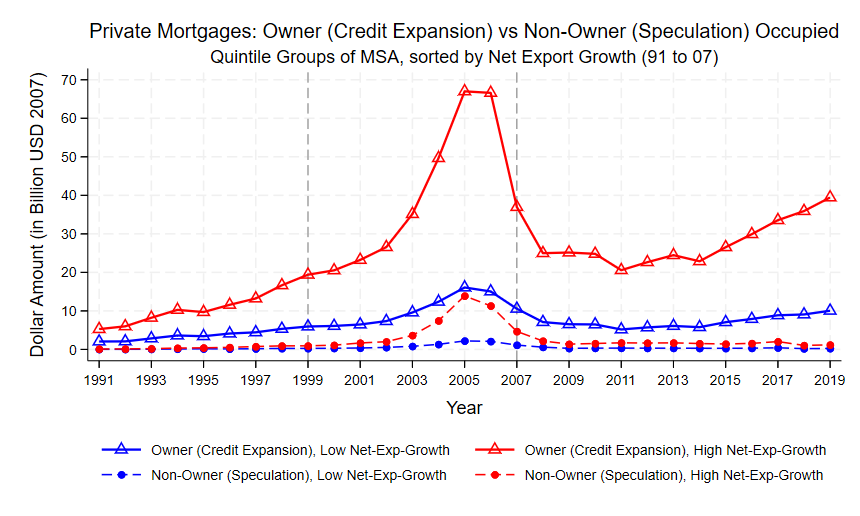}
    \caption{\textbf{Credit Expansion vs. Speculation (1991-2019) across Metropolitan Areas: High vs. Low Quintile Sorted by Net Export Growth (1991-2007).}  \smallskip \newline 
    {\footnotesize This figure shows the time series of dollar amount of ``owner-occupied" (as a measure of credit expansion) home purchase private-label mortgages (in solid lines with triangles) and ``non-owner-occupied" (speculation) home purchase private-label mortgages (in dashed lines with dots) from 1991 to 2019 for high and low quintile groups of metropolitan statistical areas (MSA). For the entire period, the quintile groups are sorted by net export growth (1991-2007) at the MSA level. The entire sample comprises 301 MSAs (712 counties) that have been consistently covered by the HMDA since 1990, owing to the limited coverage of metropolitan areas in the early years. The low quintile group comprises 61 MSAs (94 counties), and the high quintile group comprises 60 MSAs (112 counties) throughout the entire period. The dollar amount is adjusted to the 2007 USD by the Personal Consumption Expenditures Chain-type Price Index (PCEPI) from the Federal Reserve Bank of St. Louis. The red lines represent the high-quintile group, whereas the blue lines represent the low-quintile group.
        } %end of small font
    } % end of caption
    \label{fig_CreditExpansion_vs_Sepculation}
    % note that \label is given after \caption.
\end{figure}

\pagebreak 
%-----------------------------------------------------------------
%%%%%%%%%%%%%%%%%%%%%%%%%%%%%%%%%%%%
% fig_ZIP_MortGrowth_PLNJMAmt07_cbQuint_zipHalf_Quart
%%%%%%%%%%%%%%%%%%%%%%%%%%%%%%%%%%%%

%------------------------------------------------------------
% fig_ZIP_MortGrowth_PLNJMAmt07_cbQuint_zipHalf_Quart

%------------------------------------
\begin{figure}[h!] 
    \centering
    \begin{subfigure}[t]{0.9\textwidth}
        \centering
        \includegraphics[height=7.5cm]{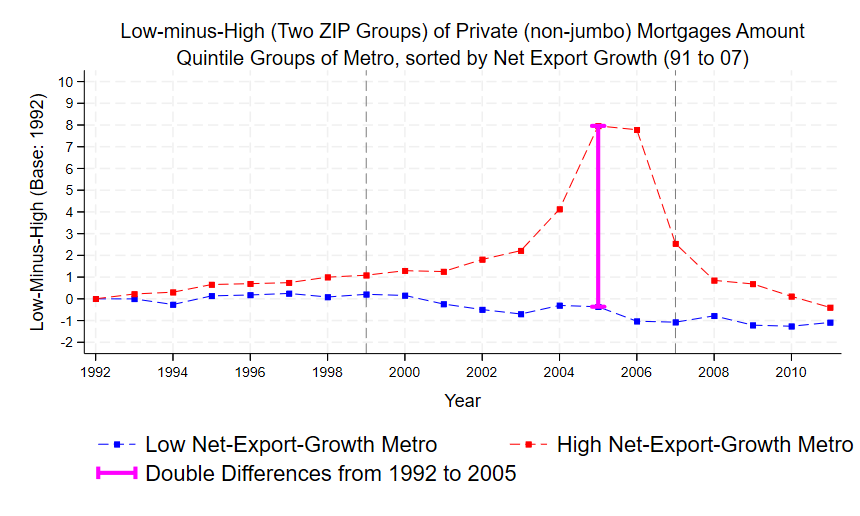}
        \caption{Income Half Groups of ZIP Codes within Metro}
    \end{subfigure}%
    \hfill 
    \begin{subfigure}[t]{0.9\textwidth}
        \centering
        \includegraphics[height=7.5cm]{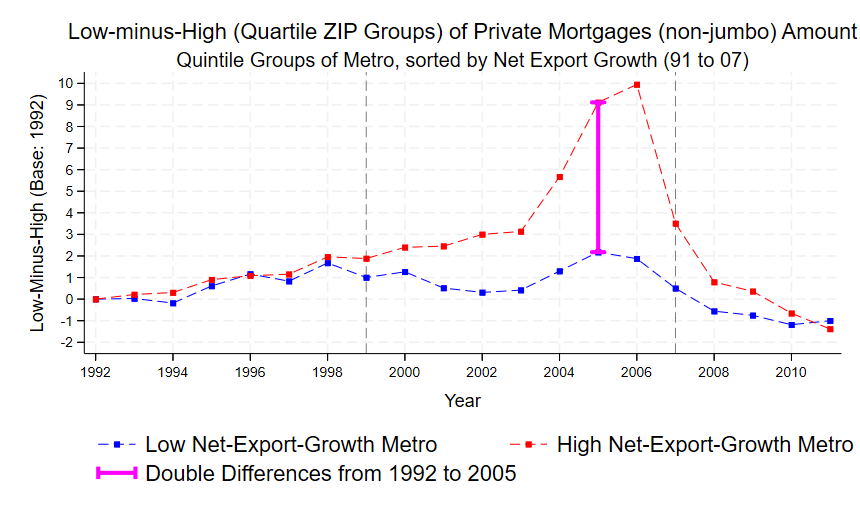}
        \caption{Income Quartile Groups of ZIP Codes within Metro}
    \end{subfigure}
    \caption{\textbf{Low-minus-High Factor for Private-Label (non-jumbo) Mortgage Amount Growth (1992-2011) Across ZIP Codes Across Metropolitan Areas.}  \smallskip  \newline 
    {\footnotesize This figure shows the time series of the low-minus-high factor for the dollar amount of private-label (non-jumbo) mortgages (PLNJM) for high (in red lines) and low (in blue lines) quintile groups of metropolitan areas (MSAs by CBSA03 code) from 1992 to 2011. Within each MSA, the low-minus-high factor is calculated as the dollar amount of private-label (non-jumbo) mortgages in the bottom-half/quartile-income ZIP codes minus that in the top-half/quartile-income ZIP codes, based on the 1998 income level. Mortgages only contain home purchase loans and are deflated to 2007 U.S. dollars. As a result, the low-minus-high factor for each MSA begins at 0 in 1992. I require that each MSA have at least five ZIP codes. The entire sample for the figure of half (quartile) ZIP groups is a balanced panel of 10,315 ZIP codes and 18 years, consistently covered by the HMDA sample. For the entire period, metropolitan areas are sorted by net export growth (1991-2007) at the MSA level. The low quintile group comprises 904 zip codes from 57 MSAs, and the high quintile group comprises 1,610 zip codes from 56 MSAs throughout the entire period. 
        } %end of small font
    } % end of caption
    \label{fig_ZIP_MortGrowth_PLNJMAmt07_cbQuint_zipHalf_Quart}
    
\end{figure} 
%------------------------------------

\pagebreak 
%-----------------------------------------------------------------
%%%%%%%%%%%%%%%%%%%%%%%%%%%%%%%%%%%%
% fig_ZIP_HPIGrowth_cbQuint_zipHalf_Quart
%%%%%%%%%%%%%%%%%%%%%%%%%%%%%%%%%%%%

%------------------------------------------------------------
% fig_ZIP_HPIGrowth_cbQuint_zipHalf_Quart

%------------------------------------
\begin{figure}[h!] 
    \centering
    \begin{subfigure}[t]{0.9\textwidth}
        \centering
        \includegraphics[height=7.5cm]{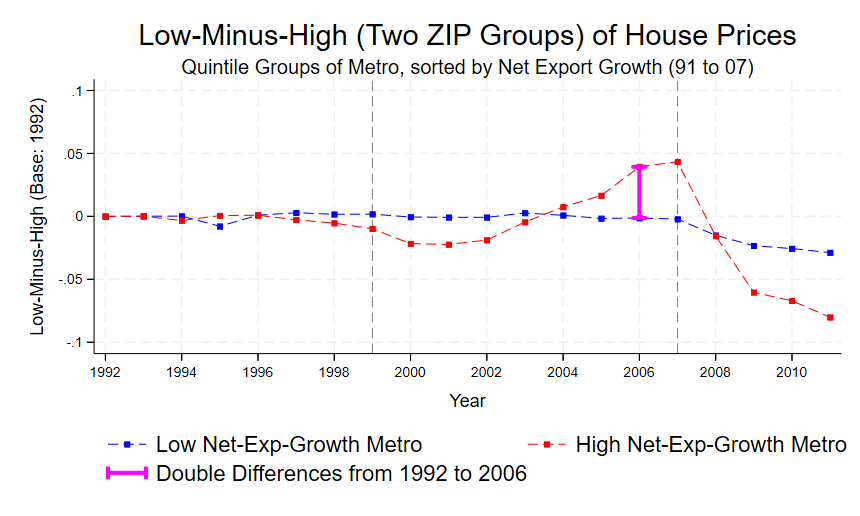}
        \caption{Half Groups of ZIP Codes within Metro}
    \end{subfigure}%
    \hfill 
    \begin{subfigure}[t]{0.9\textwidth}
        \centering
        \includegraphics[height=7.5cm]{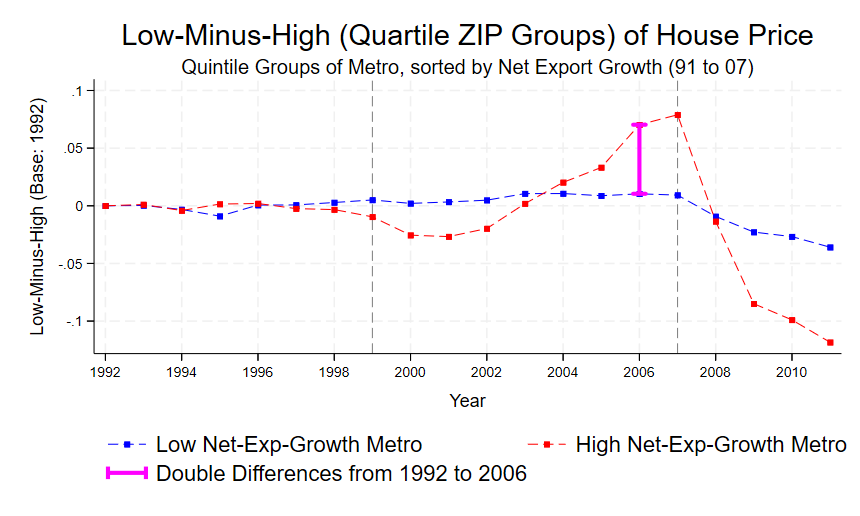}
        \caption{Quartile Groups of ZIP Codes within Metro}
    \end{subfigure}
    \caption{\textbf{Low-minus-High Factor for House Price Growth (1992-2011) Across ZIP Codes Across Metropolitan Areas}  \smallskip \newline 
    {\footnotesize This figure shows the time series of the low-minus-high factor for the weighted-average house price index for high (in red lines) and low (in blue lines) quintile groups of metropolitan areas (MSAs by CBSA03 code) from 1992 to 2011. Within each MSA, the low-minus-high factor is calculated as the simple average (without weights) of the house price index for the bottom-half/quartile-income ZIP codes minus that for the top-half/quartile-income ZIP codes, based on 1998 income levels. The house price index for the ZIP code is deflated to the 2007 value. As a result, the low-minus-high factor for each MSA begins at 0 in 1992. I require that each MSA have at least five ZIP codes. The entire sample for the figure of half (quartile) ZIP groups is a balanced panel of 7,218 ZIP codes and 18 years, consistently covered by the FHFA HPI ZIP sample. For the entire period, metropolitan areas are sorted by net export growth (1991-2007) at the MSA level. The low quintile group comprises 659 zip codes from 46 MSAs, and the high quintile group comprises 1,059 zip codes from 45 MSAs throughout the entire period. When aggregating the low-minus-high factor from MSA to the MSA quintile group, I use the number of housing units at the MSA level as the weights. 
        }%end of small font
    } % end of caption
    \label{fig_ZIP_HPIGrowth_cbQuint_zipHalf_Quart}
    
\end{figure} 
%------------------------------------

\pagebreak 
%-----------------------------------------------------------------
%%%%%%%%%%%%%%%%%%%%%%%%%%%%%%%%%%%%
% fig_ZIP_MortGrowth_PLNJMNL_cbQuint_zipHalf_Quart
%%%%%%%%%%%%%%%%%%%%%%%%%%%%%%%%%%%%

%------------------------------------------------------------
% fig_ZIP_MortGrowth_PLNJMNL_cbQuint_zipHalf_Quart

%------------------------------------
\begin{figure}[h!] 
    \centering
    \begin{subfigure}[t]{0.9\textwidth}
        \centering
        \includegraphics[height=7.5cm]{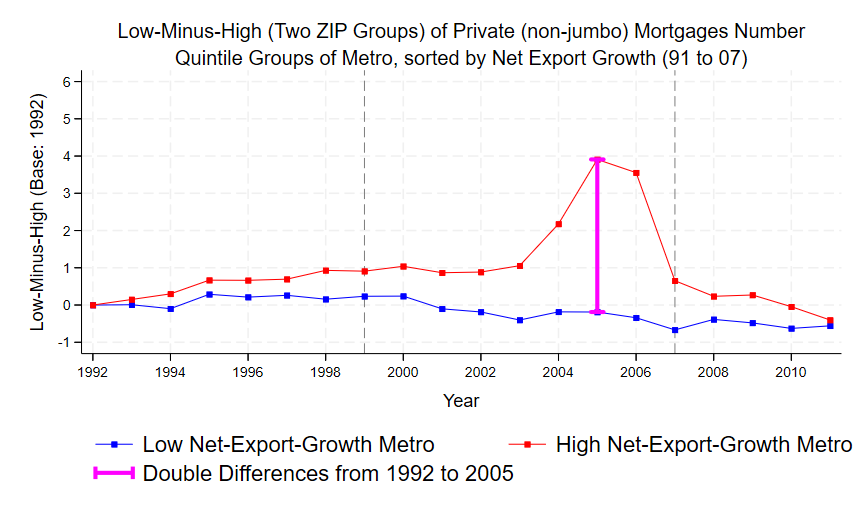}
        \caption{Income Half Groups of ZIP Codes within Metro}
    \end{subfigure}%
    \hfill 
    \begin{subfigure}[t]{0.9\textwidth}
        \centering
        \includegraphics[height=7.5cm]{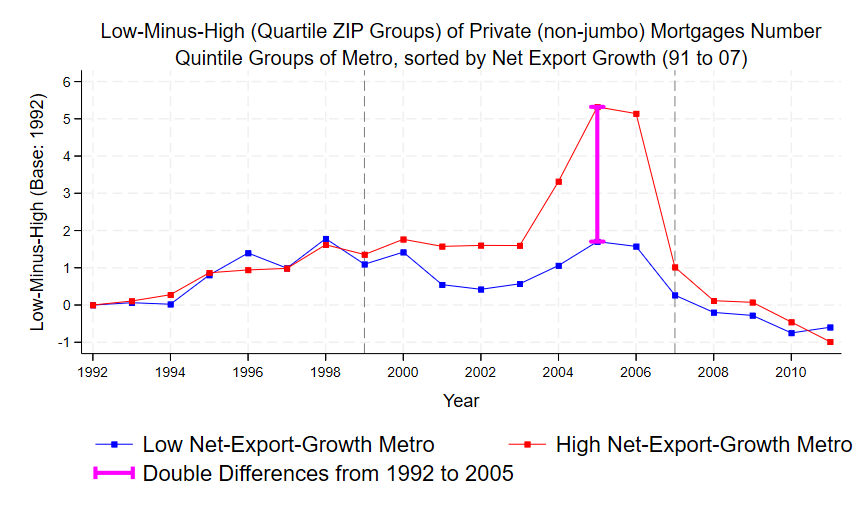}
        \caption{Income Quartile Groups of ZIP Codes within Metro}
    \end{subfigure}
    \caption{\textbf{Low-minus-High Factor for Private-Label (non-jumbo) Mortgage Number Growth (1992-2011) Across ZIP Codes Across Metropolitan Areas.}  \smallskip  \newline 
    {\footnotesize This figure shows the time series of the low-minus-high factor for the number of private-label (non-jumbo) mortgages (PLNJM) for high (in red lines) and low (in blue lines) quintile groups of metropolitan areas (MSAs by CBSA03 code) from 1992 to 2011. Within each MSA, the low-minus-high factor is calculated as the number of private-label (non-jumbo) mortgages in the bottom-half/quartile-income ZIP codes minus that in the top-half/quartile-income ZIP codes, based on the 1998 income level. Mortgages only contain home purchase loans. As a result, the low-minus-high factor for each MSA begins at 0 in 1992. I require that each MSA have at least five ZIP codes. The entire sample for the figure of half (quartile) ZIP groups is a balanced panel of 10,315 ZIP codes and 18 years, consistently covered by the HMDA sample. For the entire period, metropolitan areas are sorted by net export growth (1991-2007) at the MSA level. The low quintile group comprises 904 zip codes from 57 MSAs, and the high quintile group comprises 1,610 zip codes from 56 MSAs throughout the entire period. 
        } %end of small font
    } % end of caption
    \label{fig_ZIP_MortGrowth_PLNJMNL_cbQuint_zipHalf_Quart}
    
\end{figure} 
%------------------------------------

\pagebreak 
%-----------------------------------------------------------------
%%%%%%%%%%%%%%%%%%%%%%%%%%%%%%%%%%%%
% fig_ZIP_MortGrowth_PLNJMOwnAmt07_cbQuint_zipHalf_Quart
%%%%%%%%%%%%%%%%%%%%%%%%%%%%%%%%%%%%

%------------------------------------------------------------
% fig_ZIP_MortGrowth_PLNJMOwnAmt07_cbQuint_zipHalf_Quart

%------------------------------------
\begin{figure}[h!] 
    \centering
    \begin{subfigure}[t]{0.9\textwidth}
        \centering
        \includegraphics[height=7.5cm]{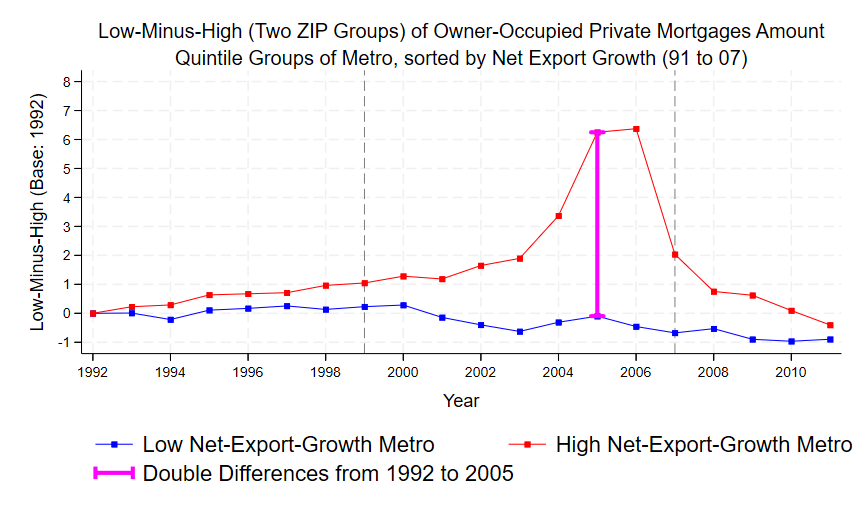}
        \caption{Income Half Groups of ZIP Codes within Metro}
    \end{subfigure}%
    \hfill 
    \begin{subfigure}[t]{0.9\textwidth}
        \centering
        \includegraphics[height=7.5cm]{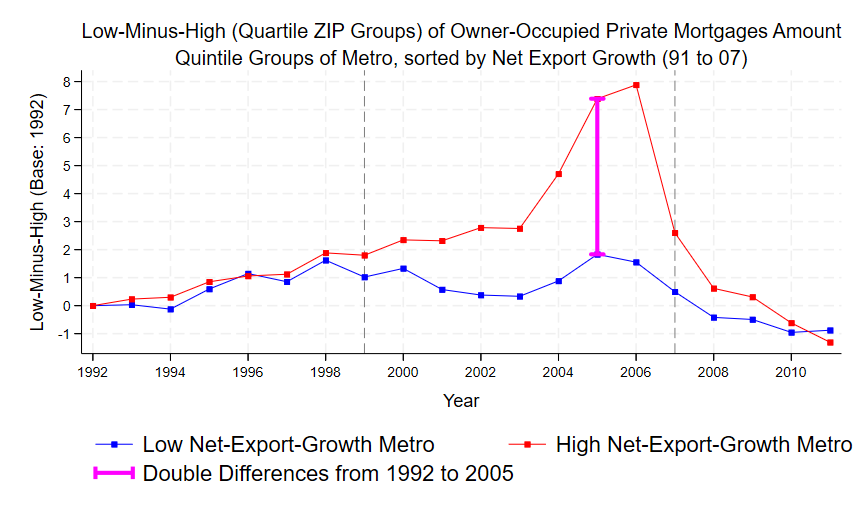}
        \caption{Income Quartile Groups of ZIP Codes within Metro}
    \end{subfigure}
    \caption{\textbf{Low-minus-High Factor for Owner-Occupied Private-Label (non-jumbo) Mortgage Amount Growth (1992-2011) Across ZIP Codes Across Metropolitan Areas.}  \smallskip  \newline 
    {\footnotesize This figure shows the time series of the low-minus-high factor for the dollar amount of owner-occupied private-label (non-jumbo) mortgages for high (in red lines) and low (in blue lines) quintile groups of metropolitan areas (MSAs by CBSA03 code) from 1992 to 2011. Within each MSA, the low-minus-high factor is calculated as the dollar amount of owner-occupied private-label (non-jumbo) mortgage in the bottom-half/quartile-income ZIP codes minus that in the top-half/quartile-income ZIP codes, based on the 1998 income level. Mortgages only contain home purchase loans and are deflated to 2007 U.S. dollars. As a result, the low-minus-high factor for each MSA begins at 0 in 1992. I require that each MSA have at least five ZIP codes. The entire sample for the figure of half (quartile) ZIP groups is a balanced panel of 10,315 ZIP codes and 18 years, consistently covered by the HMDA sample. For the entire period, metropolitan areas are sorted by net export growth (1991-2007) at the MSA level. The low quintile group comprises 904 zip codes from 57 MSAs, and the high quintile group comprises 1,610 zip codes from 56 MSAs throughout the entire period.
        } %end of small font
    } % end of caption
    \label{fig_ZIP_MortGrowth_PLNJMOwnAmt07_cbQuint_zipHalf_Quart}
    
\end{figure} 
%------------------------------------

\pagebreak 
%-----------------------------------------------------------------
%%%%%%%%%%%%%%%%%%%%%%%%%%%%%%%%%%%%
% fig_ZIP_D12t19_GSEMnPLNJM.Amt07_cbQuint_zipHalf_Quart
%%%%%%%%%%%%%%%%%%%%%%%%%%%%%%%%%%%%

%------------------------------------------------------------
% fig_ZIP_D12t19_GSEMnPLNJM.Amt07_cbQuint_zipHalf_Quart

%------------------------------------
\begin{figure}[h!] 
    \centering
    \begin{subfigure}[t]{0.9\textwidth}
        \centering
        \includegraphics[height=7.5cm]{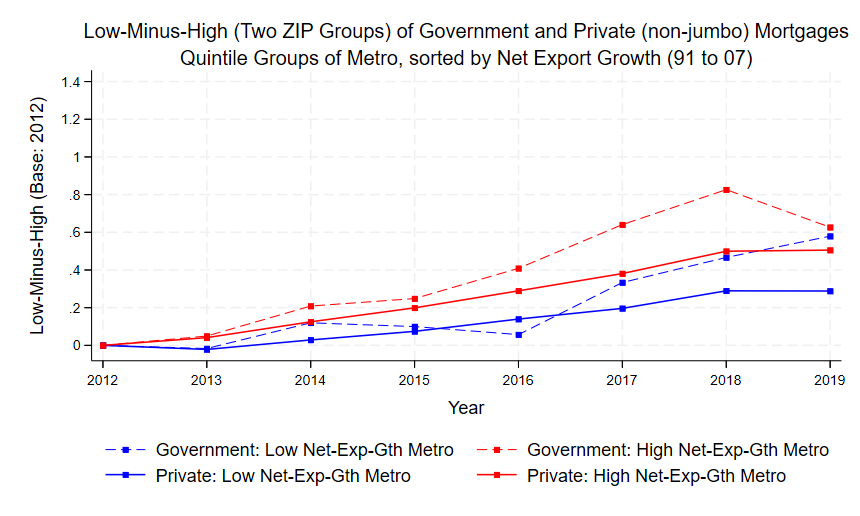}
        \caption{Income Half Groups of ZIP Codes within Metro}
    \end{subfigure}%
    \hfill 
    \begin{subfigure}[t]{0.9\textwidth}
        \centering
        \includegraphics[height=7.5cm]{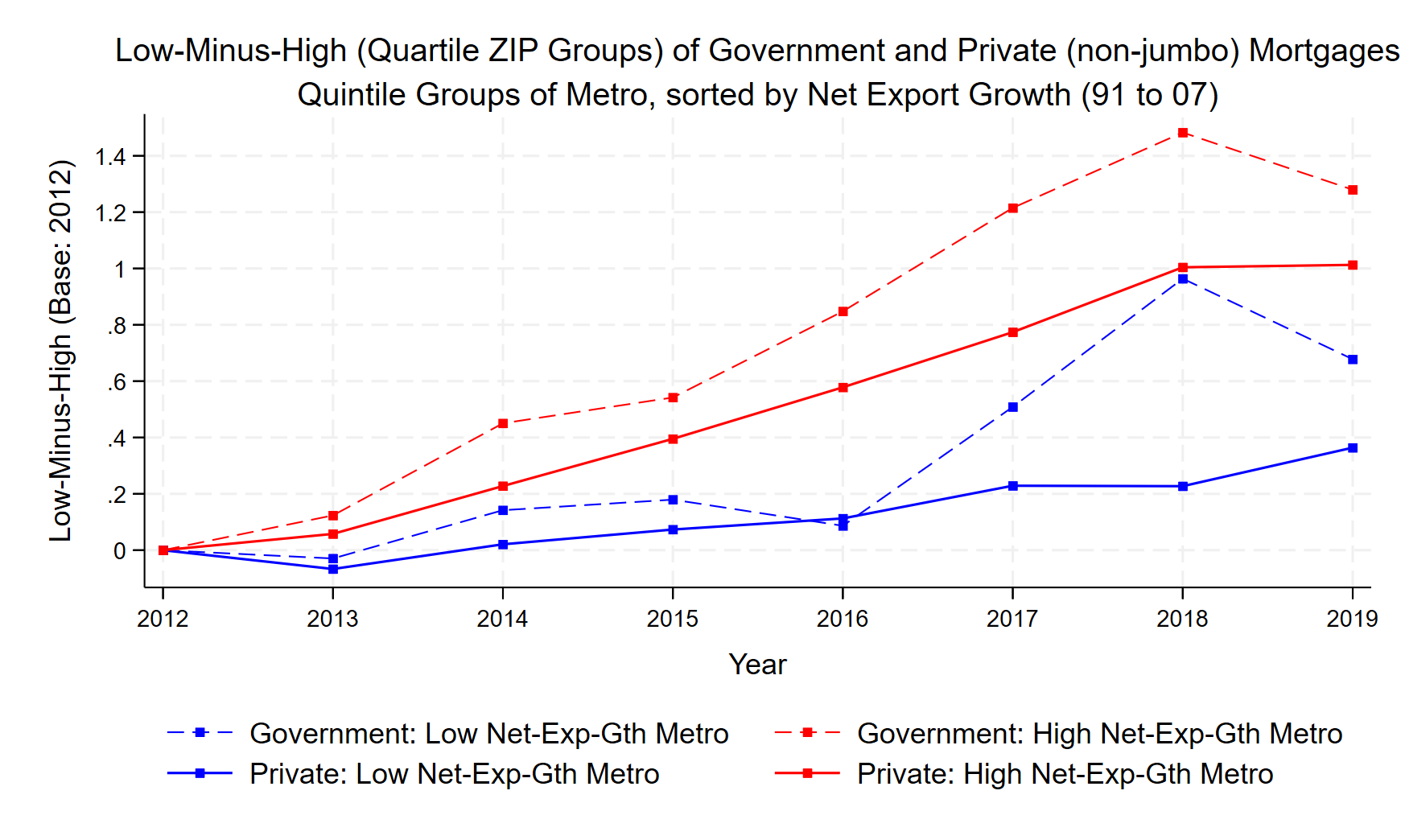}
        \caption{Income Quartile Groups of ZIP Codes within Metro}
    \end{subfigure}
    \caption{\textbf{Low-minus-High Factor for Government-Sponsored Enterprise Mortgages and Private-Label (non-jumbo) Mortgage Amount Growth (2012-2019) Across ZIP Codes Across Metropolitan Areas.}  \smallskip  \newline 
    {\footnotesize This figure shows the time series of the low-minus-high factor for the dollar amount of government-sponsored enterprise mortgages (GSEMs) and private-label (non-jumbo) mortgages (PLNJMs) for high (in red lines) and low (in blue lines) quintile groups of metropolitan areas (MSAs by CBSA09 code) from 2012 to 2019. Within each MSA, the low-minus-high factor is calculated as the dollar amount of private-label (non-jumbo) mortgages in the bottom-half/quartile-income ZIP codes minus that in the top-half/quartile-income ZIP codes, based on the 2011 income levels. Mortgages only contain home purchase loans and are deflated to 2007 U.S. dollars. As a result, the low-minus-high factor for each MSA begins at 0 in 1992. I require that each MSA have at least five ZIP codes. The entire sample for the figure of half (quartile) ZIP groups is a balanced panel of 15,724 ZIP codes and 8 years, consistently covered by the HMDA sample. For the entire period, metropolitan areas are sorted by net export growth (1991-2007) at the MSA level. The low quintile group comprises 1,385 ZIP Codes from 72 MSAs, and the high quintile group comprises 2,611 ZIP Codes from 71 MSAs for the entire period. 
        } %end of small font
    } % end of caption
    \label{fig_ZIP_D12t19_GSEMnPLNJM.Amt07_cbQuint_zipHalf_Quart}
    
\end{figure} 
%------------------------------------

\pagebreak 
%-----------------------------------------------------------------
%%%%%%%%%%%%%%%%%%%%%%%%%%%%%%%%%%%%
% fig_ZIP_D12t19_HPI_cbQuint_zipHalf_Quart.tex
%%%%%%%%%%%%%%%%%%%%%%%%%%%%%%%%%%%%

%------------------------------------------------------------
% fig_ZIP_D12t19_HPI_cbQuint_zipHalf_Quart

%------------------------------------
\begin{figure}[h!] 
    \centering
    \begin{subfigure}[t]{0.9\textwidth}
        \centering
        \includegraphics[height=7.5cm]{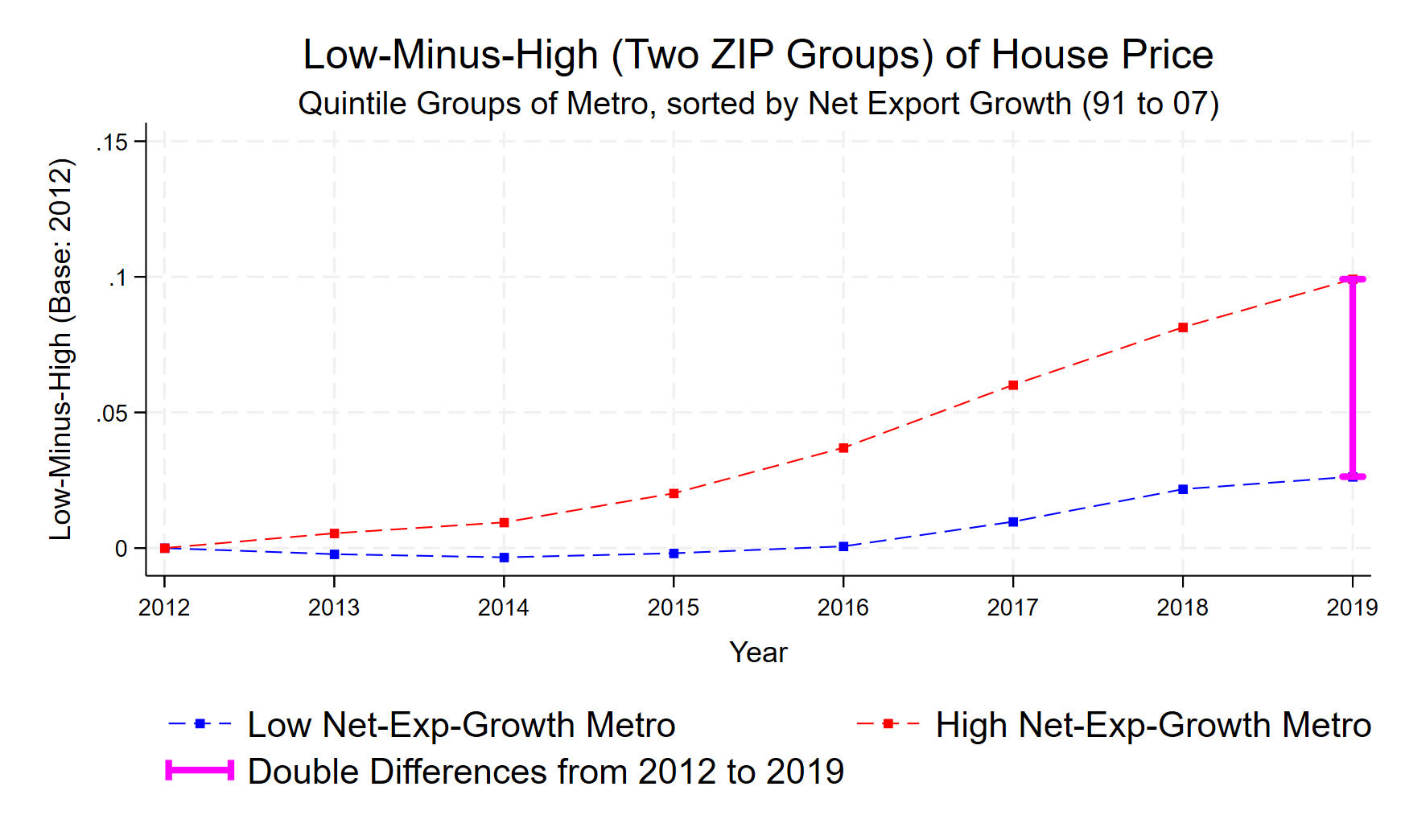}
        \caption{Income Half Groups of ZIP Codes within Metro}
    \end{subfigure}%
    \hfill 
    \begin{subfigure}[t]{0.9\textwidth}
        \centering
        \includegraphics[height=7.5cm]{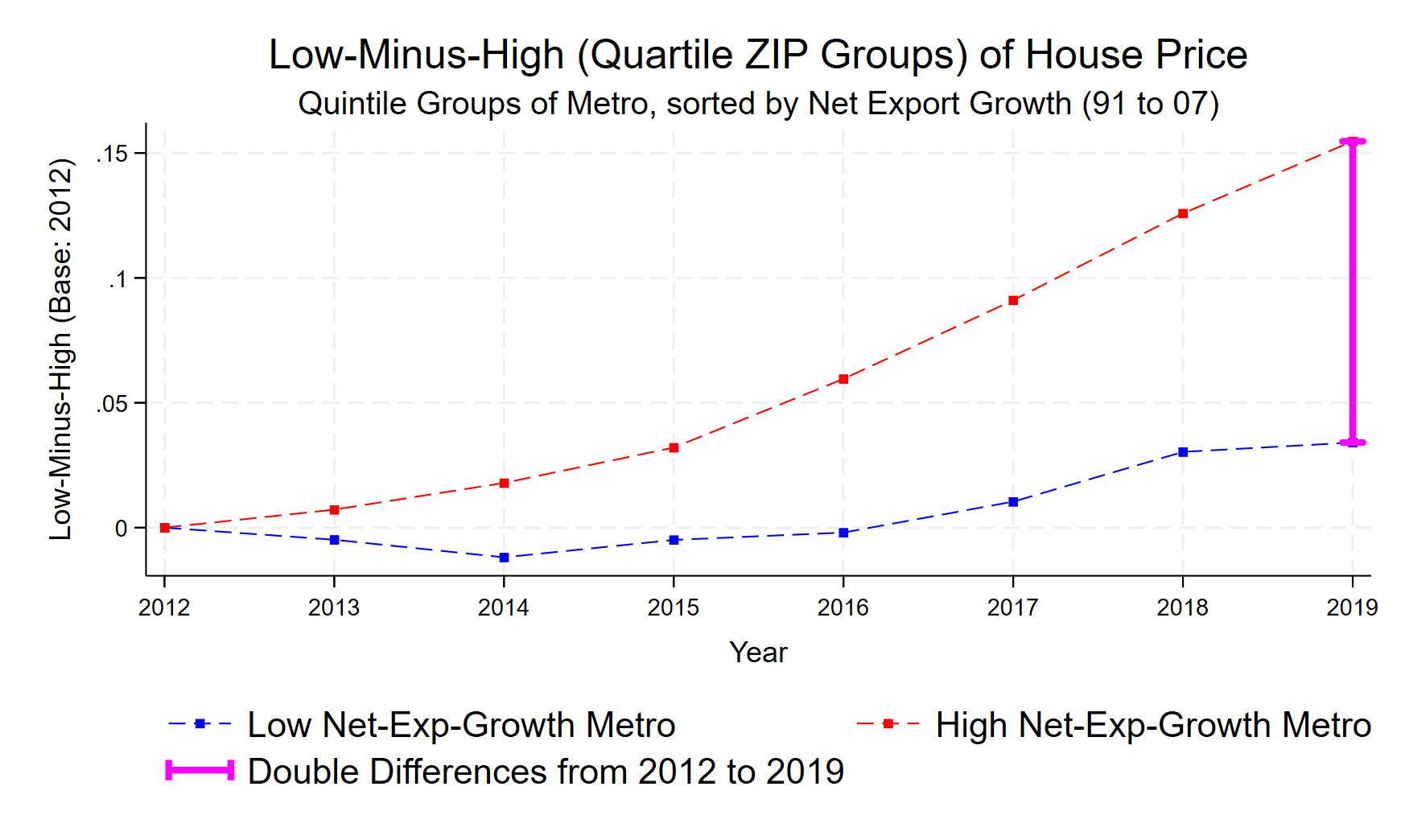}
        \caption{Income Quartile Groups of ZIP Codes within Metro}
    \end{subfigure}
    \caption{\textbf{Low-minus-High Factor for House Price Index Growth (2012-2019) Across ZIP Codes Across Metropolitan Areas.}  \smallskip  \newline 
    {\footnotesize This figure shows the time series of the low-minus-high factor for the house price index for high (in red lines) and low (in blue lines) quintile groups of metropolitan areas (MSAs by CBSA09 code) from 2012 to 2019. Within each MSA, the low-minus-high factor is calculated as the growth of the house price index in the bottom-half/quartile-income ZIP codes minus that in the top-half/quartile-income ZIP codes, based on the 2011 income levels. House price indexes are deflated to 2007 U.S. dollars. As a result, the low-minus-high factor for each MSA begins at 0 in 1992. I require that each MSA have at least five ZIP codes. The entire sample for the figure of half (quartile) ZIP groups is a balanced panel of 11,836 ZIP codes and 8 years, consistently covered by the FHFA HPI ZIP data. For the entire period, metropolitan areas are sorted by net export growth (1991-2007) at the MSA level. The low quintile group comprises 1,089 ZIP Codes from 70 MSAs, and the high quintile group comprises 1,830 ZIP Codes from 69 MSAs for the entire period. 
        } %end of small font
    } % end of caption
    \label{fig_ZIP_D12t19_HPI_cbQuint_zipHalf_Quart}
    
\end{figure} 
%------------------------------------

\pagebreak
%------------------------------------------------------------
% table 0: table_SumStat

%---------------------------------------------------------------

%%%%%%%%%%%%%%%%%%%%%%%%%%%%%%%%%%%%%%%%%%%%%%%%
% table_HPI.D02t06vsD07t09.PLNJM.4Reg
%%%%%%%%%%%%%%%%%%%%%%%%%%%%%%%%%%%%%%%%%%%%%%%%

\noindent 

\begin{table}[h!]
\centering
\caption{
\textbf{Summary Statistics} \smallskip \newline
{\scriptsize
This table reports summary statistics of variables used in regressions, separated into different periods.
} % end of small font size
} % end of caption
\label{table_SumStat}
\resizebox{0.9\columnwidth}{!}{%
\begin{tabular}{l*{6}{c}}

\toprule
Variable Names                     
            &\multicolumn{1}{c}{Count}&\multicolumn{1}{c}{Mean}&\multicolumn{1}{c}{SD}&\multicolumn{1}{c}{P25}&\multicolumn{1}{c}{P50}&\multicolumn{1}{c}{P75}\\

\midrule
\multicolumn{7}{l}{\textbf{Panel A. Prier Period (1991 - 1999)}} \\
\addlinespace
House Price Growth (07USD, 91-99, An)&         672&      0.0130&       0.016&      0.0029&      0.0146&      0.0236\\
PLNJM Growth (07USD, 91-99, An)&         705&      0.1707&      0.0845&      0.1174&      0.1638&      0.2243\\
GSEM Growth (07USD, 91-99, An)&         705&      0.1495&      0.0876&      0.0910&      0.1397&      0.1904\\
Net Export Growth (91-99, An)&         705&     -0.0018&      0.0017&     -0.0021&     -0.0013&     -0.0009\\
GIV Net Export Growth (91-99, An)&         705&     -0.0004&      0.0010&     -0.0006&     -0.0003&     -0.0001\\
Ln(Num of households, 91)&         705&     10.8667&      1.1624&     10.0189&     10.8116&     11.5926\\
Ln(household Income, 91)&         705&     10.3654&      0.1841&     10.2467&     10.3477&     10.4625\\
Ratio of Labor Force (1989)&         705&      0.6694&      0.0558&      0.6357&      0.6712&      0.7053\\
Ln(Num of House Units, 91)&         705&     10.8944&      1.1451&     10.0776&     10.8276&     11.6244\\
Housing supply elasticity&         627&      2.3940&      1.2058&      1.5618&      2.2594&      3.0016\\
House Vacancy Rate (1989)&         705&      0.0827&      0.0465&      0.0549&      0.0711&      0.0961\\
Ratio of Renters (1989)&         705&      0.3158&      0.0977&      0.2482&      0.3061&      0.3690\\
Ratio of Bachelor Educated (1989)&         705&      0.1907&      0.0821&      0.1298&      0.1754&      0.2365\\
Ratio of White Race (1989)&         705&      0.8586&      0.1281&      0.7841&      0.8983&      0.9591\\
Ratio of Immigration (80-90)&         705&      0.0165&      0.0248&      0.0030&      0.0076&      0.0172\\
Ratio of Art, Enter, and Recre Emp (1989)&         705&      0.0059&      0.0046&      0.0041&      0.0053&      0.0067\\
Ratio of Health Emp (1989)&         705&      0.0392&      0.0104&      0.0328&      0.0382&      0.0448\\
Ratio of College Students (1989)&         705&      0.0097&      0.0212&      0.0000&      0.0027&      0.0105\\
\addlinespace
\addlinespace

\midrule
\multicolumn{7}{l}{\textbf{Panel B. Boom Period (1999 - 2005)}} \\
\addlinespace
House Price Growth (07USD, 99-05, An) &         786&      0.0382&      0.0342&      0.0134&      0.0234&      0.0617\\
PLNJM Growth (07USD, 99-05, An)&         792&      0.1698&      0.0814&      0.1160&      0.1649&      0.2164\\
GSEM Growth (07USD, 99-05, An)&         792&      0.0438&      0.0711&      0.0016&      0.0389&      0.0881\\
Net Export Growth (99-05, An)&         792&     -0.0022&      0.0020&     -0.0026&     -0.0018&     -0.0012\\
GIV Net Export Growth (99-05, An)&         792&     -0.0010&      0.0011&     -0.0013&     -0.0008&     -0.0004\\
Ln(Num of households, 99)&         792&     10.8657&      1.1857&     10.0668&     10.8149&     11.6139\\
Ln(household Income, 99)&         792&     10.7197&      0.2369&     10.5625&     10.6946&     10.8402\\
Fraction of Labor Force (1999)&         792&      0.6536&      0.0552&      0.6218&      0.6563&      0.6918\\
Ln(Num of House Units, 99)&         792&     10.9005&      1.1543&     10.0936&     10.8644&     11.6262\\
Housing supply elasticity&         701&      2.4194&      1.2167&      1.6055&      2.2897&      3.0029\\
House Vacancy Rate (1999)&         792&      0.0791&      0.0494&      0.0521&      0.0687&      0.0919\\
Fraction of Renters (1999)&         792&      0.2924&      0.0977&      0.2222&      0.2795&      0.3464\\
Fraction of Bachelor Educated (1999)&         792&      0.2058&      0.0850&      0.1422&      0.1905&      0.2517\\
Fraction of White Race (1999)&         792&      0.8215&      0.1433&      0.7414&      0.8604&      0.9357\\
Fraction of Immigration (90-00)&         792&      0.0255&      0.0291&      0.0066&      0.0148&      0.0325\\
Ratio of Art, Enter, and Recre Emp (1999)&         792&      0.0076&      0.0060&      0.0049&      0.0065&      0.0085\\
Ratio of Health Emp (1999)&         792&      0.0530&      0.0114&      0.0451&      0.0519&      0.0602\\
Ratio of Tradable Service Emp (1999)&         792&      0.0591&      0.0274&      0.0396&      0.0524&      0.0719\\
Ratio of College Students (1999)&         792&      0.0092&      0.0201&      0.0000&      0.0024&      0.0109\\
\addlinespace
\addlinespace

\midrule
\multicolumn{7}{l}{\textbf{Panel C . Bust Period (2007 - 2009)}} \\
\addlinespace
House Price Growth (07USD, 07-09, An)&         786&     -0.0537&      0.0615&     -0.0701&     -0.0332&     -0.0158\\
\addlinespace

\comment{
\midrule
\multicolumn{7}{l}{\textbf{Panel D . 1992 - 2006}} \\
\addlinespace
HH Income Growth (07USD, 92-06, An)&       1,074&       0.015&       0.007&      0.0108&      0.0145&      0.0188\\
Working-age Population Growth (92-06, An)&       1,074&       0.016&       0.013&      0.0067&      0.0131&      0.0223\\
Tot (Ex Const) Emp Shr Growth (92-06, An)&       1,041&       0.002&       0.007&     -0.0009&      0.0023&      0.0055\\
Net Export Growth (92-06, An)&       1,074&      -0.002&       0.002&     -0.0024&     -0.0015&     -0.0010\\
GIV Net Export Growth (92-06, An)&       1,074&      -0.001&       0.001&     -0.0012&     -0.0008&     -0.0004\\
\addlinespace

\midrule
\multicolumn{7}{l}{\textbf{Panel E . Variables Constant Across Periods}} \\
\addlinespace
Housing supply elasticity (Saiz, 2010) &         701&       2.419&       1.217&      1.6055&      2.2897&      3.0029\\
\addlinespace
}

\bottomrule
\end{tabular}

} % end of resize box

\end{table}

%---------------------------------------------------------------------------------------
%---------------------------------------------------------------------------------------
%---------------------------------------------------------------------------------------
%---------------------------------------------------------------------------------------
% Cross-Metro Empirical: Main Tests 1
% HPI Growth in Boom (02-06) & Drop in Bust (07-09)
%---------------------------------------------------------------------------------------
%---------------------------------------------------------------------------------------
%---------------------------------------------------------------------------------------
%---------------------------------------------------------------------------------------

\pagebreak
%-----------------------------------------------------------------

%%%%%%%%%%%%%%%%%%%%%%%%%%%%%%%%%%%%
% table_HPI.D99t05.PLNJM.2SLS
%%%%%%%%%%%%%%%%%%%%%%%%%%%%%%%%%%%%

%-----------------------------------------------------------------

%%%%%%%%%%%%%%%%%%%%%%%%%%%%%%%%%%%%
% table_HPI.D99t05.PLNJM.2SLS
%%%%%%%%%%%%%%%%%%%%%%%%%%%%%%%%%%%%

\noindent 

\begin{table}[h!]
\centering
\caption{
\textbf{2SLS Regression of House Price Growth on PLNJM Growth in Boom Period (99-05)} \smallskip \newline
{\footnotesize 
This table reports the first-stage and the second-stage results of 2SLS regression $\triangle_{99,05} Ln(HPI_{c}) = \beta * \triangle_{99,05} Ln(PLNJM_{c})  + \gamma* \bm{Controls_{c}} + \alpha + \epsilon_{c}$. The left-hand-side dependent variable $\triangle_{99,05} Ln(HPI_{c})$ is the growth rate of the house price index at county $c$ 99-05, and the key variable of interest $\triangle_{99,05} Ln(PLNJM_{c})$ is the growth rate of the dollar amount of private-label (non-jumbo) mortgages (PLNJM) at county $c$ 99-05. $Controls_{c}$ indicates control variables at county $c$ in 1999. We use the gravity model-based instrumental variable ($\triangle_{99,05}\text{givNetExp}_{m}$) as IV for $\triangle_{99,05} Ln(PLNJM_{c})$. For the first-stage F-test, we report Kleibergen-Paap (2006) robust (clustered) statistics and Montiel Olea-Pflueger (2013) efficient statistics. Regression is weighted by the natural logarithm of housing units in 1999. Standard errors are clustered at the CBSA level. ***, **, and * indicate significance at the 1\%, 5\%, and 10\% levels, respectively.
} % end of small font size
} % end of caption
\label{table_HPI.D99t05.PLNJM.2SLS}

\resizebox{0.92\columnwidth}{!}{%

\begin{tabular}{l*{5}{c}}
\toprule
        &\multicolumn{1}{p{3cm}}{\centering PLNJM Growth \\ (99-05, An)}  &\multicolumn{4}{c}{\centering Housing Price Growth (07USD, 99-05, An)} \\
            \cmidrule{2-6} 
            &\multicolumn{1}{c}{(1)}&\multicolumn{1}{c}{(2)}&\multicolumn{1}{c}{(3)}&\multicolumn{1}{c}{(4)}&\multicolumn{1}{c}{(5)}\\
            
\midrule
GIV Net Export Growth (99-05, An)&   14.684***&            &            &            &            \\
               &  (4.014)   &            &            &            &            \\
\addlinespace
PLNJM Growth (07USD, 99-05, An)&            &    0.515***&    0.452***&    0.445***&    0.479***\\
               &            &  (0.110)   &  (0.118)   &  (0.139)   &  (0.156)   \\
\addlinespace
Ln(Num of households, 1999)&   -0.036   &            &    0.004*  &    0.020   &    0.020   \\
               &  (0.034)   &            &  (0.002)   &  (0.016)   &  (0.018)   \\
\addlinespace
Ln(household Income, 1999)&   -0.014   &            &    0.056***&    0.061***&    0.024   \\
               &  (0.035)   &            &  (0.013)   &  (0.015)   &  (0.018)   \\
\addlinespace
Ratio of Labor Force (1999)&    0.448***&            &   -0.176***&   -0.158***&   -0.170*  \\
               &  (0.139)   &            &  (0.044)   &  (0.046)   &  (0.093)   \\
\addlinespace
Ln(Num of House Units, 1999)&    0.042   &            &            &   -0.026   &   -0.028   \\
               &  (0.034)   &            &            &  (0.016)   &  (0.019)   \\
\addlinespace
Housing supply elasticity&   -0.006   &            &            &   -0.003   &   -0.002   \\
               &  (0.004)   &            &            &  (0.002)   &  (0.002)   \\
\addlinespace
Wharton Regulation Index&    0.024***&            &            &    0.006   &    0.004   \\
               &  (0.008)   &            &            &  (0.005)   &  (0.005)   \\
\addlinespace
House Vacancy Rate (1999)&    0.071   &            &            &    0.099** &    0.083*  \\
               &  (0.091)   &            &            &  (0.042)   &  (0.045)   \\
\addlinespace
Ratio of Renters (1999)&   -0.111*  &            &            &    0.151***&    0.081** \\
               &  (0.064)   &            &            &  (0.029)   &  (0.035)   \\
\addlinespace
Ratio of Bachelor Educated (1999)&   -0.204** &            &            &            &    0.106*  \\
               &  (0.093)   &            &            &            &  (0.062)   \\
\addlinespace
Ratio of White Race (1999)&   -0.050   &            &            &            &   -0.007   \\
               &  (0.036)   &            &            &            &  (0.019)   \\
\addlinespace
Ratio of Immigration (90-00)&    0.254   &            &            &            &    0.249** \\
               &  (0.200)   &            &            &            &  (0.103)   \\
\addlinespace
Ratio of Age 65 Above (1999)&    0.504***&            &            &            &   -0.009   \\
               &  (0.178)   &            &            &            &  (0.115)   \\
\addlinespace
Constant       &    0.035   &   -0.048** &   -0.564***&   -0.569***&   -0.144   \\
               &  (0.345)   &  (0.018)   &  (0.132)   &  (0.156)   &  (0.177)   \\
\midrule
Obs            &      712   &      786   &      786   &      707   &      707   \\
R2-adj         &    0.154   &   -0.373   &  -0.0458   &   0.0833   &   0.0134   \\
Cluster SE     &     CBSA   &     CBSA   &     CBSA   &     CBSA   &     CBSA   \\
Weight         & Ln(HU99)   & Ln(HU99)   & Ln(HU99)   & Ln(HU99)   & Ln(HU99)   \\
KP F-Stat      &  13.06          &    23.08   &    20.62   &    16.74   &    13.06   \\
MOP F-Stat     &  13.06          &    23.08   &    20.62   &    16.74   &    13.06   \\
\bottomrule

\end{tabular}

} % end of resize box

\end{table}

\pagebreak 
%---------------------------------------------------------------

%%%%%%%%%%%%%%%%%%%%%%%%%%%%%%%%%%%%%%%%%%%%%%%%
% table_HPI.D99t05.PLNJM.4Reg
%%%%%%%%%%%%%%%%%%%%%%%%%%%%%%%%%%%%%%%%%%%%%%%%

%---------------------------------------------------------------

%%%%%%%%%%%%%%%%%%%%%%%%%%%%%%%%%%%%%%%%%%%%%%%%
% table_HPI.D99t05.PLNJM.4Reg
%%%%%%%%%%%%%%%%%%%%%%%%%%%%%%%%%%%%%%%%%%%%%%%%

\noindent 

\begin{table}[h!]
\centering
\caption{
\textbf{Four Regressions of House Price Growth on PLNJM Growth in Boom Period (99-05)} \smallskip \newline
{\footnotesize 
This table reports OLS, reduced-form, first stage, and second stage results of 2SLS regression $\triangle_{99,05} Ln(HPI_{c}) = \beta * \triangle_{99,05} Ln(PLNJM_{c}) + \gamma* \bm{Controls_{c}} + \alpha + \epsilon_{c}$. The left-hand-side dependent variable $\triangle_{99,05} Ln(HPI_{c})$ is the growth rate of the house price index at county $c$ 99-05, and the key variable of interest $\triangle_{99,05} Ln(PLNJM_{c})$ is the growth rate of the dollar amount of private-label (non-jumbo) mortgages (PLNJM) at county $c$ 99-05. $Controls_{c}$ indicates control variables at county $c$ in 1999. We use the gravity model-based instrumental variable ($\triangle_{99,05}\text{givNetExp}_{m}$) as IV for $\triangle_{99,05} Ln(PLNJM_{c})$. For the first-stage F-test, we report kleibergen-Paap (2006) robust (clustered) statistics and Montiel Olea-Pflueger (2013) efficient statistics. Each regression is weighted by the natural logarithm of housing units in 1999. Standard errors are clustered at the CBSA level. ***, **, and * indicate significance at the 1\%, 5\%, and 10\% levels, respectively.
} % end of small font size
} % end of caption
\label{table_HPI.D99t05.PLNJM.4Reg}

\resizebox{\columnwidth}{!}{%

\begin{tabular}{l*{4}{c}}
\toprule
Dep Var (Panel A, B, and C)                     &\multicolumn{4}{c}{House Price Growth (1999-2005, annualized)} \\
            \cmidrule{2-5} 
            &\multicolumn{1}{c}{(1)}&\multicolumn{1}{c}{(2)}&\multicolumn{1}{c}{(3)}&\multicolumn{1}{c}{(4)}\\
            
\midrule
\multicolumn{5}{l}{\textbf{Panel A. OLS estimates}} \\
\addlinespace
PLNJM Growth (07USD, 99-05, An)&    0.184***&    0.169***&    0.131***&    0.124***\\
               &  (0.026)   &  (0.025)   &  (0.021)   &  (0.020)   \\
\addlinespace
R2-adj         &    0.165   &    0.345   &    0.544   &    0.587   \\
\addlinespace

\midrule
\multicolumn{5}{l}{\textbf{Panel B. Reduced-form estimates}} \\
\addlinespace
GIV Net Export Growth (99-05, An)&    7.877***&    6.738***&    6.402***&    6.853***\\
               &  (1.842)   &  (1.921)   &  (1.616)   &  (1.479)   \\
\addlinespace
R2-adj         &   0.0509   &    0.244   &    0.497   &    0.556   \\
\addlinespace

\midrule
\multicolumn{5}{l}{\textbf{Panel C . 2SLS estimates}} \\
\addlinespace
PLNJM Growth (07USD, 99-05, An)&    0.515***&    0.452***&    0.445***&    0.479***\\
               &  (0.110)   &  (0.118)   &  (0.139)   &  (0.156)   \\
\addlinespace

\addlinespace
\addlinespace

Dep Var (Panel D): &\multicolumn{4}{c}{PLNJM Growth (99-05, An)} \\ 
\midrule 
\multicolumn{5}{l}{\textbf{Panel D . First-stage estimates}} \\
\addlinespace
GIV Net Export Growth (99-05, An)&   15.757***&   15.312***&   14.794***&   14.684***\\
               &  (3.254)   &  (3.334)   &  (3.603)   &  (4.014)   \\
\addlinespace
KP F-Stat      &    23.08   &    20.62   &    16.74   &    13.06   \\
MOP F-Stat     &    23.08   &    20.62   &    16.74   &    13.06   \\
\addlinespace
\midrule
\multicolumn{5}{l}{\textbf{Controls (for all Panels)}} \\
Basic Controls &            &  Y   &   Y    & Y       \\
Housing Controls &           &      & Y       & Y     \\
Demographic Controls &            &      &        &  Y   \\
\midrule              
Obs            &      786   &      786   &      707   &      707   \\
Cluster SE     &     CBSA   &     CBSA   &     CBSA   &     CBSA   \\
Weight         & Ln(HU99)   & Ln(HU99)   & Ln(HU99)   & Ln(HU99)   \\
\bottomrule

\end{tabular}

} % end of resize box

\end{table}

\pagebreak 
%-----------------------------------------------------------------------
%%%%%%%%%%%%%%%%%%%%%%%%%%%%%%%%%%%%
% table_HPI.D99t05vsD07t09.PLNJM.2SLS.wide
%%%%%%%%%%%%%%%%%%%%%%%%%%%%%%%%%%%%

%-----------------------------------------------------------------------
%%%%%%%%%%%%%%%%%%%%%%%%%%%%%%%%%%%%
% table_HPI.D99t05vsD07t09.PLNJM.2SLS.wide
%%%%%%%%%%%%%%%%%%%%%%%%%%%%%%%%%%%%

\noindent 

\begin{table}[h!]
\centering
\caption{
\textbf{2SLS Stacked Regression of Housing Price Growth in Boom (99-05) and Bust (07-09) Periods on PLNJM Growth (99-05)} \smallskip \newline
{\scriptsize
This table reports 2SLS regression $\triangle_{99,05} \& \triangle_{07,09} Ln(HPI_{c}) = \beta_{99,05} * \triangle_{99,05} Ln(PLNJM_{c}) \times Dum_{99,05} + \beta_{07,09} * \triangle_{99,05} Ln(PLNJM_{c}) \times Dum_{07,09} + \gamma_{99,05}* \bm{Controls_{c}} \times Dum_{99,05} + \gamma_{07,09}* \bm{Controls_{c}} \times Dum_{07,09} + \epsilon_{period, c}$. The left-hand-side dependent variable $\triangle_{99,05} \& \triangle_{07,09} Ln(HPI_{c})$ is the stacked growth rate of the house price index at county $c$ 99-05 and 07-09. The key variable of interest $\triangle_{99,05} Ln(PLNJM_{c})$ is the growth rate of the dollar amount of private-label (non-jumbo) mortgages at county $c$ 99-05. $Controls_{c}$ indicates control variables at county $c$ in the period start year 1999. We use the gravity model-based instrumental variable $\triangle_{99,05}\text{givNetExp}_{m}$ as the IV for $\triangle_{99,05}Ln(PLNJM_{c})$. Regression is weighted by the natural logarithm of housing units in 1999.  For the first-stage F-test of two non-stacked samples, we report Kleibergen-Paap (2006) robust (clustered) statistics and Montiel Olea-Pflueger (2013) efficient statistics. Standard errors are clustered at the CBSA level. ***, **, and * indicate significance at the 1\%, 5\%, and 10\% levels, respectively.
\smallskip
} % end of small font size
} % end of caption
\label{table_HPI.D99t05vsD07t09.PLNJM.2SLS.wide}

\vspace{-2mm}

\resizebox{\columnwidth}{!}{%
\begin{tabular}{l*{8}{c}}
\toprule
\textbf{TSLS estimates}            &\multicolumn{8}{c}{Housing Price Growth (99-05 and 07-09, Annualized )} \\
            \cmidrule{2-9} 
            &\multicolumn{2}{c}{(1)}&\multicolumn{2}{c}{(2)}&\multicolumn{2}{c}{(3)}&\multicolumn{2}{c}{(4)}\\
            
\midrule
PLNJM Growth (07USD, 99-05, An) x Dum99t05&    0.515***&  (0.110)&    0.452***&  (0.118)&    0.445***&  (0.139)&    0.479***&  (0.156)\\  \addlinespace
PLNJM Growth (07USD, 99-05, An) x Dum07t09&   -0.907***&  (0.208)&   -0.839***&  (0.211)&   -0.855***&  (0.253)&   -0.889***&  (0.243)\\  \addlinespace
Dum99t05       &   -0.048** &  (0.018)&   -0.564***&  (0.132)&   -0.569***&  (0.156)&   -0.144   &  (0.177)\\  \addlinespace
Dum07t09       &    0.099***&  (0.035)&    0.605***&  (0.171)&    0.533** &  (0.251)&    0.680** &  (0.324)\\  \addlinespace
Ln(Num of HH, 99) x DumD99t05&            &         &    0.004*  &  (0.002)&    0.020   &  (0.016)&    0.020   &  (0.018)\\  \addlinespace
Ln(Num of HH, 99) x DumD07t09&            &         &   -0.008***&  (0.003)&   -0.052*  &  (0.030)&   -0.018   &  (0.029)\\  \addlinespace
Ln(HH Income, 99) x DumD99t05&            &         &    0.056***&  (0.013)&    0.061***&  (0.015)&    0.024   &  (0.018)\\  \addlinespace
Ln(HH Income, 99) x DumD07t09&            &         &   -0.057***&  (0.017)&   -0.057** &  (0.025)&   -0.078** &  (0.032)\\  \addlinespace
Ratio of Labor Force (1999) x DumD99t05&            &         &   -0.176***&  (0.044)&   -0.158***&  (0.046)&   -0.170*  &  (0.093)\\  \addlinespace
Ratio of Labor Force (1999) x DumD07t09&            &         &    0.279***&  (0.055)&    0.328***&  (0.084)&    0.318** &  (0.154)\\  \addlinespace
Ln(Num of HU, 99) x DumD99t05&            &         &            &         &   -0.026   &  (0.016)&   -0.028   &  (0.019)\\  \addlinespace
Ln(Num of HU, 99) x DumD07t09&            &         &            &         &    0.048   &  (0.031)&    0.017   &  (0.030)\\  \addlinespace
Housing supply elasticity x DumD99t05&            &         &            &         &   -0.003   &  (0.002)&   -0.002   &  (0.002)\\  \addlinespace
Housing supply elasticity x DumD07t09&            &         &            &         &    0.003   &  (0.004)&    0.001   &  (0.004)\\  \addlinespace
Wharton Regulation Index x Dum99t05&            &         &            &         &    0.006   &  (0.005)&    0.004   &  (0.005)\\  \addlinespace
Wharton Regulation Index x Dum07t09&            &         &            &         &    0.003   &  (0.009)&    0.003   &  (0.009)\\  \addlinespace
House Vacancy Rate (1999) x DumD99t05&            &         &            &         &    0.099** &  (0.042)&    0.083*  &  (0.045)\\  \addlinespace
House Vacancy Rate (1999) x DumD07t09&            &         &            &         &   -0.003   &  (0.080)&    0.003   &  (0.076)\\  \addlinespace
Ratio of Renters (1999) x DumD99t05&            &         &            &         &    0.151***&  (0.029)&    0.081** &  (0.035)\\  \addlinespace
Ratio of Renters (1999) x DumD07t09&            &         &            &         &   -0.060   &  (0.054)&    0.024   &  (0.054)\\  \addlinespace
Ratio of Bachelor Educated (1999) x DumD99t05&            &         &            &         &            &         &    0.106*  &  (0.062)\\  \addlinespace
Ratio of Bachelor Educated (1999) x DumD07t09&            &         &            &         &            &         &    0.122   &  (0.097)\\  \addlinespace
Ratio of White Race (1999) x DumD99t05&            &         &            &         &            &         &   -0.007   &  (0.019)\\  \addlinespace
Ratio of White Race (1999) x DumD07t09&            &         &            &         &            &         &    0.020   &  (0.034)\\  \addlinespace
Ratio of Immigration (90-00) x DumD99t05&            &         &            &         &            &         &    0.249** &  (0.103)\\  \addlinespace
Ratio of Immigration (90-00) x DumD07t09&            &         &            &         &            &         &   -0.635***&  (0.198)\\  \addlinespace
Ratio of Age 65 Above (1999) x Dum99t05&            &         &            &         &            &         &   -0.009   &  (0.115)\\  \addlinespace
Ratio of Age 65 Above (1999) x Dum07t09&            &         &            &         &            &         &    0.126   &  (0.200)\\  \addlinespace
\midrule
Obs            &     1576   &         &     1576   &         &     1418   &         &     1418   &         \\  
Cluster SE     &     CBSA   &         &     CBSA   &         &     CBSA   &         &     CBSA   &         \\
Weight         & Ln(HU99)   &         & Ln(HU99)   &         & Ln(HU99)   &         & Ln(HU99)   &         \\
KP F-Stat (99-05, non-stack sample)       &    23.08   &         &    20.62   &         &    16.74   &         &    13.06   &         \\
MOP F-Stat (99-05, non-stack sample)  &    23.08   &         &    20.62   &         &    16.74   &         &    13.06   &         \\
CoefEqual\_Chi2 &   23.232   &         &   17.933   &         &   12.133   &         &   12.766   &         \\
CoefEqual\_PValue&    0.000   &         &    0.000   &         &    0.000   &         &    0.000   &         \\
\bottomrule

\end{tabular}

} % end of resize box

\end{table}

\pagebreak 
%---------------------------------------------------------------

%%%%%%%%%%%%%%%%%%%%%%%%%%%%%%%%%%%%%%%%%%%%%%%%
% table_HPI.D99t05vsD07t09.PLNJM.4Reg
%%%%%%%%%%%%%%%%%%%%%%%%%%%%%%%%%%%%%%%%%%%%%%%%

%---------------------------------------------------------------

%%%%%%%%%%%%%%%%%%%%%%%%%%%%%%%%%%%%%%%%%%%%%%%%
% table_HPI.D99t05vsD07t09.PLNJM.4Reg
%%%%%%%%%%%%%%%%%%%%%%%%%%%%%%%%%%%%%%%%%%%%%%%%

\noindent 

\begin{table}[h!]
\centering
\caption{
\textbf{Four Stacked Regressions of Housing Price Growth in Boom (99-05) and Bust (07-09) Periods on PLNJM Growth (99-05)} \smallskip \newline
{\scriptsize
This table reports OLS, reduced-form, first stage, and second stages of stacked 2SLS regression $\triangle_{99,05} \quad \& \quad \triangle_{07,09} Ln(HPI_{c}) = \beta_{99,05} * \triangle_{99,05} Ln(PLNJM_{c}) \times Dum_{99,05} + \beta_{07,09} * \triangle_{99,05} Ln(PLNJM_{c}) \times Dum_{07,09} + \gamma_{99,05}* \bm{Controls_{c}} \times Dum_{99,05} + \gamma_{07,09}* \bm{Controls_{c}} \times Dum_{07,09} + \epsilon_{period, c}$. The left-hand-side dependent variable $\triangle_{99,05} \& \triangle_{07,09} Ln(HPI_{c})$ is the stacked growth rate of the house price index at county $c$ 99-05 and 07-09. The key variable of interest $\triangle_{99,05} Ln(PLNJM_{c}$ is the growth rate of the dollar amount of private-label (non-jumbo) mortgages at county $c$ 99-05. $Controls_{c}$ indicates control variables at county $c$ in the period start year 1999. We use the gravity model-based instrumental variable $\triangle_{99,05}\text{givNetExp}_{m}$ as the IV for $\triangle_{99,05} Ln(PLNJM_{c}$. Each regression is weighted by the natural logarithm of housing units in the start year (either 1999 or 2007). For the first-stage F-test of the non-stacked sample, we report kleibergen-Paap (2006) robust (clustered) statistics and Montiel Olea-Pflueger (2013) efficient statistics. Standard errors are clustered at the CBSA level. ***, **, and * indicate significance at the 1\%, 5\%, and 10\% levels, respectively.
} % end of small font size
} % end of caption
\label{table_HPI.D99t05vsD07t09.PLNJM.4Reg}
\resizebox{0.95\columnwidth}{!}{%
\begin{tabular}{l*{4}{c}}
\toprule
Dep Var (Panel A, B, and C)                      &\multicolumn{4}{c}{House Price Growth (99-05 \& 07-09, annualized)} \\
            \cmidrule{2-5} 
            &\multicolumn{1}{c}{(1)}&\multicolumn{1}{c}{(2)}&\multicolumn{1}{c}{(3)}&\multicolumn{1}{c}{(4)}\\

\midrule
\multicolumn{5}{l}{\textbf{Panel A. OLS estimates}} \\
PLNJM Growth (07USD, 99-05, An) x Dum99t05&    0.184***&    0.169***&    0.131***&    0.124***\\
               &  (0.026)   &  (0.025)   &  (0.021)   &  (0.020)   \\
\addlinespace
PLNJM Growth (07USD, 99-05, An) x Dum07t09&   -0.398***&   -0.372***&   -0.328***&   -0.303***\\
               &  (0.052)   &  (0.050)   &  (0.052)   &  (0.045)   \\
\addlinespace
R2-adj         &    0.595   &    0.667   &    0.707   &    0.747   \\
\addlinespace

\midrule
\multicolumn{5}{l}{\textbf{Panel B. Reduced-form estimates}} \\
GIV Net Export Growth (99-05, An) x Dum99t05&    7.877***&    6.738***&    6.402***&    6.853***\\
               &  (1.843)   &  (1.921)   &  (1.617)   &  (1.479)   \\
\addlinespace
GIV Net Export Growth (99-05, An) x Dum07t09&  -13.862***&  -12.500***&  -12.198***&  -12.608***\\
               &  (2.787)   &  (2.840)   &  (3.031)   &  (2.715)   \\
\addlinespace
R2-adj         &    0.501   &    0.583   &    0.653   &    0.706   \\
\addlinespace

\midrule
\multicolumn{5}{l}{\textbf{Panel C . 2SLS estimates}} \\
\addlinespace
PLNJM Growth (07USD, 99-05, An) x Dum99t05&    0.515***&    0.452***&    0.445***&    0.479***\\
               &  (0.110)   &  (0.118)   &  (0.139)   &  (0.156)   \\
\addlinespace
PLNJM Growth (07USD, 99-05, An) x Dum07t09&   -0.907***&   -0.839***&   -0.855***&   -0.889***\\
               &  (0.208)   &  (0.211)   &  (0.253)   &  (0.243)   \\
               
\addlinespace
\addlinespace

Dep Var (Panel D): &\multicolumn{4}{c}{PLNJM Growth (99-05, An)} \\ 
\midrule 
\multicolumn{5}{l}{\textbf{Panel D . First-stage estimates only for 99-05 (Non-stack sample)}} \\
\addlinespace
GIV Net Export Growth (99-05, An) x Dum99t05&   15.757***&   15.312***&   14.794***&   14.684***\\
               &  (3.254)   &  (3.334)   &  (3.603)   &  (4.014)   \\
\addlinespace
KP F-Stat &    23.08   &    20.62   &    16.74   &    13.06   \\
MOP F-Stat     &    23.08   &    20.62   &    16.74   &    13.06   \\
\addlinespace

\midrule
\multicolumn{5}{l}{\textbf{Controls (for all Panels)}} \\
DumPeriod  &    Y        &  Y   &   Y    & Y        \\
Basic Controls x DumPeriod &            &  Y   &   Y    & Y        \\
Housing Controls x DumPeriod &           &      & Y       & Y        \\
Demographic Controls x DumPeriod &            &      &        &  Y        \\

\midrule              
Obs (Panel A, B, and C)          &     1576   &     1576   &     1418   &     1418   \\
Obs (Panel D)            &      791   &      791   &      712   &      712   \\
Cluster SE     &     CBSA   &     CBSA   &     CBSA   &     CBSA     \\
Weight         & {\scriptsize Ln(HU-Start)}   & {\scriptsize Ln(HU-Start)}   &{\scriptsize Ln(HU-Start)}   &{\scriptsize Ln(HU-Start)}       \\
\bottomrule
\end{tabular}

} % end of resize box

\end{table}

%---------------------------------------------------------------------------------------
%---------------------------------------------------------------------------------------
%---------------------------------------------------------------------------------------
%---------------------------------------------------------------------------------------
% Cross-Metro Empirical: Main Tests 2 Exclusion Restriction
%---------------------------------------------------------------------------------------
%---------------------------------------------------------------------------------------
%---------------------------------------------------------------------------------------
%---------------------------------------------------------------------------------------

%------------------------------------------------------------------
%------------------------------------------------------------------
% Empirical: Main Tests 2 Exclusion Restriction
% (1) Relevance: PLNJM vs GSEM
%------------------------------------------------------------------
%------------------------------------------------------------------

\pagebreak
%---------------------------------------------------------------

%%%%%%%%%%%%%%%%%%%%%%%%%%%%%%%%%%%%
% table_GSEMvsPLNJM.D99t05.2SLS.wide
%%%%%%%%%%%%%%%%%%%%%%%%%%%%%%%%%%%%
%---------------------------------------------------------------

%%%%%%%%%%%%%%%%%%%%%%%%%%%%%%%%%%%%
% table_GSEMvsPLNJM.D99t05.2SLS.wide
%%%%%%%%%%%%%%%%%%%%%%%%%%%%%%%%%%%%

\noindent 

\begin{table}[h!]
\centering
\caption{
\textbf{2SLS Stacked Regression of GSEM and PLNJM Growth in Boom Period (99-05) on Net Export Growth (99-05)} \smallskip \newline
{\footnotesize
This table reports 2SLS stacked regression $\triangle_{99,05} Ln(PLNJM_{c}) \quad  \text{stacked with} \quad  \triangle_{99,05} Ln(GSEM_{c}) = \beta_{G} * \triangle_{99,05} \text{NetExp}_{m} \times Dum_{G} + \beta_{P} * \triangle_{99,05} \text{NetExp}_{m} \times Dum_{P} + \gamma_{G}* \bm{Controls_{c}} \times Dum_{G} + \gamma_{P}* \bm{Controls_{c}} \times Dum_{P}  + \alpha_{G} + \alpha_{P} + \epsilon_{G, c} + \epsilon_{P, c}$. The left-hand-side dependent variable $\triangle_{99,05} Ln(GSEMnPLNJM_{c})$ is the stacked growth rate of the dollar amount of either government-sponsored enterprise mortgages (GSEM) or private-label (non-jumbo) mortgages (PLNJM) at county $c$ 99-05. The key variable of interest $\triangle_{99,05} \text{NetExp}_{m}$ is the growth rate of net export at the metropolitan area (CBSA03 code) $m$ 99-05. $Controls_{c}$ indicates control variables at county $c$ in 1999. We use the gravity model-based instrumental variable ($\triangle_{99,05}\text{givNetExp}_{m}$) as IV for $\triangle_{99,05}\text{NetExp}_{m}$. For the first-stage F-test, we report Kleibergen-Paap (2006) robust (clustered) statistics and Montiel Olea-Pflueger (2013) efficient statistics. Regression is weighted by the natural logarithm of housing units in 1999. We report the statistics and p-values for the tests of coefficient equality between $\beta_{G}$ and $\beta_{P}$. Standard errors are clustered at the CBSA level. ***, **, and * indicate significance at the 1\%, 5\%, and 10\% levels, respectively.
} % end of small font size
\smallskip
} % end of caption
\label{table_GSEMvsPLNJM.D99t05.2SLS.wide}

\vspace{-2mm}

\resizebox{\columnwidth}{!}{%
\begin{tabular}{l*{8}{c}}
\toprule
\textbf{TSLS estimates}            &\multicolumn{8}{c}{GSEM and PLNJM Growth (1999-2005, annualized)} \\
            \cmidrule{2-9} 
            &\multicolumn{2}{c}{(1)}&\multicolumn{2}{c}{(2)}&\multicolumn{2}{c}{(3)}&\multicolumn{2}{c}{(4)}\\
            
\midrule
Net Export Growth (99-05, An) x Dum\_GSEM&    1.889   &  (2.539)&    4.626** &  (2.265)&    3.399   &  (2.393)&    3.603   &  (2.382)\\  \addlinespace
Net Export Growth (99-05, An) x Dum\_PLNJM&   13.694***&  (3.547)&   13.365***&  (3.643)&   13.049***&  (3.924)&   12.851***&  (3.712)\\  \addlinespace
Dum\_GSEM       &    0.046***&  (0.008)&    1.074***&  (0.187)&    0.858***&  (0.214)&    1.208***&  (0.344)\\  \addlinespace
Dum\_PLNJM      &    0.200***&  (0.009)&    0.181   &  (0.212)&    0.605** &  (0.242)&   -0.246   &  (0.359)\\  \addlinespace
Ln(Num of HH, 99) x Dum\_GSEM&            &         &   -0.015***&  (0.002)&    0.008   &  (0.037)&    0.044   &  (0.036)\\  \addlinespace
Ln(Num of HH, 99) x Dum\_PLNJM&            &         &    0.006** &  (0.003)&    0.008   &  (0.032)&   -0.016   &  (0.033)\\  \addlinespace
Ln(HH Income, 99) x Dum\_GSEM&            &         &   -0.067***&  (0.020)&   -0.052** &  (0.024)&   -0.097***&  (0.035)\\  \addlinespace
Ln(HH Income, 99) x Dum\_PLNJM&            &         &   -0.004   &  (0.022)&   -0.050** &  (0.025)&    0.009   &  (0.035)\\  \addlinespace
Ratio of Labor Force (1999) x Dum\_GSEM&            &         &   -0.214***&  (0.075)&   -0.194** &  (0.097)&   -0.182*  &  (0.107)\\  \addlinespace
Ratio of Labor Force (1999) x Dum\_PLNJM&            &         &   -0.021   &  (0.107)&    0.116   &  (0.107)&    0.493***&  (0.144)\\  \addlinespace
Ln(Num of HU, 99) x Dum\_GSEM&            &         &            &         &   -0.021   &  (0.038)&   -0.055   &  (0.037)\\  \addlinespace
Ln(Num of HU, 99) x Dum\_PLNJM&            &         &            &         &    0.001   &  (0.033)&    0.023   &  (0.033)\\  \addlinespace
Housing supply elasticity x Dum\_GSEM&            &         &            &         &   -0.002   &  (0.004)&   -0.003   &  (0.003)\\  \addlinespace
Housing supply elasticity x Dum\_PLNJM&            &         &            &         &   -0.004   &  (0.005)&   -0.004   &  (0.005)\\  \addlinespace
Wharton Regulation Index x Dum\_GSEM&            &         &            &         &   -0.011** &  (0.005)&   -0.012** &  (0.005)\\  \addlinespace
Wharton Regulation Index x Dum\_PLNJM&            &         &            &         &    0.022***&  (0.008)&    0.020** &  (0.008)\\  \addlinespace
House Vacancy Rate (1999) x Dum\_GSEM&            &         &            &         &    0.224** &  (0.104)&    0.223** &  (0.106)\\  \addlinespace
House Vacancy Rate (1999) x Dum\_PLNJM&            &         &            &         &    0.111   &  (0.097)&    0.059   &  (0.093)\\  \addlinespace
Ratio of Renters (1999) x Dum\_GSEM&            &         &            &         &   -0.018   &  (0.053)&    0.041   &  (0.061)\\  \addlinespace
Ratio of Renters (1999) x Dum\_PLNJM&            &         &            &         &   -0.130** &  (0.059)&   -0.104   &  (0.064)\\  \addlinespace
Ratio of Bachelor Educated (1999) x Dum\_GSEM&            &         &            &         &            &         &    0.147*  &  (0.081)\\  \addlinespace
Ratio of Bachelor Educated (1999) x Dum\_PLNJM&            &         &            &         &            &         &   -0.296***&  (0.101)\\  \addlinespace
Ratio of White Race (1999) x Dum\_GSEM&            &         &            &         &            &         &    0.066***&  (0.025)\\  \addlinespace
Ratio of White Race (1999) x Dum\_PLNJM&            &         &            &         &            &         &   -0.056   &  (0.037)\\  \addlinespace
Ratio of Immigration (90-00) x Dum\_GSEM&            &         &            &         &            &         &   -0.219   &  (0.160)\\  \addlinespace
Ratio of Immigration (90-00) x Dum\_PLNJM&            &         &            &         &            &         &    0.298   &  (0.205)\\  \addlinespace
Ratio of Age 65 Above (1999) x Dum\_GSEM&            &         &            &         &            &         &    0.179   &  (0.127)\\  \addlinespace
Ratio of Age 65 Above (1999) x Dum\_PLNJM&            &         &            &         &            &         &    0.638***&  (0.181)\\  \addlinespace
\midrule
Obs            &     1584   &         &     1584   &         &     1426   &         &     1426   &         \\
Cluster SE     &     CBSA   &         &     CBSA   &         &     CBSA   &         &     CBSA   &         \\
Weight         & Ln(HU99)   &         & Ln(HU99)   &         & Ln(HU99)   &         & Ln(HU99)   &         \\
KP F-Stat (Non-Stack Sample)    &   23.36  &  &   23.11  &   &   16.17  &   &   18.39   &  \\
MOP F-Stat (Non-Stack Sample) &   23.36  &  &   23.11  &   &   16.17  &   &   18.39   &  \\
CoefEqual\_Chi2 &    8.165   &         &    7.083   &         &    7.005   &         &    8.222   &         \\
CoefEqual\_PValue&    0.004   &         &    0.008   &         &    0.008   &         &    0.004   &         \\
\bottomrule

\end{tabular}

} % end of resize box

\end{table}

\pagebreak 
%---------------------------------------------------------------

%%%%%%%%%%%%%%%%%%%%%%%%%%%%%%%%%%%%%%%%%%%%%%%%
% table_GSEMvsPLNJM.D99t05.4Reg
%%%%%%%%%%%%%%%%%%%%%%%%%%%%%%%%%%%%%%%%%%%%%%%%
%---------------------------------------------------------------

%%%%%%%%%%%%%%%%%%%%%%%%%%%%%%%%%%%%%%%%%%%%%%%%
% table_GSEMvsPLNJM.D99t05.4Reg
%%%%%%%%%%%%%%%%%%%%%%%%%%%%%%%%%%%%%%%%%%%%%%%%

\noindent 

\begin{table}[h!]
\centering
\caption{
\textbf{Four Stacked Regressions of GSEM and PLNJM Growth in Boom Period (99-05) on Net Export Growth (99-05)} \smallskip \newline
{\scriptsize
This table reports OLS, reduced-form, and the second stage of stacked 2SLS regression $\triangle_{99,05} Ln(GSEMnPLNJM_{c}) = \beta_{G} * \triangle_{99,05} \text{NetExp}_{m} \times Dum_{G} + \beta_{P} * \triangle_{99,05} \text{NetExp}_{m} \times Dum_{P} + \gamma_{G}* \bm{Controls_{c}} \times Dum_{G} + \gamma_{P}* \bm{Controls_{c}} \times Dum_{P}  + \alpha_{G} + \alpha_{P} + \epsilon_{G, c} + \epsilon_{P, c}$. The left-hand-side dependent variable $\triangle_{99,05} Ln(GSEMnPLNJM_{c})$ is the stacked growth rate of the dollar amount of either government-sponsored enterprise mortgages (GSEM) or private-label (non-jumbo) mortgages (PLNJM) at county $c$ 99-05. The key variable of interest $\triangle_{99,05} \text{NetExp}_{m}$ is the growth rate of net export at the metropolitan area (CBSA03 code) $m$ 99-05. $Controls_{c}$ indicates control variables at county $c$ in 1999. We use the gravity model-based instrumental variable ($\triangle_{99,05}\text{givNetExp}_{m}$) as IV for $\triangle_{99,05}\text{NetExp}_{m}$. For the first-stage F-test, we report Kleibergen-Paap (2006) robust (clustered) statistics and Montiel Olea-Pflueger (2013) efficient statistics. Each regression is weighted by the natural logarithm of housing units in 1999. Standard errors are clustered at the CBSA level. ***, **, and * indicate significance at the 1\%, 5\%, and 10\% levels, respectively.
} % end of small font size
} % end of caption
\label{table_GSEMvsPLNJM.D99t05.4Reg}

\resizebox{0.95\columnwidth}{!}{%

\begin{tabular}{l*{4}{c}}
\toprule
Dep Var               &\multicolumn{4}{c}{GSEM and PLNJM Growth (99-05, annualized)} \\
            \cmidrule{2-5} 
            &\multicolumn{1}{c}{(1)}&\multicolumn{1}{c}{(2)}&\multicolumn{1}{c}{(3)}&\multicolumn{1}{c}{(4)}\\
            
\midrule
\multicolumn{5}{l}{\textbf{Panel A. OLS estimates}} \\
\addlinespace
Net Export Growth (99-05, An) x Dum\_GSEM&    2.928** &    4.552***&    3.678***&    3.584***\\
               &  (1.360)   &  (1.380)   &  (1.361)   &  (1.339)   \\
\addlinespace
Net Export Growth (99-05, An) x Dum\_PLNJM&    8.391***&    8.110***&    6.759***&    8.237***\\
               &  (1.552)   &  (1.575)   &  (1.670)   &  (1.864)   \\
\addlinespace
R2-adj         &    0.733   &    0.761   &    0.774   &    0.782   \\
\addlinespace

\midrule
\multicolumn{5}{l}{\textbf{Panel B. Reduced-form estimates}} \\
\addlinespace
GIV Net Export Growth (99-05, An) x Dum\_GSEM&    2.173   &    5.288** &    3.845   &    4.113   \\
               &  (2.987)   &  (2.567)   &  (2.809)   &  (2.724)   \\
\addlinespace
GIV Net Export Growth (99-05, An) x Dum\_PLNJM&   15.753***&   15.278***&   14.763***&   14.670***\\
               &  (3.256)   &  (3.332)   &  (3.625)   &  (4.025)   \\
\addlinespace
R2-adj         &    0.733   &    0.760   &    0.774   &    0.781   \\
\addlinespace

\midrule
\multicolumn{5}{l}{\textbf{Panel C . 2SLS estimates}} \\
\addlinespace
Net Export Growth (99-05, An) x Dum\_GSEM&    1.889   &    4.626** &    3.399   &    3.603   \\
               &  (2.539)   &  (2.265)   &  (2.393)   &  (2.382)   \\
\addlinespace
Net Export Growth (99-05, An) x Dum\_PLNJM&   13.694***&   13.365***&   13.049***&   12.851***\\
               &  (3.547)   &  (3.643)   &  (3.924)   &  (3.712)   \\
\addlinespace
\addlinespace

Dep Var (Panel D): &\multicolumn{4}{c}{Net Export Growth (99-05, An)} \\ 
\midrule 
\multicolumn{5}{l}{\textbf{Panel D . First-stage estimates for non-stack sample}} \\
\addlinespace
GIV Net Export Growth (99-05, An)  &1.150***&    1.143***&    1.131***&    1.142***\\
                &  (0.238)   &  (0.238)   &  (0.281)   &  (0.266)   \\
\addlinespace
KP F-Stat      &    23.36   &    23.11   &    16.17   &    18.39   \\
MOP F-Stat     &    23.36   &    23.11   &    16.17   &    18.39   \\
\addlinespace

\midrule
\multicolumn{5}{l}{\textbf{Controls (for all Panels)}} \\
Dum\_MortgageType  &    Y        &  Y   &   Y    & Y       \\
Basic Controls x Dum\_MortgageType &            &  Y   &   Y    & Y     \\
Housing Controls x Dum\_MortgageType &           &      & Y       & Y    \\
Demographic Controls x Dum\_MortgageType &            &      &        &  Y     \\

\midrule          
Obs (Panel A, B, \& C)        &     1584   &     1584   &     1426   &     1426   \\
Obs (Panel D)        &      792   &      792   &      713   &      713   \\
Cluster SE     &     CBSA   &     CBSA   &     CBSA   &     CBSA      \\
Weight         & Ln(HU99)   & Ln(HU99)   & Ln(HU99)   & Ln(HU99)     \\
\bottomrule

\end{tabular}

} % end of resize box

\end{table}

%---------------------------------------------------------------
%---------------------------------------------------------------
% Empirical: Main Tests 2: Exclusion Restriction
% (2) Fully Controlling Demand via GSEM in Boom (99-05)
%---------------------------------------------------------------
%---------------------------------------------------------------

\pagebreak 
%----------------------------------------------------------------

%%%%%%%%%%%%%%%%%%%%%%%%%%%%%%%%%%%%
% table_HPI.D91t99vsD99t05.2SLS.wide
%%%%%%%%%%%%%%%%%%%%%%%%%%%%%%%%%%%%
%-----------------------------------------------------------------

%%%%%%%%%%%%%%%%%%%%%%%%%%%%%%%%%%%%
% table_HPI.D99t05.PLNJM_m_GSEM.2SLS
%%%%%%%%%%%%%%%%%%%%%%%%%%%%%%%%%%%%

\noindent 

\begin{table}[h!]
\centering
\caption{
\textbf{2SLS Regression of House Price Growth on Differential Growth (PLNJM Minus GSEM) in Boom Period (99-05)} \smallskip \newline
{\footnotesize 
This table reports the first-stage and the second-stage results of 2SLS regression $\triangle_{99,05} Ln(HPI_{c}) = \beta * (\triangle_{99,05} Ln(PLNJM_{c}) - \triangle_{99,05} Ln(GSEM_{c})) + \gamma* \bm{Controls_{c}} + \alpha + \epsilon_{c}$. The left-hand-side dependent variable $\triangle_{99,05} Ln(HPI_{c})$ is the growth rate of the house price index at county $c$ 99-05, and the key variable of interest $(\triangle_{99,05} Ln(PLNJM_{c}) - \triangle_{99,05} Ln(GSEM_{c}))$ is the differential growth rate between (1) the dollar amount of private-label (non-jumbo) mortgages (PLNJM) and (2) the dollar amount of government-sponsored enterprise mortgage (GSEM) at county $c$ 99-05. $Controls_{c}$ indicates control variables at county $c$ in 1999. We use the gravity model-based instrumental variable ($\triangle_{99,05}\text{givNetExp}_{m}$) as IV for the above differential growth rate. For the first-stage F-test, we report Kleibergen-Paap (2006) robust (clustered) statistics and Montiel Olea-Pflueger (2013) efficient statistics. Regression is weighted by the natural logarithm of housing units in 1999. Standard errors are clustered at the CBSA level. ***, **, and * indicate significance at the 1\%, 5\%, and 10\% levels, respectively.
} % end of small font size
} % end of caption
\label{table_HPI.D99t05.PLNJM_m_GSEM.2SLS}

\resizebox{0.9\columnwidth}{!}{%

\begin{tabular}{l*{5}{c}}
\toprule
        &\multicolumn{1}{p{3cm}}{\centering Diff Growth \\ (99-05, An)}  &\multicolumn{4}{c}{\centering Housing Price Growth (07USD, 99-05, An)} \\
            \cmidrule{2-6} 
            &\multicolumn{1}{c}{(1)}&\multicolumn{1}{c}{(2)}&\multicolumn{1}{c}{(3)}&\multicolumn{1}{c}{(4)}&\multicolumn{1}{c}{(5)}\\
            
\midrule
GIV Net Export Growth (99-05, An)&   10.546***&            &            &            &            \\
               &  (3.559)   &            &            &            &            \\
\addlinespace
Diff Growth (PLNJM minus GSEM)&            &    0.622***&    0.741***&    0.648***&    0.712***\\
               &            &  (0.165)   &  (0.249)   &  (0.227)   &  (0.271)   \\
\addlinespace
Ln(Num of households, 1999)&   -0.048   &            &   -0.009   &    0.008   &    0.040   \\
               &  (0.039)   &            &  (0.006)   &  (0.024)   &  (0.031)   \\
\addlinespace
Ln(household Income, 1999)&    0.088** &            &    0.003   &    0.036** &   -0.045   \\
               &  (0.043)   &            &  (0.025)   &  (0.014)   &  (0.039)   \\
\addlinespace
Ratio of Labor Force (1999)&    0.593***&            &   -0.297***&   -0.269***&   -0.365** \\
               &  (0.151)   &            &  (0.083)   &  (0.088)   &  (0.184)   \\
\addlinespace
Ln(Num of House Units, 1999)&    0.067*  &            &            &   -0.025   &   -0.058*  \\
               &  (0.039)   &            &            &  (0.025)   &  (0.034)   \\
\addlinespace
Housing supply elasticity&   -0.002   &            &            &   -0.003   &   -0.003   \\
               &  (0.005)   &            &            &  (0.003)   &  (0.003)   \\
\addlinespace
Wharton Regulation Index&    0.036***&            &            &   -0.005   &   -0.009   \\
               &  (0.008)   &            &            &  (0.009)   &  (0.010)   \\
\addlinespace
House Vacancy Rate (1999)&   -0.137   &            &            &    0.213***&    0.221** \\
               &  (0.106)   &            &            &  (0.076)   &  (0.087)   \\
\addlinespace
Ratio of Renters (1999)&   -0.161*  &            &            &    0.175***&    0.146*  \\
               &  (0.096)   &            &            &  (0.050)   &  (0.078)   \\
\addlinespace
Ratio of Bachelor Educated (1999)&   -0.362***&            &            &            &    0.263** \\
               &  (0.098)   &            &            &            &  (0.122)   \\
\addlinespace
Ratio of White Race (1999)&   -0.113***&            &            &            &    0.049   \\
               &  (0.039)   &            &            &            &  (0.042)   \\
\addlinespace
Ratio of Immigration (90-00)&    0.482** &            &            &            &    0.029   \\
               &  (0.212)   &            &            &            &  (0.197)   \\
\addlinespace
Ratio of Age 65 Above (1999)&    0.355*  &            &            &            &   -0.013   \\
               &  (0.181)   &            &            &            &  (0.159)   \\
\addlinespace
Constant       &   -1.219***&   -0.040*  &    0.210   &   -0.132   &    0.723*  \\
               &  (0.410)   &  (0.021)   &  (0.294)   &  (0.146)   &  (0.431)   \\
\midrule
Obs            &      712   &      786   &      786   &      707   &      707   \\
R2-adj         &    0.303   &   -1.359   &   -1.947   &   -1.164   &   -1.484   \\
Cluster SE     &     CBSA   &     CBSA   &     CBSA   &     CBSA   &     CBSA   \\
Weight         & Ln(HU99)   & Ln(HU99)   & Ln(HU99)   & Ln(HU99)   & Ln(HU99)   \\
CD F-Stat      &  10.41          &    15.54   &    9.411   &    10.41   &    10.41   \\
KP F-Stat      &  8.050          &    12.34   &    8.961   &    10.36   &    8.050   \\
MOP F-Stat     &  8.050          &    12.34   &    8.961   &    10.36   &    8.050   \\
\bottomrule

\end{tabular}

} % end of resize box

\end{table}

\pagebreak 
%----------------------------------------------------------------
%%%%%%%%%%%%%%%%%%%%%%%%%%%%%%%%%%%%%%%%%%%%%%%%
% table_HPI.D91t99vsD99t05.4Reg
%%%%%%%%%%%%%%%%%%%%%%%%%%%%%%%%%%%%%%%%%%%%%%%%
%---------------------------------------------------------------

%%%%%%%%%%%%%%%%%%%%%%%%%%%%%%%%%%%%%%%%%%%%%%%%
% table_HPI.D99t05.PLNJM_m_GSEM.4Reg
%%%%%%%%%%%%%%%%%%%%%%%%%%%%%%%%%%%%%%%%%%%%%%%%

\noindent 

\begin{table}[h!]
\centering
\caption{
\textbf{Four Regressions of House Price Growth on Differential Growth Rate (PLNJM Minus GSEM) in Boom Period (99-05)} \smallskip \newline
{\footnotesize 
This table reports OLS, reduced-form, first stage, and second stage results of 2SLS regression $\triangle_{99,05} Ln(HPI_{c}) = \beta * (\triangle_{99,05} Ln(PLNJM_{c}) - \triangle_{99,05} Ln(GSEM_{c})) + \gamma* \bm{Controls_{c}} + \alpha + \epsilon_{c}$. The left-hand-side dependent variable $\triangle_{99,05} Ln(HPI_{c})$ is the growth rate of the house price index at county $c$ 99-05, and the key variable of interest $(\triangle_{99,05} Ln(PLNJM_{c}) - \triangle_{99,05} Ln(GSEM_{c}))$ is the differential growth rate between (1) the dollar amount of private-label (non-jumbo) mortgages (PLNJM) and (2) the dollar amount of government-sponsored enterprise mortgage (GSEM) at county $c$ 99-05. $Controls_{c}$ indicates control variables at county $c$ in 1999. We use the gravity model-based instrumental variable ($\triangle_{99,05}\text{givNetExp}_{m}$) as IV for the above differential growth rate. For the first-stage F-test, we report Kleibergen-Paap (2006) robust (clustered) statistics and Montiel Olea-Pflueger (2013) efficient statistics. Regression is weighted by the natural logarithm of housing units in 1999. Standard errors are clustered at the CBSA level. ***, **, and * indicate significance at the 1\%, 5\%, and 10\% levels, respectively.
} % end of small font size
} % end of caption
\label{table_HPI.D99t05.PLNJM_m_GSEM.4Reg}

\resizebox{0.95\columnwidth}{!}{%

\begin{tabular}{l*{4}{c}}
\toprule
Dep Var (Panel A, B, and C)                     &\multicolumn{4}{c}{House Price Growth (1999-2005, annualized)} \\
            \cmidrule{2-5} 
            &\multicolumn{1}{c}{(1)}&\multicolumn{1}{c}{(2)}&\multicolumn{1}{c}{(3)}&\multicolumn{1}{c}{(4)}\\
            
\midrule
\multicolumn{5}{l}{\textbf{Panel A. OLS estimates}} \\
\addlinespace
Diff Growth (PLNJM minus GSEM)&    0.154***&    0.120***&    0.094***&    0.083***\\
               &  (0.021)   &  (0.023)   &  (0.019)   &  (0.017)   \\
\addlinespace
R2-adj         &    0.163   &    0.290   &    0.512   &    0.552   \\
\addlinespace

\midrule
\multicolumn{5}{l}{\textbf{Panel B. Reduced-form estimates}} \\
\addlinespace
GIV Net Export Growth (99-05, An)&    7.877***&    6.738***&    6.402***&    6.853***\\
               &  (1.842)   &  (1.921)   &  (1.616)   &  (1.479)   \\
\addlinespace
R2-adj         &   0.0509   &    0.244   &    0.497   &    0.556   \\
\addlinespace

\midrule
\multicolumn{5}{l}{\textbf{Panel C . 2SLS estimates}} \\
\addlinespace
Diff Growth (PLNJM minus GSEM)&    0.622***&    0.741***&    0.648***&    0.712***\\
               &  (0.165)   &  (0.249)   &  (0.227)   &  (0.271)   \\
\addlinespace

\addlinespace
\addlinespace

Dep Var (Panel D): &\multicolumn{4}{c}{PLNJM Growth (99-05, An)} \\ 
\midrule 
\multicolumn{5}{l}{\textbf{Panel D . First-stage estimates}} \\
\addlinespace
GIV Net Export Growth (99-05, An)&   13.578***&    9.944***&   10.874***&   10.546***\\
               &  (3.902)   &  (3.221)   &  (3.319)   &  (3.559)   \\
\addlinespace
CD F-Stat      &    15.54   &    9.411   &    10.41   &    10.41   \\
KP F-Stat      &    12.34   &    8.961   &    10.36   &    8.050   \\
MOP F-Stat     &    12.34   &    8.961   &    10.36   &    8.050   \\
\addlinespace
\midrule
\multicolumn{5}{l}{\textbf{Controls (for all Panels)}} \\
Basic Controls &            &  Y   &   Y    & Y       \\
Housing Controls &           &      & Y       & Y     \\
Demographic Controls &            &      &        &  Y   \\
\midrule              
Obs            &      791   &      791   &      712   &      712   \\
Cluster SE     &     CBSA   &     CBSA   &     CBSA   &     CBSA   \\
Weight         & Ln(HU99)   & Ln(HU99)   & Ln(HU99)   & Ln(HU99)   \\
\bottomrule

\end{tabular}

} % end of resize box

\end{table}

%---------------------------------------------------------------
%---------------------------------------------------------------
% Empirical: Main Tests 2: Exclusion Restriction
% (3) Prior (91-99) vs. Boom (99-05) 
%---------------------------------------------------------------
%---------------------------------------------------------------

\pagebreak 
%----------------------------------------------------------------

%%%%%%%%%%%%%%%%%%%%%%%%%%%%%%%%%%%%
% table_HPI.D91t99vsD99t05.2SLS.wide
%%%%%%%%%%%%%%%%%%%%%%%%%%%%%%%%%%%%
%-----------------------------------------------------------------------
%%%%%%%%%%%%%%%%%%%%%%%%%%%%%%%%%%%%
% table_HPI.D91t99vsD99t05.2SLS.wide
%%%%%%%%%%%%%%%%%%%%%%%%%%%%%%%%%%%%

\noindent 

\begin{table}[h!]
\centering
\caption{
\textbf{2SLS Stacked Regression of House Price Growth on Net Export Growth in Prior (91-99) and Boom (99-05) Periods} \smallskip \newline
{\scriptsize 
This table reports 2SLS stacked regression $\triangle_{91,99} \& \triangle_{99,05} Ln(HPI_{c}) = \beta_{91,99} * \triangle_{91,99} \text{NetExp}_{m} \times Dum_{91,99} + \beta_{99,05} * \triangle_{99,05} \text{NetExp}_{m} \times Dum_{99,05} + \gamma_{91,99}* \bm{Controls_{c}} \times Dum_{91,99} + \gamma_{99,05}* \bm{Controls_{c}} \times Dum_{99,05}  + \alpha_{91,99} + \alpha_{99,05} + \epsilon_{period, c}$. The left-hand-side dependent variable $\triangle_{91,99} \& \triangle_{99,05} Ln(HPI_{c})$ is the stacked growth rate of house price index at county $c$ 91-99 and 99-05. The key variable of interest $\triangle_{91,99} \text{NetExp}_{m}$ and $\triangle_{99,05} \text{NetExp}_{m}$ are the growth rate of net export at the metropolitan area (CBSA03 code) $m$ 91-99 and 99-05, respectively. $Controls_{c}$ indicates control variables at county $c$ in the period start year either 1991 or 1999. We use the gravity model-based instrumental variable $\triangle_{91,99}\text{givNetExp}_{m}$ and $\triangle_{99,05}\text{givNetExp}_{m}$ as IVs for $\triangle_{91,99}\text{NetExp}_{m}$ and $\triangle_{99,05}\text{NetExp}_{m}$. Regression is weighted by the natural logarithm of housing units in start year (either 1991 or 1999). We report the statistics and p-values for the tests of coefficient equality between $\beta_{91,99}$ and $\beta_{99,05}$. For the first-stage F-test of two non-stacked samples, we report Kleibergen-Paap (2006) robust (clustered) statistics and Montiel Olea-Pflueger (2013) efficient statistics. We report Standard errors are clustered at the CBSA level. ***, **, and * indicate significance at the 1\%, 5\%, and 10\% levels, respectively.
\smallskip
} % end of small font size
} % end of caption
\label{table_HPI.D91t99vsD99t05.2SLS.wide}

\vspace{-2mm}

\resizebox{\columnwidth}{!}{%

\begin{tabular}{l*{8}{c}}
\toprule
\textbf{TSLS estimates}            &\multicolumn{8}{c}{House Price Growth (07USD, 91-99 \& 99-05, annualized)} \\
            \cmidrule{2-9} 
            &\multicolumn{2}{c}{(1)}&\multicolumn{2}{c}{(2)}&\multicolumn{2}{c}{(3)}&\multicolumn{2}{c}{(4)}\\
            
\midrule
Net Export Growth (91-99, An) x Dum91t99&   -3.708** &  (1.737)&   -3.275*  &  (1.740)&   -3.169   &  (1.988)&   -2.972   &  (2.009)\\ \addlinespace
Net Export Growth (99-05, An) x Dum99t05&    6.827***&  (2.246)&    5.878** &  (2.307)&    5.630** &  (2.277)&    5.979***&  (1.996)\\ \addlinespace
Dummry 91-99   &    0.007*  &  (0.003)&    0.192***&  (0.047)&    0.172***&  (0.056)&    0.275***&  (0.083)\\ \addlinespace
Dummry 99-05   &    0.056***&  (0.006)&   -0.467***&  (0.071)&   -0.291***&  (0.070)&   -0.254** &  (0.113)\\ \addlinespace
Ln(Num of HH, 1991) x Dum91t99&            &         &   -0.003***&  (0.001)&   -0.036***&  (0.010)&   -0.027***&  (0.009)\\ \addlinespace
Ln(Num of HH, 1999) x Dum99t05&            &         &    0.006***&  (0.002)&    0.014   &  (0.011)&   -0.001   &  (0.009)\\ \addlinespace
Ln(HH Income, 1991) x Dum91t99&            &         &   -0.019***&  (0.005)&   -0.018***&  (0.006)&   -0.029***&  (0.009)\\ \addlinespace
Ln(HH Income, 1999) x Dum99t05&            &         &    0.053***&  (0.007)&    0.038***&  (0.007)&    0.028***&  (0.011)\\ \addlinespace
Ratio of Labor Force (1989) x Dum91t99&            &         &    0.052** &  (0.021)&    0.068***&  (0.026)&    0.032   &  (0.028)\\ \addlinespace
Ratio of Labor Force (1999) x Dum99t05&            &         &   -0.187***&  (0.046)&   -0.104** &  (0.046)&    0.073   &  (0.054)\\ \addlinespace
Ln(Num of HU, 1991) x Dum91t99&            &         &            &         &    0.036***&  (0.010)&    0.027***&  (0.009)\\ \addlinespace
Ln(Num of HU, 1999) x Dum99t05&            &         &            &         &   -0.016   &  (0.011)&   -0.004   &  (0.009)\\ \addlinespace
Housing supply elasticity x Dum91t99&            &         &            &         &   -0.001   &  (0.001)&   -0.002   &  (0.001)\\ \addlinespace
Housing supply elasticity x Dum99t05&            &         &            &         &   -0.005***&  (0.002)&   -0.005***&  (0.002)\\ \addlinespace
Wharton Regulation Index x Dum91t99&            &         &            &         &   -0.005*  &  (0.003)&   -0.005*  &  (0.003)\\ \addlinespace
Wharton Regulation Index x Dum99t05&            &         &            &         &    0.016***&  (0.003)&    0.014***&  (0.003)\\ \addlinespace
House Vacancy Rate (1989) x Dum91t99&            &         &            &         &   -0.084***&  (0.025)&   -0.072***&  (0.025)\\ \addlinespace
House Vacancy Rate (1999) x Dum99t05&            &         &            &         &    0.139***&  (0.033)&    0.099***&  (0.033)\\ \addlinespace
Ratio of Renters (1989) x Dum91t99&            &         &            &         &   -0.019** &  (0.009)&   -0.006   &  (0.013)\\ \addlinespace
Ratio of Renters (1999) x Dum99t05&            &         &            &         &    0.094***&  (0.019)&    0.028   &  (0.023)\\ \addlinespace
Ratio of Bachelor Educated (1989) x Dum91t99&            &         &            &         &            &         &    0.035** &  (0.016)\\ \addlinespace
Ratio of Bachelor Educated (1999) x Dum99t05&            &         &            &         &            &         &   -0.037   &  (0.032)\\ \addlinespace
Ratio of White Race (1989) x Dum91t99&            &         &            &         &            &         &    0.019** &  (0.008)\\ \addlinespace
Ratio of White Race (1999) x Dum99t05&            &         &            &         &            &         &   -0.038** &  (0.016)\\ \addlinespace
Ratio of Immigration (80-90) x Dum91t99&            &         &            &         &            &         &   -0.115***&  (0.044)\\ \addlinespace
Ratio of Immigration (90-00) x Dum99t05&            &         &            &         &            &         &    0.397***&  (0.075)\\ \addlinespace
Ratio of Age 65 Above (1989) x Dum91t99&            &         &            &         &            &         &   -0.057*  &  (0.033)\\ \addlinespace
Ratio of Age 65 Above (1999) x Dum99t05&            &         &            &         &            &         &    0.296***&  (0.060)\\ \addlinespace
\midrule
Obs            &     1459   &         &     1459   &         &     1315   &         &     1315   &         \\
Cluster SE     &     CBSA   &         &     CBSA   &         &     CBSA   &         &     CBSA   &         \\
Weight         & {\scriptsize Ln(HU-Start)}   &    & {\scriptsize Ln(HU-Start)}   &    &{\scriptsize Ln(HU-Start)}   &    &{\scriptsize Ln(HU-Start)}   &   \\
KP F-Stat (91-99, Non-Stack Sample)      &    36.93   &         &    38.81   &         &    29.20   &         &    28.53   &         \\
MOP F-Stat (91-99, Non-Stack Sample)  &    36.93   &         &    38.81   &         &    29.20   &         &    28.53   &         \\
KP F-Stat (99-05, Non-Stack Sample)       &    23.25   &         &    23.00   &         &    16.18   &         &    18.32   &         \\
MOP F-Stat (99-05, Non-Stack Sample) &    23.25   &         &    23.00   &         &    16.18   &         &    18.32   &         \\
CoefEqual\_Chi2 &   12.057   &         &    8.189   &         &    6.914   &         &    8.416   &         \\
CoefEqual\_PValue&    0.001   &         &    0.004   &         &    0.009   &         &    0.004   &         \\
\bottomrule

\end{tabular}

} % end of resize box

\end{table}

\pagebreak 
%----------------------------------------------------------------
%%%%%%%%%%%%%%%%%%%%%%%%%%%%%%%%%%%%%%%%%%%%%%%%
% table_HPI.D91t99vsD99t05.4Reg
%%%%%%%%%%%%%%%%%%%%%%%%%%%%%%%%%%%%%%%%%%%%%%%%
%---------------------------------------------------------------

%%%%%%%%%%%%%%%%%%%%%%%%%%%%%%%%%%%%%%%%%%%%%%%%
% table_HPI.D91t99vsD99t05.4Reg
%%%%%%%%%%%%%%%%%%%%%%%%%%%%%%%%%%%%%%%%%%%%%%%%

\noindent 

\begin{table}[h!]
\centering
\caption{
\textbf{Four Stacked Regressions of House Price Growth on Net Export Growth in Prior (91-99) and Boom (99-05) Periods} \smallskip \newline
{\scriptsize
This table reports OLS, reduced-form, first stage and second stages of stacked 2SLS regression $\triangle_{91,99} \& \triangle_{99,05} Ln(HPI_{c}) = \beta_{91,99} * \triangle_{91,99} \text{NetExp}_{m} \times Dum_{91,99} + \beta_{99,05} * \triangle_{99,05} \text{NetExp}_{m} \times Dum_{99,05} + \gamma_{91,99}* \bm{Controls_{c}} \times Dum_{91,99} + \gamma_{99,05}* \bm{Controls_{c}} \times Dum_{99,05}  + \alpha_{91,99} + \alpha_{99,05} + \epsilon_{period, c}$. The left-hand-side dependent variable $\triangle_{91,99} \& \triangle_{99,05} Ln(HPI_{c})$ is the stacked growth rate of house price index at county $c$ 91-99 and 99-05. The key variable of interest $\triangle_{91,99} \text{NetExp}_{m}$ and $\triangle_{99,05} \text{NetExp}_{m}$ are the growth rate of net export at the metropolitan area (CBSA03 code) $m$ 91-99 and 99-05, respectively. $Controls_{c}$ indicates control variables at county $c$ in the period start year either 1991 or 1999. We use the gravity model-based instrumental variable $\triangle_{91,99}\text{givNetExp}_{m}$ and $\triangle_{99,05}\text{givNetExp}_{m}$ as IVs for $\triangle_{91,99}\text{NetExp}_{m}$ and $\triangle_{99,05}\text{NetExp}_{m}$. Each regression is weighted by the natural logarithm of housing units in the start year (either 1991 or 1999). For the first-stage F-test of two non-stacked samples, we report Kleibergen-Paap (2006) robust (clustered) statistics and Montiel Olea-Pflueger (2013) efficient statistics. Standard errors are clustered at the CBSA level. ***, **, and * indicate significance at the 1\%, 5\%, and 10\% levels, respectively.
} % end of small font size
} % end of caption
\label{table_HPI.D91t99vsD99t05.4Reg}
\resizebox{0.78\columnwidth}{!}{%
\begin{tabular}{l*{4}{c}}
\toprule
Dep Var (Panel A, B, and C)                      &\multicolumn{4}{c}{House Price Growth (91-99 \& 99-05, annualized)} \\
            \cmidrule{2-5} 
            &\multicolumn{1}{c}{(1)}&\multicolumn{1}{c}{(2)}&\multicolumn{1}{c}{(3)}&\multicolumn{1}{c}{(4)}\\

\midrule
\multicolumn{5}{l}{\textbf{Panel A. OLS estimates}} \\
Net Export Growth (91-99, An) x Dum91t99&   -0.875   &   -0.280   &    0.136   &    0.464   \\
               &  (1.037)   &  (0.975)   &  (0.984)   &  (0.967)   \\
\addlinespace
Net Export Growth (99-05, An) x Dum99t05&    4.441***&    3.720***&    1.938** &    2.389***\\
               &  (1.030)   &  (1.058)   &  (0.823)   &  (0.757)   \\
\addlinespace
R2-adj        &    0.560   &    0.641   &    0.734   &    0.760   \\
\addlinespace

\midrule
\multicolumn{5}{l}{\textbf{Panel B. Reduced-form estimates}} \\
GIV Net Export Growth (91-99, An) x Dum91t99&   -3.894** &   -3.453** &   -3.155*  &   -2.933*  \\
               &  (1.592)   &  (1.619)   &  (1.733)   &  (1.741)   \\
\addlinespace
GIV Net Export Growth (99-05, An) x Dum99t05&    7.877***&    6.738***&    6.402***&    6.853***\\
               &  (1.843)   &  (1.922)   &  (1.618)   &  (1.481)   \\
\addlinespace
R2-adj         &    0.559   &    0.640   &    0.744   &    0.770   \\
\addlinespace

\midrule
\multicolumn{5}{l}{\textbf{Panel C . 2SLS estimates}} \\
\addlinespace
Net Export Growth (91-99, An) x Dum91t99&   -3.708** &   -3.275*  &   -3.169   &   -2.972   \\
               &  (1.737)   &  (1.740)   &  (1.988)   &  (2.009)   \\
\addlinespace
Net Export Growth (99-05, An) x Dum99t05&    6.827***&    5.878** &    5.630** &    5.979***\\
               &  (2.246)   &  (2.307)   &  (2.277)   &  (1.996)   \\
\addlinespace
\addlinespace

Dep Var (Panel D): &\multicolumn{4}{c}{Net Export Growth (91-99, An)} \\ 
\midrule 
\multicolumn{5}{l}{\textbf{Panel D . First-stage estimates only for 91-99 (non-stack sample)}} \\
\addlinespace
GIV Net Export Growth (91-99, An)&    1.050***&    1.055***&    0.996***&    0.987***\\
               &  (0.173)   &  (0.169)   &  (0.184)   &  (0.185)   \\
\addlinespace
KP F-Stat      &    36.93   &    38.81   &    29.20   &    28.53   \\
MOP F-Stat &    36.93   &    38.81   &    29.20   &    28.53   \\

\addlinespace
\addlinespace

Dep Var (Panel E): &\multicolumn{4}{c}{Net Export Growth (99-05, An)} \\ 
\midrule 
\multicolumn{5}{l}{\textbf{Panel E . First-stage estimates only for 99-05 (non-stack sample)}} \\
\addlinespace
GIV Net Export Growth (99-05, An)&    1.150***&    1.143***&    1.132***&    1.142***\\
               &  (0.238)   &  (0.238)   &  (0.281)   &  (0.266)   \\
\addlinespace
KP F-Stat      &    23.25   &    23.00   &    16.18   &    18.32   \\
MOP F-Stat &    23.25   &    23.00   &    16.18   &    18.32   \\

\midrule
\multicolumn{5}{l}{\textbf{Controls (for all Panels)}} \\
Basic Controls x DumPeriod &            &  Y   &   Y    & Y        \\
Housing Controls x DumPeriod &           &      & Y       & Y        \\
Demographic Controls x DumPeriod &            &      &        &  Y        \\
\midrule              
Obs (Panel A, B, and C)         &     1459   &     1459   &     1315   &     1315   \\
Obs (Panel D)           &      673   &      673   &      608   &      608   \\
Obs (Panel E)          &      791   &      791   &      712   &      712   \\
Cluster SE     &     CBSA   &     CBSA   &     CBSA   &     CBSA   \\    
Weight         & {\scriptsize Ln(HU-Start)}   & {\scriptsize Ln(HU-Start)}   &{\scriptsize Ln(HU-Start)}   &{\scriptsize Ln(HU-Start)}       \\
\bottomrule
\end{tabular}

} % end of resize box

\end{table}

\pagebreak 
%----------------------------------------------------------------
%%%%%%%%%%%%%%%%%%%%%%%%%%%%%%%%%%%%
% table_PLNJM.D91t99vsD99t05.2SLS.wide
%%%%%%%%%%%%%%%%%%%%%%%%%%%%%%%%%%%%
%----------------------------------------------------------------

%%%%%%%%%%%%%%%%%%%%%%%%%%%%%%%%%%%%
% table_PLNJM.D91t99vsD99t05.2SLS.wide
%%%%%%%%%%%%%%%%%%%%%%%%%%%%%%%%%%%%

\noindent 

\begin{table}[h!]
\centering
\caption{
\textbf{2SLS Stacked Regression of PLNJM Growth in Prior (91-99) and Boom (99-05) Periods on Net Export Growth (99-05)} \smallskip \newline
{\scriptsize 
This table reports 2SLS stacked regression $\triangle_{91,99} \& \triangle_{99,05} Ln(PLNJM_{c}) = \beta_{91,99} * \triangle_{91,99} \text{NetExp}_{m} \times Dum_{91,99} + \beta_{99,05} * \triangle_{99,05} \text{NetExp}_{m} \times Dum_{99,05} + \gamma_{91,99}* \bm{Controls_{c}} \times Dum_{91,99} + \gamma_{99,05}* \bm{Controls_{c}} \times Dum_{99,05}  + \alpha_{91,99} + \alpha_{99,05} + \epsilon_{period, c}$. The left-hand-side dependent variable $\triangle_{91,99} \& \triangle_{99,05} Ln(PLNJM_{c})$ is the stacked growth rate of the dollar amount of private-label (non-jumbo) mortgages (PLNJM) at county $c$ 91-99 and 99-05. The key variable of interest $\triangle_{91,99} \text{NetExp}_{m}$ and $\triangle_{99,05} \text{NetExp}_{m}$ are the growth rate of net export at the metropolitan area (CBSA03 code) $m$ 91-99 and 99-05, respectively. $Controls_{c}$ indicates control variables at county $c$ in the period start year either 1991 or 1999. We use the gravity model-based instrumental variable $\triangle_{91,99}\text{givNetExp}_{m}$ and $\triangle_{99,05}\text{givNetExp}_{m}$ as IVs for $\triangle_{91,99}\text{NetExp}_{m}$ and $\triangle_{99,05}\text{NetExp}_{m}$. Regression is weighted by the natural logarithm of housing units in the start year (either 1991 or 1999). We report the statistics and p-values for the tests of coefficient equality between $\beta_{91,99}$ and $\beta_{99,05}$. For the first-stage F-test of two separate non-stack samples, we report kleibergen-Paap (2006) robust (clustered) statistics and Montiel Olea-Pflueger (2013) efficient statistics. Standard errors are clustered at the CBSA level. ***, **, and * indicate significance at the 1\%, 5\%, and 10\% levels, respectively.
\smallskip
} % end of small font size
} % end of caption
\label{table_PLNJM.D91t99vsD99t05.2SLS.wide}

\vspace{-2mm}

\resizebox{\columnwidth}{!}{%

\begin{tabular}{l*{8}{c}}
\toprule
\textbf{TSLS estimates}            &\multicolumn{8}{c}{PLNJM Growth (91-99 and 99-05, annualized)} \\
            \cmidrule{2-9} 
            &\multicolumn{2}{c}{(1)}&\multicolumn{2}{c}{(2)}&\multicolumn{2}{c}{(3)}&\multicolumn{2}{c}{(4)}\\
            
\midrule
Net Export Growth (91-99, An) x Dum91t99&   -1.730   &  (4.373)&   -2.067   &  (4.312)&   -9.389   &  (6.769)&  -10.197   &  (6.243)\\  \addlinespace
Net Export Growth (99-05, An) x Dum99t05&   14.682***&  (3.716)&   14.336***&  (3.830)&   14.181***&  (4.174)&   14.519***&  (3.794)\\  \addlinespace
Dummry 91-99   &    0.167***&  (0.009)&    0.350   &  (0.266)&    0.362   &  (0.346)&    0.329   &  (0.549)\\  \addlinespace
Dummry 99-05   &    0.202***&  (0.009)&    0.200   &  (0.212)&    0.638***&  (0.243)&   -0.282   &  (0.353)\\  \addlinespace
Ln(Num of HH, 1991) x Dum91t99&            &         &    0.000   &  (0.004)&   -0.021   &  (0.064)&   -0.059   &  (0.068)\\  \addlinespace
Ln(Num of HH, 1999) x Dum99t05&            &         &    0.007** &  (0.003)&    0.015   &  (0.033)&   -0.008   &  (0.033)\\  \addlinespace
Ln(HH Income, 1991) x Dum91t99&            &         &   -0.014   &  (0.029)&   -0.017   &  (0.037)&    0.012   &  (0.057)\\  \addlinespace
Ln(HH Income, 1999) x Dum99t05&            &         &   -0.006   &  (0.022)&   -0.054** &  (0.025)&    0.011   &  (0.035)\\  \addlinespace
Ratio of Labor Force (1989) x Dum91t99&            &         &   -0.058   &  (0.102)&    0.022   &  (0.102)&   -0.160   &  (0.132)\\  \addlinespace
Ratio of Labor Force (1999) x Dum99t05&            &         &   -0.019   &  (0.107)&    0.120   &  (0.107)&    0.514***&  (0.146)\\  \addlinespace
Ln(Num of HU, 1991) x Dum91t99&            &         &            &         &    0.017   &  (0.064)&    0.054   &  (0.069)\\  \addlinespace
Ln(Num of HU, 1999) x Dum99t05&            &         &            &         &   -0.005   &  (0.034)&    0.016   &  (0.034)\\  \addlinespace
Housing supply elasticity x Dum91t99&            &         &            &         &   -0.012** &  (0.005)&   -0.007   &  (0.004)\\  \addlinespace
Housing supply elasticity x Dum99t05&            &         &            &         &   -0.005   &  (0.005)&   -0.005   &  (0.005)\\  \addlinespace
Wharton Regulation Index x Dum91t99&            &         &            &         &    0.004   &  (0.008)&    0.005   &  (0.007)\\  \addlinespace
Wharton Regulation Index x Dum99t05&            &         &            &         &    0.021***&  (0.008)&    0.019** &  (0.008)\\  \addlinespace
House Vacancy Rate (1989) x Dum91t99&            &         &            &         &    0.336** &  (0.139)&    0.320** &  (0.131)\\  \addlinespace
House Vacancy Rate (1999) x Dum99t05&            &         &            &         &    0.118   &  (0.098)&    0.064   &  (0.094)\\  \addlinespace
Ratio of Renters (1989) x Dum91t99&            &         &            &         &   -0.007   &  (0.083)&   -0.183** &  (0.086)\\  \addlinespace
Ratio of Renters (1999) x Dum99t05&            &         &            &         &   -0.135** &  (0.059)&   -0.103   &  (0.063)\\  \addlinespace
Ratio of Bachelor Educated (1989) x Dum91t99&            &         &            &         &            &         &   -0.082   &  (0.097)\\  \addlinespace
Ratio of Bachelor Educated (1999) x Dum99t05&            &         &            &         &            &         &   -0.320***&  (0.099)\\  \addlinespace
Ratio of White Race (1989) x Dum91t99&            &         &            &         &            &         &   -0.050   &  (0.042)\\  \addlinespace
Ratio of White Race (1999) x Dum99t05&            &         &            &         &            &         &   -0.056   &  (0.037)\\  \addlinespace
Ratio of Immigration (80-90) x Dum91t99&            &         &            &         &            &         &    1.068***&  (0.227)\\  \addlinespace
Ratio of Immigration (90-00) x Dum99t05&            &         &            &         &            &         &    0.303   &  (0.205)\\  \addlinespace
Ratio of Age 65 Above (1989) x Dum91t99&            &         &            &         &            &         &   -0.411** &  (0.176)\\  \addlinespace
Ratio of Age 65 Above (1999) x Dum99t05&            &         &            &         &            &         &    0.669***&  (0.185)\\  \addlinespace
Constant       &            &         &            &         &            &         &            &         \\  \addlinespace
\midrule
Obs            &     1489   &         &     1489   &         &     1341   &         &     1341   &         \\
Cluster SE     &     CBSA   &         &     CBSA   &         &     CBSA   &         &     CBSA   &         \\
Weight         & {\scriptsize Ln(HU-Start)}  &   & {\scriptsize Ln(HU-Start)}  &   &{\scriptsize Ln(HU-Start)}  &   &{\scriptsize Ln(HU-Start)}  &  \\
KP F-Stat (91-99, Non-stack Sample)    &    40.61   &         &    41.60   &         &    21.65   &         &    21.58   &         \\
MOP F-Stat (91-99, Non-stack Sample) &    40.61   &         &    41.60   &         &    21.65   &         &    21.58   &         \\
KP F-Stat (99-05, Non-stack Sample)     &    20.24   &         &    20.13   &         &    13.70   &         &    15.70   &         \\
MOP F-Stat (99-05, Non-stack Sample) &    20.24   &         &    20.13   &         &    13.70   &         &    15.70   &         \\
CoefEqual\_Chi2 &    9.508   &         &    9.162   &         &    9.527   &         &    12.29   &         \\
CoefEqual\_PValue&  0.00205   &         &  0.00247   &         &  0.00202   &         & 0.000454   &         \\
\bottomrule

\end{tabular}

} % end of resize box

\end{table}

\pagebreak 
%---------------------------------------------------------------

%%%%%%%%%%%%%%%%%%%%%%%%%%%%%%%%%%%%%%%%%%%%%%%%
% table_PLNJM.D91t99vsD99t05.4Reg
%%%%%%%%%%%%%%%%%%%%%%%%%%%%%%%%%%%%%%%%%%%%%%%%

%---------------------------------------------------------------

%%%%%%%%%%%%%%%%%%%%%%%%%%%%%%%%%%%%%%%%%%%%%%%%
% table_PLNJM.D91t99vsD99t05.4Reg
%%%%%%%%%%%%%%%%%%%%%%%%%%%%%%%%%%%%%%%%%%%%%%%%

\noindent 

\begin{table}[h!]
\centering
\caption{
\textbf{Four Stacked Regressions of PLNJM Growth in Prior (91-99) and Boom (99-05) Periods on Net Export Growth (99-05)} \smallskip \newline
{\scriptsize
This table reports OLS, reduced-form, first stage and second stages of 2SLS stacked regression $\triangle_{91,99} \quad \&  \quad \triangle_{99,05}  Ln(PLNJM_{c}) = \beta_{91,99} * \triangle_{91,99} \text{NetExp}_{m} \times Dum_{91,99} + \beta_{99,05} * \triangle_{99,05} \text{NetExp}_{m} \times Dum_{99,05} + \gamma_{91,99}* \bm{Controls_{c}} \times Dum_{91,99} + \gamma_{99,05}* \bm{Controls_{c}} \times Dum_{99,05}  + \alpha_{91,99} + \alpha_{99,05} + \epsilon_{period, c}$. The left-hand-side dependent variable $\triangle_{91,99} \& \triangle_{99,05} Ln(PLNJM_{c})$ is the stacked growth rate of the dollar amount of private-label (non-jumbo) mortgages (PLNJM) at county $c$ 91-99 and 99-05. The key variable of interest $\triangle_{91,99} \text{NetExp}_{m}$ and $\triangle_{99,05} \text{NetExp}_{m}$ are the growth rate of net export at the metropolitan area (CBSA03 code) $m$ 91-99 and 99-05, respectively. $Controls_{c}$ indicates control variables at county $c$ in the period start year either 1991 or 1999. We use the gravity model-based instrumental variable $\triangle_{91,99}\text{givNetExp}_{m}$ and $\triangle_{99,05}\text{givNetExp}_{m}$ as IVs for $\triangle_{91,99}\text{NetExp}_{m}$ and $\triangle_{99,05}\text{NetExp}_{m}$. Each regression is weighted by the natural logarithm of housing units in the start year (either 1991 or 1999). For the first-stage F-test of two non-stack samples, we report Kleibergen-Paap (2006) robust (clustered) statistics and Montiel Olea-Pflueger (2013) efficient statistics. We report Standard errors are clustered at the CBSA level. ***, **, and * indicate significance at the 1\%, 5\%, and 10\% levels, respectively.
} % end of small font size
} % end of caption
\label{table_PLNJM.D91t99vsD99t05.4Reg}
\resizebox{0.74\columnwidth}{!}{%
\begin{tabular}{l*{4}{c}}
\toprule
Dep Var (Panel A, B, and C)                      &\multicolumn{4}{c}{PLNJM Growth (91-99 \& 99-05, annualized)} \\
            \cmidrule{2-5} 
            &\multicolumn{1}{c}{(1)}&\multicolumn{1}{c}{(2)}&\multicolumn{1}{c}{(3)}&\multicolumn{1}{c}{(4)}\\

\midrule
\multicolumn{5}{l}{\textbf{Panel A. OLS estimates}} \\
Net Export Growth (91-99, An) x Dum91t99&    4.119   &    4.540   &    1.529   &   -0.332   \\
               &  (3.236)   &  (3.213)   &  (3.908)   &  (3.748)   \\
\addlinespace
Net Export Growth (99-05, An) x Dum99t05&    8.675***&    8.421***&    7.183***&    9.121***\\
               &  (1.689)   &  (1.721)   &  (1.863)   &  (1.966)   \\
\addlinespace
R2-adj         &    0.817   &    0.818   &    0.831   &    0.841   \\
\addlinespace

\midrule
\multicolumn{5}{l}{\textbf{Panel B. Reduced-form estimates}} \\
GIV Net Export Growth (91-99, An) x Dum91t99&   -1.982   &   -2.372   &   -9.653   &  -10.465*  \\
               &  (4.934)   &  (4.857)   &  (6.119)   &  (5.672)   \\
\addlinespace
GIV Net Export Growth (99-05, An) x Dum99t05&   16.584***&   16.078***&   15.569***&   15.868***\\
               &  (3.363)   &  (3.464)   &  (3.833)   &  (4.255)   \\
\addlinespace
R2-adj         &    0.817   &    0.817   &    0.832   &    0.841   \\
\addlinespace

\midrule
\multicolumn{5}{l}{\textbf{Panel C . 2SLS estimates}} \\
\addlinespace
Net Export Growth (91-99, An) x Dum91t99&   -1.730   &   -2.067   &   -9.389   &  -10.197   \\
               &  (4.373)   &  (4.312)   &  (6.769)   &  (6.243)   \\
\addlinespace
Net Export Growth (99-05, An) x Dum99t05&   14.682***&   14.336***&   14.181***&   14.519***\\
               &  (3.716)   &  (3.830)   &  (4.174)   &  (3.794)   \\
\addlinespace
\addlinespace

Dep Var (Panel D): &\multicolumn{4}{c}{Net Export Growth (91-99, An)} \\ 
\midrule 
\multicolumn{5}{l}{\textbf{Panel D . First-stage estimates only for 91-99 (non-stack sample)}} \\
\addlinespace
GIV NEG (91-99, An)&    1.146***&    1.148***&    1.028***&    1.026***\\
               &  (0.180)   &  (0.178)   &  (0.221)   &  (0.221)   \\
\addlinespace
KP F-Stat      &    40.61   &    41.60   &    21.65   &    21.58   \\
MOP F-Stat     &    40.61   &    41.60   &    21.65   &    21.58   \\

\addlinespace
\addlinespace

Dep Var (Panel E): &\multicolumn{4}{c}{Net Export Growth (99-05, An) x Dum92t02 } \\ 
\midrule 
\multicolumn{5}{l}{\textbf{Panel E. First-stage estimates only for 99-05 (non-stack sample)}} \\
\addlinespace
GIV NEG (99-05, An)&    1.130***&    1.122***&    1.098***&    1.093***\\
               &  (0.251)   &  (0.250)   &  (0.297)   &  (0.276)   \\
\addlinespace
KP F-Stat     &    20.24   &    20.13   &    13.70   &    15.70   \\
MOP F-Stat    &    20.24   &    20.13   &    13.70   &    15.70   \\

\midrule
\multicolumn{5}{l}{\textbf{Controls (for all Panels)}} \\
DumPeriod  &    Y        &  Y   &   Y    & Y      \\
Basic Controls x DumPeriod &            &  Y   &   Y    & Y      \\
Housing Controls x DumPeriod &           &      & Y       & Y     \\
Demographic Controls x DumPeriod &            &      &        &  Y     \\

\midrule              
Obs (Panel A, B, and C)         &     1489   &     1489   &     1341   &     1341   \\
Obs (Panel D)    &     701   &      701   &      632   &      632   \\
Obs (Panel E)   &      788   &      788   &      709   &      709   \\
Cluster SE     &     CBSA   &     CBSA   &     CBSA   &     CBSA   \\
Weight         & {\scriptsize Ln(HU-Start)}   & {\scriptsize Ln(HU-Start)}   &{\scriptsize Ln(HU-Start)}   &{\scriptsize Ln(HU-Start)}      \\
\bottomrule
\end{tabular}

} % end of resize box

\end{table}

%---------------------------------------------------------------
%---------------------------------------------------------------
% Empirical: Main Tests 2 Exclusion Restriction
% (4) Overidentification tests
%---------------------------------------------------------------
%---------------------------------------------------------------

\pagebreak 
%----------------------------------------------------------------

%%%%%%%%%%%%%%%%%%%%%%%%%%%%%%%%%%%%
% table_HPI.D99t05.PLNJM.2SLS_OverID.tex
%%%%%%%%%%%%%%%%%%%%%%%%%%%%%%%%%%%%

%-----------------------------------------------------------------

%%%%%%%%%%%%%%%%%%%%%%%%%%%%%%%%%%%%
% table_HPI.D99t05.PLNJM.2SLS.overID
%%%%%%%%%%%%%%%%%%%%%%%%%%%%%%%%%%%%

\noindent 

\begin{table}[h!]
\centering
\caption{
\textbf{Over-Identification Test: 2SLS Regression of House Price Growth on PLNJM Growth with Two Instruments in Boom Period (99-05)}} 
\vspace{-4mm}
\begin{tablenotes}
\setstretch{1} \footnotesize  
\item \noindent This table reports the first-stage and the second-stage results of 2SLS regression $\triangle_{99,05} Ln(HPI_{c}) = \beta * \triangle_{99,05} Ln(PLNJM_{c})  + \gamma* \bm{Controls_{c}} + \alpha + \epsilon_{c}$. The left-hand-side dependent variable $\triangle_{99,05} Ln(HPI_{c})$ is the growth rate of the house price index at county $c$ 99-05, and the key variable of interest $\triangle_{99,05} Ln(PLNJM_{c})$ is the growth rate of the dollar amount of private-label (non-jumbo) mortgages (PLNJM) at county $c$ 99-05. $Controls_{c}$ indicates control variables at county $c$ in 1999. We use the gravity model-based instrumental variable ($\triangle_{99,05}\text{givNetExp}_{m}$) and its square as two IVs for $\triangle_{99,05} Ln(PLNJM_{c})$. For the first-stage F-test, we report Kleibergen-Paap (2006) robust (clustered) statistics and Montiel Olea-Pflueger (2013) efficient statistics. Regression is weighted by the natural logarithm of housing units in 1999. Standard errors are clustered at the CBSA level. ***, **, and * indicate significance at the 1\%, 5\%, and 10\% levels, respectively.
\end{tablenotes} 
\vspace{2mm}

\label{table_HPI.D99t05.PLNJM.2SLS.overID}

\resizebox{0.9\columnwidth}{!}{%

\begin{tabular}{l*{5}{c}}
\toprule
        &\multicolumn{1}{p{3cm}}{\centering PLNJM Growth \\ (99-05, An)}  &\multicolumn{4}{c}{\centering Housing Price Growth (07USD, 99-05, An)} \\
            \cmidrule{2-6} 
            &\multicolumn{1}{c}{(1)}&\multicolumn{1}{c}{(2)}&\multicolumn{1}{c}{(3)}&\multicolumn{1}{c}{(4)}&\multicolumn{1}{c}{(5)}\\
            
\midrule
GIV Net Export Growth (99-05, An)&   18.284** &            &            &            &            \\
               &  (8.767)   &            &            &            &            \\
\addlinespace
GIV Net Export Growth Suqare (99-05, An)&  701.650   &            &            &            &            \\
               &(1500.819)   &            &            &            &            \\
\addlinespace
PLNJM Growth (07USD, 99-05, An)&            &    0.462***&    0.427***&    0.478***&    0.514***\\
               &            &  (0.103)   &  (0.119)   &  (0.137)   &  (0.148)   \\
\addlinespace
Ln(Num of households, 1999)&   -0.038   &            &    0.005** &    0.020   &    0.021   \\
               &  (0.034)   &            &  (0.002)   &  (0.016)   &  (0.018)   \\
\addlinespace
Ln(household Income, 1999)&   -0.014   &            &    0.056***&    0.063***&    0.026   \\
               &  (0.035)   &            &  (0.013)   &  (0.015)   &  (0.018)   \\
\addlinespace
Ratio of Labor Force (1999)&    0.446***&            &   -0.184***&   -0.166***&   -0.192** \\
               &  (0.138)   &            &  (0.045)   &  (0.046)   &  (0.089)   \\
\addlinespace
Ln(Num of House Units, 1999)&    0.044   &            &            &   -0.026   &   -0.029   \\
               &  (0.034)   &            &            &  (0.017)   &  (0.019)   \\
\addlinespace
Housing supply elasticity&   -0.006   &            &            &   -0.003   &   -0.002   \\
               &  (0.004)   &            &            &  (0.002)   &  (0.002)   \\
\addlinespace
Wharton Regulation Index&    0.024***&            &            &    0.005   &    0.003   \\
               &  (0.008)   &            &            &  (0.004)   &  (0.005)   \\
\addlinespace
House Vacancy Rate (1999)&    0.067   &            &            &    0.090** &    0.078*  \\
               &  (0.092)   &            &            &  (0.042)   &  (0.045)   \\
\addlinespace
Ratio of Renters (1999)&   -0.113*  &            &            &    0.157***&    0.087** \\
               &  (0.064)   &            &            &  (0.029)   &  (0.034)   \\
\addlinespace
Ratio of Bachelor Educated (1999)&   -0.207** &            &            &            &    0.112*  \\
               &  (0.094)   &            &            &            &  (0.062)   \\
\addlinespace
Ratio of White Race (1999)&   -0.050   &            &            &            &   -0.004   \\
               &  (0.035)   &            &            &            &  (0.019)   \\
\addlinespace
Ratio of Immigration (90-00)&    0.252   &            &            &            &    0.234** \\
               &  (0.198)   &            &            &            &  (0.102)   \\
\addlinespace
Ratio of Age 65 Above (1999)&    0.499***&            &            &            &   -0.036   \\
               &  (0.179)   &            &            &            &  (0.109)   \\
\addlinespace
Constant       &    0.031   &   -0.038** &   -0.565***&   -0.592***&   -0.159   \\
               &  (0.345)   &  (0.017)   &  (0.136)   &  (0.158)   &  (0.179)   \\
\midrule
Obs            &      712   &      786   &      786   &      707   &      707   \\
Cluster SE     &     CBSA   &     CBSA   &     CBSA   &     CBSA   &     CBSA   \\
Weight         & Ln(HU99)   & Ln(HU99)   & Ln(HU99)   & Ln(HU99)   & Ln(HU99)   \\
KP F-Stat      & 6.679           &    13.23   &    12.07   &    8.886   &    6.679   \\
MOP F-Stat     & 5.826           &    10.97   &    10.17   &    7.828   &    5.826   \\
Hansen J-Test Stat&            &    1.570   &    1.774   &    0.701   &    0.366   \\
Hansen J-Test P-value&            &    0.210   &    0.183   &    0.402   &    0.545   \\
\bottomrule
\end{tabular}

} % end of resize box

\end{table}

\pagebreak 
%----------------------------------------------------------------
%%%%%%%%%%%%%%%%%%%%%%%%%%%%%%%%%%%%%%%%%%%%%%%%
% table_HPI.D99t05.PLNJM.4Reg.OverID.tex
%%%%%%%%%%%%%%%%%%%%%%%%%%%%%%%%%%%%%%%%%%%%%%%%
%---------------------------------------------------------------

%%%%%%%%%%%%%%%%%%%%%%%%%%%%%%%%%%%%%%%%%%%%%%%%
% table_HPI.D99t05.PLNJM.4Reg.OverID
%%%%%%%%%%%%%%%%%%%%%%%%%%%%%%%%%%%%%%%%%%%%%%%%

\noindent 

\begin{table}[h!]
\centering
\caption{
\textbf{Over-Identification Test: Four Regressions of House Price Growth on PLNJM Growth in Boom Period (99-05)} }
\vspace{-4mm}
\begin{tablenotes}
\setstretch{1} \footnotesize  
\item \noindent This table reports OLS, reduced-form, first stage, and second stage results of 2SLS regression $\triangle_{99,05} Ln(HPI_{c}) = \beta * \triangle_{99,05} Ln(PLNJM_{c})  + \gamma* \bm{Controls_{c}} + \alpha + \epsilon_{c}$. The left-hand-side dependent variable $\triangle_{99,05} Ln(HPI_{c})$ is the growth rate of the house price index at county $c$ 99-05, and the key variable of interest $\triangle_{99,05} Ln(PLNJM_{c})$ is the growth rate of the dollar amount of private-label (non-jumbo) mortgages (PLNJM) at county $c$ 99-05. $Controls_{c}$ indicates control variables at county $c$ in 1999. We use the gravity model-based instrumental variable ($\triangle_{99,05}\text{givNetExp}_{m}$) and its square as two IVs for $\triangle_{99,05} Ln(PLNJM_{c})$. For the first-stage F-test, we report Kleibergen-Paap (2006) robust (clustered) statistics and Montiel Olea-Pflueger (2013) efficient statistics. Regression is weighted by the natural logarithm of housing units in 1999. Standard errors are clustered at the CBSA level. ***, **, and * indicate significance at the 1\%, 5\%, and 10\% levels, respectively.
\end{tablenotes} 
\vspace{2mm}

\label{table_HPI.D99t05.PLNJM.4Reg.overID}

\resizebox{\columnwidth}{!}{%

\begin{tabular}{l*{4}{c}}
\toprule
Dep Var (Panel A, B, and C)                     &\multicolumn{4}{c}{House Price Growth (1999-2005, annualized)} \\
            \cmidrule{2-5} 
            &\multicolumn{1}{c}{(1)}&\multicolumn{1}{c}{(2)}&\multicolumn{1}{c}{(3)}&\multicolumn{1}{c}{(4)}\\
            
\midrule
\multicolumn{5}{l}{\textbf{Panel A. OLS estimates}} \\
\addlinespace
PLNJM Growth (07USD, 99-05, An)&    0.184***&    0.169***&    0.131***&    0.124***\\
               &  (0.026)   &  (0.025)   &  (0.021)   &  (0.020)   \\
\addlinespace
R2-adj         &    0.165   &    0.345   &    0.544   &    0.587   \\
\addlinespace

\midrule
\multicolumn{5}{l}{\textbf{Panel B. Reduced-form estimates}} \\
\addlinespace
GIV Net Export Growth (99-05, An)&   13.649***&   12.934***&   11.070***&   10.911***\\
               &  (4.488)   &  (4.583)   &  (3.213)   &  (2.878)   \\
\addlinespace
GIV Net Export Growth Suqare (99-05, An)& 1130.863*  & 1214.625** &  904.606** &  788.374*  \\
               &(591.457)   &(597.750)   &(445.203)   &(409.426)   \\
\addlinespace
R2-adj         &   0.0587   &    0.254   &    0.503   &    0.560    \\
\addlinespace

\midrule
\multicolumn{5}{l}{\textbf{Panel C . 2SLS estimates}} \\
\addlinespace
PLNJM Growth (07USD, 99-05, An)&    0.462***&    0.427***&    0.478***&    0.514***\\
               &  (0.103)   &  (0.119)   &  (0.137)   &  (0.148)   \\
\addlinespace

\addlinespace
\addlinespace

Dep Var (Panel D): &\multicolumn{4}{c}{PLNJM Growth (99-05, An)} \\ 
\midrule 
\multicolumn{5}{l}{\textbf{Panel D . First-stage estimates}} \\
\addlinespace
GIV Net Export Growth (99-05, An)&   18.905** &   19.331** &   18.543** &   18.284** \\
               &  (7.855)   &  (8.023)   &  (8.395)   &  (8.767)   \\
\addlinespace
GIV Net Export Growth Square (99-05, An)&  618.919   &  790.426   &  729.041   &  701.650   \\
               &(1183.483)   &(1191.748)   &(1288.572)   &(1500.819)   \\
\addlinespace
KP F-Stat      &    13.23   &    12.07   &    8.886   &    6.679   \\
MOP F-Stat     &    10.97   &    10.17   &    7.828   &    5.826   \\
\addlinespace
\midrule
\multicolumn{5}{l}{\textbf{Controls (for all Panels)}} \\
Basic Controls &            &  Y   &   Y    & Y        \\
Housing Controls &           &      & Y       & Y        \\
Demographic Controls &            &      &        &  Y       \\
\midrule              
Obs            &      786   &      786   &      707   &      707   \\
Cluster SE     &     CBSA   &     CBSA   &     CBSA   &     CBSA   \\
Weight         & Ln(HU99)   & Ln(HU99)   & Ln(HU99)   & Ln(HU99)   \\
\bottomrule

\end{tabular}

} % end of resize box

\end{table}

%\pagebreak
%-----------------------------------------------------------------

%---------------------------------------------------------------------------------------
%---------------------------------------------------------------------------------------
% Cross-Metro Empirical: Main Tests 3 Recovery and Long-Term Trend
%---------------------------------------------------------------------------------------
%---------------------------------------------------------------------------------------

\pagebreak 
%----------------------------------------------------------------

%%%%%%%%%%%%%%%%%%%%%%%%%%%%%%%%%%%%
% table_HPI.D91t19.NEG.D91t07.2SLS
%%%%%%%%%%%%%%%%%%%%%%%%%%%%%%%%%%%%
%-----------------------------------------------------------------

%%%%%%%%%%%%%%%%%%%%%%%%%%%%%%%%%%%%
% table_HPI.D91t19.NEG.D91t07.2SLS
%%%%%%%%%%%%%%%%%%%%%%%%%%%%%%%%%%%%

\noindent 

\begin{table}[h!]
\centering
\caption{
\textbf{2SLS Regression of House Price Growth (91-19) on Net Export Growth (91-07)} \smallskip \newline
{\footnotesize 
This table reports 2SLS regression $\triangle_{91,19} Ln(HPI_{c}) = \beta * \triangle_{91,07} \text{NetExp}_{m} + \gamma* \bm{Controls_{c}} + \alpha + \epsilon_{c}$. The left-hand-side dependent variable $\triangle_{91,19} Ln(HPI_{c})$ is the growth rate of the house price index at county $c$ 91-19, and the key variable of interest $\triangle_{91,07} \text{NetExp}_{m}$ is the growth rate of net export at the metropolitan area (CBSA09 code) $m$ 91-07. $Controls_{c}$ indicates control variables at county $c$ in 1991. We use the gravity model-based instrumental variable ($\triangle_{91,07}\text{givNetExp}_{m}$) as IV for $\triangle_{91,07}\text{NetExp}_{m}$. To reduce the impact of outliers, I winsorize at 2\% and 98\% for $\triangle_{91,07}\text{NetExp}_{m}$ and $\triangle_{91,07}\text{givNetExp}_{m}$. For the first-stage F-test, we report Kleibergen-Paap (2006) robust (clustered) statistics and Montiel Olea-Pflueger (2013) efficient statistics. Regression is weighted by the natural logarithm of housing units in 1991. Standard errors are clustered at the CBSA09 level. ***, **, and * indicate significance at the 1\%, 5\%, and 10\% levels, respectively.
} % end of small font size
\smallskip \newline
} % end of caption
\label{table_HPI.D91t19.NEG.D91t07.2SLS}

\resizebox{0.95\columnwidth}{!}{%

\begin{tabular}{l*{5}{c}}
\toprule
            &\multicolumn{5}{c}{House Price Growth (1991-2019, 07USD, annualized)} \\
            \cmidrule{2-6} 
            &\multicolumn{1}{c}{(1)}&\multicolumn{1}{c}{(2)}&\multicolumn{1}{c}{(3)}&\multicolumn{1}{c}{(4)}&\multicolumn{1}{c}{(5)}\\
            
\midrule
GIV Net Export Growth (91-07, An, 07USD)&    1.023***&            &            &            &            \\
               &  (0.099)   &            &            &            &            \\
\addlinespace
Net Export Growth (91-07, An, 07USD)&            &    1.210***&    1.229***&    1.273***&    0.971** \\
               &            &  (0.367)   &  (0.370)   &  (0.313)   &  (0.413)   \\
\addlinespace
Ln(Num of HH, 1991)&   -0.001   &            &    0.000   &   -0.001** &   -0.017***\\
               &  (0.000)   &            &  (0.000)   &  (0.000)   &  (0.005)   \\
\addlinespace
Ln(HH Income, 1991)&    0.000   &            &   -0.002   &   -0.015***&   -0.015***\\
               &  (0.001)   &            &  (0.003)   &  (0.003)   &  (0.004)   \\
\addlinespace
Ratio of Labor Force (1989)&   -0.000   &            &    0.010   &    0.021** &    0.039***\\
               &  (0.001)   &            &  (0.008)   &  (0.009)   &  (0.010)   \\
\addlinespace
Ratio of Bachelor Educated (1989)&    0.001   &            &            &    0.034***&    0.030***\\
               &  (0.001)   &            &            &  (0.006)   &  (0.007)   \\
\addlinespace
Ratio of White Race (1989)&   -0.000   &            &            &    0.009***&    0.009***\\
               &  (0.000)   &            &            &  (0.003)   &  (0.003)   \\
\addlinespace
Ratio of Immigration (80-90)&    0.005** &            &            &    0.115***&    0.107***\\
               &  (0.002)   &            &            &  (0.015)   &  (0.018)   \\
\addlinespace
Ratio of Age 65 Above (1989)&   -0.001   &            &            &    0.026** &    0.019*  \\
               &  (0.001)   &            &            &  (0.010)   &  (0.011)   \\
\addlinespace
Ln(Num of HU, 1991)&    0.001   &            &            &            &    0.016***\\
               &  (0.000)   &            &            &            &  (0.005)   \\
\addlinespace
House Vacancy Rate (1989)&    0.001   &            &            &            &    0.003   \\
               &  (0.001)   &            &            &            &  (0.009)   \\
\addlinespace
Ratio of Renters (1989)&   -0.001   &            &            &            &    0.009*  \\
               &  (0.001)   &            &            &            &  (0.005)   \\
\addlinespace
Housing supply elasticity&   -0.000   &            &            &            &   -0.002***\\
               &  (0.000)   &            &            &            &  (0.000)   \\
\addlinespace
Wharton Regulation Index&   -0.000   &            &            &            &    0.000   \\
               &  (0.000)   &            &            &            &  (0.001)   \\
\addlinespace
Constant       &   -0.002   &    0.013***&    0.024   &    0.140***&    0.139***\\
               &  (0.005)   &  (0.001)   &  (0.023)   &  (0.029)   &  (0.034)   \\
\midrule
Obs            &      740   &      877   &      877   &      877   &      740   \\
R2-adj         &    0.527   &            &            &            &            \\
Cluster SE     &     CBSA09   &     CBSA09   &     CBSA09   &     CBSA09   &     CBSA09   \\
Weight         & Ln(HU91)   & Ln(HU91)   & Ln(HU91)   & Ln(HU91)   & Ln(HU91)   \\
KP F-Stat      &   105.6    &    163.7   &    164.3   &    175.8   &    105.6   \\
MOP F-Stat     &   105.6    &    163.7   &    164.3   &    175.8   &    105.6   \\
\bottomrule

\end{tabular}

} % end of resize box

\end{table}

\pagebreak 
%----------------------------------------------------------------
%%%%%%%%%%%%%%%%%%%%%%%%%%%%%%%%%%%%%%%%%%%%%%%%
% table_HPI.D91t19.NEG.D91t07.4Reg
%%%%%%%%%%%%%%%%%%%%%%%%%%%%%%%%%%%%%%%%%%%%%%%%
%---------------------------------------------------------------

%%%%%%%%%%%%%%%%%%%%%%%%%%%%%%%%%%%%%%%%%%%%%%%%
% table_HPI.D91t19.NEG.D91t07.4Reg
%%%%%%%%%%%%%%%%%%%%%%%%%%%%%%%%%%%%%%%%%%%%%%%%

\noindent 

\begin{table}[h!]
\centering
\caption{
\textbf{Four Regression of House Price Growth (91-19) on Net Export Growth (91-07)} \smallskip \newline
{\footnotesize 
This table reports OLS, reduced-form, first stage, and second stage results of 2SLS regression $\triangle_{91,19} Ln(HPI_{c}) = \beta * \triangle_{91,07} \text{NetExp}_{m} + \gamma* \bm{Controls_{c}} + \alpha + \epsilon_{c}$. The left-hand-side dependent variable $\triangle_{91,19} Ln(HPI_{c})$ is the growth rate of the house price index at county $c$ 91-19, and the key variable of interest $\triangle_{91,07} \text{NetExp}_{m}$ is the growth rate of net export at the metropolitan area (CBSA09 code) $m$ 91-07. $Controls_{c}$ indicates control variables at county $c$ in 1991. We use the gravity model-based instrumental variable ($\triangle_{91,07}\text{givNetExp}_{m}$) as IV for $\triangle_{91,07}\text{NetExp}_{m}$. To reduce the impact of outliers, I winsorize at 2\% and 98\% for $\triangle_{91,07}\text{NetExp}_{m}$ and $\triangle_{91,07}\text{givNetExp}_{m}$. For the first-stage F-test, we report Kleibergen-Paap (2006) robust (clustered) statistics and Montiel Olea-Pflueger (2013) efficient statistics. Regression is weighted by the natural logarithm of the number of housing units in 1991. Standard errors are clustered at the CBSA level. ***, **, and * indicate significance at the 1\%, 5\%, and 10\% levels, respectively.
} % end of small font size
} % end of caption
\label{table_HPI.D91t19.NEG.D91t07.4Reg}

\resizebox{\columnwidth}{!}{%

\begin{tabular}{l*{4}{c}}
\toprule
Dep Var (Panel A, B, and C)                     &\multicolumn{4}{c}{\small House Price Index Growth  (1991-2019, 07USD, annualized)} \\
            \cmidrule{2-5} 
            &\multicolumn{1}{c}{(1)}&\multicolumn{1}{c}{(2)}&\multicolumn{1}{c}{(3)}&\multicolumn{1}{c}{(4)}\\
            
\midrule
\multicolumn{5}{l}{\textbf{Panel A. OLS estimates}} \\
\addlinespace
Net Export Growth (91-07, An, 07USD)&    1.679***&    1.698***&    1.516***&    1.374***\\
               &  (0.252)   &  (0.258)   &  (0.234)   &  (0.280)   \\
\addlinespace
R2-adj         &   0.0685   &   0.0716   &    0.237   &    0.359   \\
\addlinespace

\midrule
\multicolumn{5}{l}{\textbf{Panel B. Reduced-form estimates}} \\
\addlinespace
GIV Net Export Growth (91-07, An, 07USD)&    1.352***&    1.354***&    1.399***&    0.993** \\
               &  (0.453)   &  (0.449)   &  (0.370)   &  (0.466)   \\
\addlinespace
R2-adj         &   0.0184   &   0.0216   &    0.203   &    0.332   \\
\addlinespace

\midrule
\multicolumn{5}{l}{\textbf{Panel C . 2SLS estimates}} \\
\addlinespace
Net Export Growth (91-07, An, 07USD)&    1.210***&    1.229***&    1.273***&    0.971** \\
               &  (0.367)   &  (0.370)   &  (0.313)   &  (0.413)   \\

\addlinespace
\addlinespace

Dep Var (Panel D): &\multicolumn{4}{c}{Net Export Growth (91-07, An)} \\ 
\midrule 
\multicolumn{5}{l}{\textbf{Panel D . First-stage estimates}} \\
\addlinespace
GIV Net Export Growth (91-07, An, 07USD)&    1.117***&    1.101***&    1.099***&    1.023***\\
               &  (0.087)   &  (0.086)   &  (0.083)   &  (0.099)   \\
\addlinespace
KP F-Stat      &    163.7   &    164.3   &    175.8   &    105.6   \\
MOP F-Stat     &    163.7   &    164.3   &    175.8   &    105.6   \\

\addlinespace
\midrule
\multicolumn{5}{l}{\textbf{Controls (for all Panels)}} \\
Basic Controls &            &  Y   &   Y    & Y       \\
Demographic Controls  &         &      & Y      &  Y      \\
Housing Controls  &           &      &        & Y       \\
\midrule              
Obs            &      877   &      877   &      877   &      740   \\
Cluster SE     &     CBSA09   &     CBSA09   &     CBSA09   &     CBSA09   \\
Weight         & Ln(HU91)   & Ln(HU91)   & Ln(HU91)   & Ln(HU91)    \\
\bottomrule

\end{tabular}

} % end of resize box

\end{table}

\pagebreak
%-----------------------------------------------------------------
%%%%%%%%%%%%%%%%%%%%%%%%%%%%%%%%%%%%%%%%%%%%%%%%
% table_GSEMvsPLNJM.D91t19.4Reg
%%%%%%%%%%%%%%%%%%%%%%%%%%%%%%%%%%%%%%%%%%%%%%%%
%---------------------------------------------------------------

%%%%%%%%%%%%%%%%%%%%%%%%%%%%%%%%%%%%%%%%%%%%%%%%
% table_GSEMvsPLNJM.D91t19.4Reg.NEG.D91t07
%%%%%%%%%%%%%%%%%%%%%%%%%%%%%%%%%%%%%%%%%%%%%%%%

\noindent 

\begin{table}[h!]
\centering
\caption{
\textbf{Four Stacked Regressions of GSEM and PLNJM Growth in Long Term (91-19) on Net Export Growth (91-07)} \smallskip \newline
{\scriptsize
This table reports OLS, reduced-form, and the second stage of stacked 2SLS regression $\triangle_{91,19} Ln(\text{GSEM or PLNJM}_{c}) = \beta_{G} * \triangle_{91,07} \text{NetExp}_{m} \times Dum_{G} + \beta_{P} * \triangle_{91,07} \text{NetExp}_{m} \times Dum_{P} + \gamma_{G}* \bm{Controls_{c}} \times Dum_{G} + \gamma_{P}* \bm{Controls_{c}} \times Dum_{P}  + \alpha_{G} + \alpha_{P} + \epsilon_{G, c} + \epsilon_{P, c}$. The left-hand-side dependent variable $\triangle_{91,19} Ln(\text{GSEM or PLNJM}_{c})$ is the stacked growth rate of the dollar amount of either government-sponsored enterprise mortgages (GSEM) or private-label (non-jumbo) mortgages (PLNJM) at county $c$ 91-19. The key variable of interest $\triangle_{91,07} \text{NetExp}_{m}$ is the growth rate of net export at the metropolitan area (CBSA03 code) $m$ 91-07. $Controls_{c}$ indicates control variables at county $c$ in 1991. We use the gravity model-based instrumental variable ($\triangle_{91,07}\text{givNetExp}_{m}$) as IV for $\triangle_{91,07}\text{NetExp}_{m}$. For the first-stage F-test, we report Kleibergen-Paap (2006) robust (clustered) statistics and Montiel Olea-Pflueger (2013) efficient statistics. Each regression is weighted by the natural logarithm of housing units in 1991. Standard errors are clustered at the CBSA03 level. ***, **, and * indicate significance at the 1\%, 5\%, and 10\% levels, respectively.
} % end of small font size
} % end of caption
\label{table_GSEMvsPLNJM.D91t19.4Reg.NEG.D91t07}

\resizebox{0.90\columnwidth}{!}{%

\begin{tabular}{l*{4}{c}}
\toprule
Dep Var               &\multicolumn{4}{c}{GSEM and PLNJM Growth (99-05, annualized)} \\
            \cmidrule{2-5} 
            &\multicolumn{1}{c}{(1)}&\multicolumn{1}{c}{(2)}&\multicolumn{1}{c}{(3)}&\multicolumn{1}{c}{(4)}\\
            
\midrule
\multicolumn{5}{l}{\textbf{Panel A. OLS estimates}} \\
\addlinespace
Net Export Growth (91-07, An) x Dum\_GSEM&   -0.169   &    1.684** &    1.486*  &    0.479   \\
               &  (0.880)   &  (0.775)   &  (0.794)   &  (0.859)   \\
\addlinespace
Net Export Growth (91-07, An) x Dum\_PLMNJ&    1.839   &    2.125*  &    2.140*  &    0.677   \\
               &  (1.245)   &  (1.258)   &  (1.205)   &  (1.288)   \\
\addlinespace
\addlinespace
R2-adj         &    0.824   &    0.850   &    0.853   &    0.865   \\
\addlinespace

\midrule
\multicolumn{5}{l}{\textbf{Panel B. Reduced-form estimates}} \\
\addlinespace
GIV Net Export Growth (91-07, An) x Dum\_GSEM&    0.265   &    1.385   &    1.224   &    0.907   \\
               &  (0.882)   &  (0.842)   &  (0.859)   &  (1.016)   \\
\addlinespace
GIV Net Export Growth (91-07, An) x Dum\_PLMNJ&    0.461   &    0.631   &    0.712   &   -1.281   \\
               &  (1.405)   &  (1.406)   &  (1.368)   &  (1.559)   \\
\addlinespace
R2-adj         &    0.823   &    0.849   &    0.852   &    0.865   \\
\addlinespace

\midrule
\multicolumn{5}{l}{\textbf{Panel C . 2SLS estimates}} \\
\addlinespace
Net Export Growth (91-07, An) x Dum\_GSEM&    0.279   &    1.475*  &    1.318   &    0.955   \\
               &  (0.930)   &  (0.860)   &  (0.890)   &  (1.057)   \\
\addlinespace
Net Export Growth (91-07, An) x Dum\_PLMNJ&    0.485   &    0.672   &    0.767   &   -1.349   \\
               &  (1.470)   &  (1.484)   &  (1.457)   &  (1.671)   \\
\addlinespace
\addlinespace

Dep Var (Panel D): &\multicolumn{4}{c}{Net Export Growth (91-07, An)} \\ 
\midrule 
\multicolumn{5}{l}{\textbf{Panel D . First-stage estimates for non-stack sample}} \\
\addlinespace
GIV Net Export Growth (91-07, An)  &    0.952***&    0.939***&    0.928***&    0.950***\\
               &  (0.122)   &  (0.117)   &  (0.114)   &  (0.135)   \\
\addlinespace
KP F-Stat      &    61.07   &    64.65   &    66.87   &    49.25   \\
MOP F-Stat     &    61.07   &    64.65   &    66.87   &    49.25   \\
\addlinespace

\midrule
\multicolumn{5}{l}{\textbf{Controls (for all Panels)}} \\
Dum\_MortgageType  &    Y        &  Y   &   Y    & Y       \\
Basic Controls x Dum\_MortgageType &            &  Y   &   Y    & Y     \\
Demographic Controls x Dum\_MortgageType &           &      & Y       & Y    \\
Housing Controls x Dum\_MortgageType &            &      &        &  Y     \\

\midrule          
Obs (Panel A, B, \& C)        &     1424   &     1424   &     1424   &     1284   \\
Obs (Panel D)        &      712   &      712   &      712   &      712   \\
Cluster SE     &     CBSA03   &     CBSA03   &     CBSA03   &     CBSA03      \\
Weight         & Ln(HU91)   & Ln(HU91)   & Ln(HU91)   & Ln(HU91)     \\
\bottomrule

\end{tabular}

} % end of resize box

\end{table}

\pagebreak 
%----------------------------------------------------------------

%%%%%%%%%%%%%%%%%%%%%%%%%%%%%%%%%%%%
% table_HPI.D12t19.NEG.D91t07.2SLS
%%%%%%%%%%%%%%%%%%%%%%%%%%%%%%%%%%%%
%-----------------------------------------------------------------

%%%%%%%%%%%%%%%%%%%%%%%%%%%%%%%%%%%%
% table_HPI.D12t19.NEG.D91t07.2SLS
%%%%%%%%%%%%%%%%%%%%%%%%%%%%%%%%%%%%

\noindent 

\begin{table}[h!]
\centering
\caption{
\textbf{2SLS Regression of House Price Growth (12-19) on Net Export Growth (91-07)} \smallskip \newline
{\footnotesize 
This table reports 2SLS regression $\triangle_{12,19} Ln(HPI_{c}) = \beta * \triangle_{91,07} \text{NetExp}_{m} + \gamma* \bm{Controls_{c}} + \alpha + \epsilon_{c}$. The left-hand-side dependent variable $\triangle_{12,19} Ln(HPI_{c})$ is the growth rate of the house price index at county $c$ 12-19, and the key variable of interest $\triangle_{91,07} \text{NetExp}_{m}$ is the growth rate of net export at the metropolitan area (CBSA03 code) $m$ 12-07. $Controls_{c}$ indicates control variables at county $c$ in 1999. We use the gravity model-based instrumental variable ($\triangle_{91,07}\text{givNetExp}_{m}$) as IV for $\triangle_{91,07}\text{NetExp}_{m}$. To reduce the impact of outliers, I winsorize at 4\% and 96\% for $\triangle_{91,07}\text{NetExp}_{m}$ and $\triangle_{91,07}\text{givNetExp}_{m}$. For the first-stage F-test, we report Kleibergen-Paap (2006) robust (clustered) statistics and Montiel Olea-Pflueger (2013) efficient statistics. Regression is weighted by the natural logarithm of the number of housing units in 1999. Standard errors are clustered at the CBSA level. ***, **, and * indicate significance at the 1\%, 5\%, and 10\% levels, respectively.
} % end of small font size
\smallskip \newline
} % end of caption
\label{table_HPI.D12t19.NEG.D91t07.2SLS}

\resizebox{0.95\columnwidth}{!}{%

\begin{tabular}{l*{5}{c}}
\toprule
            &\multicolumn{5}{c}{House Price Growth (2012-2019, 07USD, annualized)} \\
            \cmidrule{2-6} 
            &\multicolumn{1}{c}{(1)}&\multicolumn{1}{c}{(2)}&\multicolumn{1}{c}{(3)}&\multicolumn{1}{c}{(4)}&\multicolumn{1}{c}{(5)}\\
            
\midrule
GIV Net Export Growth (91-07, An, 07USD)&    1.123***&            &            &            &            \\
               &  (0.081)   &            &            &            &            \\
\addlinespace
Net Export Growth (91-07, An, 07USD)&            &    4.070***&    3.208** &    3.205***&    3.028** \\
               &            &  (1.503)   &  (1.324)   &  (1.182)   &  (1.397)   \\
\addlinespace
Ln(Num of households, 1999)&   -0.001** &            &    0.007***&    0.005***&   -0.020** \\
               &  (0.000)   &            &  (0.001)   &  (0.001)   &  (0.009)   \\
\addlinespace
Ln(household Income, 1999)&   -0.000   &            &   -0.000   &    0.014   &    0.010   \\
               &  (0.000)   &            &  (0.009)   &  (0.012)   &  (0.013)   \\
\addlinespace
Fraction of Labor Force (1999)&   -0.001   &            &    0.032   &    0.038   &    0.055   \\
               &  (0.001)   &            &  (0.036)   &  (0.034)   &  (0.038)   \\
\addlinespace
Fraction of Bachelor Educated (1999)&    0.002** &            &            &   -0.091***&   -0.096***\\
               &  (0.001)   &            &            &  (0.027)   &  (0.031)   \\
\addlinespace
Fraction of White Race (1999)&   -0.000   &            &            &    0.018** &    0.010   \\
               &  (0.000)   &            &            &  (0.009)   &  (0.012)   \\
\addlinespace
Fraction of Immigration (90-00)&    0.002   &            &            &    0.326***&    0.359***\\
               &  (0.002)   &            &            &  (0.074)   &  (0.077)   \\
\addlinespace
Ratio of Age 65 Above (1999)&   -0.004** &            &            &   -0.040   &   -0.056   \\
               &  (0.002)   &            &            &  (0.065)   &  (0.068)   \\
\addlinespace
Ln(Num of House Units, 1999)&    0.001** &            &            &            &    0.025***\\
               &  (0.000)   &            &            &            &  (0.010)   \\
\addlinespace
House Vacancy Rate (1999)&    0.001   &            &            &            &   -0.006   \\
               &  (0.001)   &            &            &            &  (0.034)   \\
\addlinespace
Fraction of Renters (1999)&   -0.001*  &            &            &            &   -0.032   \\
               &  (0.001)   &            &            &            &  (0.026)   \\
\addlinespace
Housing supply elasticity&   -0.000   &            &            &            &   -0.001   \\
               &  (0.000)   &            &            &            &  (0.001)   \\
\addlinespace
Wharton Regulation Index&   -0.000   &            &            &            &    0.003   \\
               &  (0.000)   &            &            &            &  (0.003)   \\
\addlinespace
Constant       &    0.001   &    0.035***&   -0.062   &   -0.197*  &   -0.149   \\
               &  (0.003)   &  (0.004)   &  (0.079)   &  (0.110)   &  (0.133)   \\
\midrule
Obs            &      640   &      709   &      709   &      709   &      639   \\
R2-adj         &    0.569   &            &            &            &            \\
Cluster SE     &     CBSA03 &     CBSA03   &     CBSA03   &     CBSA03   &     CBSA03   \\
Weight         & Ln(HU99)   & Ln(HU99)   & Ln(HU99)   & Ln(HU99)   & Ln(HU99)   \\
KP F-Stat      &  191.5     &    223.4   &    226.4   &    237.4   &    191.5   \\
MOP F-Stat     &  191.5     &    223.4   &    226.4   &    237.4   &    191.5   \\
\bottomrule

\end{tabular}

} % end of resize box

\end{table}

\pagebreak 
%----------------------------------------------------------------
%%%%%%%%%%%%%%%%%%%%%%%%%%%%%%%%%%%%%%%%%%%%%%%%
% table_HPI.D12t19.NEG.D91t07.4Reg
%%%%%%%%%%%%%%%%%%%%%%%%%%%%%%%%%%%%%%%%%%%%%%%%
%---------------------------------------------------------------

%%%%%%%%%%%%%%%%%%%%%%%%%%%%%%%%%%%%%%%%%%%%%%%%
% table_HPI.D12t19.NEG.D91t07.4Reg
%%%%%%%%%%%%%%%%%%%%%%%%%%%%%%%%%%%%%%%%%%%%%%%%

\noindent 

\begin{table}[h!]
\centering
\caption{
\textbf{Four Regression of House Price Growth (12-19) on Net Export Growth (91-07)} \smallskip \newline
{\footnotesize 
This table reports OLS, reduced-form, first stage, and second stage results of 2SLS regression $\triangle_{12,19} Ln(HPI_{c}) = \beta * \triangle_{91,07} \text{NetExp}_{m} + \gamma* \bm{Controls_{c}} + \alpha + \epsilon_{c}$. The left-hand-side dependent variable $\triangle_{12,19} Ln(HPI_{c})$ is the growth rate of the house price index at county $c$ 12-19, and the key variable of interest $\triangle_{91,07} \text{NetExp}_{m}$ is the growth rate of net export at the metropolitan area (CBSA03 code) $m$ 12-07. $Controls_{c}$ indicates control variables at county $c$ in 1999. We use the gravity model-based instrumental variable ($\triangle_{91,07}\text{givNetExp}_{m}$) as IV for $\triangle_{91,07}\text{NetExp}_{m}$. To reduce the impact of outliers, I winsorize at 4\% and 96\% for $\triangle_{91,07}\text{NetExp}_{m}$ and $\triangle_{91,07}\text{givNetExp}_{m}$. For the first-stage F-test, we report Kleibergen-Paap (2006) robust (clustered) statistics and Montiel Olea-Pflueger (2013) efficient statistics. Regression is weighted by the natural logarithm of the number of housing units in 1999. Standard errors are clustered at the CBSA level. ***, **, and * indicate significance at the 1\%, 5\%, and 10\% levels, respectively.
} % end of small font size
} % end of caption
\label{table_HPI.D12t19.NEG.D91t07.4Reg}

\resizebox{\columnwidth}{!}{%

\begin{tabular}{l*{4}{c}}
\toprule
Dep Var (Panel A, B, and C)                     &\multicolumn{4}{c}{\small House Price Index Growth  (2012-2019, 07USD, annualized)} \\
            \cmidrule{2-5} 
            &\multicolumn{1}{c}{(1)}&\multicolumn{1}{c}{(2)}&\multicolumn{1}{c}{(3)}&\multicolumn{1}{c}{(4)}\\
            
\midrule
\multicolumn{5}{l}{\textbf{Panel A. OLS estimates}} \\
\addlinespace
Net Export Growth (91-07, An, 07USD)&    3.981***&    2.921***&    2.735***&    2.474** \\
               &  (1.043)   &  (0.922)   &  (0.944)   &  (1.054)   \\
\addlinespace
R2-adj         &   0.0362   &    0.147   &    0.266   &    0.309   \\
\addlinespace

\midrule
\multicolumn{5}{l}{\textbf{Panel B. Reduced-form estimates}} \\
\addlinespace
GIV Net Export Growth (91-07, An, 07USD)&    4.907***&    3.815** &    3.779***&    3.399** \\
               &  (1.846)   &  (1.587)   &  (1.395)   &  (1.598)   \\
\addlinespace
R2-adj         &   0.0213   &    0.141   &    0.262   &    0.306   \\
\addlinespace

\midrule
\multicolumn{5}{l}{\textbf{Panel C . 2SLS estimates}} \\
\addlinespace
Net Export Growth (91-07, An, 07USD)&    4.070***&    3.208** &    3.205***&    3.028** \\
               &  (1.503)   &  (1.324)   &  (1.182)   &  (1.397)   \\

\addlinespace
\addlinespace

Dep Var (Panel D): &\multicolumn{4}{c}{Net Export Growth (91-07, An)} \\ 
\midrule 
\multicolumn{5}{l}{\textbf{Panel D . First-stage estimates}} \\
\addlinespace
GIV Net Export Growth (91-07, An, 07USD)&    1.206***&    1.189***&    1.180***&    1.123***\\
               &  (0.081)   &  (0.079)   &  (0.077)   &  (0.081)   \\
\addlinespace
KP F-Stat      &    223.4   &    226.4   &    237.4   &    191.5   \\
MOP F-Stat     &    223.4   &    226.4   &    237.4   &    191.5   \\

\addlinespace
\midrule
\multicolumn{5}{l}{\textbf{Controls (for all Panels)}} \\
Basic Controls &            &  Y   &   Y    & Y       \\
Demographic Controls  &         &      & Y      &  Y      \\
Housing Controls  &           &      &        & Y       \\
\midrule              
Obs            &      709   &      709   &      709   &      639   \\
Cluster SE     &     CBSA03   &     CBSA03   &     CBSA03   &     CBSA03   \\
Weight         & Ln(HU99)   & Ln(HU99)   & Ln(HU99)   & Ln(HU99)   \\
\bottomrule

\end{tabular}

} % end of resize box

\end{table}

%---------------------------------------------------------------------------------------
%---------------------------------------------------------------------------------------
% Empirical: Against Speculation
%---------------------------------------------------------------------------------------
%---------------------------------------------------------------------------------------

\pagebreak 
%---------------------------------------------------------------

%%%%%%%%%%%%%%%%%%%%%%%%%%%%%%%%%%%%%%%%%%%%%%%%
% table_PLNJMNonOwn.D99t05.PLNJMOwn.4Reg
%%%%%%%%%%%%%%%%%%%%%%%%%%%%%%%%%%%%%%%%%%%%%%%%

%---------------------------------------------------------------

%%%%%%%%%%%%%%%%%%%%%%%%%%%%%%%%%%%%
% table_PLNJMNonOwn.D99t05.PLNJMOwn.4Reg
%%%%%%%%%%%%%%%%%%%%%%%%%%%%%%%%%%%%

\noindent 

\begin{table}[h!]
\centering
\caption{
\textbf{2SLS Regression of PLNJM (Non-Owner-Occupied) Growth on PLNJM (Owner-Occupied) Growth in Boom Period (99-05)} \smallskip \newline
{\footnotesize
This table reports OLS, reduced-form, first stage, and second stage results of 2SLS regression $\triangle_{99,05} Ln(PLNJM\_NonOwn_{c}) = \beta * \triangle_{99,05} Ln(PLNJM\_Own_{c}) + \gamma* \bm{Controls_{c}} + \alpha + \epsilon_{c}$. The left-hand-side dependent variable $\triangle_{99,05} Ln(PLNJM\_NonOwn_{c})$ is the growth rate of the dollar amount of non-owner-occupied private-label (non-jumbo) mortgages at county $c$ 99-05 and the key variable of interest $\triangle_{99,05} Ln(PLNJM\_Own_{c})$ is the growth rate of the dollar amount of owner-occupied private-label (non-jumbo) mortgages at county $c$ 99-05. $Controls_{c}$ indicates control variables at county $c$ in 1999. We use the gravity model-based instrumental variable ($\triangle_{99,05}\text{givNetExp}_{m}$) as IV for $\triangle_{99,05} Ln(PLNJM\_Own_{c})$. For the first-stage F-test, we report Kleibergen-Paap (2006) robust (clustered) statistics and Montiel Olea-Pflueger (2013) efficient statistics. Regression is weighted by the natural logarithm of housing units in 1999 as the weight for the housing market. Standard errors are clustered at the CBSA level. ***, **, and * indicate significance at the 1\%, 5\%, and 10\% levels, respectively.
} % end of small font size
} % end of caption
\label{table_PLNJMNonOwn.D99t05.PLNJMOwn.4Reg}

\resizebox{\columnwidth}{!}{%

\begin{tabular}{l*{4}{c}}
\toprule
         &\multicolumn{4}{c}{PLNJM (Non-Owner-Occupied) Growth (07USD, 99-05)} \\
            \cmidrule{2-5} 
            &\multicolumn{1}{c}{(1)}&\multicolumn{1}{c}{(2)}&\multicolumn{1}{c}{(3)}&\multicolumn{1}{c}{(4)}\\
            
\midrule
\multicolumn{5}{l}{\textbf{Panel A. OLS estimates}} \\
\addlinespace
PLNJM (Owner-Occupied) Growth (99-05)&    1.134***&    1.125***&    1.122***&    1.140***\\
               &  (0.073)   &  (0.066)   &  (0.071)   &  (0.067)   \\
\addlinespace
R2-adj        &    0.377   &    0.407   &    0.429   &    0.469   \\
\addlinespace

\midrule
\multicolumn{5}{l}{\textbf{Panel B. Reduced-form estimates}} \\
\addlinespace
GIV Net Export Growth (99-05)&   17.499***&   15.974** &   19.425***&   19.398***\\
               &  (6.704)   &  (6.453)   &  (6.578)   &  (5.944)   \\
\addlinespace
R2-adj         &   0.0155   &   0.0501   &    0.109   &    0.146   \\
\addlinespace

\midrule
\multicolumn{5}{l}{\textbf{Panel C . 2SLS estimates}} \\
\addlinespace
PLNJM (Owner-Occupied) Growth (99-05)&    1.177***&    1.106***&    1.376***&    1.395***\\
               &  (0.349)   &  (0.331)   &  (0.337)   &  (0.319)   \\
\addlinespace

\addlinespace
\addlinespace

Dep Var (Panel D): &\multicolumn{4}{c}{PLNJM (Owner-Occupied) Growth (07USD, 99-05)} \\ 
\midrule 
\multicolumn{5}{l}{\textbf{Panel D . First-stage estimates}} \\
\addlinespace
GIV Net Export Growth (99-05)&   14.868***&   14.440***&   14.115***&   13.908***\\
               &  (3.077)   &  (3.139)   &  (3.473)   &  (3.931)   \\
\addlinespace
KP F-Stat      &    23.34   &    21.16   &    16.52   &    12.52   \\
MOP F-Eff      &    23.34   &    21.16   &    16.52   &    12.52   \\
\addlinespace
\midrule
\multicolumn{5}{l}{\textbf{Controls (for all Panels)}} \\
Basic Controls &            &  Y   &   Y    & Y        \\
Housing Controls &           &      & Y       & Y        \\
Demographic Controls &            &      &        &  Y       \\
\midrule              
Obs            &      774   &      774   &      695   &      695   \\
Cluster SE     &     CBSA   &     CBSA   &     CBSA   &     CBSA   \\
Weight         & Ln(HU99)   & Ln(HU99)   & Ln(HU99)   & Ln(HU99)   \\
\bottomrule

\end{tabular}

} % end of resize box

\end{table}

\pagebreak 
%-----------------------------------------------------------------------
%%%%%%%%%%%%%%%%%%%%%%%%%%%%%%%%%%%%
% table_HPI.D99t05.PLNJMOwn_vs_NonOwnIndependent.4Reg
%%%%%%%%%%%%%%%%%%%%%%%%%%%%%%%%%%%%

%-----------------------------------------------------------------

%%%%%%%%%%%%%%%%%%%%%%%%%%%%%%%%%%%%
% table_HPI.D99t05.PLNJMOwn_vs_NonOwnIndependent.4Reg
%%%%%%%%%%%%%%%%%%%%%%%%%%%%%%%%%%%%

\noindent 

\begin{table}[h!]
\centering
\caption{
\textbf{2SLS Regression of House Price Growth on Owner-Occupied PLNJM Growth and Non-Owner-Occupied PLNJM Growth in Boom Period (99-05)} \smallskip \newline
{\footnotesize 
This table reports OLS, reduced-form, first stage, and second stage results of 2SLS regression $\triangle_{99,05} Ln(HPI_{c}) = \beta * \triangle_{99,05} Ln(PLNJM\_Own_{c}) + \theta* \text{Credit-Independent Speculation (c,99-05)}  +  \gamma \bm{Controls_{c}} + \epsilon_{c}$. The left-hand-side dependent variable $\triangle_{99,05} Ln(HPI_{c})$ is the growth rate of the house price index at county $c$ 99-05. As a proxy for pure credit expansion, $\triangle_{99,05} Ln(PLNJM\_Own_{c})$ is the growth rate of the dollar amount (07USD) of owner-occupied private-label (non-jumbo) mortgages (PLNJM) at county $c$ 99-05. As a proxy for credit-independent speculation at county $c$ 99-05, $\text{Credit-Independent Speculation (c,99-05)}$ is derived from the regression Eq (\ref{eq:PLNJMNonOwn_on_PLNJMOwn}), which is a part of growth rate of the non-owner-occupied private-label (non-jumbo) mortgages that cannot be explained by the growth rate of owner-occupied private-label (non-jumbo) mortgages. $Controls_{c}$ indicates control variables at county $c$ in 1999. We use the gravity model-based instrumental variable ($\triangle_{99,05}\text{givNetExp}_{m}$) as IV for $\triangle_{99,05} Ln(PLNJM\_Own_{c})$. For the first-stage F-test, we report Kleibergen-Paap (2006) robust (clustered) statistics and Montiel Olea-Pflueger (2013) efficient statistics. Regression is weighted by the natural logarithm of housing units in 1999. Standard errors are clustered at the CBSA level. ***, **, and * indicate significance at the 1\%, 5\%, and 10\% levels, respectively.
} % end of small font size
\smallskip 
} % end of caption
\label{table_HPI.D99t05.PLNJMOwn_vs_NonOwnIndependent.4Reg}

\resizebox{0.95\columnwidth}{!}{%

\begin{tabular}{l*{4}{c}}
\toprule
       &\multicolumn{4}{c}{House Price Growth (07USD, 99-05)} \\
       \cmidrule{2-5} 
        &\multicolumn{1}{c}{(1)}   &\multicolumn{1}{c}{(2)}   &\multicolumn{1}{c}{(3)}   &\multicolumn{1}{c}{(4)}   \\

\midrule
\multicolumn{5}{l}{\textbf{Panel A. OLS estimates}} \\
\addlinespace
PLNJM Own Dollar Growth (Credit Expansion, 99-05)&    0.184***&    0.176***&    0.141***&    0.131***\\
               &  (0.030)   &  (0.028)   &  (0.024)   &  (0.022)   \\
\addlinespace
PLNJM NonOwn Dollar Growth (Credit Independent, 99-05)&    0.040*  &    0.055***&    0.043***&    0.033***\\
               &  (0.022)   &  (0.019)   &  (0.016)   &  (0.011)   \\
\addlinespace
R2-adj         &    0.145   &    0.344   &    0.549   &    0.587   \\
\addlinespace

\midrule
\multicolumn{5}{l}{\textbf{Panel B. Reduced-form estimates}} \\
\addlinespace
GIV Net Export Growth (99-05)&    8.080***&    7.049***&    6.481***&    6.877***\\
               &  (1.861)   &  (1.953)   &  (1.621)   &  (1.475)   \\
\addlinespace
PLNJM NonOwn Dollar Growth (Credit Independent, 99-05)&    0.019   &    0.034*  &    0.024   &    0.016   \\
               &  (0.023)   &  (0.019)   &  (0.015)   &  (0.010)   \\
\addlinespace
R2-adj         &   0.0546   &    0.253   &    0.502   &    0.556   \\
\addlinespace

\midrule
\multicolumn{5}{l}{\textbf{Panel C . 2SLS estimates}} \\
\addlinespace
PLNJM Own Dollar Growth (Credit Expansion, 99-05)&    0.579***&    0.523***&    0.476***&    0.509***\\
               &  (0.119)   &  (0.134)   &  (0.147)   &  (0.163)   \\
\addlinespace
PLNJM NonOwn Dollar Growth (Credit Independent, 99-05)&    0.092***&    0.103***&    0.088***&    0.085***\\
               &  (0.026)   &  (0.026)   &  (0.025)   &  (0.025)   \\
\addlinespace

\addlinespace
\addlinespace

Dep Var (Panel D): &\multicolumn{4}{c}{PLNJM Own Growth (07USD, 99-05)} \\ 
\midrule 
\multicolumn{5}{l}{\textbf{Panel D . First-stage estimates}} \\
\addlinespace
GIV Net Export Growth (99-05)&   14.495***&   13.927***&   14.082***&   13.908***\\
               &  (3.162)   &  (3.224)   &  (3.485)   &  (3.804)   \\
\addlinespace
PLNJM NonOwn Dollar Growth (Credit Independent, 99-05)&   -0.115***&   -0.123***&   -0.124***&   -0.122***\\
               &  (0.032)   &  (0.029)   &  (0.029)   &  (0.028)   \\
\addlinespace
KP F-Stat      &    20.58   &    18.10   &    16.23   &    13.14   \\
MOP F-Eff      &    20.58   &    18.10   &    16.23   &    13.14   \\
\addlinespace
\midrule
\multicolumn{5}{l}{\textbf{Controls (for all Panels)}} \\
Basic Controls &            &  Y   &   Y    & Y        \\
Housing Controls &           &      & Y       & Y        \\
Demographic Controls &            &      &        &  Y        \\

\midrule              
Obs            &      769   &      769   &      690   &      690   \\
Cluster SE     &     CBSA   &     CBSA   &     CBSA   &     CBSA   \\
Weight         & Ln(HU99)   & Ln(HU99)   & Ln(HU99)   & Ln(HU99)   \\
\bottomrule

\end{tabular}

} % end of resize box

\end{table}

\pagebreak 
%---------------------------------------------------------------

%%%%%%%%%%%%%%%%%%%%%%%%%%%%%%%%%%%%%%%%%%%%%%%%
% table_PLNJMNonOwn.D91t99vsD99t05.4Reg
%%%%%%%%%%%%%%%%%%%%%%%%%%%%%%%%%%%%%%%%%%%%%%%%
%----------------------------------------------------------------

%%%%%%%%%%%%%%%%%%%%%%%%%%%%%%%%%%%%
% table_PLNJMNonOwn.D91t99vsD99t05.4Reg
%%%%%%%%%%%%%%%%%%%%%%%%%%%%%%%%%%%%

\noindent 

\begin{table}[h!]
\centering
\caption{
\textbf{2SLS Stacked Regression of Non-Owner-Occupied PLNJM on Net Export Growth in Prior (91-99) and Boom (99-05) Periods} \smallskip \newline
{\scriptsize 
This table reports OLS, reduced-form, first stage, and second stage results of 2SLS regression $\triangle_{91,99} \& \triangle_{99,05} Ln(PLNJM\_NonOwn_{c}) = \beta_{91,99} * \triangle_{91,99} \text{NetExp}_{m} \times Dum_{91,99} + \beta_{99,05} * \triangle_{99,05} \text{NetExp}_{m} \times Dum_{99,05} + \gamma_{91,99}* \bm{Controls_{c}} \times Dum_{91,99} + \gamma_{99,05}* \bm{Controls_{c}} \times Dum_{99,05}  + \epsilon_{period, c}$. The left-hand-side dependent variable $\triangle_{91,99} \& \triangle_{99,05} Ln(PLNJM\_NonOwn_{c})$ is the stacked growth rate of the dollar amount (07USD) of non-owner-occupied private-label (non-jumbo) mortgages (PLNJM) at county $c$ 91-99 and 99-05. The key variable of interest $\triangle_{91,99} \text{NetExp}_{m}$ and $\triangle_{99,05} \text{NetExp}_{m}$ are the growth rate of net export at the metropolitan area (CBSA03 code) $m$ 91-99 and 99-05, respectively. $Controls_{c}$ indicates control variables at county $c$ in the period start year, either 1991 or 1999. We use the gravity model-based instrumental variable $\triangle_{91,99}\text{givNetExp}_{m}$ and $\triangle_{99,05}\text{givNetExp}_{m}$ as IVs for $\triangle_{91,99}\text{NetExp}_{m}$ and $\triangle_{99,05}\text{NetExp}_{m}$. We report the statistics and p-values for the tests of coefficient equality between $\beta_{91,99}$ and $\beta_{99,05}$. For the first-stage F-test of two separate non-stack samples, we report kleibergen-Paap (2006) robust (clustered) statistics and Montiel Olea-Pflueger (2013) efficient statistics. Regression is weighted by the natural logarithm of housing units in the start year (either 1991 or 1999). Standard errors are clustered at the CBSA level. ***, **, and * indicate significance at the 1\%, 5\%, and 10\% levels, respectively.
} % end of small font size
} % end of caption
\label{table_PLNJMNonOwn.D91t99vsD99t05.4Reg}

\resizebox{0.82\columnwidth}{!}{%

\begin{tabular}{l*{4}{c}}
\toprule
\textbf{TSLS estimates}            &\multicolumn{4}{c}{Private-label Mortgage (Non-Owner) Growth (91-99 or 99-05, An)} \\
            \cmidrule{2-5} 
            &\multicolumn{1}{c}{(1)}&\multicolumn{1}{c}{(2)}&\multicolumn{1}{c}{(3)}&\multicolumn{1}{c}{(4)}\\
            
\midrule
\multicolumn{5}{l}{\textbf{Panel A. OLS estimates}} \\
Net Export Growth (91-99, An) x Dum91t99&    2.704   &    1.157   &   -6.329   &   -8.105   \\
               &  (5.496)   &  (5.061)   &  (6.034)   &  (5.839)   \\
\addlinespace
Net Export Growth (99-05, An) x Dum99t05&   12.087***&   11.561***&   12.149***&   12.703***\\
               &  (3.418)   &  (3.169)   &  (3.071)   &  (2.850)   \\
\addlinespace
R2-adj         &    0.848   &    0.855   &    0.862   &    0.867   \\
\addlinespace

\midrule
\multicolumn{5}{l}{\textbf{Panel B. Reduced-form estimates}} \\
GIV Net Export Growth (91-99, An) x Dum91t99&    5.928   &    4.449   &   -3.850   &   -8.336   \\
               &  (9.956)   &  (8.720)   & (10.661)   & (10.789)   \\
\addlinespace
GIV Net Export Growth (99-05, An) x Dum99t05&   17.499***&   15.974** &   19.425***&   19.398***\\
               &  (6.708)   &  (6.459)   &  (6.591)   &  (5.959)   \\
\addlinespace
R2-adj         &    0.847   &    0.854   &    0.861   &    0.866   \\
\addlinespace

\midrule
\multicolumn{5}{l}{\textbf{Panel C . 2SLS estimates}} \\
\addlinespace
Net Export Growth (91-99, An) x Dum91t99&    6.259   &    4.708   &   -5.127   &  -11.153   \\
               & (10.131)   &  (8.990)   & (14.568)   & (15.376)   \\
\addlinespace
Net Export Growth (99-05, An) x Dum99t05&   15.270** &   14.024** &   17.254** &   17.058***\\
               &  (6.508)   &  (5.991)   &  (6.821)   &  (5.598)   \\
CoefEqual\_Chi2 &    0.544   &    0.717   &    1.976   &    3.014   \\
CoefEqual\_PValue&    0.461   &    0.397   &    0.160   &   0.0826   \\            
\addlinespace
\addlinespace

Dep Var (Panel D): &\multicolumn{4}{c}{Net Export Growth (91-99, An)} \\ 
\midrule 
\multicolumn{5}{l}{\textbf{Panel D. First-stage estimates only for 91-99 (non-stack sample)}} \\
\addlinespace
GIV NEG (91-99, An)&    1.077***&    1.074***&    0.979***&    0.971***\\
               &  (0.128)   &  (0.125)   &  (0.136)   &  (0.133)   \\
\addlinespace
KP F-Stat       &    35.62   &    37.16   &    25.09   &    25.75   \\
MOP F-Stat       &    35.62   &    37.16   &    25.09   &    25.75   \\

\addlinespace
\addlinespace

Dep Var (Panel E): &\multicolumn{4}{c}{Net Export Growth (99-05, An) } \\ 
\midrule 
\multicolumn{5}{l}{\textbf{Panel E. First-stage estimates only for 99-05 (non-stack sample)}} \\
\addlinespace
GIV NEG (99-05, An)&    1.150***&    1.143***&    1.131***&    1.142***\\
               &  (0.238)   &  (0.238)   &  (0.281)   &  (0.266)   \\
\addlinespace
KP F-Stat      &    22.72   &    22.40   &    15.46   &    17.70   \\
MOP F-Stat      &    22.72   &    22.40   &    15.46   &    17.70   \\

\addlinespace
\addlinespace

\midrule
\multicolumn{5}{l}{\textbf{Controls (DumPeriod for stacked sample and no DumPeriod for non-stacked sample)}} \\
DumPeriod  &    Y        &  Y   &   Y    & Y    \\
Basic Controls x DumPeriod &            &  Y   &   Y    & Y    \\
Housing Controls x DumPeriod &           &      & Y       & Y    \\
Demographic Controls x DumPeriod &            &      &        &  Y    \\

\midrule              
Obs (Panel A, B, and C)         &     1266   &     1266   &     1139   &     1139   \\
Obs (Panel D)           &      492   &      492   &      444   &      444   \\
Obs (Panel E)           &      774   &      774   &      695   &      695   \\
Cluster SE     &     CBSA   &     CBSA   &     CBSA   &     CBSA     \\
Weight         & {\scriptsize Ln(HU-Start)}   & {\scriptsize Ln(HU-Start)}   &{\scriptsize Ln(HU-Start)}   &{\scriptsize Ln(HU-Start)}    \\
\bottomrule

\end{tabular}

} % end of resize box

\end{table}

%---------------------------------------------------------------------------------------
%---------------------------------------------------------------------------------------
%---------------------------------------------------------------------------------------
%---------------------------------------------------------------------------------------
% Model-based New Prediction - Double Differences
%---------------------------------------------------------------------------------------
%---------------------------------------------------------------------------------------
%---------------------------------------------------------------------------------------
%---------------------------------------------------------------------------------------

\pagebreak 
%---------------------------------------------------------------
%%%%%%%%%%%%%%%%%%%%%%%%%%%%%%%%%%%%%%%%%%%%%%%%
% table_ZIP.LMH.PLNJM.D99t05vsD05t08.4Reg
%%%%%%%%%%%%%%%%%%%%%%%%%%%%%%%%%%%%%%%%%%%%%%%%
%---------------------------------------------------------------

%%%%%%%%%%%%%%%%%%%%%%%%%%%%%%%%%%%%%%%%%%%%%%%%
% table_ZIP.LMH.PLNJM.D99t05vsD05t08.4Reg
%%%%%%%%%%%%%%%%%%%%%%%%%%%%%%%%%%%%%%%%%%%%%%%%

\noindent 

\begin{table}[h!]
\centering
\caption{
\textbf{Four stacked Regressions of Low-minus-High PLNJM Growth in Boom (99-05) and Bust (05-08) on Net Export Growth (99-05) } \smallskip \newline
{\footnotesize 
This table reports OLS, reduced-form, first stage, and second stage results of 2SLS stacked regression $\text{LMH}\triangle_{99,05} \&  \triangle_{05,08} Ln(PLNJM_{m}) = \beta_{99,05} * \triangle_{99,05} \text{NetExp}_{m} \times Dum_{99,05} + \beta_{05,08} * \triangle_{99,05} \text{NetExp}_{m} \times Dum_{05,08} + \gamma_{99,05}* Dum_{99,05} + \gamma_{05,08}* Dum_{05,08} + \epsilon_{period, m}$. The left-hand-side dependent variable $\text{LMH}\triangle_{99,05} \&  \triangle_{05,08} Ln(PLNJM_{m})$ is the differential growth rate of the dollar amount (07USD) of private-label (non-jumbo) mortgages (PLNJM) at metro $m$ between the low- and high-income ZIP codes 99-05 and 05-08. The dependent variables in column (1) to (4) are the above ``low-minus-high" PLNJM growth based on two, tertile, quartile, and quintile income groups of ZIP codes within metropolitan areas. The key variable of interest $\triangle_{99,05} \text{NetExp}_{m}$ is the growth rate of net export at the metropolitan area (CBSA03 code) $m$ 99-05. We use the gravity model-based instrumental variable ($\triangle_{99,05}\text{givNetExp}_{m}$) as IV for $\triangle_{99,05}\text{NetExp}_{m}$. For the first-stage F-test, we report kleibergen-Paap (2006) robust (clustered) statistics and Montiel Olea-Pflueger (2013) efficient statistics. Each regression is weighted by the natural logarithm of housing units in metro $m$ in 1999. Standard errors are clustered at the CBSA level. ***, **, and * indicate significance at the 1\%, 5\%, and 10\% levels, respectively. ``An" means annualized.  ``An" means annualized variable. 
} % end of small font size
} % end of caption
\label{table_ZIP.LMH.PLNJM.D99t05vsD05t08.4Reg}

\resizebox{\columnwidth}{!}{

\begin{tabular}{l*{4}{c}}
\toprule
Dep Var (Panel A, B, and C)                     &\multicolumn{4}{c}{Low-minus-High PLNJM Growth (99-05 \& 05-08, An, 07USD)} \\
            \cmidrule{2-5} 
            &\multicolumn{1}{c}{(1)}&\multicolumn{1}{c}{(2)}&\multicolumn{1}{c}{(3)}&\multicolumn{1}{c}{(4)} \\
            
            &\multicolumn{1}{c}{Two Groups}&\multicolumn{1}{c}{Tertile Groups}&\multicolumn{1}{c}{Quartile Groups}&\multicolumn{1}{c}{Quintile Groups}\\
            
\midrule
\multicolumn{5}{l}{\textbf{Panel A. OLS estimates}} \\
\addlinespace
Net Export Growth (99-05, An) x Dum99t05&    3.861*  &    5.928** &    6.716** &    6.965** \\
               &  (2.019)   &  (2.478)   &  (2.634)   &  (3.217)   \\
\addlinespace
Net Export Growth (99-05, An) x Dum05t08&   -7.331***&   -9.383***&   -8.909** &   -9.485** \\
               &  (2.573)   &  (3.093)   &  (4.191)   &  (4.712)   \\
\addlinespace
R2-adj         &    0.224   &    0.256   &    0.257   &    0.245   \\
\addlinespace

\midrule
\multicolumn{5}{l}{\textbf{Panel B. Reduced-form estimates}} \\
\addlinespace
GIV Net Export Growth (99-05, An) x Dum99t05&    6.859*  &    9.984** &   13.131** &   14.882** \\
               &  (3.728)   &  (4.980)   &  (6.034)   &  (6.970)   \\
\addlinespace
GIV Net Export Growth (99-05, An) x Dum05t08&   -6.997   &  -17.342***&  -18.368***&  -25.099***\\
               &  (6.460)   &  (5.400)   &  (6.583)   &  (8.211)   \\
\addlinespace
R2-adj         &    0.216   &    0.256   &    0.260   &    0.252   \\
\addlinespace

\midrule
\multicolumn{5}{l}{\textbf{Panel C . 2SLS estimates}} \\
\addlinespace
Net Export Growth (99-05, An) x Dum99t05&    7.926*  &   11.537** &   15.168** &   17.184** \\
               &  (4.466)   &  (5.843)   &  (6.692)   &  (7.616)   \\
\addlinespace
Net Export Growth (99-05, An) x Dum05t08&   -8.085   &  -20.040***&  -21.217** &  -28.992***\\
               &  (6.479)   &  (7.034)   &  (8.330)   & (11.154)   \\
\addlinespace
CoefEqual\_Chi2 &    3.017   &    8.540   &    8.334   &    8.438   \\
CoefEqual\_PValue&    0.082   &    0.003   &    0.004   &    0.004   \\

\addlinespace 
\addlinespace

Dep Var (Panel D): &\multicolumn{4}{c}{Net Export Growth (99-05, An)} \\ 
\midrule 
\multicolumn{5}{l}{\textbf{Panel D . First-stage estimates only for 99-05 (Non-stack sample)}} \\
\addlinespace
GIV Net Export Growth (99-05, An) x Dum99t05&    0.865***&    0.865***&    0.865***&    0.865***\\
               &  (0.227)   &  (0.227)   &  (0.227)   &  (0.227)   \\
\addlinespace
KP F-Stat      &    14.54   &    14.54   &    14.55   &    14.50   \\
MOP F-Stat     &    14.54   &    14.54   &    14.55   &    14.50   \\
\addlinespace

\midrule  
Obs (Panel A, B, and C)          &      600   &      600   &      598   &      597   \\
Obs (Panel D)           &      300   &      300   &      299   &      298   \\
Cluster SE     &     CBSA   &     CBSA   &     CBSA   &     CBSA     \\
Weight         &{\scriptsize Ln(CBSAHU99)}   &{\scriptsize Ln(CBSAHU99)}   &{\scriptsize Ln(CBSAHU99)}   &{\scriptsize Ln(CBSAHU99)}   \\
\bottomrule

\end{tabular}
} % end of resize box

\end{table}

\pagebreak 
%---------------------------------------------------------------
%%%%%%%%%%%%%%%%%%%%%%%%%%%%%%%%%%%%%%%%%%%%%%%%
% table_ZIP.LMH.GSEM.D99t05vsD05t08.4Reg
%%%%%%%%%%%%%%%%%%%%%%%%%%%%%%%%%%%%%%%%%%%%%%%%
%---------------------------------------------------------------

%%%%%%%%%%%%%%%%%%%%%%%%%%%%%%%%%%%%%%%%%%%%%%%%
% table_ZIP.LMH.GSEM.D99t05vsD05t08.4Reg
%%%%%%%%%%%%%%%%%%%%%%%%%%%%%%%%%%%%%%%%%%%%%%%%

\noindent 

\begin{table}[h!]
\centering
\caption{
\textbf{Four stacked Regressions of Low-minus-High GSEM Growth in Boom (99-05) and Bust (05-08) on Net Export Growth (99-05) } \smallskip \newline
{\footnotesize 
This table reports OLS, reduced-form, first stage, and second stage results of 2SLS stacked regression $\text{LMH}\triangle_{99,05} \&  \triangle_{05,08} Ln(GSEM_{m}) = \beta_{99,05} * \triangle_{99,05} \text{NetExp}_{m} \times Dum_{99,05} + \beta_{05,08} * \triangle_{99,05} \text{NetExp}_{m} \times Dum_{05,08} + \gamma_{99,05}* Dum_{99,05} + \gamma_{05,08}* Dum_{05,08} + \epsilon_{period, m}$. The left-hand-side dependent variable $\text{LMH}\triangle_{99,05} \&  \triangle_{05,08} Ln(GSEM_{m})$ is the differential growth rate of the dollar amount (07USD) of government-sponsored enterprise mortgages (GSEM) at metro $m$ between the low- and high-income ZIP codes 99-05 and 05-08. The dependent variables in column (1) to (4) are the above ``low-minus-high" GSEM growth based on two, tertile, quartile, and quintile income groups of ZIP codes within metropolitan areas. The key variable of interest $\triangle_{99,05} \text{NetExp}_{m}$ is the growth rate of net export at the metropolitan area (CBSA03 code) $m$ 99-05. We use the gravity model-based instrumental variable ($\triangle_{99,05}\text{givNetExp}_{m}$) as IV for $\triangle_{99,05}\text{NetExp}_{m}$. For the first-stage F-test, we report kleibergen-Paap (2006) robust (clustered) statistics and Montiel Olea-Pflueger (2013) efficient statistics. Each regression is weighted by the natural logarithm of housing units in metro $m$ in 1999. Standard errors are clustered at the CBSA level. ***, **, and * indicate significance at the 1\%, 5\%, and 10\% levels, respectively. ``An" means annualized variable. 
} % end of small font size
} % end of caption
\label{table_ZIP.LMH.GSEM.D99t05vsD05t08.4Reg}

\resizebox{\columnwidth}{!}{

\begin{tabular}{l*{4}{c}}
\toprule
Dep Var (Panel A, B, and C)                     &\multicolumn{4}{c}{Low-minus-High GSEM Growth (99-05 \& 05-08, An, 07USD)} \\
            \cmidrule{2-5} 
            &\multicolumn{1}{c}{(1)}&\multicolumn{1}{c}{(2)}&\multicolumn{1}{c}{(3)}&\multicolumn{1}{c}{(4)} \\
            
            &\multicolumn{1}{c}{Two Groups}&\multicolumn{1}{c}{Tertile Groups}&\multicolumn{1}{c}{Quartile Groups}&\multicolumn{1}{c}{Quintile Groups}\\
            
\midrule
\multicolumn{5}{l}{\textbf{Panel A. OLS estimates}} \\
\addlinespace
Net Export Growth (99-05, An) x Dum99t05&    1.176   &    2.843   &    3.519   &    2.203   \\
               &  (1.748)   &  (2.264)   &  (2.457)   &  (2.744)   \\
\addlinespace
Net Export Growth (99-05, An) x Dum05t08&   -4.750** &   -4.100*  &   -5.851** &   -5.395   \\
               &  (2.047)   &  (2.432)   &  (2.864)   &  (3.483)   \\
\addlinespace
R2-adj         &  0.00811   &   0.0194   &   0.0316   &   0.0277   \\
\addlinespace

\midrule
\multicolumn{5}{l}{\textbf{Panel B. Reduced-form estimates}} \\
\addlinespace
GIV Net Export Growth (99-05, An) x Dum99t05&   -0.417   &    0.312   &    2.668   &    2.414   \\
               &  (2.553)   &  (3.417)   &  (4.125)   &  (5.253)   \\
\addlinespace
GIV Net Export Growth (99-05, An) x Dum05t08&   -1.359   &   -0.214   &   -3.645   &   -6.711   \\
               &  (5.268)   &  (4.692)   &  (5.655)   &  (7.095)   \\
\addlinespace
R2-adj         &-0.000784   &   0.0134   &   0.0244   &   0.0259   \\
\addlinespace

\midrule
\multicolumn{5}{l}{\textbf{Panel C . 2SLS estimates}} \\
\addlinespace
Net Export Growth (99-05, An) x Dum99t05 &   -0.482   &    0.361   &    3.081   &    2.788   \\
               &  (2.949)   &  (3.916)   &  (4.674)   &  (6.029)   \\
\addlinespace
Net Export Growth (99-05, An) x Dum05t08 &   -1.571   &   -0.247   &   -4.212   &   -7.754   \\
               &  (5.834)   &  (5.370)   &  (5.973)   &  (7.534)   \\
\addlinespace
CoefEqual\_Chi2 &    0.022   &    0.006   &    0.677   &    0.814   \\
CoefEqual\_PValue&    0.883   &    0.941   &    0.411   &    0.367   \\

\addlinespace 
\addlinespace

Dep Var (Panel D): &\multicolumn{4}{c}{Net Export Growth (99-05, An)} \\ 
\midrule 
\multicolumn{5}{l}{\textbf{Panel D . First-stage estimates only for 99-05 (Non-stack sample)}} \\
\addlinespace
GIV Net Export Growth (99-05, An) x Dum99t05&    0.865***&    0.865***&    0.865***&    0.865***\\
               &  (0.227)   &  (0.227)   &  (0.227)   &  (0.227)   \\
\addlinespace
KP F-Stat      &    14.54   &    14.54   &    14.55   &    14.55   \\
MOP F-Stat     &    14.54   &    14.54   &    14.55   &    14.55   \\
\addlinespace

\midrule  
Obs (Panel A, B, and C)           &      600   &      600   &      599   &      599   \\
Obs (Panel D)           &      300   &      300   &      300   &      300   \\
Cluster SE     &     CBSA   &     CBSA   &     CBSA   &     CBSA     \\
Weight         &{\scriptsize Ln(CBSAHU99)}   &{\scriptsize Ln(CBSAHU99)}   &{\scriptsize Ln(CBSAHU99)}   &{\scriptsize Ln(CBSAHU99)}   \\
\bottomrule

\end{tabular}
} % end of resize box

\end{table}

\pagebreak 
%---------------------------------------------------------------
%%%%%%%%%%%%%%%%%%%%%%%%%%%%%%%%%%%%%%%%%%%%%%%%
% table_ZIP.LMH.PLNJM.D92t99vsD99t05.4Reg.tex
%%%%%%%%%%%%%%%%%%%%%%%%%%%%%%%%%%%%%%%%%%%%%%%%
%---------------------------------------------------------------

%%%%%%%%%%%%%%%%%%%%%%%%%%%%%%%%%%%%%%%%%%%%%%%%
% table_ZIP.LMH.PLNJM.D92t99vsD99t05.4Reg
%%%%%%%%%%%%%%%%%%%%%%%%%%%%%%%%%%%%%%%%%%%%%%%%

\noindent 

\begin{table}[h!]
\centering
\caption{
\textbf{Four stacked Regressions of Low-minus-High PLNJM Growth on Net Export Growth in Prior (92-99) and Boom (99-05)} \smallskip \newline
{\footnotesize 
This table reports OLS, reduced-form, first stage, and second stage results of 2SLS stacked regression $\text{LMH}\triangle_{92,99} \&  \triangle_{99,05} Ln(PLNJM_{m}) = \beta_{92,99} * \triangle_{92,99} \text{NetExp}_{m} \times Dum_{92,99} + \beta_{99,05} * \triangle_{99,05} \text{NetExp}_{m} \times Dum_{99,05} + \gamma_{92,99}* Dum_{92,99} + \gamma_{99,05}* Dum_{99,05} + \epsilon_{period, m}$. The left-hand-side dependent variable $\text{LMH}\triangle_{92,99} \&  \triangle_{99,05} Ln(PLNJM_{m})$ is the differential growth rate of the dollar amount (07USD) of private-label (non-jumbo) mortgages (PLNJM) at metro $m$ between the low- and high-income ZIP codes 92-99 and 99-05. The dependent variables in columns (1) to (4) are the above ``low-minus-high" PLNJM growth based on two, tertile, quartile, and quintile income groups of ZIP codes within metropolitan areas in 1998. The key variable of interest $\triangle_{92,99} \&  \triangle_{99,05}\text{NetExp}_{m}$ is the growth rate of net export at the metropolitan area (CBSA03 code) $m$ 92-99 and 99-05. We use the gravity model-based instrumental variable ($\triangle_{92,99}\text{givNetExp}_{m}$ and $\triangle_{99,05}\text{givNetExp}_{m}$) as IV for $\triangle_{92,99}\text{NetExp}_{m}$ and $\triangle_{99,05}\text{NetExp}_{m}$. For the first-stage F-test, we report kleibergen-Paap (2006) robust (clustered) statistics and Montiel Olea-Pflueger (2013) efficient statistics. Each regression is weighted by the natural logarithm of housing units in metro $m$ in 1992 for 92-99 period and in 1999 for 99-05 period. Standard errors are clustered at the CBSA level. ***, **, and * indicate significance at the 1\%, 5\%, and 10\% levels, respectively. ``An" means annualized variable. 
} % end of small font size
} % end of caption
\label{table_ZIP.LMH.PLNJM.D92t99vsD99t05.4Reg}

\resizebox{0.88\columnwidth}{!}{

\begin{tabular}{l*{4}{c}}
\toprule
Dep Var (Panel A, B, and C)                     &\multicolumn{4}{c}{Low-minus-High PLNJM Growth (92-99 \& 99-05, An, 07USD)} \\
            \cmidrule{2-5} 
            &\multicolumn{1}{c}{(1)}&\multicolumn{1}{c}{(2)}&\multicolumn{1}{c}{(3)}&\multicolumn{1}{c}{(4)} \\
            
            &\multicolumn{1}{c}{Two Groups}&\multicolumn{1}{c}{Tertile Groups}&\multicolumn{1}{c}{Quartile Groups}&\multicolumn{1}{c}{Quintile Groups}\\
            
\midrule
\multicolumn{5}{l}{\textbf{Panel A. OLS estimates}} \\
\addlinespace
Net Export Growth (92-99, An) x Dum92t99&    2.053   &    4.439   &    3.574   &    2.709   \\
               &  (2.121)   &  (2.817)   &  (2.536)   &  (2.669)   \\
\addlinespace
Net Export Growth (99-05, An) x Dum99t05&    3.861*  &    5.928** &    6.716** &    6.965** \\
               &  (2.019)   &  (2.478)   &  (2.634)   &  (3.217)   \\
\addlinespace
R2-adj         &    0.106   &    0.109   &    0.106   &   0.0932   \\
\addlinespace

\midrule
\multicolumn{5}{l}{\textbf{Panel B. Reduced-form estimates}} \\
\addlinespace
GIV Net Export Growth (92-99, An) x Dum92t99&    1.325   &    7.495   &    5.478   &    1.810   \\
               &  (3.783)   &  (5.481)   &  (4.991)   &  (4.630)   \\
\addlinespace
GIV Net Export Growth (99-05, An) x Dum99t05&    6.859*  &    9.984** &   13.131** &   14.882** \\
               &  (3.728)   &  (4.981)   &  (6.035)   &  (6.971)   \\
\addlinespace
R2-adj         &    0.104   &    0.109   &    0.108   &   0.0954   \\
\addlinespace

\midrule
\multicolumn{5}{l}{\textbf{Panel C . 2SLS estimates}} \\
\addlinespace
Net Export Growth (92-99, An) x Dum92t99 &    1.176   &    6.650   &    4.844   &    1.604   \\
               &  (3.320)   &  (4.807)   &  (4.362)   &  (4.077)   \\
\addlinespace
Net Export Growth (99-05, An) x Dum99t05 &    7.926*  &   11.537** &   15.168** &   17.184** \\
               &  (4.466)   &  (5.843)   &  (6.692)   &  (7.616)   \\
\addlinespace
CoefEqual\_Chi2 &    1.184   &    0.385   &    1.445   &    2.754   \\
CoefEqual\_PValue&    0.276   &    0.535   &    0.229   &    0.097   \\

\addlinespace 
\addlinespace

Dep Var (Panel D): &\multicolumn{4}{c}{Net Export Growth (92-99, An)} \\ 
\midrule 
\multicolumn{5}{l}{\textbf{Panel D . First-stage estimates only for 92-99 (Non-stack sample)}} \\
\addlinespace
GIV Net Export Growth (92-99, An) x Dum92t99&    1.128***&    1.128***&    1.128***&    1.128***\\
               &  (0.100)   &  (0.100)   &  (0.100)   &  (0.100)   \\
\addlinespace
KP F-Stat      &    127.5   &    127.5   &    128.8   &    125.6   \\
MOP F-Stat     &    127.5   &    127.5   &    128.8   &    125.6   \\
\addlinespace
\addlinespace

Dep Var (Panel E): &\multicolumn{4}{c}{Net Export Growth (99-05, An)} \\ 
\midrule 
\multicolumn{5}{l}{\textbf{Panel D . First-stage estimates only for 99-05 (Non-stack sample)}} \\
\addlinespace
Net Export Growth (99-05, An) x Dum99t05&    7.926*  &   11.537** &   15.168** &   17.184** \\
               &  (4.466)   &  (5.843)   &  (6.692)   &  (7.616)   \\
\addlinespace
KP F-Stat      &    14.54   &    14.54   &    14.55   &    14.50   \\
MOP F-Stat     &    14.54   &    14.54   &    14.55   &    14.50   \\
\addlinespace

\midrule  
Obs (Panel A, B, and C)         &      580   &      580   &      578   &      575   \\
Obs (Panel D)            &      280   &      280   &      279   &      277   \\
Obs (Panel E)          &      300   &      300   &      299   &      298   \\
Cluster SE     &     CBSA   &     CBSA   &     CBSA   &     CBSA     \\
Weight         &{\scriptsize Ln(CBSAHU,start)}   &{\scriptsize Ln(CBSAHU,start)}   &{\scriptsize Ln(CBSAHU,start)}   &{\scriptsize Ln(CBSAHU,start)}   \\
\bottomrule

\end{tabular}
} % end of resize box

\end{table}

\pagebreak 
%---------------------------------------------------------------
%%%%%%%%%%%%%%%%%%%%%%%%%%%%%%%%%%%%%%%%%%%%%%%%
% table_ZIP.LMH.HPI.D00t06vsD07t09.PLNJM.4Reg
%%%%%%%%%%%%%%%%%%%%%%%%%%%%%%%%%%%%%%%%%%%%%%%%
%---------------------------------------------------------------
%%%%%%%%%%%%%%%%%%%%%%%%%%%%%%%%%%%%%%%%%%%%%%%%
% table_ZIP.LMH.HPI.D00t06vsD07t09.PLNJM.4Reg
%%%%%%%%%%%%%%%%%%%%%%%%%%%%%%%%%%%%%%%%%%%%%%%%
\noindent 

\begin{table}[h!]
\centering
\caption{
\textbf{Four Stacked Regressions of Low-minus-High Housing Price Growth in Boom (00-06) and Bust (07-09) Periods on Low-minus-High PLNJM Growth (99-05)} \smallskip \newline
{\footnotesize
This table reports OLS, reduced-form, first stage, and second stages results of stacked 2SLS regression $\text{LMH}\triangle_{00,06} \quad \& \quad \text{LMH}\triangle_{07,09} Ln(HPI_{m})  = \beta_{00,06} * \text{LMH}\triangle_{99,05} Ln(PLNJM_{m}) \times Dum_{00,06} + \beta_{07,09} * \text{LMH}\triangle_{99,05} Ln(PLNJM_{m}) \times Dum_{07,09} + \alpha_{00,06} + \alpha_{07,09} + \epsilon_{period, m}$. The left-hand-side dependent variable $\text{LMH}\triangle_{00,06} \& \text{LMH}\triangle_{07,09} Ln(HPI_{m})$ is the stacked differential growth rate of the house price index between low-income and high-income ZIP codes within metro $m$ 00-06 and 07-09. The key variable of interest $\text{LMH}\triangle_{99,05} Ln(PLNJM_{m}$ is the growth rate of the dollar amount of private-label (non-jumbo) mortgages between low-income and high-income ZIP codes within metro $m$ 99-05. We use the gravity model-based instrumental variable $\triangle_{99,05}\text{givNetExp}_{m}$ as the IV for $\text{LMH}\triangle_{99,05} Ln(PLNJM_{m}$. Each regression is weighted by the natural logarithm of housing units in 1999. For the first-stage F-test of the non-stacked sample, we report kleibergen-Paap (2006) robust (clustered) statistics and Montiel Olea-Pflueger (2013) efficient statistics. Standard errors are clustered at the CBSA level. ***, **, and * indicate significance at the 1\%, 5\%, and 10\% levels, respectively. ``An" means annualized variable. 
} % end of small font size
} % end of caption
\label{table_ZIP.LMH.HPI.D00t06vsD07t09.PLNJM.4Reg}
\resizebox{\columnwidth}{!}{%
\begin{tabular}{l*{4}{c}}
\toprule
Dep Var (Panel A, B, and C)                      &\multicolumn{4}{c}{Low-minus-High House Price Growth (00-06 \& 07-09, An, 07USD)} \\
            \cmidrule{2-5} 
            &\multicolumn{1}{c}{(1)}&\multicolumn{1}{c}{(2)}&\multicolumn{1}{c}{(3)}&\multicolumn{1}{c}{(4)} \\
            
            &\multicolumn{1}{c}{Two Groups}&\multicolumn{1}{c}{Tertile Groups}&\multicolumn{1}{c}{Quartile Groups}&\multicolumn{1}{c}{Quintile Groups}\\

\midrule
\multicolumn{5}{l}{\textbf{Panel A. OLS estimates}} \\
Low-minus-High PLNJM Growth (07USD, 99-05, An) x Dum00t06 &    0.053***&0.050***     &0.048***     &0.041***     \\
               &(0.010)      &(0.014)     &(0.011)      &(0.010)      \\
\addlinespace
Low-minus-High PLNJM Growth (07USD, 99-05, An) x Dum07t09 &-0.218***    &-0.219***    &-0.206***    &-0.172***    \\
               &(0.045)      & (0.046)     &(0.044)      &(0.035)      \\
\addlinespace
R2-adj         &    0.296   &    0.313   &    0.298   &    0.285   \\
\addlinespace

\midrule
\multicolumn{5}{l}{\textbf{Panel B. Reduced-form estimates}} \\
GIV Net Export Growth (99-05, An) x Dum00t06&    0.681*  &    1.234***&    1.608***&    1.731***\\
               &  (0.349)   &  (0.465)   &  (0.570)   &  (0.557)   \\
\addlinespace
GIV Net Export Growth (99-05, An) x Dum07t09&   -5.044***&   -6.860***&   -7.995***&   -8.206***\\
               &  (1.486)   &  (1.909)   &  (2.173)   &  (2.386)   \\
\addlinespace
R2-adj         &    0.175   &    0.179   &    0.182   &    0.191   \\
\addlinespace

\midrule
\multicolumn{5}{l}{\textbf{Panel C . 2SLS estimates}} \\
\addlinespace
Low-minus-High PLNJM Growth (07USD, 99-05, An) x Dum00t06 &0.096     &0.117*     &0.134**     &0.122*       \\
               &(0.060)      &(0.060)      &(0.064)      &(0.063)     \\
\addlinespace
Low-minus-High PLNJM Growth (07USD, 99-05, An) x Dum07t09 &-0.711**    &-0.650**    &-0.665***    &-0.579**      \\
               & (0.341)     &(0.274)    &(0.251)    &(0.248)     \\
\addlinespace
CoefEqual\_Chi2 &    4.332   &    5.604   &    6.979   &    5.381   \\
CoefEqual\_PValue&    0.037   &    0.018   &    0.008   &    0.020   \\          
\addlinespace
\addlinespace

Dep Var (Panel D): &\multicolumn{4}{c}{Low-minus-High PLNJM Growth (07 USD, 99-05, An)} \\ 
\midrule 
\multicolumn{5}{l}{\textbf{Panel D . First-stage estimates only for 00-06 (Non-stack sample)}} \\
\addlinespace
GIV Net Export Growth (99-05, An) x Dum00t06&    7.090** &   10.554** &   12.028** &   14.173** \\
               &  (3.446)   &  (4.551)   &  (4.820)   &  (6.253)   \\
\addlinespace
KP F-Stat      &    4.233   &    5.379   &    6.228   &    5.138   \\
MOP F-Stat     &    4.233   &    5.379   &    6.228   &    5.138   \\
\addlinespace

\midrule         
Obs (Panel A, B, and C)          &      540   &      540   &      540   &      540   \\
Obs (Panel D)            &      270   &      270   &      270   &      270   \\
Cluster SE     &     CBSA   &     CBSA   &     CBSA   &     CBSA    \\
Weight         & {\scriptsize Ln(CBSAHU99)}   & {\scriptsize Ln(CBSAHU99)}   &{\scriptsize Ln(CBSAHU99)}   &{\scriptsize Ln(CBSAHU99)}   \\
\bottomrule
\end{tabular}

}  % end of resize box

\end{table}

\pagebreak 
%---------------------------------------------------------------
%%%%%%%%%%%%%%%%%%%%%%%%%%%%%%%%%%%%%%%%%%%%%%%%
% table_ZIP.LMH.HPI.D00t06vsD07t09.NEG.4Reg
%%%%%%%%%%%%%%%%%%%%%%%%%%%%%%%%%%%%%%%%%%%%%%%%
%---------------------------------------------------------------

%%%%%%%%%%%%%%%%%%%%%%%%%%%%%%%%%%%%%%%%%%%%%%%%
% table_ZIP.LMH.HPI.D92t99vsD00t06.NEG.4Reg
%%%%%%%%%%%%%%%%%%%%%%%%%%%%%%%%%%%%%%%%%%%%%%%%

\noindent 

\begin{table}[h!]
\centering
\caption{
\textbf{Four stacked Regressions of Low-minus-High House Price Growth on Net Export Growth in Prior (92-99) and Boom (00-06)} \smallskip \newline
{\footnotesize 
This table reports OLS, reduced-form, first stage, and second stage results of 2SLS stacked regression $\text{LMH}\triangle_{92,99} \&  \triangle_{00,06} Ln(HPI_{m}) = \beta_{92,99} * \triangle_{92,99} \text{NetExp}_{m} \times Dum_{92,99} + \beta_{00,06} * \triangle_{99,05} \text{NetExp}_{m} \times Dum_{00,06} + Dum_{92,99} + Dum_{00,06}  + \epsilon_{period, m}$. The left-hand-side dependent variable $\text{LMH}\triangle_{92,99} \&  \triangle_{00,06} Ln(HPI_{m})$ is the differential growth rate of the house price index at metro $m$ between the low- and high-income ZIP codes 92-99 and 99-05. The dependent variables in columns (1) to (4) are the above ``low-minus-high" house price growth based on two, tertile, quartile, and quintile income groups of ZIP codes within metropolitan areas in 1998. The key variable of interest $\triangle_{92,99} \&  \triangle_{99,05}\text{NetExp}_{m}$ is the growth rate of net export at the metropolitan area (CBSA03 code) $m$ 92-99 and 99-05. We use the gravity model-based instrumental variable ($\triangle_{92,99}\text{givNetExp}_{m}$ and $\triangle_{99,05}\text{givNetExp}_{m}$) as IV for $\triangle_{92,99}\text{NetExp}_{m}$ and $\triangle_{99,05}\text{NetExp}_{m}$. For the first-stage F-test, we report kleibergen-Paap (2006) robust (clustered) statistics and Montiel Olea-Pflueger (2013) efficient statistics. Each regression is weighted by the natural logarithm of housing units in metro $m$ in 1992 for 92-99 period and in 1999 for 00-06 period. Standard errors are clustered at the CBSA level. ***, **, and * indicate significance at the 1\%, 5\%, and 10\% levels, respectively. ``An" means annualized variable. 
} % end of small font size
} % end of caption
\label{table_ZIP.LMH.HPI.D92t99vsD00t06.NEG.4Reg}

\resizebox{0.88\columnwidth}{!}{

\begin{tabular}{l*{4}{c}}
\toprule
Dep Var (Panel A, B, and C)                     &\multicolumn{4}{c}{Low-minus-High House Price Growth (92-99 \& 00-06, An, 07USD)} \\
            \cmidrule{2-5} 
            &\multicolumn{1}{c}{(1)}&\multicolumn{1}{c}{(2)}&\multicolumn{1}{c}{(3)}&\multicolumn{1}{c}{(4)} \\
            
            &\multicolumn{1}{c}{Two Groups}&\multicolumn{1}{c}{Tertile Groups}&\multicolumn{1}{c}{Quartile Groups}&\multicolumn{1}{c}{Quintile Groups}\\
            
\midrule
\multicolumn{5}{l}{\textbf{Panel A. OLS estimates}} \\
\addlinespace
Net Export Growth (92-99, An) x Dum92t99&   -0.316*  &   -0.487** &   -0.498** &   -0.569** \\
               &  (0.178)   &  (0.210)   &  (0.242)   &  (0.246)   \\
\addlinespace
Net Export Growth (99-05, An) x Dum00t06&    0.295   &    0.598** &    0.610*  &    0.727** \\
               &  (0.179)   &  (0.262)   &  (0.328)   &  (0.315)   \\
\addlinespace
R2-adj         &   0.0315   &   0.0341   &   0.0401   &   0.0420   \\
\addlinespace

\midrule
\multicolumn{5}{l}{\textbf{Panel B. Reduced-form estimates}} \\
\addlinespace
GIV Net Export Growth (92-99, An) x Dum92t99&   -0.219   &   -0.482   &   -0.490   &   -0.602   \\
               &  (0.269)   &  (0.314)   &  (0.377)   &  (0.393)   \\
\addlinespace
GIV Net Export Growth (99-05, An) x Dum00t06&    0.681*  &    1.234***&    1.608***&    1.731***\\
               &  (0.349)   &  (0.465)   &  (0.570)   &  (0.557)   \\
\addlinespace
R2-adj         &   0.0307   &   0.0331   &   0.0438   &   0.0443   \\
\addlinespace

\midrule
\multicolumn{5}{l}{\textbf{Panel C . 2SLS estimates}} \\
\addlinespace
Net Export Growth (92-99, An) x Dum92t99&   -0.188   &   -0.415   &   -0.421   &   -0.517   \\
               &  (0.230)   &  (0.265)   &  (0.319)   &  (0.330)   \\
\addlinespace
Net Export Growth (99-05, An) x Dum00t06&    0.680*  &    1.231** &    1.604***&    1.727***\\
               &  (0.366)   &  (0.506)   &  (0.609)   &  (0.612)   \\
\addlinespace
CoefEqual\_Chi2 &    3.319   &    6.771   &    7.690   &    8.857   \\
CoefEqual\_PValue &    0.068   &    0.009   &    0.006   &    0.003   \\

\addlinespace 
\addlinespace

Dep Var (Panel D): &\multicolumn{4}{c}{GIV Net Export Growth (92-99, An, 07USD)} \\ 
\midrule 
\multicolumn{5}{l}{\textbf{Panel D. First-stage estimates only for 92-99 (Non-stack sample)}} \\
\addlinespace
Net Export Growth (92-99, An) x Dum92t99 &   -0.188   &   -0.415   &   -0.421   &   -0.517   \\
               &  (0.230)   &  (0.265)   &  (0.319)   &  (0.330)   \\
\addlinespace
KP F-Stat      &    128.8   &    128.8   &    128.8   &    128.8   \\
MOP F-Stat     &    128.8   &    128.8   &    128.8   &    128.8   \\
\addlinespace
\addlinespace

Dep Var (Panel E): &\multicolumn{4}{c}{GIV Net Export Growth (99-05, An, 07USD)} \\ 
\midrule 
\multicolumn{5}{l}{\textbf{Panel D. First-stage estimates only for 00-06 (Non-stack sample)}} \\
\addlinespace
Net Export Growth (99-05, An) x Dum00t06&    0.680*  &    1.231** &    1.604***&    1.727***\\
               &  (0.366)   &  (0.506)   &  (0.609)   &  (0.612)   \\
\addlinespace
KP F-Stat      &    20.95   &    20.95   &    20.95   &    20.95   \\
MOP F-Stat     &    20.95   &    20.95   &    20.95   &    20.95   \\
\addlinespace

\midrule  
Obs (Panel A, B, and C)         &      495   &      495   &      495   &      495   \\
Obs (Panel D)           &      225   &      225   &      225   &      225   \\
Obs (Panel E)          &      270   &      270   &      270   &      270   \\
Cluster SE     &     CBSA   &     CBSA   &     CBSA   &     CBSA     \\
Weight         &{\scriptsize Ln(CBSAHU,start)}   &{\scriptsize Ln(CBSAHU,start)}   &{\scriptsize Ln(CBSAHU,start)}   &{\scriptsize Ln(CBSAHU,start)}   \\
\bottomrule

\end{tabular}
} % end of resize box

\end{table}

%----------------------------------------------------------------------------
% Double Differences: Recovery Period (2012-2019)
%----------------------------------------------------------------------------

\pagebreak 
%---------------------------------------------------------------
%%%%%%%%%%%%%%%%%%%%%%%%%%%%%%%%%%%%%%%%%%%%%%%%
% table_ZIP.LMH.PLNJM.D12t19
%%%%%%%%%%%%%%%%%%%%%%%%%%%%%%%%%%%%%%%%%%%%%%%%
%---------------------------------------------------------------

%%%%%%%%%%%%%%%%%%%%%%%%%%%%%%%%%%%%%%%%%%%%%%%%
% table_ZIP.LMH.PLNJM.D12t19
%%%%%%%%%%%%%%%%%%%%%%%%%%%%%%%%%%%%%%%%%%%%%%%%

\noindent 

\begin{table}[h!]
\centering
\caption{
\textbf{Four stacked Regressions of Low-minus-High PLNJM Growth in Recovery (12-19) on Net Export Growth (91-07) } \smallskip \newline
{\footnotesize 
This table reports OLS, reduced-form, first stage, and second stage results of 2SLS stacked regression $\text{LMH}\triangle_{12,19} Ln(PLNJM_{m}) = \beta * \triangle_{91,07} \text{NetExp}_{m} + \text{Constant} + \epsilon_{c}$. The left-hand-side dependent variable $\text{LMH}\triangle_{12,19} Ln(PLNJM_{m})$ is the differential growth rate of the dollar amount (07USD) of private-label (non-jumbo) mortgages (PLNJM) at metro $m$ between the low- and high-income ZIP codes 12-19 . The dependent variables in column (1) to (4) are the above ``low-minus-high" PLNJM growth based on two, tertile, quartile, and quintile income groups of ZIP codes within metropolitan areas. The key variable of interest $\triangle_{91,07} \text{NetExp}_{m}$ is the growth rate of net export at the metropolitan area (CBSA09 code) $m$ 91-07. We use the gravity model-based instrumental variable ($\triangle_{91,07}\text{givNetExp}_{m}$) as IV for $\triangle_{91,07}\text{NetExp}_{m}$. For the first-stage F-test, we report kleibergen-Paap (2006) robust (clustered) statistics and Montiel Olea-Pflueger (2013) efficient statistics. Each regression is weighted by the natural logarithm of housing units in metro $m$ in 2011. Standard errors are clustered at the CBSA level. ***, **, and * indicate significance at the 1\%, 5\%, and 10\% levels, respectively. ``An" means annualized.  ``An" means annualized variable. 
} % end of small font size
} % end of caption
\label{table_ZIP.LMH.PLNJM.D12t19}

\resizebox{\columnwidth}{!}{

\begin{tabular}{l*{4}{c}}
\toprule
Dep Var (Panel A, B, and C)                     &\multicolumn{4}{c}{Low-minus-High PLNJM Growth (12-19, An, 07USD)} \\
            \cmidrule{2-5} 
            &\multicolumn{1}{c}{(1)}&\multicolumn{1}{c}{(2)}&\multicolumn{1}{c}{(3)}&\multicolumn{1}{c}{(4)} \\
            
            &\multicolumn{1}{c}{Two Groups}&\multicolumn{1}{c}{Tertile Groups}&\multicolumn{1}{c}{Quartile Groups}&\multicolumn{1}{c}{Quintile Groups}\\
            
\midrule
\multicolumn{5}{l}{\textbf{Panel A. OLS estimates}} \\
\addlinespace
Net Export Growth (91-07, An)&    3.211***&    5.137***&    6.231***&    5.714***\\
               &  (0.994)   &  (1.598)   &  (1.771)   &  (1.961)   \\
\addlinespace
R2-adj         &   0.0217   &   0.0264   &   0.0288   &   0.0195   \\
\addlinespace

\midrule
\multicolumn{5}{l}{\textbf{Panel B. Reduced-form estimates}} \\
\addlinespace
GIV Net Export Growth (91-07, An)&    2.684*  &    5.045***&    5.885***&    6.048** \\
               &  (1.396)   &  (1.942)   &  (2.112)   &  (2.379)   \\
\addlinespace
R2-adj         &  0.00683   &   0.0130   &   0.0130   &   0.0113   \\
\addlinespace

\midrule
\multicolumn{5}{l}{\textbf{Panel C . 2SLS estimates}} \\
\addlinespace
Net Export Growth (91-07, An)&    2.788** &    5.242***&    6.115***&    6.285***\\
               &  (1.329)   &  (1.916)   &  (2.056)   &  (2.412)   \\

\addlinespace 
\addlinespace

Dep Var (Panel D): &\multicolumn{4}{c}{Net Export Growth (91-07, An)} \\ 
\midrule 
\multicolumn{5}{l}{\textbf{Panel D . First-stage estimates}} \\
\addlinespace
GIV Net Export Growth (91-07, An)&    0.962***&    0.962***&    0.962***&    0.962***\\
               &  (0.116)   &  (0.116)   &  (0.116)   &  (0.116)   \\
\addlinespace
KP F-Stat      &    68.90   &    68.90   &    68.90   &    68.87   \\
MOP F-Stat     &    68.90   &    68.90   &    68.90   &    68.87   \\
\addlinespace

\midrule  
Obs           &      359   &      359   &      359   &      358   \\
Cluster SE     &     CBSA09   &     CBSA09   &     CBSA09   &     CBSA09     \\
Weight         & Ln(HU11)   & Ln(HU11)   & Ln(HU11)   & Ln(HU11)   \\
\bottomrule

\end{tabular}
} % end of resize box

\end{table}

\pagebreak 
%---------------------------------------------------------------
%%%%%%%%%%%%%%%%%%%%%%%%%%%%%%%%%%%%%%%%%%%%%%%%
% table_ZIP.LMH.GSEM.D12t19
%%%%%%%%%%%%%%%%%%%%%%%%%%%%%%%%%%%%%%%%%%%%%%%%
%---------------------------------------------------------------

%%%%%%%%%%%%%%%%%%%%%%%%%%%%%%%%%%%%%%%%%%%%%%%%
% table_ZIP.LMH.GSEM.D12t19
%%%%%%%%%%%%%%%%%%%%%%%%%%%%%%%%%%%%%%%%%%%%%%%%

\noindent 

\begin{table}[h!]
\centering
\caption{
\textbf{Four stacked Regressions of Low-minus-High GSEM Growth in Recovery (12-19) on Net Export Growth (91-07) } \smallskip \newline
{\footnotesize 
This table reports OLS, reduced-form, first stage, and second stage results of 2SLS stacked regression $\text{LMH}\triangle_{12,19} Ln(GSEM_{m}) = \beta * \triangle_{91,07} \text{NetExp}_{m} + \text{Constant} + \epsilon_{c}$. The left-hand-side dependent variable $\text{LMH}\triangle_{12,19} Ln(GSEM_{m})$ is the differential growth rate of the dollar amount (07USD) of government-sponsored enterprise mortgages at metro $m$ between the low- and high-income ZIP codes 12-19. The dependent variables in column (1) to (4) are the above ``low-minus-high" GSEM growth based on two, tertile, quartile, and quintile income groups of ZIP codes within metropolitan areas. The key variable of interest $\triangle_{91,07} \text{NetExp}_{m}$ is the growth rate of net export at the metropolitan area (CBSA09 code) $m$ 91-07. We use the gravity model-based instrumental variable ($\triangle_{91,07}\text{givNetExp}_{m}$) as IV for $\triangle_{91,07}\text{NetExp}_{m}$. For the first-stage F-test, we report kleibergen-Paap (2006) robust (clustered) statistics and Montiel Olea-Pflueger (2013) efficient statistics. Each regression is weighted by the natural logarithm of housing units in metro $m$ in 2011. Standard errors are clustered at the CBSA level. ***, **, and * indicate significance at the 1\%, 5\%, and 10\% levels, respectively. ``An" means annualized.  ``An" means annualized variable. 
} % end of small font size
} % end of caption
\label{table_ZIP.LMH.GSEM.D12t19}

\resizebox{\columnwidth}{!}{

\begin{tabular}{l*{4}{c}}
\toprule
Dep Var (Panel A, B, and C)                     &\multicolumn{4}{c}{Low-minus-High GSEM Growth (12-19, An, 07USD)} \\
            \cmidrule{2-5} 
            &\multicolumn{1}{c}{(1)}&\multicolumn{1}{c}{(2)}&\multicolumn{1}{c}{(3)}&\multicolumn{1}{c}{(4)} \\
            
            &\multicolumn{1}{c}{Two Groups}&\multicolumn{1}{c}{Tertile Groups}&\multicolumn{1}{c}{Quartile Groups}&\multicolumn{1}{c}{Quintile Groups}\\
            
\midrule
\multicolumn{5}{l}{\textbf{Panel A. OLS estimates}} \\
\addlinespace
Net Export Growth (91-07, An)&    1.230   &    4.173***&    4.957***&    6.008***\\
               &  (1.158)   &  (1.547)   &  (1.853)   &  (2.203)   \\
\addlinespace
R2-adj         &-0.000167   &   0.0150   &   0.0128   &   0.0157   \\
\addlinespace

\midrule
\multicolumn{5}{l}{\textbf{Panel B. Reduced-form estimates}} \\
\addlinespace
GIV Net Export Growth (91-07, An)&    0.644   &    2.556   &    5.214** &    6.012** \\
               &  (1.538)   &  (2.259)   &  (2.285)   &  (2.715)   \\
\addlinespace
R2-adj         & -0.00240   & 0.000961   &  0.00689   &  0.00764   \\
\addlinespace

\midrule
\multicolumn{5}{l}{\textbf{Panel C . 2SLS estimates}} \\
\addlinespace
Net Export Growth (91-07, An)&    0.669   &    2.656   &    5.418** &    6.247** \\
               &  (1.586)   &  (2.241)   &  (2.311)   &  (2.662)   \\

\addlinespace 
\addlinespace

Dep Var (Panel D): &\multicolumn{4}{c}{Net Export Growth (91-07, An)} \\ 
\midrule 
\multicolumn{5}{l}{\textbf{Panel D . First-stage estimates}} \\
\addlinespace
GIV Net Export Growth (91-07, An)&    0.962***&    0.962***&    0.962***&    0.962***\\
               &  (0.116)   &  (0.116)   &  (0.116)   &  (0.116)   \\
\addlinespace
KP F-Stat      &    68.90   &    68.90   &    68.90   &    68.90   \\
MOP F-Stat     &    68.90   &    68.90   &    68.90   &    68.90   \\
\addlinespace

\midrule  
Obs           &      359   &      359   &      359   &      359   \\
Cluster SE     &     CBSA09   &     CBSA09   &     CBSA09   &     CBSA09     \\
Weight         & Ln(HU11)   & Ln(HU11)   & Ln(HU11)   & Ln(HU11)   \\
\bottomrule

\end{tabular}
} % end of resize box

\end{table}

\pagebreak 
%---------------------------------------------------------------
%%%%%%%%%%%%%%%%%%%%%%%%%%%%%%%%%%%%%%%%%%%%%%%%
% table_ZIP.LMH.PLNJM_m_GSEM.D12t19
%%%%%%%%%%%%%%%%%%%%%%%%%%%%%%%%%%%%%%%%%%%%%%%%
%---------------------------------------------------------------

%%%%%%%%%%%%%%%%%%%%%%%%%%%%%%%%%%%%%%%%%%%%%%%%
% table_ZIP.LMH.PLNJM_m_GSEM.D12t19
%%%%%%%%%%%%%%%%%%%%%%%%%%%%%%%%%%%%%%%%%%%%%%%%

\noindent 

\begin{table}[h!]
\centering
\caption{
\textbf{Four stacked Regressions of Difference Between Low-minus-High PLNJM Growth and Low-minus-High GSEM Growth in Recovery (12-19) on Net Export Growth (91-07) } \smallskip \newline
{\footnotesize 
This table reports OLS, reduced-form, first stage, and second stage results of 2SLS stacked regression $\text{LMH}\triangle_{12,19} Ln(PLNJM_{m}) - \text{LMH}\triangle_{12,19} Ln(GSEM_{m}) = \beta * \triangle_{91,07} \text{NetExp}_{m} + \text{Constant} + \epsilon_{c}$. The left-hand-side dependent variable is the difference between the low-minus-high private-label (non-jumbo) mortgages growth and the low-minus-high government-sponsored enterprise mortgages growth at metro $m$ 12-19. The dependent variables in column (1) to (4) are the above difference based on two, tertile, quartile, and quintile income groups of ZIP codes within metropolitan areas. The key variable of interest $\triangle_{91,07} \text{NetExp}_{m}$ is the growth rate of net export at the metropolitan area (CBSA09 code) $m$ 91-07. We use the gravity model-based instrumental variable ($\triangle_{91,07}\text{givNetExp}_{m}$) as IV for $\triangle_{91,07}\text{NetExp}_{m}$. For the first-stage F-test, we report kleibergen-Paap (2006) robust (clustered) statistics and Montiel Olea-Pflueger (2013) efficient statistics. Each regression is weighted by the natural logarithm of housing units in metro $m$ in 2011. Standard errors are clustered at the CBSA level. ***, **, and * indicate significance at the 1\%, 5\%, and 10\% levels, respectively. ``An" means annualized.  ``An" means annualized variable. 
} % end of small font size
} % end of caption
\label{table_ZIP.LMH.PLNJM_m_GSEM.D12t19}

\resizebox{\columnwidth}{!}{

\begin{tabular}{l*{4}{c}}
\toprule
Dep Var (Panel A, B, and C)                     &\multicolumn{4}{c}{LMH PLNJM Growth minus LMH GSEM Growth (12-19, An, 07USD)} \\
            \cmidrule{2-5} 
            &\multicolumn{1}{c}{(1)}&\multicolumn{1}{c}{(2)}&\multicolumn{1}{c}{(3)}&\multicolumn{1}{c}{(4)} \\
            
            &\multicolumn{1}{c}{Two Groups}&\multicolumn{1}{c}{Tertile Groups}&\multicolumn{1}{c}{Quartile Groups}&\multicolumn{1}{c}{Quintile Groups}\\
            
\midrule
\multicolumn{5}{l}{\textbf{Panel A. OLS estimates}} \\
\addlinespace
Net Export Growth (91-07, An)&    1.981   &    0.964   &    1.274   &   -0.218   \\
               &  (1.215)   &  (1.964)   &  (2.232)   &  (2.718)   \\
\addlinespace
R2-adj         &  0.00380   & -0.00190   & -0.00173   & -0.00278   \\
\addlinespace

\midrule
\multicolumn{5}{l}{\textbf{Panel B. Reduced-form estimates}} \\
\addlinespace
GIV Net Export Growth (91-07, An)&    2.040   &    2.489   &    0.671   &    0.081   \\
               &  (1.582)   &  (1.985)   &  (2.056)   &  (2.622)   \\
\addlinespace
R2-adj         &  0.00113   & 0.000565   & -0.00263   & -0.00281   \\
\addlinespace

\midrule
\multicolumn{5}{l}{\textbf{Panel C . 2SLS estimates}} \\
\addlinespace
Net Export Growth (91-07, An)&    2.120   &    2.586   &    0.697   &    0.084   \\
               &  (1.585)   &  (2.118)   &  (2.123)   &  (2.719)   \\

\addlinespace 
\addlinespace

Dep Var (Panel D): &\multicolumn{4}{c}{Net Export Growth (91-07, An)} \\ 
\midrule 
\multicolumn{5}{l}{\textbf{Panel D . First-stage estimates}} \\
\addlinespace
GIV Net Export Growth (91-07, An)&    0.962***&    0.962***&    0.962***&    0.962***\\
               &  (0.116)   &  (0.116)   &  (0.116)   &  (0.116)   \\
\addlinespace
KP F-Stat      &    68.90   &    68.90   &    68.90   &    68.87   \\
MOP F-Stat     &    68.90   &    68.90   &    68.90   &    68.87   \\
\addlinespace

\midrule  
Obs           &      359   &      359   &      359   &      358   \\
Cluster SE     &     CBSA09   &     CBSA09   &     CBSA09   &     CBSA09     \\
Weight         & Ln(HU11)   & Ln(HU11)   & Ln(HU11)   & Ln(HU11)   \\
\bottomrule

\end{tabular}
} % end of resize box

\end{table}

\pagebreak 
%---------------------------------------------------------------
%%%%%%%%%%%%%%%%%%%%%%%%%%%%%%%%%%%%%%%%%%%%%%%%
% table_ZIP.LMH.HPI.D12t19
%%%%%%%%%%%%%%%%%%%%%%%%%%%%%%%%%%%%%%%%%%%%%%%%
%---------------------------------------------------------------

%%%%%%%%%%%%%%%%%%%%%%%%%%%%%%%%%%%%%%%%%%%%%%%%
% table_ZIP.LMH.HPI.D12t19
%%%%%%%%%%%%%%%%%%%%%%%%%%%%%%%%%%%%%%%%%%%%%%%%

\noindent 

\begin{table}[h!]
\centering
\caption{
\textbf{Four stacked Regressions of Low-minus-High House Price Growth in Recovery (12-19) on Net Export Growth (91-07) } \smallskip \newline
{\footnotesize 
This table reports OLS, reduced-form, first stage, and second stage results of 2SLS stacked regression $\text{LMH}\triangle_{12,19} Ln(HPI_{m}) = \beta * \triangle_{91,07} \text{NetExp}_{m} + \text{Constant} + \epsilon_{c}$. The left-hand-side dependent variable $\text{LMH}\triangle_{12,19} Ln(HPI_{m})$ is the differential growth rate of the house price index at metro $m$ between the low- and high-income ZIP codes 12-19. The dependent variables in column (1) to (4) are the above ``low-minus-high" house price growth based on two, tertile, quartile, and quintile income groups of ZIP codes within metropolitan areas. The key variable of interest $\triangle_{91,07} \text{NetExp}_{m}$ is the growth rate of net export at the metropolitan area (CBSA09 code) $m$ 91-07. We use the gravity model-based instrumental variable ($\triangle_{91,07}\text{givNetExp}_{m}$) as IV for $\triangle_{91,07}\text{NetExp}_{m}$. For the first-stage F-test, we report kleibergen-Paap (2006) robust (clustered) statistics and Montiel Olea-Pflueger (2013) efficient statistics. Each regression is weighted by the natural logarithm of housing units in metro $m$ in 2011. Standard errors are clustered at the CBSA level. ***, **, and * indicate significance at the 1\%, 5\%, and 10\% levels, respectively. ``An" means annualized.  ``An" means annualized variable. 
} % end of small font size
} % end of caption
\label{table_ZIP.LMH.HPI.D12t19}

\resizebox{\columnwidth}{!}{

\begin{tabular}{l*{4}{c}}
\toprule
Dep Var (Panel A, B, and C)                     &\multicolumn{4}{c}{Low-minus-High House Price Growth (12-19, An, 07USD)} \\
            \cmidrule{2-5} 
            &\multicolumn{1}{c}{(1)}&\multicolumn{1}{c}{(2)}&\multicolumn{1}{c}{(3)}&\multicolumn{1}{c}{(4)} \\
            
            &\multicolumn{1}{c}{Two Groups}&\multicolumn{1}{c}{Tertile Groups}&\multicolumn{1}{c}{Quartile Groups}&\multicolumn{1}{c}{Quintile Groups}\\
            
\midrule
\multicolumn{5}{l}{\textbf{Panel A. OLS estimates}} \\
\addlinespace
Net Export Growth (91-07, An)&    0.866***&    1.128***&    1.424***&    1.588***\\
               &  (0.200)   &  (0.245)   &  (0.297)   &  (0.330)   \\
\addlinespace
R2-adj         &   0.0371   &   0.0354   &   0.0422   &   0.0420   \\
\addlinespace

\midrule
\multicolumn{5}{l}{\textbf{Panel B. Reduced-form estimates}} \\
\addlinespace
GIV Net Export Growth (91-07, An)&    0.500*  &    0.739*  &    0.822*  &    0.748   \\
               &  (0.303)   &  (0.390)   &  (0.465)   &  (0.504)   \\
\addlinespace
R2-adj         &  0.00404   &  0.00569   &  0.00496   &  0.00230   \\
\addlinespace

\midrule
\multicolumn{5}{l}{\textbf{Panel C . 2SLS estimates}} \\
\addlinespace
Net Export Growth (91-07, An)&    0.476*  &    0.705*  &    0.784*  &    0.713   \\
               &  (0.279)   &  (0.362)   &  (0.423)   &  (0.458)   \\

\addlinespace 
\addlinespace

Dep Var (Panel D): &\multicolumn{4}{c}{Net Export Growth (91-07, An)} \\ 
\midrule 
\multicolumn{5}{l}{\textbf{Panel D . First-stage estimates}} \\
\addlinespace
GIV Net Export Growth (91-07, An)&    1.049***&    1.049***&    1.049***&    1.049***\\
               &  (0.093)   &  (0.093)   &  (0.093)   &  (0.093)   \\
\addlinespace
KP F-Stat      &    127.5   &    127.5   &    127.5   &    127.5   \\
MOP F-Stat     &    127.5   &    127.5   &    127.5   &    127.5   \\
\addlinespace

\midrule  
Obs           &      346   &      346   &      346   &      346   \\
Cluster SE     &     CBSA09   &     CBSA09   &     CBSA09   &     CBSA09     \\
Weight         & Ln(HU11)   & Ln(HU11)   & Ln(HU11)   & Ln(HU11)   \\
\bottomrule

\end{tabular}
} % end of resize box

\end{table}

%---------------------------------------------------------------------------------------
% Double Differences: Robustness
%---------------------------------------------------------------------------------------

\pagebreak 
%---------------------------------------------------------------
%%%%%%%%%%%%%%%%%%%%%%%%%%%%%%%%%%%%%%%%%%%%%%%%
% table_ZIP.LMH.nL.PLNJM.D99t05vsD05t08.4Reg
%%%%%%%%%%%%%%%%%%%%%%%%%%%%%%%%%%%%%%%%%%%%%%%%
%---------------------------------------------------------------

%%%%%%%%%%%%%%%%%%%%%%%%%%%%%%%%%%%%%%%%%%%%%%%%
% table_ZIP.LMH.nL.PLNJM.D99t05vsD05t08.4Reg
%%%%%%%%%%%%%%%%%%%%%%%%%%%%%%%%%%%%%%%%%%%%%%%%

\noindent 

\begin{table}[h!]
\centering
\caption{
\textbf{Four stacked Regressions of Low-minus-High PLNJM Number Growth in Boom (99-05) and Bust (05-08) on Net Export Growth (99-05) } \smallskip \newline
{\footnotesize 
This table reports OLS, reduced-form, first stage, and second stage results of 2SLS stacked regression $\text{LMH}\triangle_{99,05} \&  \triangle_{05,08} Ln(PLNJM_{m}) = \beta_{99,05} * \triangle_{99,05} \text{NetExp}_{m} \times Dum_{99,05} + \beta_{05,08} * \triangle_{99,05} \text{NetExp}_{m} \times Dum_{05,08} + + \gamma_{99,05}* Dum_{99,05} + \gamma_{05,08}* Dum_{05,08} + \epsilon_{period, m}$. The left-hand-side dependent variable $\text{LMH}\triangle_{99,05} \&  \triangle_{05,08} Ln(PLNJM_{m})$ is the differential growth rate of the number of private-label (non-jumbo) mortgages (PLNJM) at metro $m$ between the low- and high-income ZIP codes 99-05 and 05-08. The dependent variables in columns (1) to (4) are the above ``low-minus-high" PLNJM growth based on two, tertile, quartile, and quintile income groups of ZIP codes within metropolitan areas. The key variable of interest $\triangle_{99,05} \text{NetExp}_{m}$ is the growth rate of net export at the metropolitan area (CBSA03 code) $m$ 99-05. We use the gravity model-based instrumental variable ($\triangle_{99,05}\text{givNetExp}_{m}$) as IV for $\triangle_{99,05}\text{NetExp}_{m}$. For the first-stage F-test, we report kleibergen-Paap (2006) robust (clustered) statistics and Montiel Olea-Pflueger (2013) efficient statistics. Each regression is weighted by the natural logarithm of housing units in metro $m$ in 1999. Standard errors are clustered at the CBSA level. ***, **, and * indicate significance at the 1\%, 5\%, and 10\% levels, respectively. ``An" means annualized variable. 
} % end of small font size
} % end of caption
\label{table_ZIP.LMH.nL.PLNJM.D99t05vsD05t08.4Reg}

\resizebox{\columnwidth}{!}{

\begin{tabular}{l*{4}{c}}
\toprule
Dep Var (Panel A, B, and C)                     &\multicolumn{4}{c}{Low-minus-High PLNJM Number Growth (99-05 \& 05-08, An, 07USD)} \\
            \cmidrule{2-5} 
            &\multicolumn{1}{c}{(1)}&\multicolumn{1}{c}{(2)}&\multicolumn{1}{c}{(3)}&\multicolumn{1}{c}{(4)} \\
            
            &\multicolumn{1}{c}{Two Groups}&\multicolumn{1}{c}{Tertile Groups}&\multicolumn{1}{c}{Quartile Groups}&\multicolumn{1}{c}{Quintile Groups}\\
            
\midrule
\multicolumn{5}{l}{\textbf{Panel A. OLS estimates}} \\
\addlinespace
Net Export Growth (99-05, An) x Dum99t05&    2.933   &    5.407** &    5.959** &    6.340** \\
               &  (2.046)   &  (2.428)   &  (2.592)   &  (3.078)   \\
\addlinespace
Net Export Growth (99-05, An) x Dum05t08&   -6.325** &   -8.357***&   -8.804** &   -8.951** \\
               &  (2.639)   &  (2.860)   &  (3.855)   &  (4.538)   \\
\addlinespace
R2-adj         &    0.193   &    0.219   &    0.215   &    0.202   \\
\addlinespace

\midrule
\multicolumn{5}{l}{\textbf{Panel B. Reduced-form estimates}} \\
\addlinespace
GIV Net Export Growth (99-05, An) x Dum99t05&    6.387*  &   10.583** &   14.201** &   16.688***\\
               &  (3.716)   &  (4.548)   &  (5.710)   &  (6.402)   \\
\addlinespace
GIV Net Export Growth (99-05, An) x Dum05t08&   -4.687   &  -15.332***&  -17.859***&  -23.741***\\
               &  (6.673)   &  (5.081)   &  (6.465)   &  (8.697)   \\
\addlinespace
R2-adj         &    0.186   &    0.221   &    0.220   &    0.212   \\
\addlinespace

\midrule
\multicolumn{5}{l}{\textbf{Panel C . 2SLS estimates}} \\
\addlinespace
Net Export Growth (99-05, An) x Dum99t05&    7.380   &   12.229** &   16.404** &   19.270** \\
               &  (4.563)   &  (5.920)   &  (6.960)   &  (7.583)   \\
\addlinespace
Net Export Growth (99-05, An) x Dum05t08&   -5.417   &  -17.717***&  -20.629** &  -27.424** \\
               &  (7.002)   &  (6.570)   &  (8.123)   & (11.562)   \\
\addlinespace
CoefEqual\_Chi2 &    1.700   &    7.357   &    7.564   &    7.713   \\
CoefEqual\_PValue&    0.192   &    0.007   &    0.006   &    0.005   \\

\addlinespace 
\addlinespace

Dep Var (Panel D): &\multicolumn{4}{c}{Net Export Growth (99-05, An)} \\ 
\midrule 
\multicolumn{5}{l}{\textbf{Panel D . First-stage estimates only for 99-05 (Non-stack sample)}} \\
\addlinespace
GIV Net Export Growth (99-05, An) x Dum99t05&    0.865***&    0.865***&    0.865***&    0.865***\\
               &  (0.227)   &  (0.227)   &  (0.227)   &  (0.227)   \\
\addlinespace
KP F-Stat      &    14.54   &    14.54   &    14.55   &    14.50   \\
MOP F-Stat     &    14.54   &    14.54   &    14.55   &    14.50   \\
\addlinespace

\midrule  
Obs (Panel A, B, and C)           &      600   &      600   &      598   &      597   \\
Obs (Panel D)           &      300   &      300   &      299   &      298   \\
Cluster SE     &     CBSA   &     CBSA   &     CBSA   &     CBSA     \\
Weight         &{\scriptsize Ln(CBSAHU99)}   &{\scriptsize Ln(CBSAHU99)}   &{\scriptsize Ln(CBSAHU99)}   &{\scriptsize Ln(CBSAHU99)}   \\
\bottomrule

\end{tabular}
} % end of resize box

\end{table}

\pagebreak 
%---------------------------------------------------------------
%%%%%%%%%%%%%%%%%%%%%%%%%%%%%%%%%%%%%%%%%%%%%%%%
% table_ZIP.LMH.PLNJM.Own.D99t05vsD05t08.4Reg
%%%%%%%%%%%%%%%%%%%%%%%%%%%%%%%%%%%%%%%%%%%%%%%%
%---------------------------------------------------------------

%%%%%%%%%%%%%%%%%%%%%%%%%%%%%%%%%%%%%%%%%%%%%%%%
% table_ZIP.LMH.PLNJM.Own.D99t05vsD05t08.4Reg
%%%%%%%%%%%%%%%%%%%%%%%%%%%%%%%%%%%%%%%%%%%%%%%%

\noindent 

\begin{table}[h!]
\centering
\caption{
\textbf{Four stacked Regressions of Low-minus-High PLNJM (Owner-Occupied) Growth in Boom (99-05) and Bust (05-08) on Net Export Growth (99-05) } \smallskip \newline
{\footnotesize 
This table reports OLS, reduced-form, first stage, and second stage results of 2SLS stacked regression $\text{LMH}\triangle_{99,05} \&  \triangle_{05,08} Ln(PLNJM\_Own_{m}) = \beta_{99,05} * \triangle_{99,05} \text{NetExp}_{m} \times Dum_{99,05} + \beta_{05,08} * \triangle_{99,05} \text{NetExp}_{m} \times Dum_{05,08} + + \gamma_{99,05}* Dum_{99,05} + \gamma_{05,08}* Dum_{05,08} + \epsilon_{period, m}$. The left-hand-side dependent variable $\text{LMH}\triangle_{99,05} \&  \triangle_{05,08} Ln(PLNJM\_Own_{m})$ is the differential growth rate of the dollar amount (07USD) of private-label (non-jumbo) mortgages (PLNJM) (Owner-Occupied) at metro $m$ between the low- and high-income ZIP codes 99-05 and 05-08. The dependent variables in column (1) to (4) are the above ``low-minus-high" PLNJM growth based on two, tertile, quartile, and quintile income groups of ZIP codes within metropolitan areas. The key variable of interest $\triangle_{99,05} \text{NetExp}_{m}$ is the growth rate of net export at the metropolitan area (CBSA03 code) $m$ 99-05. We use the gravity model-based instrumental variable ($\triangle_{99,05}\text{givNetExp}_{m}$) as IV for $\triangle_{99,05}\text{NetExp}_{m}$. For the first-stage F-test, we report kleibergen-Paap (2006) robust (clustered) statistics and Montiel Olea-Pflueger (2013) efficient statistics. Each regression is weighted by the natural logarithm of housing units in metro $m$ in 1999. Standard errors are clustered at the CBSA level. ***, **, and * indicate significance at the 1\%, 5\%, and 10\% levels, respectively. ``An" means annualized variable. 
} % end of small font size
} % end of caption
\label{table_ZIP.LMH.PLNJM.Own.D99t05vsD05t08.4Reg}

\resizebox{\columnwidth}{!}{

\begin{tabular}{l*{4}{c}}
\toprule
Dep Var (Panel A, B, and C)                     &\multicolumn{4}{c}{Low-minus-High PLNJM (Owner) Growth (99-05 \& 05-08, An, 07USD)} \\
            \cmidrule{2-5} 
            &\multicolumn{1}{c}{(1)}&\multicolumn{1}{c}{(2)}&\multicolumn{1}{c}{(3)}&\multicolumn{1}{c}{(4)} \\
            
            &\multicolumn{1}{c}{Two Groups}&\multicolumn{1}{c}{Tertile Groups}&\multicolumn{1}{c}{Quartile Groups}&\multicolumn{1}{c}{Quintile Groups}\\
            
\midrule
\multicolumn{5}{l}{\textbf{Panel A. OLS estimates}} \\
\addlinespace
Net Export Growth (99-05, An) x Dum99t05&    3.564*  &    5.581** &    6.296** &    6.798** \\
               &  (2.024)   &  (2.466)   &  (2.586)   &  (3.228)   \\
\addlinespace
Net Export Growth (99-05, An) x Dum05t08&   -6.653***&   -8.034***&   -7.235*  &   -8.865*  \\
               &  (2.500)   &  (3.023)   &  (4.252)   &  (4.806)   \\
\addlinespace
R2-adj         &    0.179   &    0.216   &    0.223   &    0.205   \\
\addlinespace

\midrule
\multicolumn{5}{l}{\textbf{Panel B. Reduced-form estimates}} \\
\addlinespace
GIV Net Export Growth (99-05, An) x Dum99t05&    6.533*  &   10.002** &   12.951** &   15.008** \\
               &  (3.772)   &  (4.796)   &  (5.735)   &  (6.796)   \\
\addlinespace
GIV Net Export Growth (99-05, An) x Dum05t08&   -6.518   &  -16.683***&  -16.775***&  -22.690***\\
               &  (6.476)   &  (5.169)   &  (6.453)   &  (7.933)   \\
\addlinespace
R2-adj         &    0.173   &    0.218   &    0.227   &    0.212   \\
\addlinespace

\midrule
\multicolumn{5}{l}{\textbf{Panel C . 2SLS estimates}} \\
\addlinespace
Net Export Growth (99-05, An) x Dum99t05&    7.550*  &   11.557** &   14.960** &   17.330** \\
               &  (4.473)   &  (5.780)   &  (6.484)   &  (7.555)   \\
\addlinespace
Net Export Growth (99-05, An) x Dum05t08&   -7.532   &  -19.278***&  -19.377** &  -26.210** \\
               &  (6.560)   &  (7.013)   &  (8.156)   & (10.644)   \\
\addlinespace
CoefEqual\_Chi2 &    2.591   &    7.989   &    7.954   &    8.211   \\
CoefEqual\_PValue&    0.107   &    0.005   &    0.005   &    0.004   \\

\addlinespace 
\addlinespace

Dep Var (Panel D): &\multicolumn{4}{c}{Net Export Growth (99-05, An)} \\ 
\midrule 
\multicolumn{5}{l}{\textbf{Panel D . First-stage estimates only for 99-05 (Non-stack sample)}} \\
\addlinespace
GIV Net Export Growth (99-05, An) x Dum99t05&    0.865***&    0.865***&    0.865***&    0.865***\\
               &  (0.227)   &  (0.227)   &  (0.227)   &  (0.227)   \\
\addlinespace
KP F-Stat      &    14.54   &    14.54   &    14.55   &    14.50   \\
MOP F-Stat     &    14.54   &    14.54   &    14.55   &    14.50   \\
\addlinespace

\midrule  
Obs (Panel A, B, and C)           &      600   &      600   &      598   &      597   \\
Obs (Panel D)           &      300   &      300   &      299   &      298   \\
Cluster SE     &     CBSA   &     CBSA   &     CBSA   &     CBSA     \\
Weight         &{\scriptsize Ln(CBSAHU99)}   &{\scriptsize Ln(CBSAHU99)}   &{\scriptsize Ln(CBSAHU99)}   &{\scriptsize Ln(CBSAHU99)}   \\
\bottomrule

\end{tabular}
} % end of resize box

\end{table}

\pagebreak 
%---------------------------------------------------------------
%%%%%%%%%%%%%%%%%%%%%%%%%%%%%%%%%%%%%%%%%%%%%%%%
% table_ZIP.LMH.GSEM.D99t05vsD05t08.4Reg
%%%%%%%%%%%%%%%%%%%%%%%%%%%%%%%%%%%%%%%%%%%%%%%%
%---------------------------------------------------------------

%%%%%%%%%%%%%%%%%%%%%%%%%%%%%%%%%%%%%%%%%%%%%%%%
% table_ZIP.LMH.PLNJM.D99t05vsD05t08.FullContr.4Reg
%%%%%%%%%%%%%%%%%%%%%%%%%%%%%%%%%%%%%%%%%%%%%%%%

\noindent 

\begin{table}[h!]
\centering
\caption{
\textbf{Four stacked Regressions of Low-minus-High PLNJM Growth in Boom (99-05) and Bust (05-08) on Net Export Growth (99-05) with full set of controls } \smallskip \newline
{\footnotesize 
This table reports OLS, reduced-form, first stage, and second stage results of 2SLS stacked regression $\text{LMH}\triangle_{99,05} \&  \triangle_{05,08} Ln(PLNJM_{m}) = \beta_{99,05} * \triangle_{99,05} \text{NetExp}_{m} \times Dum_{99,05} + \beta_{05,08} * \triangle_{99,05} \text{NetExp}_{m} \times Dum_{05,08} + \gamma_{99,05}* Dum_{99,05} + \gamma_{05,08}* Dum_{05,08} + \theta_{99,05}* \text{Controls} + \theta_{05,08}* \text{Controls} + \epsilon_{period, m}$. The left-hand-side dependent variable $\text{LMH}\triangle_{99,05} \&  \triangle_{05,08} Ln(PLNJM_{m})$ is the differential growth rate of the dollar amount (07USD) of private-label (non-jumbo) mortgages (PLNJM) at metro $m$ between the low- and high-income ZIP codes 99-05 and 05-08. The dependent variables in column (1) to (4) are the above ``low-minus-high" PLNJM growth based on two, tertile, quartile, and quintile income groups of ZIP codes within metropolitan areas. The key variable of interest $\triangle_{99,05} \text{NetExp}_{m}$ is the growth rate of net export at the metropolitan area (CBSA03 code) $m$ 99-05. We use the gravity model-based instrumental variable ($\triangle_{99,05}\text{givNetExp}_{m}$) as IV for $\triangle_{99,05}\text{NetExp}_{m}$. Controls variables include basic, housing, and demographic controls. For the first-stage F-test, we report kleibergen-Paap (2006) robust (clustered) statistics and Montiel Olea-Pflueger (2013) efficient statistics. Each regression is weighted by the natural logarithm of housing units in metro $m$ in 1999. Standard errors are clustered at the CBSA level. ***, **, and * indicate significance at the 1\%, 5\%, and 10\% levels, respectively. ``An" means annualized.  ``An" means annualized variable. 
} % end of small font size
} % end of caption
\label{table_ZIP.LMH.PLNJM.D99t05vsD05t08.FullContr.4Reg}

\resizebox{\columnwidth}{!}{

\begin{tabular}{l*{4}{c}}
\toprule
Dep Var (Panel A, B, and C)                     &\multicolumn{4}{c}{Low-minus-High PLNJM Growth (99-05 \& 05-08, An, 07USD)} \\
            \cmidrule{2-5} 
            &\multicolumn{1}{c}{(1)}&\multicolumn{1}{c}{(2)}&\multicolumn{1}{c}{(3)}&\multicolumn{1}{c}{(4)} \\
            
            &\multicolumn{1}{c}{Two Groups}&\multicolumn{1}{c}{Tertile Groups}&\multicolumn{1}{c}{Quartile Groups}&\multicolumn{1}{c}{Quintile Groups}\\
            
\midrule
\multicolumn{5}{l}{\textbf{Panel A. OLS estimates}} \\
\addlinespace
Net Export Growth (99-05, An) x Dum99t05&    2.108   &    4.392*  &    4.922*  &    5.797*  \\
               &  (1.869)   &  (2.470)   &  (2.618)   &  (3.387)   \\
\addlinespace
Net Export Growth (99-05, An) x Dum05t08&   -3.351   &   -7.644** &   -7.877** &   -9.656** \\
               &  (2.495)   &  (3.011)   &  (3.600)   &  (4.474)   \\
\addlinespace
R2-adj         &    0.412   &    0.461   &    0.473   &    0.402   \\
\addlinespace

\midrule
\multicolumn{5}{l}{\textbf{Panel B. Reduced-form estimates}} \\
\addlinespace
GIV Net Export Growth (99-05, An) x Dum99t05&    5.232   &   10.237** &   13.008** &   16.061** \\
               &  (3.306)   &  (4.379)   &  (5.404)   &  (6.995)   \\
\addlinespace
GIV Net Export Growth (99-05, An) x Dum05t08&    0.924   &  -13.126** &  -13.437*  &  -20.567** \\
               &  (5.898)   &  (6.126)   &  (7.203)   &  (9.138)   \\
\addlinespace
R2-adj         &    0.411   &    0.462   &    0.476   &    0.407   \\
\addlinespace

\midrule
\multicolumn{5}{l}{\textbf{Panel C . 2SLS estimates}} \\
\addlinespace
Net Export Growth (99-05, An) x Dum99t05&    6.138   &   12.009** &   15.210** &   18.724** \\
               &  (4.295)   &  (6.056)   &  (6.696)   &  (8.414)   \\
\addlinespace
Net Export Growth (99-05, An) x Dum05t08&    1.084   &  -15.398** &  -15.711*  &  -24.048** \\
               &  (6.885)   &  (7.609)   &  (8.531)   & (11.831)   \\
\addlinespace
CoefEqual\_Chi2   &    0.346   &    5.931   &    6.394   &    6.335   \\
CoefEqual\_PValue   &    0.556   &    0.015   &    0.011   &    0.012   \\

\addlinespace 
\addlinespace

Dep Var (Panel D): &\multicolumn{4}{c}{Net Export Growth (99-05, An)} \\ 
\midrule 
\multicolumn{5}{l}{\textbf{Panel D. First-stage estimates only for 99-05 (Non-stack sample)}} \\
\addlinespace
GIV Net Export Growth (99-05, An) x Dum99t05&    0.852***&    0.852***&    0.852***&    0.852***\\
               &  (0.252)   &  (0.252)   &  (0.252)   &  (0.252)   \\
\addlinespace
KP F-Stat      &    11.47   &    11.47   &    11.50   &    11.56   \\
MOP F-Stat     &    11.47   &    11.47   &    11.50   &    11.56   \\
\addlinespace

\midrule  
Obs (Panel A, B, and C)         &      516   &      516   &      514   &      513   \\
Obs (Panel D)           &      258   &      258   &      257   &      256   \\
Cluster SE     &     CBSA   &     CBSA   &     CBSA   &     CBSA     \\
Weight         &{\scriptsize Ln(CBSAHU99)}   &{\scriptsize Ln(CBSAHU99)}   &{\scriptsize Ln(CBSAHU99)}   &{\scriptsize Ln(CBSAHU99)}   \\
\bottomrule

\end{tabular}
} % end of resize box

\end{table}

\pagebreak 
%---------------------------------------------------------------
%%%%%%%%%%%%%%%%%%%%%%%%%%%%%%%%%%%%%%%%%%%%%%%%
% table_ZIP.LMH.PLNJM.D99t05vsD05t08.FullContr.2SLS.wide.tex
%%%%%%%%%%%%%%%%%%%%%%%%%%%%%%%%%%%%%%%%%%%%%%%%
%---------------------------------------------------------------

%%%%%%%%%%%%%%%%%%%%%%%%%%%%%%%%%%%%%%%%%%%%%%%%
% table_ZIP.LMH.PLNJM.D99t05vsD05t08.FullContr.2SLS.wide
%%%%%%%%%%%%%%%%%%%%%%%%%%%%%%%%%%%%%%%%%%%%%%%%

\noindent 

\begin{table}[h!]
\centering
\caption{
\textbf{2SLS Regressions of Low-minus-High PLNJM Growth in Boom (99-05) and Bust (05-08) on Net Export Growth (99-05) with full set of controls } \smallskip \newline
{\footnotesize 
This table reports results of 2SLS stacked regression $\text{LMH}\triangle_{99,05} \&  \triangle_{05,08} Ln(PLNJM_{m}) = \beta_{99,05} * \triangle_{99,05} \text{NetExp}_{m} \times Dum_{99,05} + \beta_{05,08} * \triangle_{99,05} \text{NetExp}_{m} \times Dum_{05,08} + \gamma_{99,05}* Dum_{99,05} + \gamma_{05,08}* Dum_{05,08} + \theta_{99,05}* \text{Controls} + \theta_{05,08}* \text{Controls} + \epsilon_{period, m}$. The left-hand-side dependent variable $\text{LMH}\triangle_{99,05} \&  \triangle_{05,08} Ln(PLNJM_{m})$ is the differential growth rate of the dollar amount (07USD) of private-label (non-jumbo) mortgages (PLNJM) at metro $m$ between the low- and high-income ZIP codes 99-05 and 05-08. The dependent variables in column (1) to (4) are the above ``low-minus-high" PLNJM growth based on two, tertile, quartile, and quintile income groups of ZIP codes within metropolitan areas. The key variable of interest $\triangle_{99,05} \text{NetExp}_{m}$ is the growth rate of net export at the metropolitan area (CBSA03 code) $m$ 99-05. We use the gravity model-based instrumental variable ($\triangle_{99,05}\text{givNetExp}_{m}$) as IV for $\triangle_{99,05}\text{NetExp}_{m}$. Controls variables include basic, housing, and demographic controls. For the first-stage F-test, we report kleibergen-Paap (2006) robust (clustered) statistics and Montiel Olea-Pflueger (2013) efficient statistics. Each regression is weighted by the natural logarithm of housing units in metro $m$ in 1999. Standard errors are clustered at the CBSA level. ***, **, and * indicate significance at the 1\%, 5\%, and 10\% levels, respectively. ``An" means annualized.  ``An" means annualized variable. 
} % end of small font size
} % end of caption
\label{table_ZIP.LMH.PLNJM.D99t05vsD05t08.FullContr.2SLS.wide}

\resizebox{\columnwidth}{!}{

\begin{tabular}{l*{4}{cc}}
\toprule
\textbf{TSLS estimates}        &\multicolumn{8}{c}{Low-minus-High PLNJM Growth (99-05 \& 05-08, An, 07USD)} \\
                \cmidrule{2-9} 
               &\multicolumn{2}{c}{(1)}&\multicolumn{2}{c}{(2)}&\multicolumn{2}{c}{(3)}&\multicolumn{2}{c}{(4)}\\
               &\multicolumn{2}{c}{Two Groups}&\multicolumn{2}{c}{Tertile Groups}&\multicolumn{2}{c}{Quartile Groups}&\multicolumn{2}{c}{Quintile Groups}\\
\midrule
Net Export Growth (99-05, An) x Dum99t05&    6.138   &  (4.295)&   12.009** &  (6.056)&   15.210** &  (6.696)&   18.724** &  (8.414)\\  \addlinespace
Net Export Growth (99-05, An) x Dum05t08&    1.084   &  (6.885)&  -15.398** &  (7.609)&  -15.711*  &  (8.531)&  -24.048** & (11.831)\\  \addlinespace
Dummry 99-05   &   -1.745***&  (0.510)&   -2.379***&  (0.675)&   -2.910***&  (0.727)&   -2.867***&  (0.889)\\  \addlinespace
Dummry 05-08   &    3.371***&  (0.681)&    4.681***&  (0.950)&    5.200***&  (1.075)&    6.543***&  (1.266)\\  \addlinespace
Ln(Num of HH, 99) x Dum99t05&    0.054   &  (0.102)&    0.232*  &  (0.125)&    0.281*  &  (0.153)&    0.206   &  (0.190)\\  \addlinespace
Ln(Num of HH, 99) x Dum05t08&   -0.123   &  (0.145)&   -0.288   &  (0.176)&   -0.391*  &  (0.202)&   -0.348   &  (0.270)\\  \addlinespace
Ln(HH Income, 99) x Dum99t05&    0.137***&  (0.052)&    0.198***&  (0.069)&    0.245***&  (0.074)&    0.227** &  (0.095)\\  \addlinespace
Ln(HH Income, 99) x Dum05t08&   -0.340***&  (0.072)&   -0.459***&  (0.101)&   -0.506***&  (0.112)&   -0.642***&  (0.131)\\  \addlinespace
Ratio of Labor Force (1999) x Dum99t05&   -0.011   &  (0.165)&   -0.085   &  (0.214)&   -0.138   &  (0.257)&   -0.033   &  (0.348)\\  \addlinespace
Ratio of Labor Force (1999) x Dum05t08&    0.728***&  (0.236)&    0.570*  &  (0.308)&    0.708*  &  (0.368)&    0.759*  &  (0.440)\\  \addlinespace
Ln(Num of HU, 99) x Dum99t05&   -0.050   &  (0.102)&   -0.231*  &  (0.125)&   -0.274*  &  (0.153)&   -0.191   &  (0.191)\\  \addlinespace
Ln(Num of HU, 99) x Dum05t08&    0.115   &  (0.148)&    0.278   &  (0.178)&    0.372*  &  (0.205)&    0.329   &  (0.269)\\  \addlinespace
Housing supply elasticity x Dum99t05&   -0.003   &  (0.003)&   -0.001   &  (0.004)&    0.002   &  (0.006)&    0.002   &  (0.007)\\  \addlinespace
Housing supply elasticity x Dum05t08&   -0.002   &  (0.006)&   -0.005   &  (0.008)&   -0.007   &  (0.011)&   -0.016   &  (0.014)\\  \addlinespace
Wharton Regulation Index x Dum99t05&    0.004   &  (0.006)&    0.005   &  (0.008)&    0.007   &  (0.011)&   -0.002   &  (0.016)\\  \addlinespace
Wharton Regulation Index x Dum05t08&   -0.001   &  (0.010)&   -0.002   &  (0.013)&    0.004   &  (0.015)&    0.003   &  (0.018)\\  \addlinespace
House Vacancy Rate (1999) x Dum99t05&    0.051   &  (0.243)&    0.465*  &  (0.246)&    0.630** &  (0.276)&    0.677*  &  (0.364)\\  \addlinespace
House Vacancy Rate (1999) x Dum05t08&   -0.340   &  (0.240)&   -0.609** &  (0.295)&   -0.848** &  (0.345)&   -0.955** &  (0.486)\\  \addlinespace
Ratio of Renters (1999) x Dum99t05&    0.381***&  (0.126)&    0.455***&  (0.158)&    0.485***&  (0.174)&    0.505** &  (0.221)\\  \addlinespace
Ratio of Renters (1999) x Dum05t08&   -0.490***&  (0.181)&   -0.507** &  (0.219)&   -0.620** &  (0.244)&   -0.810***&  (0.290)\\  \addlinespace
Ratio of Bachelor Educated (1999) x Dum99t05&   -0.203   &  (0.140)&   -0.199   &  (0.192)&   -0.298   &  (0.228)&   -0.283   &  (0.290)\\  \addlinespace
Ratio of Bachelor Educated (1999) x Dum05t08&    0.239   &  (0.171)&    0.556** &  (0.220)&    0.567** &  (0.256)&    1.003***&  (0.376)\\  \addlinespace
Ratio of White Race (1999) x Dum99t05&    0.108** &  (0.055)&    0.095   &  (0.066)&    0.096   &  (0.073)&    0.078   &  (0.094)\\  \addlinespace
Ratio of White Race (1999) x Dum05t08&   -0.070   &  (0.070)&   -0.011   &  (0.095)&    0.040   &  (0.121)&    0.078   &  (0.144)\\  \addlinespace
Ratio of Immigration (90-00) x Dum99t05&    0.343   &  (0.252)&    0.371   &  (0.328)&    0.332   &  (0.346)&    0.278   &  (0.514)\\  \addlinespace
Ratio of Immigration (90-00) x Dum05t08&    0.094   &  (0.356)&   -0.245   &  (0.433)&   -0.002   &  (0.490)&    0.073   &  (0.648)\\  \addlinespace
Ratio of Age 65 Above (1999) x Dum99t05&    0.620** &  (0.263)&    0.788** &  (0.336)&    0.709*  &  (0.389)&    0.663   &  (0.498)\\  \addlinespace
Ratio of Age 65 Above (1999) x Dum05t08&    0.081   &  (0.328)&   -0.264   &  (0.420)&   -0.313   &  (0.503)&   -0.243   &  (0.613)\\  \addlinespace
\midrule
Obs            &      516   &         &      516   &         &      514   &         &      513   &         \\
Cluster SE     &     CBSA   &         &     CBSA   &         &     CBSA   &         &     CBSA   &         \\
Weight         & Ln(HU99)   &         & Ln(HU99)   &         & Ln(HU99)   &         & Ln(HU99)   &         \\
CD F-Stat      &    91.14   &         &    88.71   &         &    65.06   &         &    76.06   &         \\
KP F-Stat      &    7.257   &         &    6.926   &         &    4.490   &         &    5.857   &         \\
CoefEqual\_Chi2 &    3.017   &         &    7.663   &         &    6.537   &         &    6.381   &         \\
CoefEqual\_PValue&    0.082   &         &    0.006   &         &    0.011   &         &    0.012   &         \\
\bottomrule

\end{tabular}
} % end of resize box

\end{table}

%---------------------------------------------------------------
%---------------------------------------------------------------
% end of this tex file
%---------------------------------------------------------------
%---------------------------------------------------------------

%----------------------------------------------------------------------

%----------------------------------------------------------------------
% section 10: Appendix 

\clearpage
\pagenumbering{arabic}% resets `page` counter to 1
\renewcommand*{\thepage}{A\arabic{page}}
% renew page numbering in the appendix 

\counterwithin{figure}{section}
\counterwithin{table}{section}
% count figures and tables within Appendix section

\appendix

%------------------------------------------------------------
%------------------------------------------------------------
\clearpage
%------------------------------------------------------------

%------------------------------------------------------------
\section{Appendix}

%--------------------------------------------------------------------------------------
%\subsect{Appendix for Data Details}
%--------------------------------------------------------------------------------------

%------------------------------------------------------------
%------------------------------------------------------------

%------------------------------------------------------------
\subsection{Appendix for Data Details}\label{subsec:App_Data}

%------------------------------------------------------------
%------------------------------------------------------------
\subsubsection{HMDA Mortgage Data Processing Procedures at ZIP level}
In this section, I illustrate the processing details regarding the mortgage data at the ZIP code level. I will describe the processes of crosswalking from census tracts to ZIP codes, retaining data in the 1990-2000-unchanged ZIP codes, linking ZIP codes to primary counties and metropolitan areas, and grouping ZIP codes into income groups within metropolitan areas. 

\paragraph{Crosswalk from Census Tracts to ZIP Codes} 
First, I use the 712 or 800 consistent counties across time to filter HMDA data, depending on the period of interest. Second, HMDA mortgage data uses various versions of census tracts across time. 1992-2002, 2003-2011, 2012-2021 HMDA datasets use 1990, 2000, and 2010 version of the census tract, respectively. From the Missouri Census Data Center \footnote{Missouri Census Data Center's website is here: \url{https://mcdc.missouri.edu/applications/geocorr.html}}, I obtain the crosswalk file that links 1990 (2000 and 2010) census tracts to 1990 (2000 and 2010) ZIP codes, respectively. I use population weight in the crosswalk since multiple census tracts can match multiple ZIP codes. I use 1992-1999 as the prior period for comparison because I intend to retain a reasonably large number of ZIP codes in a balanced panel for comparison. To prevent within-ZIP code changes from contaminating my analysis, I only keep ZIP codes whose 90\% of the population is covered by HMDA reporting. The deleted ZIP codes are primarily those on the boundary of metropolitan areas, as HMDA mandates reporting in metropolitan areas. 

ZIP codes are reasonably large areas within metropolitan areas. One ZIP code typically contains several census blocks, which are designed to provide more statistical uniformity with an average population of 4,000 or more. In the 1992-2011 balanced mortgage panel data for 10,315 ZIP codes, the population sizes in 1990 at the 25\%, 50\%, and 75\% quantiles are 3,463, 10,448, and 23,234, respectively. \footnote{For more information comparing census tracts and ZIP codes, see \url{https://proximityone.com/tracts_zips.htm}.}

\paragraph{Retaining Data in the 1990-2000-unchanged ZIP codes}
According to \cite{bailey2023national}, the ZIP code geographic boundary can change in relation to population growth and decline across time. For example, their detailed work, based on the U.S. Postal Service's Postal Bulletin, shows that 3,015 changes occurred between 1990 and 2000. This number of changes is not small, given that 29,401 distinct ZIP codes were in operation at the beginning of 1990 (by the crosswalk file from the Missouri Census Data Center described above). Typical changes include merge, split, and boundary alteration. These changes can result in measurement error (matching observations to the wrong units) or missing data (due to an observation reporting a ZIP code that was not in operation at the beginning of the sample period). 

To avoid the potential measurement error and missing data, I get a list of 1990-2000-unchanged ZIP codes by following steps. First, from the Missouri Census Data Center, I get the crosswalk file from 1990 ZIP codes to 1990 counties with population weights. Second, I delete any ZIP codes that may experience a change during 1990-2000 from the 1990 ZIP codes list. Specifically, I delete ZIP codes that appear in the following variables in the following files by \cite{bailey2023national}: (1) variable ``oldzipcode" or ``newzipcode" in file ``ZIP\_Code\_Changes\_1990-2000\_Documentation.xls" and (2) ``zipcode" in file ``ZIP\_xwalk\_90-00.dta". Out of the 29,401 ZIP codes in the year 1990, 26,577 ZIP codes did not experience any change. Third, I match these ZIP codes to ZIP codes in the crosswalk file from 2000 ZIP codes to 2000 counties with population weights. I retain only matched ZIP codes. The above three steps result in 26,577 1990-2000-unchanged ZIP codes in the U.S. (including metropolitan and rural areas). 

\paragraph{Link ZIP codes to Their Primary Counties and Metropolitan Areas}
A ZIP code can be associated with multiple counties (metropolitan areas) when it lies along the border. For each match zip code, I only keep its primary matched county that contains the largest population in the zip code. From the Missouri Census Data Center, I get the crosswalk file that links 2000 ZIP codes to 2000 counties. I retain counties in the 2003 version of metropolitan areas in U.S. mainland (excluding Hawaii and Alaska).\footnote{Metropolitan areas definition (CBSA code) of 2003 can be found here: \url{https://www2.census.gov/programs\%2Dsurveys/metro\%2Dmicro/geographies/reference\%2Dfiles/2003/historical\%2Ddelineation\%2Dfiles/030606omb\%2Dnecta\%2Dcnecta.xls}.}

The above steps yield 13,513 1990-2000-unchanged ZIP codes that are matched to primary counties within metropolitan areas (2003 version CBSA code) in the U.S. mainland. I use these ZIP codes to filter HMDA ZIP-level data.

\paragraph{Group ZIP codes Within Metropolitan Areas}
To conduct the analysis in Section (\ref{sec:Empirical_DoubleDifference}) for model-based new predictions, I need to aggregate mortgage data by ZIP code into half-, tertile-, quartile-, and quintile-income groups within metropolitan areas. To ensure each metropolitan statistical area (MSA) contains sufficient ZIP codes for the calculations, I require that each MSA include at least five ZIP codes. Since ZIP codes are relatively large areas, I believe five ZIP codes constitute a reasonable minimum for the calculation. For descriptive figures, there are 10,315 ZIP codes from 284 metros (676 counties) for the 1992-2011 period and 15,724 ZIP codes from 359 metros (1083 counties) for the 2012-2019 period. For regression analysis on mortgages of ``double differences", there are 10,315 ZIP codes from 284 metros (676 counties) for the 1992-1999 period, 11,165 ZIP codes from 303 metros (768 counties) for the 1999-2011 period, and 15,724 ZIP codes from 359 metros (1083 counties) for the 2012-2019 period.

%------------------------------------------------------------
%------------------------------------------------------------
\subsubsection{House Price Data at ZIP level}
The annual house price index at the ZIP level based on repeat sales comes from the Federal Housing Financing Agency.\footnote{The data website of the Federal Housing Financing Agency is here: \url{https://www.fhfa.gov/DataTools/Downloads/Pages/House-Price-Index-Datasets.aspx\#qexe}.} This database has reasonably good coverage of the ZIP codes of US metropolitan areas in and after 1992 (11,556 ZIP codes in 1992), but a little bit less coverage than the HMDA mortgage data.

\paragraph{Merge House Price and Mortgage Data at ZIP Level}
For the Figure (\ref{fig_ZIP_HPIGrowth_cbQuint_zipHalf_Quart}) and (\ref{fig_ZIP_D12t19_HPI_cbQuint_zipHalf_Quart}) of ``double differences" on house prices, I only require that each MSA contains at least 5 ZIP codes, resulting in a balanced panel of 7,218 ZIP codes from 1992 to 2011 and 11,836 ZIP codes from 2012 to 2019. For regression analyses of house prices and mortgages, I also require that the ZIP codes are covered by both the house price and mortgage data, as in Section \ref{subsec:DoubleDifferences_HousePrice_Empirical}. The merged dataset contains fewer ZIP codes than the mortgage data, as the house price data covers fewer ZIP codes, resulting in 7,099 ZIP codes in 227 metropolitan areas for the 1992-2011 sample, 8,578 ZIP codes in 272 metropolitan areas for the 1999-2011 sample, and 11,717 ZIP codes in 346 metropolitan areas for the 2012-2019 sample.

\paragraph{Interpolating House Units At the County Level}
I collect the number of housing units at the ZIP level from the Decennial Census Summary Files for 1900 and 2000. Since Census data are collected as of March in 1990 and 2000, I treat the number of house units as the one at the end of the last year (1989 and 1999, respectively). I interpolate the number of house units from 1990 to 2000 using a linear function. For county-level housing units between 2001 and 2020, I use annual estimates from the American Community Survey from the U.S. Census. Then, I aggregate county-level housing units to metropolitan areas based on the CBSA03 code. County-level housing units in the initial year are used as weights for cross-metro (county) regression analysis, and metro-level housing units in the initial year are used as weights for "double differences" (cross ZIP codes cross metro) regression analysis. 

\comment{
Lastly, I use the number of house units as weights to aggregate house prices across zip code groups within metropolitan areas. 
}

%\subsubsection{Six Metropolitan Areas Severely Affected by 2005 Hurricanes}
\noindent \textbf{Six Metropolitan Areas Severely Affected by 2005 Hurricanes}  Following Section \ref{subsec:2005Hurricanes}, I delete six metropolitan areas in our regression analysis for ``double differences" in boom and bust periods that contain the twelve counties that were severely affected by 2005 Hurricanes. These six metropolitan areas are Abbeville (LA, CBSA03 code: 10020), Gulfport-Biloxi (MS, 25060), Key West-Marathon (FL, 28580), Lake Charles (LA, 29340), New Orleans-Metairie-Kenner (LA, 35380), and Pascagoula (MS, 37700).

%--------------------------------------------------------------------------------------
%\subsection{Appendix: GIV for Imports}
%--------------------------------------------------------------------------------------

%-----------------------------------------------------
%-----------------------------------------------------
%-----------------------------------------------------
%-----------------------------------------------------

\subsection{Gravity Model-based IV: US Imports}\label{subsec:GIV_imports}

I have illustrated the key idea of the gravity model-based instrument for US exports by \cite{feenstra2019us} in Section \ref{subsec:GIV_exports}. For completeness, I show how they construct IV for US imports here. The gravity-based IV for US imports starts from a simple symmetric constant-elasticity equation by \cite{romalis2007nafta}:
\vspace{-1mm}
\begin{equation}{\label{eq:imp_gravity}}
    \frac{X^{j,US}_{s,v,t}}{X^{j,i}_{s,v,t}} = \Bigg( \frac{w^{j}_{s,t}d^{j,US}\tau^{j,US}_{s,t}}{w^{j}_{s,t}d^{j,i}\tau^{j,i}_{s,t}} \Bigg) ^{1-\sigma} \frac{(P^{US}_{s,t})^{\sigma-1}E^{US}_{s,t}}{(P^{i}_{s,t})^{\sigma-1}E^{i}_{s,t}} = \Bigg(\frac{d^{j,US}\tau^{j,US}_{s,t}}{d^{j,i}\tau^{j,i}_{s,t}} \Bigg) ^{1-\sigma} \frac{(P^{US}_{s,t})^{\sigma-1}E^{US}_{s,t}}{(P^{i}_{s,t})^{\sigma-1}E^{i}_{s,t}}
\end{equation}
In the above formula, $X^{j,US}_{s,v,t}$ is country $j$'s export to US in product variant $v$ in industry $s$ in year $t$. $X^{j,i}_{s,v,t}$ is the similar term but representing country $j$'s export to country $i$. $w^{j}_{s,t}$ is the relative marginal cost of production in industry $s$ in country $j$, which is canceled out in the above equation. $\tau^{j,US}_{s,t}$ and $\tau^{j,i}_{s,t}$ are the \textit{ad valorem} total import tariff on country $j$'s export to the US and country $i$. $d^{j,US}$ and $d^{j,i}$ are the bilateral distance and other fixed trade costs from country $j$ to the US and to country $i$. $P^{US}_{s,t}$ and $P^{i}_{s,t}$ are the aggregate price index in the US and country $i$; $E^{US}_{s,t}$ and $E^{i}_{s,t}$ are the total expenditure in the US and country $i$. Lastly, $\sigma$ denotes the constant elasticity of substitution ($\sigma>1$). 

Like the one for export, the intuition of this gravity-style model for import is quite straightforward. The ratio of country $i$'s export to the US relative to country $j$ is declining with the ratio of bilateral distance and the ratio of \textit{ad valorem} import tariff, but rising with the ratio of aggregate price index and total expenditure.

Assume that there are $N^{j}_{s,t}$ identical product varieties exported by country $j$ in industry $s$ and year $t$, one can re-arrange the above equation, multiply both sides with $N^{j}_{s,t}$, and sum over countries $i \neq US$:
\vspace{-1mm}
\begin{equation*}
    N^{j}_{s,t}X^{j,US}_{s,v,t}*\sum_{i\neq US} \big[ ( d^{j,i})^{1-\sigma} (P^{i}_{s,t})^{\sigma-1}E^{i}_{s,t} \big] = (d^{j,US}\tau^{j,US}_{s,t})^{1-\sigma} (P^{US}_{s,t})^{\sigma-1}E^{US}_{s,t}* \sum_{i \neq US} \big[ N^{j}_{s,t}X^{j,i}_{s,v,t} (\tau^{j,i}_{s,t})^{\sigma-1} \big ]
\end{equation*}

Like before, as the above equation holds for any countries $i \neq US$, one can choose the set of countries that have similar economic conditions with the US (so that they are buyers competing with US buyers when country $j$ considers its export) to make my prediction more accurate. \cite{feenstra2019us} use the eight high-income countries proposed by \cite{autor2013china}. 

I denote the sectoral export by country $j$ to the US and to country $i$  as $X^{j,US}_{s,t} \equiv X^{j,US}_{s,v,t}*N^{j}_{s,t}$ and $X^{j,i}_{s,t} \equiv X^{j,i}_{s,v,t}*N^{j}_{s,t}$. Then I can get 
\vspace{-1mm}
\begin{equation*}
    X^{j,US}_{s,t}*\sum_{i\neq US} \big[ ( d^{j,i})^{1-\sigma} (P^{i}_{s,t})^{\sigma-1}E^{i}_{s,t} \big] = (d^{j,US}\tau^{j,US}_{s,t})^{1-\sigma} (P^{US}_{s,t})^{\sigma-1}E^{US}_{s,t}* \sum_{i \neq US} \big[ X^{j,i}_{s,t} (\tau^{j,i}_{s,t})^{\sigma-1} \big ]
\end{equation*}

After a few re-arrangement, I can get the formula for $ X^{j,US}_{s,t}$:
\vspace{-1mm}
\begin{equation}
    X^{j,US}_{s,t} =   \frac{(d^{j,US}\tau^{j,US}_{s,t})^{1-\sigma}(P^{US}_{s,t})^{\sigma-1}E^{US}_{s,t}}{\sum_{i\neq US} \big[ (d^{j,i})^{1-\sigma}(P^{i}_{s,t})^{\sigma-1}E^{i}_{s,t} \big]}  
    * \bigg( \sum_{k\neq US} X^{j,k}_{s,t} \bigg)  * \Bigg\{  \sum_{i\neq US} \bigg[ \frac{ X^{j,i}_{s,t} }{\sum_{k\neq US} X^{j,k}_{s,t}} (\tau^{j,i}_{s,t})^{\sigma -1} \bigg] \Bigg\}
\end{equation}
 
Note in the above formula, I multiply and divide by $\sum_{k\neq US} X^{j,k}_{s,t}$ to prepare for the regression setup in the next step. Now I can take logs of the above equation and move the term $\lnb{\sum_{k\neq US} X^{j,k}_{s,t}}$ to the left-hand side of the equation to get the regression-style formula:
\vspace{-1mm}
\begin{equation} \label{eq:imp_gravityRegression}
\resizebox{0.92\textwidth}{!}{%
\begin{math}
\begin{aligned}
\lnb{X^{j,US}_{s,t}} & = \underbrace{ \lnb{\sum_{k\neq US}X^{j,k}_{s,t}} }_{\text{Term 0}} + \underbrace{ \lnb{(P^{US}_{s,t})^{\sigma-1}E^{US}_{s,t}} }_{\text{Ind-Year FE: } \gamma^{US}_{s,t}} + \underbrace{(1-\sigma)\lnb{d^{j,US}}}_{\text{Exporting-country FE: } \delta^{j,US}} \\
& + \underbrace{(1-\sigma)\lnb{\tau^{j,US}_{s,t}}}_{\text{Term 1}} + \underbrace{(\sigma-1) \lnb{ \Bigg\{  \sum_{i\neq US} \bigg[ \frac{ X^{j,i}_{s,t} }{\sum_{k\neq US} X^{j,k}_{s,t}} (\tau^{j,i}_{s,t})^{\sigma -1} \bigg] \Bigg\}^{\frac{1}{\sigma-1}} }}_{\text{Term 2: } (\sigma-1) \lnb{T^{j}_{s,t}}} + \epsilon^{j}_{s,t} \\
\end{aligned}
\end{math}
} %end of \scalemath \resizebos
\end{equation}
Now, the US import from country $j$ in the industry $s$ year $t$ can be decomposed into six terms. ``Term 0'' is the other eight high-income countries' import from country $j$, which reflects the world supply. The second term $\gamma^{US}_{s,t}$ reflects the US demand shocks, which is potentially endogenous. I remove this term by the US industry-by-year fixed effects. The third term $\delta^{j,US}$ reflects the distance from country $j$ to the US and all other industry- and year-invariant trade costs. Since this term is predetermined rather than a shock, I remove it by the exporting-country fixed effects. ``Term 1" is the tariff on country $j$'s exports imposed by the US, which is out of the control of exporting firms. I keep this term to capture the shock from tariffs. ``Term 2" is the weighted average tariffs on country $j$'s exports imposed by other eight high-income countries. Intuitively, when this weighted average tariffs on country $j$'s exports rise, destination country $j$ will export to the US as a substitution. I keep this term to capture this substitution effect. The last term $\epsilon^{j}_{s,t} = - \lnb{ \sum_{i\neq US} [ (d^{j,i})^{1-\sigma} (P^{i}_{s,t})^{\sigma-1}E^{i}_{s,t} ] } $ is unobserved and remains in the regression error term. 

After the above regression, I can isolate predicted US imports that are presumably exogenous:
\begin{equation} \label{eq:imp_gravityPreUSImp}
 \lnb{ \widehat{X^{j,US}_{s,t}} } = \lnb{\sum_{k\neq US}X^{j,k}_{s,t}} + \hat{\beta_1} *\lnb{\tau^{j,US}_{s,t}} + \hat{\beta_2}* \lnb{T^{j}_{s,t}}
\end{equation}

%--------------------------------------------------------------------------------------
%\subsection{Appendixe: Empirical for Credit Expansion}
%--------------------------------------------------------------------------------------

%------------------------------------------------------------
%------------------------------------------------------------
\subsection{Appendix for the Empirical Results Supporting Credit Expansion}\label{subsec:App_EmpCreditExpansion}

\subsubsection{Exclusion Restriction: House Price Growth Prior vs. Boom}

In section \ref{sec:ExclusionRestriction} Table (\ref{table_HPI.D91t99vsD99t05.2SLS.wide}) and (\ref{table_HPI.D91t99vsD99t05.4Reg}), in which I use prior period to test the exclusion restriction of my instrumental variable approach indirectly, I winsorize net export growth and its GIV in the prior period (1991-1999). In this appendix, I will compare results with and without winsorization to show that a few outliers can make the coefficient of net export growth significantly negative. 

First, for the prior period (1991-1999), the 2SLS with and without winsorization can be seen from the tiny table below. We can see that 2SLS estimates without winsorization is -1.809 and significant at the 1\% level. However, 2SLS estimates with winsorization at the 3\% and 97\% level is -1.902 but not significant. Comparison between two results show that a few ourliers can make the impact of net export growth on house price growth significantly negative. 

%------------------------------------
\begin{table}[H]
\vspace{-2mm}
\centering 
\caption{
\textbf{2SLS Regression of House Price Growth on Net Export Growth in Prior (91-99) With and Without Winsorization}
}
\resizebox{0.9\columnwidth}{!}{%

\begin{tabular}{l*{4}{c}}
\toprule
\textbf{TSLS estimates}            &\multicolumn{4}{c}{House Price Growth (07USD, 91-99, annualized)} \\
            \cmidrule{2-5} 
            &\multicolumn{1}{c}{(1)}&\multicolumn{1}{c}{(2)}&\multicolumn{1}{c}{(3)}&\multicolumn{1}{c}{(4)}\\
            
\midrule
\multicolumn{5}{l}{\textbf{Panel A. Sample with Wonsorization}} \\
Net Export Growth (91-99, An) x Dum91t99&   -3.708** &   -3.275*  &   -3.169   &   -2.972   \\
               &  (1.737)   &  (1.740)   &  (1.988)   &  (2.009)   \\
\addlinespace
\multicolumn{5}{l}{\textbf{Panel B. Sample without Wonsorization}} \\
Net Export Growth (91-99, An) x Dum91t99&   -2.419***&   -2.256***&   -2.603***&   -2.529***\\
               &  (0.718)   &  (0.743)   &  (0.966)   &  (0.922)   \\
\addlinespace
\midrule
\multicolumn{5}{l}{\textbf{Controls}} \\
Basic Controls  &            &  Y   &   Y    & Y        \\
Housing Controls  &           &      & Y       & Y        \\
Demographic Controls &            &      &        &  Y        \\
\midrule          
Obs for Both Panels  &      673   &      673   &      608   &      608   \\
Cluster SE     &     CBSA   &     CBSA   &     CBSA   &     CBSA  \\
Weight         & {\scriptsize Ln(HU91)}   & {\scriptsize Ln(HU91)}   &{\scriptsize Ln(HU91)}   &{\scriptsize Ln(HU91)}    \\
\bottomrule

\end{tabular}

} % end of resize box

\end{table}
%------------------------------------

\begin{figure}[H] 
    \centering
    
    \begin{subfigure}[t]{0.48\textwidth}
        \centering
        \includegraphics[height=4.8cm]{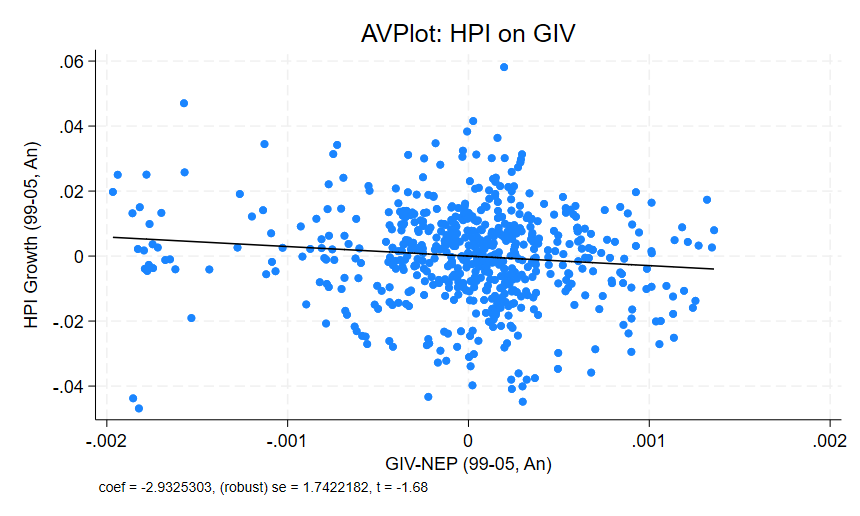}
        \caption{With Winsorization}
    \end{subfigure}
    \begin{subfigure}[t]{0.48\textwidth}
        \centering
        \includegraphics[height=4.8cm]{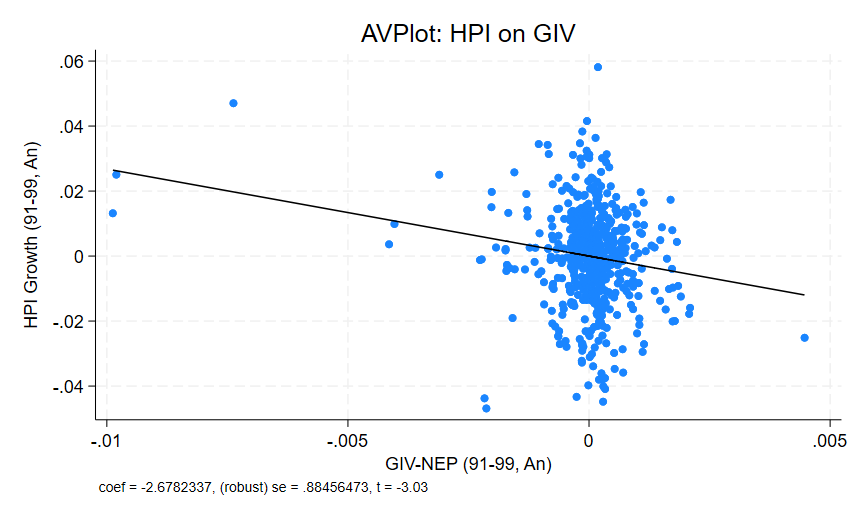}
        \caption{Without Winsorization}
    \end{subfigure}
    \caption{Reduced-form Regression of House Price Growth on Net Export Growth in Prior Period (1991-1999) With and Without Winsorization at 3\% and 97\%}
    \label{fig_RedFormReg_HPG_NEG_with&wtihout_Winsorization}
\end{figure}

Another way to illustrate the above effect of a few outliers is to compare the added-variable plots (AV-plot) of two reduced-form regressions. In Figure (\ref{fig_RedFormReg_HPG_NEG_with&wtihout_Winsorization}), the left subfigure shows the AV-plot of the reduced-form regression for the sample with winsorization and the right subfigure is for the sample without winsorization. Without winsorization, a few outliers make the impact of net export growth on house price grow significantly negative in the right subfigure. By winsorization, the left subfigure shows that, for most observations, such negative is insignificant.

%------------------------------------
%------------------------------------

\subsubsection{Exclusion Restriction: PLNJM Growth Prior vs. Boom}

In section \ref{sec:ExclusionRestriction} Table (\ref{table_PLNJM.D91t99vsD99t05.2SLS.wide}) and (\ref{table_PLNJM.D91t99vsD99t05.4Reg}), in which I use the prior period to provide suggestive validation for the exclusion restriction of my instrumental variable approach, I drop four outliers based on the GIV of net export growth in the prior period (1991-1999). These four outliers are Howard County (county FIPS: 18067), Tipton County (18159), Durham County (37063), and Orange County (37135). In this appendix, I will compare results with and without dropping outliers to show that a few outliers can make the coefficient of net export growth significantly negative. 

First, for the prior period (1991-1999), the 2SLS with and without dropping four outliers can be seen from the tiny table below. We can see that 2SLS estimates without dropping is -11.720 and significant at the 1\% level. However, 2SLS estimates with dropping the three outliers is -7.784 but not significant. Comparison between two results show that a few outliers can make the impact of net export growth on private-label (non-jumbo) mortgages growth significantly negative.

%------------------------------------
\begin{table}[h!]
\vspace{-2mm}
\centering 
\caption{
\textbf{2SLS Regression of private-label (non-jumbo) mortgage Growth on Net Export Growth in Prior (91-99) With and Without Dropping Outliers}
}
\resizebox{0.9\columnwidth}{!}{%

\begin{tabular}{l*{4}{c}}
\toprule
\textbf{TSLS estimates}            &\multicolumn{4}{c}{private-label (non-jumbo) mortgage Growth (07USD, 91-99, annualized)} \\
            \cmidrule{2-5} 
            &\multicolumn{1}{c}{(1)}&\multicolumn{1}{c}{(2)}&\multicolumn{1}{c}{(3)}&\multicolumn{1}{c}{(4)}\\
            
\midrule
\multicolumn{5}{l}{\textbf{Panel A. Sample with Dropping Four Outliers}} \\
Net Export Growth (91-99, An)&   -1.730   &   -2.067   &   -9.389   &  -10.197   \\
              &  (4.373)   &  (4.312)   &  (6.769)   &  (6.243)   \\
\addlinespace
\multicolumn{5}{l}{\textbf{Panel B. Sample without Dropping Four Outliers}} \\
Net Export Growth (91-99, An)&   -6.108   &   -6.404   &  -12.865***&  -13.517***\\
               &  (4.297)   &  (4.226)   &  (4.835)   &  (4.380)   \\
\addlinespace
\midrule
\multicolumn{5}{l}{\textbf{Controls}} \\
Basic Controls  &            &  Y   &   Y    & Y       \\
Housing Controls  &           &      & Y       & Y        \\
Demographic Controls &            &      &        &  Y       \\
\midrule          
Obs for Panel A           &      701   &      701   &      632   &      632   \\
Obs for Panel B           &      705   &      705   &      636   &      636   \\
Cluster SE     &     CBSA   &     CBSA   &     CBSA   &     CBSA     \\
Weight         & Ln(HU91)   & Ln(HU91)   & Ln(HU91)   & Ln(HU91)   \\\\
\bottomrule

\end{tabular}

} % end of resize box

\end{table}
%------------------------------------

\begin{figure}[h!] 
    \centering
    \caption{Reduced-form Regression of private-label (non-jumbo) mortgage Growth on Net Export Growth in Prior Period (1991-1999) With and Without Dropping the Three Outliers}
    \label{fig_RedFormReg_PLNJMG_NEG_with&wtihout_drooping}
    \begin{subfigure}[t]{0.48\textwidth}
        \centering
        \includegraphics[height=4.8cm]{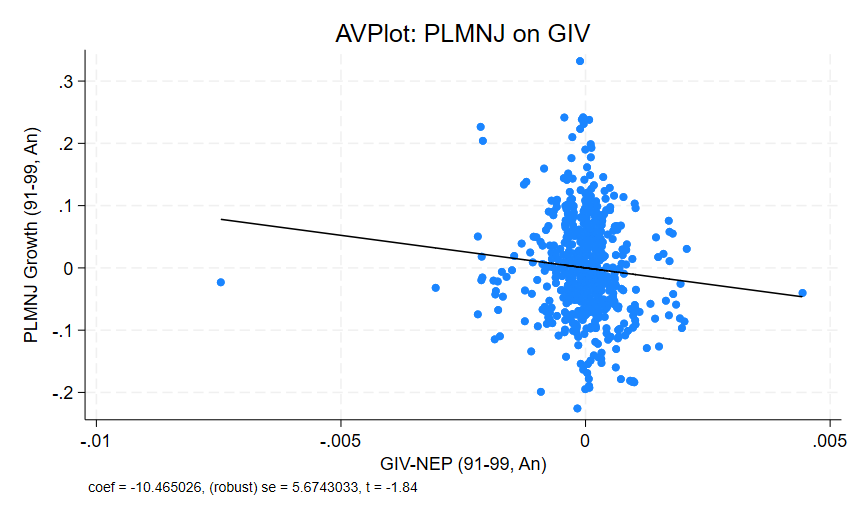}
        \caption{With Dropping Outliers}
    \end{subfigure}%
    ~ 
    \begin{subfigure}[t]{0.48\textwidth}
        \centering
        \includegraphics[height=4.8cm]{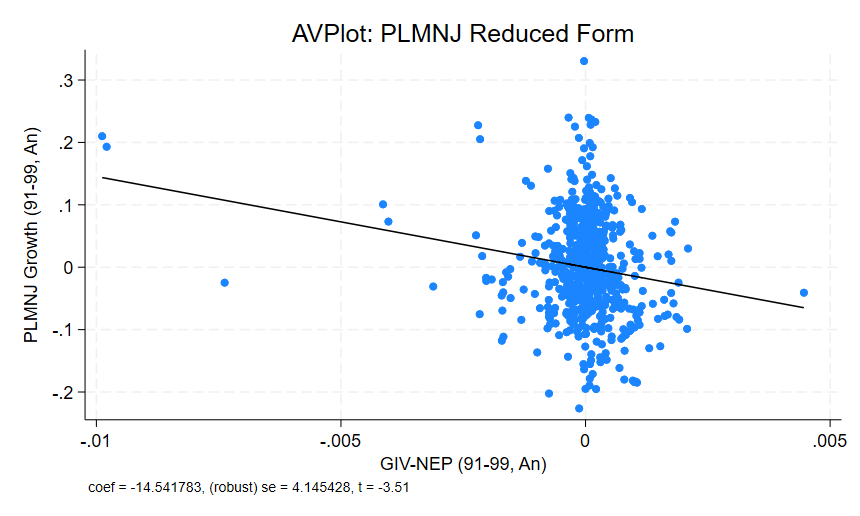}
        \caption{Without Dropping Outliers}
    \end{subfigure}
\end{figure}

Another way to illustrate the above effect of a few outliers is to compare the added-variable plots (AV-plot) of two reduced-form regressions. In Figure (\ref{fig_RedFormReg_PLNJMG_NEG_with&wtihout_drooping}), the left subfigure shows the AV-plot of the reduced-form regression for the sample with dropping the four outliers and the right subfigure is for the sample without dropping outliers. Without dropping, a few outliers make the impact of net export growth on private-label mortgage grow significantly negative in the right subfigure. By dropping the three outliers, the left subfigure shows that, for most observations, such a negative impact is insignificant.

%------------------------------------
%------------------------------------

%--------------------------------------------------------------------------------------
%\subsection{Empirical.Robustness}
%--------------------------------------------------------------------------------------

%------------------------------------------------------------
%------------------------------------------------------------
\subsection{Robustness: State Heterogeneity}\label{subsec:Empirical.Robustness}
%------------------------------------------------------------
%------------------------------------------------------------
In this subsection, I perform robustness tests to support the main conclusion that credit expansion in private-label mortgages induced by net export growth causes the 1999-2009 U.S. house price boom and bust across metropolitan areas. I show that this main conclusion is robust to state-level differences in anti-predatory lending laws \citep{di2017credit}, recourse laws \citep{ghent2011recourse}, judicial requirement in foreclosure \citep{mian2015foreclosures}, distinction between sand states and other states \citep{choi2016sand}, and state capital gain tax \citep{gao2020economic}.

%------------------------------------------------------------
\subsubsection{Anti-Predatory-Lending States vs. Other States}
%------------------------------------------------------------

In this subsection, my main conclusion holds after controlling for the cross-state difference in anti-predatory lending laws. In addition, I show that growth in private-label (non-jumbo) mortgages leads to a stronger housing boom and a stronger bust in states with anti-predatory lending laws before 2004.

According to \citet{di2017credit}, prior to 2004, a dozen of states had already implemented anti-predatory laws to protect mortgage borrowers against unfair and deceptive practices. However, on January 7th, 2004, the Office of the Comptroller of the Currency (OCC) preempted national banks (rather than state-chartered depository institutions and independent mortgage companies) from state antipredatory-lending law (APL law). They show that such deregulation results in the national bank’s credit expansion (relative to state-chartered depository institutions and independent mortgage companies), house price growth, and nontradable employment rise in 2004-2006, but a sharp decline subsequently in these states relative to other states.\footnote{This preemption is controversial afterward, especially since its spirit is somewhat against Section 1044 of Dodd-Frank \citep{sykes2019federal}. However, in 2011, OCC reaffirmed the 2004 preemption while deleting some controversial language from its preemption rules. OCC's formal document can be found here: \url{https://www.federalregister.gov/documents/2011/05/26/2011-12859/office-of-thrift-supervision-integration-dodd-frank-act-implementation}}. Thus, in the period 2004-2005, all states that had implemented APL laws before could not apply such laws to national banks. Appendix Table 1 from \cite{di2017credit} summarizes the list of APL states before 2004. Since I only have annual HMDA data, I restrict my APL states to the states that implemented APL at least half a year before 2004, resulting in eleven APL states.\footnote{According to \cite{di2017credit}, these eleven APL states (including DC) are California, Connecticut, District of Columbia, Georgia, Maryland, Michigan, Minnesota, New York, North Carolina, Texas, and West Virginia.}

My stacked 2SLS regression is 
\begin{equation}\label{eq:HPIBoomBustonPLNJM_APLvsNone}
\resizebox{0.92\textwidth}{!}{$
\begin{aligned}
\triangle_{99,05} \& \triangle_{07,09} Ln(HPI_{c}) & = \beta_{Boom} * \triangle_{99,05} Ln(PLNJM_{c}) \times Dum_{99,05} + \beta_{Bust} * \triangle_{99,05} Ln(PLNJM_{c}) \times Dum_{07,09} \\
 & + \beta_{APL, Boom} * \triangle_{99,05} Ln(PLNJM_{c}) \times Dum_{99,05} \times Dum_{APL} + \beta_{APL, Bust} * \triangle_{99,05} Ln(PLNJM_{c}) \times Dum_{07,09} \times Dum_{APL} \\
 & + \gamma_{Boom} * \bm{Controls_{c}} \times Dum_{99,05} + \gamma_{Bust} * \bm{Controls_{c}} \times Dum_{07,09} + \epsilon_{c}
\end{aligned}
$} %end of \resizebox
\end{equation}
Controls, weight, and standard errors are the same as Eq (\ref{eq:HPI_reg_PLNJM}). 

Table (\ref{table_Robust.APLvsNone.HPI.D99t05.D07t09}) reports the above 2SLS results and shows two key conclusions. First, compared to non-APL states, APL states experience a differentially stronger house price boom (99-05) and a differentially stronger bust (07-09) in the high-net-export-growth metropolitan areas, caused by the growth of private-label (non-jumbo) mortgages 1999-2005. Second, after controlling the differential trend in APL-states, for all metropolitan areas, the growth of private-label (non-jumbo) mortgages 1999-2005 leads to a stronger house price boom (99-05) and a stronger bust (07-09) in the high-net-export-growth metropolitan areas.

Please note that in either boom or bust period, there are two endogenous variables here. $\triangle_{99,05} Ln(PLNJM_{c})$ is instrumented by $\triangle_{99,05}\text{givNetExp}_{m}$ and $\triangle_{99,05} Ln(PLNJM_{c}) \times Dum_{APL}$ is instrumented by $\triangle_{99,05}\text{givNetExp}_{m} \times Dum_{APL}$. For each of the two separate first-stage regression F-tests, I report Sanderson-Windmeijer robust (clustered) F-statistics \citep{sanderson2016weak}. The SW F-statistics is 12.94 for PLNJM growth and is 16.25 for the interaction between PLNJM growth and dummy of APL-states in column (4), meaning each F-stage regression is significant, and each instrument is strong for its endogenous variable. To evaluate the overall strength of two instruments, I report the p-value of robust (clustered) Kleibergen-Paap test statistic calculated by \citep{windmeijer2021testing}. This p-value is 0.0037 in column (4), meaning the two separate instruments are jointly strong for the two endogenous variables.

%------------------------------------------------------------
\subsubsection{Non-Recourse vs. Recourse States}
%------------------------------------------------------------

In this subsection, I show that growth in private-label mortgages leads to a stronger housing bust in non-recourse states. In addition, my main results are significant within the non-recourse states and within the recourse states.

\cite{ghent2011recourse} document that, in 39 states of the U.S., mortgages are recourse loans.\footnote{39 recourse states can be found in Table 1 in \cite{ghent2011recourse}. 11 non-recourse states are Alaska, Arizona, California, Iowa, Minnesota, Montana, North Carolina, North Dakota, Oregon, Washington, and Wisconsin.} In these states, lenders could go after the borrower's other assets to recover the mortgage loss not covered by the proceedings from a foreclosure sale through obtaining a deficiency judgment. They find that in recourse states, borrowers are less sensitive to negative equity, and defaults are more likely to proceed via a lender-friendly procedure. I expect that non-recourse mortgages may discourage lenders from expanding mortgage credit but induce households to be more willing to default, given falling house prices. I need to test whether my main conclusion still holds after controlling for such a potential trend.

Regression specification is the same as Eq (\ref{eq:HPIBoomBustonPLNJM_APLvsNone}) except that the dummy variable is for non-recourse states. Table (\ref{table_Robust.NRCvsRC.HPI.D99t05.D07t09}) reports the above 2SLS results and shows two key conclusions. First, compared to recourse states, non-recourse states experience a stronger house price bust (07-09), caused by the growth of private-label (non-jumbo) mortgages from 1999 to 2005. This evidence is consistent with the notion that non-recourse mortgages induce households to be more willing to default given falling house prices. Second, after controlling for the differential trend in the non-recourse states, for all metropolitan areas, the growth of private-label (non-jumbo) mortgages 1999-2005 leads to a stronger house price boom (99-05) and a stronger bust (07-09) in the high-net-export-growth metropolitan areas. The SW F-statistics and p-value of robust (clustered) Kleibergen-Paap test statistics both indicate that instruments are both separately and jointly strong for the two endogenous variables.

%------------------------------------------------------------
\subsubsection{Non-Judicial vs Judicial States}
%------------------------------------------------------------

In this subsection, I show that growth in private-label mortgages leads to a stronger housing bust in non-judicial states. In addition, my main results are significant within the non-judicial states and within the judicial states.

\cite{mian2015foreclosures} shows that, in 20 states of the U.S., foreclosures of a delinquent property need judicial judgment.\footnote{20 judicial states can be found in Figure 2 in \cite{mian2015foreclosures} and \url{https://www.realtytrac.com/real-estate-guides/foreclosure-laws/}. The twenty judicial states are Connecticut, Delaware, Florida, Illinois, Indiana, Kansas, Kentucky, Louisiana, Maine, Maryland, Massachusetts, Nebraska, New Jersey, New Mexico, New York, North Dakota, Ohio, Pennsylvania, South Carolina, and Vermont.} In these states, in order to sell a delinquent property through foreclosure, lenders must file a notice with a judge providing evidence of the delinquency of the mortgage and get court approval. In contrast, in non-judicial states, the foreclosure process is much easier and does not need court approval. For more details, see \cite{mian2015foreclosures}. \cite{mian2015foreclosures} find that lenders in non-judicial states are twice as likely to foreclose on delinquent mortgages. I expect non-judicial states may experience a stronger housing bust, caused by growth in private-label mortgages in the boom period (99-05). I need to test whether my main conclusion still holds after controlling for this state-level difference.

Regression specification is the same as Eq (\ref{eq:HPIBoomBustonPLNJM_APLvsNone}) except that the dummy variable is for non-judicial states. Table (\ref{table_Robust.NJDvsJD.HPI.D99t05.D07t09}) reports the above 2SLS results and shows two key conclusions. First, compared to judicial states, non-judicial states experience a stronger house price bust (07-09). This evidence is consistent with the notion that the absence of judicial procedure eases the process of foreclosure \citep{mian2015foreclosures}. Second, after controlling the differential trend in non-recourse states, for all metropolitan areas, the growth of private-label (non-jumbo) mortgages 1999-2005 leads to a stronger house price boom (99-05) and a stronger bust (07-09) in the high-net-export-growth metropolitan areas. The SW F-statistics and p-value of robust (clustered) Kleibergen-Paap test statistics both show that instruments are both separately and jointly strong for the two endogenous variables.

%------------------------------------------------------------
\subsubsection{Sand vs Other States}
%------------------------------------------------------------

In this subsection, I show that growth in private-label mortgages leads to a stronger housing price boom and bust cycle in sand states. In addition, my main results are significant within both the sand states and other states.

Many studies highlight that sand states (Arizona, California, Florida, and Nevada) experienced phenomenal housing cycles in comparison to the rest of the United States \citep{choi2016sand}. I expect that, in my setting, sand states may experience a stronger housing bust, caused by growth in private-label mortgages in the boom period (99-05). 

For regression specification, I add an interaction of a dummy variable for sand states and a period dummy to Eq (\ref{eq:HPIBoomBustonPLNJM}). I do not use an interaction of three terms because the limited number of metropolitan areas and counties presents a weak IV concern. These four sand states have 48 metropolitan areas and 73 counties.\footnote{If I were to use the interaction of three terms $PLNJM Growth (99-05) \times DumPeriod \times DumSand$, The SW F-statistics are 7.005 and 7.924 for the first-stage regression of PLNJM Growth and $PLNJM Growth \times DumSand$.} Since here, my primary focus is on the differential housing boom and bust of sand states, a dummy variable can serve this purpose. In order to study the different impact of the growth of private-label mortgages, more granular data is necessary.

Table (\ref{table_Robust.SandvsNone.HPI.D99t05.D07t09}) reports the above 2SLS results and shows two key conclusions. First, compared to other states, sand states experience a stronger house price boom (99-05) and a stronger bust (07-09), shown in the coefficients of interaction terms of sand dummy and period dummy. Second, after controlling for the differential trend in the sand states, I only use the within-sand-states and within-other-states differences across metropolitan areas. The 2SLS regression shows that the growth of private-label (non-jumbo) mortgages 1999-2005 leads to a stronger house price boom (99-05) and a stronger bust (07-09) in the high-net-export-growth metropolitan areas. Since the cross-group differences between sand states and other states are removed by the interaction terms of sand dummy and period dummy, my 2SLS shall be interpreted as evidence strongly supporting my main conclusion: induced by net export growth, private-label mortgage causes the housing price boom (99-05) and a stronger bust (07-09) across all metropolitan areas. Meanwhile, the reduced cross-metro variation due to the sand-state dummy reduces the F-statistics: the kleibergen-Paap (2006) robust (clustered) statistics and Montiel Olea-Pflueger (2013) efficient F-statistics are both 9.242, a number slightly lower than 10.

%\subsection{Robustness}

%------------------------------------------------------------
\subsubsection{State Capital Gain Tax}

In this section, I show that my major conclusion is robust to the inclusion of state capital gain tax as a control variable. \cite{gao2020economic} find that speculation measured by non-owner-occupied purchase mortgages is discouraged by the state capital gain tax, and such speculation contributes to the housing boom and bust. Based on their findings, I expect the state capital gain tax to discourage the housing boom and bust. 

For regression specification, I add interaction of state capital gain tax rate and period dummy to Eq (\ref{eq:HPIBoomBustonPLNJM}).\footnote{I would like to thank Zhenyu Gao, Michael Sockin, and Wei Xiong for sharing the state capital gain tax rate as of 2005.} Table (\ref{table_Robust.StCapGainTax.HPI.D99t05.D07t09}) reports the above 2SLS results and shows two key conclusions. First, the state capital gain tax rate contributes positively to the housing boom (99-05) and bust (07-09). This result contradicts with the discouraging effect of capital gain tax on housing speculation \citep{gao2020economic}. My result may reflect the owner-occupied housing exclusion by state capital gain tax, which encourages owner-occupied home purchase. According to Tax Foundation\footnote{The Tax Foundation Report: \url{https://taxfoundation.org/research/all/federal/capital-gains-taxes/}.}, homeowners may exclude up to \$250,000 (\$500,000 per couple) of capital gain if the homeowners had lived in the home for at least two of the previous five years. The exemption may be taken only once every two years and applies at both the federal and state levels. Thus, my results may sum the above two opposing effects of state capital gain on home purchase.

After controlling for state capital gain, my major conclusion holds: the growth of private-label (non-jumbo) mortgages 1999-2005 leads to the house price boom (99-05) and bust (07-09) across all metropolitan areas.

%------------------------------------------------------------
%\subsubsection{Purchase and Refinance}

%------------------------------------------------------------
%\subsubsection{Lags in Employment Share}

%--------------------------------------------------------------------------------------
%\subsection{Empirical: Verification Tests}
%--------------------------------------------------------------------------------------

%------------------------------------------------------------
%------------------------------------------------------------
%------------------------------------------------------------
\subsection{Mechanisms: Net Export Growth and Local Economic Conditions}\label{subsec:NEG_LocalEconConditions}
In this section, I verify the assumption that net export growth causes a differential higher increase in local employment growth, household income growth, and population growth in the high-net-export-growth metropolitan areas. This assumption is based on the economic base theory and links the causal relation between net export growth to credit expansion in private-label mortgages (PLM). The growth in employment, household income, and population are three channels through which net export growth induces credit expansion in PLM from 1999 to 2005. The analysis in this subsection includes all counties in the metropolitan areas (2003 CBSA code), since we do not rely on the HMDA mortgage data.

I do not claim these three channels are my contribution since there is ample evidence in the literature. Researchers show evidence of local employment changes due to imports \citep{autor2013china, pierce2016surprisingly} and exports \citep{feenstra2005world}. Studies also document the local household income changes due to imports \citep{barrot2022import} and exports \cite{minot1998export}. For changes in population growth (including migration), \cite{greenland2019import} provides causal evidence on imports, and \cite{li2023globalization} provides causal evidence on exports. 

I conduct regression analysis with the following specifications. 
%\vspace{-3mm}
\begin{equation}\label{eq:LocalEconCond_reg_NEG}
    \triangle_{91,07} Ln(LocalEconConditions_{c}) = \beta * \triangle_{91,07} \text{NetExp}_{m} + \gamma* \bm{Controls_{c}} + \alpha + \epsilon_{c}
\end{equation}
$\triangle_{91,07} Ln(LocalEconConditions_{c})$ represents the growth of local economic condition at the county $c$, which can be employment, household income, and population. I use the gravity-model-based instrumental variable ($\triangle_{91,07}\text{givNetExp}_{m}$) as IV for $\triangle_{91,07}\text{NetExp}_{m}$. Since these three variables determine the housing market in the long-term, or these three variables and the local housing market are jointly determined, I only include basic controls and demographic controls.

\comment{I include control variables at county $c$ only in or before the year 1991, which prevents any impact from net export growth. To take into consideration the county size, the regression is weighted by the natural logarithm of the number of households in 1991. Logarithms rather than the absolute number of households are chosen to prevent regression results from being dominated by a few super populous counties. To take into account that households might commute to work across counties within a metropolitan area, I measure net export growth in the metropolitan area.\footnote{According to US Census, "the general concept of a metropolitan statistical area is that of a core area containing a substantial population nucleus, together with adjacent communities having a high degree of economic and social integration with that core." (\url{https://www.census.gov/programs-surveys/metro-micro/about.html}} Additionally, I cluster Standard errors at the metropolitan area (MA) level.
} % end of comment

%------------------------------------------------------------
%\subsection{Net Export and Employment Growth}
%------------------------------------------------------------
\subsubsection{Net Export Growth and Employment Growth}

In this test, I use the total employment share of the working-age population (ages 15-64) excluding the construction industry to avoid the impact of mortgage credit on construction employment between 1999 and 2007. 

Table (\ref{table_BEATotExConstEmp.D91t07.4Reg}) reports OLS, reduced-form, second stage, and first stage of the results of total employment growth. First, panel A shows the positive and significant impact of net export growth (91-07, annualized) on total (excluding construction) employment share growth (91-07, annualized). Panel B reports the reduced-form estimates and shows that gravity-model-based net export growth increases household income growth. This significant impact is quite stable to the inclusion of various controls. First-stage estimates in panel D show that the strong positive correlation between the GIV and net export growth is quite stable across various specifications. The first-stage robust (clustered) Kleibergen-Paap F-statistic and the Montiel Olea-Pflueger Efficient F-statistics are both 94.55 in column (3) with all control variables. As emphasized by \cite{jiang2017have}, it is important to compare 2SLS and OLS estimates, given that, on average, 255 papers in the "Big Three" finance journals report 2SLS estimates nine times the size of OLS estimates. The key concerns are the bias from a weak IV and publication bias in searching for "blown-up" IV estimates. Besides my two F-statistics, I also confirm that 2SLS estimates are all less than twice OLS estimates throughout various specifications, where my baseline results in column (3) indicate a ratio of $0.326/0.175=1.86$, much lower than nine. Thus, it is very unlikely that my estimates are biased by a weak instrument. 

Panel C reports 2SLS estimates. the 2SLS estimates are quite stable as I introduce the controls in columns (2)-(3). According to column (3), one standard cross-sectional difference in annualized net export growth results in $ 0.326 \times 0.164\% = 0.053\%$ difference in employment share increase annually, translating into $0.853\%$ difference from 1991 to 2007 at the metro level. Such an effect is statistically significant at the one percent level.

\comment{$0.093\%$ cross-sectional difference in annual income growth rate is so small that it is very likely ignored by GSEM policy. }

%------------------------------------------------------------
%\subsection{Net Export and Household Income}
%------------------------------------------------------------
\subsubsection{Net Export Growth and Household Income Gorwth}

Table (\ref{table_IRSHHIncome.D91t07.4Reg}) reports OLS, reduced-form, second stage, and first stage of the results of household income growth. First, panel A shows the positive and significant impact of net export growth (91-07, annualized) on household income growth (91-07, annualized), with a significance level at 1\%. A similar pattern appears in reduced-form and 2SLS estimates. First-stage estimates in panel D show a strong positive correlation between the GIV and net export growth. The two first-stage F-statistics are very high (with both KP F-stat and MOP F-stat being 93.73 in column (3). Additionally, the 2SLS estimates are very close to OLS estimates (column (3) indicates a ratio of $1.036/0.993=1.04$). Thus, my estimates are free from the concern of a weak instrument. 

As for economic meaning, the 2SLS estimate in Column (3) in panel C shows that one standard cross-sectional difference in annualized net export growth results in $1.036 \times 0.165\% = 0.171\%$ difference in household income growth, translating into $2.742\%$ difference from 1991 to 2007 at the metro level.

%------------------------------------------------------------
%\subsection{Net Export and Population}
%------------------------------------------------------------
\subsubsection{Net Export Growth and Population Growth}

Table (\ref{table_WorkAgePop.D91t07.4Reg}) reports OLS, reduced-form, second stage, and first stage of the results of working-age population (age 15-64) growth and total population growth. I do not include housing controls, as population growth is the most important determinant of housing growth and housing regulation. First, panel A shows the positive and significant impact of net export growth (91-07, annualized) on working-age and total population growth (91-07, annualized), all with significance level at 1\%. A similar pattern shows up in reduced-form estimates and 2SLS estimates. First-stage estimates in panel D show a strong positive correlation between the net export growth and the gravity model-based instrumental variable. The first two-stage statistics are very high (with both KP F-stat and MOP F-stat being 93.75 in columns (3) and (6)). In addition, the 2SLS estimates are very close to OLS estimates, with column (3) indicating a ratio of $0.750/0.761=0.99$ and column (6) indicating a ratio of $1.059/1.093=0.969$. Thus, my estimates are very unlikely to be biased by a weak instrument. 
 
In terms of economic meaning, Column (3) and (6) in panel C shows that one standard cross-sectional difference in annualized net export growth results in $ 0.750\times 0.165\%  = 0.124\%$ difference in total population growth and $ 1.074\times 0.165\%  = 0.178\%$ difference in working-age population growth, translating into $1.985\%$ and $2.842\%$ differences from 1991 to 2007 at the metro level.

%------------------------------------------------------------
% end of the entire subsection
%------------------------------------------------------------

%--------------------------------------------------------------------------------------
%\subsection{Appendix: Omitted Derivations in the Model.tex}
%--------------------------------------------------------------------------------------

%-----------------------------------------------------
%-----------------------------------------------------
%-----------------------------------------------------
%-----------------------------------------------------
\subsection{Omitted Derivations in the Model}

\subsubsection{Boundary Conditions for income-to-house ratios}\label{subsec:app_model}

\noindent \textbf{PLM in HNEG area}

The assumption that all households can pay the full mortgage amount in up-state means that even a household with the lowest income can pay the full mortgage amount: $\underline{I_{H,0}} ( 1+g_{u} \textcolor{red}{+ \mathit{\Delta}g} - c ) \geq H_{H,0} ( 1 + FCR^{*}_{P} + R_{P,H} + TF)$. This inequality implies 
\begin{equation*}
\frac{\underline{I_{H,0}}}{H_{H,0}} \geq \frac{ 1 + FCR^{*}_{P} + R_{P,H} + TF}{1+g_{u} \textcolor{red}{+ \mathit{\Delta}g} - c} \coloneqq \underline{i_{P,H}}
\end{equation*}
The assumption that all households default in down-state means that even a household with the highest income could default: $\overline{I_{H,0}}(1+g_{d} \textcolor{red}{+ \mathit{\Delta}g} - c) < H_{H,0} ( 1 + FCR^{*}_{P} + R_{P,H} + TF) $. This inequality implies 
\begin{equation*}
\frac{\overline{I_{H,0}}}{H_{H,0}} < \frac{ 1 + FCR^{*}_{P} + R_{P,H} + TF}{1+g_{d} \textcolor{red}{+ \mathit{\Delta}g} - c} \coloneqq \overline{i_{P,H}}
\end{equation*}
Therefore, I only assume that household have income-to-house ratio in time 0: $i_0 \in [\underline{i_{P,H}}, \overline{i_{P,H}}  )$. This assumption matches the practice. For the households that would default with probability 1 ($i_0<\underline{i_{P,H}}$ in my setting), lenders would not grant mortgages at all. For the households that would not default at all ($i_0>\overline{i_{P,H}}$ in my setting), they could always get mortgages by paying the lowest rate (risk-free rate $r_f$), thus having no variation in mortgage rates.  In either case, there is no economic trade-off in the mortgage market that is key to my model demonstration. Again, I purposefully ignore these two extreme groups of households to make my model simple and focus my attention on the key economic mechanism of interest.

\noindent \textbf{PLM in LNEG area}

The assumption that all households can pay the full mortgage amount in up-state means that even a household with the lowest income can pay the full mortgage amount: $\underline{I_{L,0}}(1+g_{u} \textcolor{blue}{- \mathit{\Delta}g} - c) \geq H_{L,0} ( 1 + FCR^{*}_{P} + R_{P,L} + TF) $. This inequality implies 
\begin{equation*}
\frac{\underline{I_{L,0}}}{H_{L,0}} \geq \frac{ 1 + FCR^{*}_{P} + R_{P,L} + TF}{1+g_{u} \textcolor{blue}{- \mathit{\Delta}g} - c} \coloneqq \underline{i_{P,L}}
\end{equation*}
The assumption that all households default in down-state means that even a household with the highest income could default: $\overline{I_{L,0}}(1+g_{d} \textcolor{blue}{\textcolor{blue}{- \mathit{\Delta}g}} - c) < H_{L,0} ( 1 + FCR^{*}_{P} + R_{P,L} + TF)$. This inequality implies 
\begin{equation*}
\frac{\overline{I_{L,0}}}{H_{L,0}} < \frac{ 1 + FCR^{*}_{P} + R_{P,L} + TF}{1+g_{d} \textcolor{blue}{\textcolor{blue}{- \mathit{\Delta}g}} - c} \coloneqq \overline{i_{P,L}}
\end{equation*}
Therefore, I only assume that household have income-to-house ratio in time 0: $i_0 \in [\underline{i_{P,L}}, \overline{i_{P,L}}  )$. This assumption matches the practice. For the households that would default with probability 1 ($i_0<\underline{i_{P,L}}$ in my setting), lenders would not grant mortgages at all. For the households that would not default at all ($i_0>\overline{i_{P,L}}$ in my setting), they could always get mortgages by paying the lowest rate (risk-free rate $r_f$), thus having no variation in mortgage rates.  In either case, there is no economic trade-off in the mortgage market that is key to my model demonstration. Again, I purposefully ignore these two extreme groups of households to make my model simple and focus my attention on the key economic mechanism of interest.

\subsubsection{When Up- and Down-states Have Unequal Probability}\label{sec:App_Model_P}
In the main part of the model, for easy exposition, I assume that both up- and down-states have the probability of $\frac{1}{2}$, as a way to model uncertainty at the household level. In a more general and realistic case, the up-state can have a probability $p>\frac{1}{2}$ (say around 95\%). I will show in this appendix that my main prediction still holds that credit expansion in PLM is stronger in HNEG area than the one in LNEG area when PLM funding cost declines. This prediction means $i^{**}_{P,H}< i^{**}_{P,L}$ in math form.

First, let us derive $i^{**}_{P,H}$ for PLM in the HNEG area. The break-even condition for PLM in the HNEG area shall be modified from Equation (\ref{eq:BE_P_H}) as follows: 
\begin{equation}\label{eq:BE_P_H_app}
\resizebox{0.9\textwidth}{!}{$
    p*H_{H,0} \underbrace{( 1 + FCR^{*}_{P} + R_{P,H} + TF)}_{ MR^{*}_{P,H}} + (1-p) [\underbrace{RR* H_{H,0}(1\textcolor{red}{+ \mathit{\Delta}g + \mathit{\Delta}e + \mathit{\Delta}p })}_{\text{Recovery Value}} + \underbrace{I_{H,0}(1+g_{d}\textcolor{red}{+ \mathit{\Delta}g} -c)}_{\text{Disposable Income}}] = H_{H,0}(1+FCR^{*}_{P} + TF)
$} %end of resizebox
\end{equation}
Then the mortgage rate $ MR^{*}_{P,L} $ is set to make the equality hold: 
\begin{equation}\label{eq:MR_P_H_app}
\resizebox{0.9\textwidth}{!}{$
    MR^{*}_{P,H} = \frac{1}{p} [ (1 + FCR^{*}_{P} + TF) - (1-p)RR(1\textcolor{red}{+ \mathit{\Delta}g + \mathit{\Delta}e + \mathit{\Delta}p }) - (1-p)(1+g_{d} \textcolor{red}{+ \mathit{\Delta}g}-c)*i_0 ] 
$} %end of resizebox
\end{equation}
Then I can derive the lower bound of the income-to-house ratio that can get PLM when the funding cost of PLM declines to $FCR^{**}_P$:
\begin{equation}\label{eq:i_newPH_app}
%\resizebox{0.9\textwidth}{!}{$
    i^{**}_{P,H} = \frac{ (1 + FCR^{**}_P + TF) - (1-p)RR(1 \textcolor{red}{+\mathit{\Delta}g + \mathit{\Delta}e  + \mathit{\Delta}p  } ) - p\overline{MR_P}}{ (1-p) (1+g_{d} \textcolor{red}{+ \mathit{\Delta}g} - c )}
%$} %end of resizebox
\end{equation}

Second, let us derive $i^{**}_{P,L}$ for PLM in the LNEG area. The break-even condition for PLM in the LNEG area shall be modified from Equation (\ref{eq:BE_P_L}) as follows: 
\begin{equation}\label{eq:BE_P_L_app}
\resizebox{0.9\textwidth}{!}{$
    p*H_{L,0} \underbrace{( 1 + FCR^{*}_{P} + R_{P,L} + TF)}_{ MR^{*}_{P,L}} + (1-p) [\underbrace{RR* H_{L,0}(1\textcolor{blue}{- \mathit{\Delta}g - \mathit{\Delta}e - \mathit{\Delta}p })}_{\text{Recovery Value}} + \underbrace{I_{L,0}(1+g_{d}\textcolor{blue}{- \mathit{\Delta}g} -c)}_{\text{Disposable Income}}] = H_{L,0}(1+FCR^{*}_{P} + TF)
$} %end of resizebox
\end{equation}

Then the mortgage rate $ MR^{*}_{P,L} $ is set to make the equality hold:  
\begin{equation}\label{eq:MR_P_L_app}
\resizebox{0.9\textwidth}{!}{$
    MR^{*}_{P,L} = \frac{1}{p} [ (1 + FCR^{*}_{P} + TF) - (1-p)RR(1\textcolor{blue}{- \mathit{\Delta}g - \mathit{\Delta}e - \mathit{\Delta}p }) - (1-p)(1+g_{d} \textcolor{blue}{- \mathit{\Delta}g}-c)*i_0 ] 
$} %end of resizebox
\end{equation}

Then I can derive the lower bound of the income-to-house ratio $i^{**}_{P,L}$ that can get PLM when the funding cost of PLM declines to $FCR^{**}_P$:
\begin{equation}\label{eq:i_newPL_app}
%\resizebox{0.9\textwidth}{!}{$
    i^{**}_{P,L} = \frac{ (1 + FCR^{**}_P + TF) - (1-p)RR(1 \textcolor{blue}{-\mathit{\Delta}g - \mathit{\Delta}e  - \mathit{\Delta}p  } ) - p\overline{MR_P}}{ (1-p) (1+g_{d} \textcolor{blue}{- \mathit{\Delta}g} - c )}
%$} %end of resizebox
\end{equation}

By comparing both numerators and denominators, I can easily know $i^{**}_{P,H}< i^{**}_{P,L}$. Therefore, my major prediction still holds that credit expansion in PLM is stronger in HNEG area than one in LNEG area when PLM funding cost declines. Again, the necessary economic conditions for the above prediction are that there are additional positive growth rates in household income, employment, and net migration inflow in the high-net-export-growth metropolitan areas.

\clearpage 
%-----------------------------------------------------
\subsection{Appendix Figures and Tables}

%\pagebreak
%------------------------------------------------------------
%------------------------------------------------------------
% figure 3: fig_NEP_D91t07_FirstStage

%------------------------------------------------------------
% figure 4:fig_NEP_D91t07_FirstStage

\begin{figure}[h!] 
    \centering
    \includegraphics[width=\textwidth]{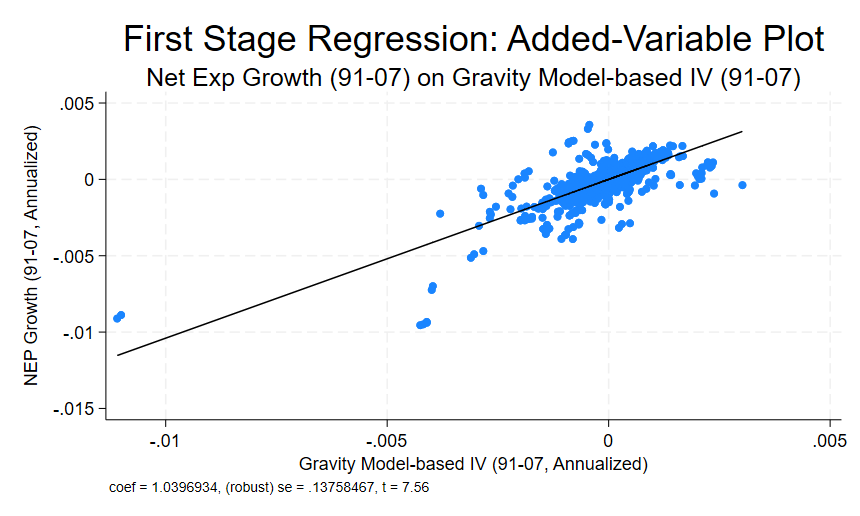}
    \caption{\textbf{First Stage Regression: Net Export Growth (91-07) on Gravity Model-based Net Export Growth (91-07)} \smallskip \newline 
    {\footnotesize This figure displays the added-variable plot of first stage regression: net export expansion growth (91-07) on gravity model-based instrumental variable growth (91-07) with full controls in column (4) in Table (\ref{table_BEATotExConstEmp.D91t07.4Reg}). Regression observations are weighted by logarithmic of number of total population in 1991 at county level. Standard errors are clustered at metropolitan areas (CBSA03) level. The coefficient is 0.941, clustered standard error is 0.124, and the t-statistic is 7.62. 
        } %end of small font
    } % end of caption
    \label{fig_NEP_D91t07_FirstStage}
    % note that \label is given after \caption.
\end{figure}

%---------------------------------------------------------------

%%%%%%%%%%%%%%%%%%%%%%%%%%%%%%%%%%%%%%%%%%%%%%%%
% table_Robust.APLvsNone.HPI.D99t05.D07t09
%%%%%%%%%%%%%%%%%%%%%%%%%%%%%%%%%%%%%%%%%%%%%%%%

%---------------------------------------------------------------

%%%%%%%%%%%%%%%%%%%%%%%%%%%%%%%%%%%%%%%%%%%%%%%%
% table_Robust.APLvsNone.HPI.D99t05.D07t09
%%%%%%%%%%%%%%%%%%%%%%%%%%%%%%%%%%%%%%%%%%%%%%%%

\noindent 

\begin{table}[h!]
\centering
\caption{
\textbf{Robustness Test for Anti-Predatory Lending States vs. Non-Anti-Predatory Lending States.} \\
Stacked 2SLS Regressions of House Price Growth in Boom (99-05) and Bust (07-09) Periods on PLNJM Growth (99-05)  \smallskip \newline
{\scriptsize
This table reports OLS, reduced-form, first stage, and second stages of stacked 2SLS regression $\triangle_{99,05} \& \triangle_{07,09} Ln(HPI_{c})  = \beta_{Boom} * \triangle_{99,05} Ln(PLNJM_{c}) \times Dum_{99,05} + \beta_{Bust} * \triangle_{99,05} Ln(PLNJM_{c}) \times Dum_{07,09} + \beta_{APL, Boom} * \triangle_{99,05} Ln(PLNJM_{c}) \times Dum_{99,05} \times Dum_{APL} + \beta_{APL, Bust} * \triangle_{99,05} Ln(PLNJM_{c}) \times Dum_{07,09} \times Dum_{APL} + \gamma_{Boom} * \bm{Controls_{c}} \times Dum_{99,05} + \gamma_{Bust} * \bm{Controls_{c}} \times Dum_{07,09} + \epsilon_{c}$. The left-hand-side dependent variable $\triangle_{99,05} \& \triangle_{07,09} Ln(HPI_{c})$ is the stacked growth rate of the house price index at county $c$ 99-05 and 07-09. The key variable of interest $\triangle_{99,05} Ln(PLNJM_{c}$ is the growth rate of the dollar amount of private-label (non-jumbo) mortgages at county $c$ 99-05. $Dum_{APL}$ is the dummy variable for counties in states with anti-predatory lending laws. $Controls_{c}$ indicates control variables at county $c$ in the period start year 1999. In either boom or bust period, we have two endogenous variables here: $\triangle_{99,05} Ln(PLNJM_{c})$ is instrumented by $\triangle_{99,05}\text{givNetExp}_{m}$ and $\triangle_{99,05} Ln(PLNJM_{c}) \times Dum_{APL}$ is instrumented by $\triangle_{99,05}\text{givNetExp}_{m} \times Dum_{APL}$. For each of the first-stage F-tests of two endogenous variables, we report Sanderson-Windmeijer (2016) robust (clustered) statistics. To evaluate the overall strength of instruments, we report the p-value of robust (clustered) Kleibergen-Paap test statistics calculated by Windmeijer (2021). Each regression is weighted by the natural logarithm of housing units in 1999. Standard errors are clustered at the CBSA level. ***, **, and * indicate significance at the 1\%, 5\%, and 10\% levels, respectively.
} % end of small font size
} % end of caption
\label{table_Robust.APLvsNone.HPI.D99t05.D07t09}

\resizebox{\columnwidth}{!}{%

\begin{tabular}{l*{4}{c}}
\toprule
\textbf{TSLS Estimates}              &\multicolumn{4}{c}{House Price Growth (07USD, 99-05 or 07-09, An)} \\
            \cmidrule{2-5} 
            &\multicolumn{1}{c}{(1)}&\multicolumn{1}{c}{(2)}&\multicolumn{1}{c}{(3)}&\multicolumn{1}{c}{(4)}\\
            
\midrule
PLNJM Growth (99-05, An) x Dum99t05&    0.445***&    0.395***&    0.412***&    0.444***\\
               &  (0.102)   &  (0.114)   &  (0.140)   &  (0.153)   \\
\addlinespace
PLNJM Growth (99-05, An) x Dum07t09&   -0.789***&   -0.737***&   -0.782***&   -0.826***\\
               &  (0.177)   &  (0.189)   &  (0.232)   &  (0.223)   \\
\addlinespace
PLNJM Growth (99-05) x Dum99t05 x DumAPL&    0.172***&    0.146***&    0.115** &    0.118** \\
               &  (0.055)   &  (0.049)   &  (0.047)   &  (0.052)   \\
\addlinespace
PLNJM Growth (99-05) x Dum07t09 x DumAPL&   -0.292***&   -0.261***&   -0.249***&   -0.210** \\
               &  (0.080)   &  (0.069)   &  (0.079)   &  (0.085)   \\
\addlinespace
\midrule
DumPeriod  &    Y        &  Y   &   Y    & Y      \\
Basic Controls x DumPeriod &            &  Y   &   Y    & Y     \\
Housing Controls x DumPeriod &           &      & Y       & Y       \\
Demographic Controls x DumPeriod &            &      &        &  Y       \\

\midrule
Obs            &     1576   &     1576   &     1418   &     1418   \\
Cluster SE     &     CBSA   &     CBSA   &     CBSA   &     CBSA   \\
Weight         & Ln(HU99)   & Ln(HU99)   & Ln(HU99)   & Ln(HU99)   \\
SW F-Stat: PLNJM 1st-Stage&    21.85   &    19.08   &    15.17   &    12.49   \\
SW F-Stat: PLNJMxDumAPL 1st-Stage&    33.34   &    30.04   &    18.63   &    16.25   \\
KP Robust UnderID P-Value&   0.0004   &   0.0004   &   0.0022   &   0.0037   \\
CoefEqual\_Chi2 &   22.498   &   15.796   &   11.123   &   12.261   \\
CoefEqual\_PValue&    0.000   &    0.000   &    0.001   &    0.000   \\

\bottomrule

\end{tabular}

} % end of resize box

\end{table}

\clearpage 
%---------------------------------------------------------------

%%%%%%%%%%%%%%%%%%%%%%%%%%%%%%%%%%%%%%%%%%%%%%%%
% table_Robust.NRCvsRC.HPI.D99t05.D07t09
%%%%%%%%%%%%%%%%%%%%%%%%%%%%%%%%%%%%%%%%%%%%%%%%

%---------------------------------------------------------------

%%%%%%%%%%%%%%%%%%%%%%%%%%%%%%%%%%%%%%%%%%%%%%%%
% table_Robust.NRCvsRC.HPI.D99t05.D07t09
%%%%%%%%%%%%%%%%%%%%%%%%%%%%%%%%%%%%%%%%%%%%%%%%

\noindent 

\begin{table}[h!]
\centering
\caption{
\textbf{Robustness Test for Non-Recourse States vs. Recourse States.} \\
Stacked 2SLS Regressions of House Price Growth in Boom (99-05) and Bust (07-09) Periods on PLNJM Growth (99-05)  \smallskip \newline
{\scriptsize
This table reports stacked 2SLS regression $\triangle_{99,05} \& \triangle_{07,09} Ln(HPI_{c}) = \beta_{Boom} * \triangle_{99,05} Ln(PLNJM_{c}) \times Dum_{99,05} + \beta_{Bust} * \triangle_{99,05} Ln(PLNJM_{c}) \times Dum_{07,09}  + \beta_{NRC, Boom} * \triangle_{99,05} Ln(PLNJM_{c}) \times Dum_{99,05} \times Dum_{NRC} + \beta_{NRC, Bust} * \triangle_{99,05} Ln(PLNJM_{c}) \times Dum_{07,09} \times Dum_{NRC} + \gamma_{Boom} * \bm{Controls_{c}} \times Dum_{99,05} + \gamma_{Bust} * \bm{Controls_{c}} \times Dum_{07,09} + \epsilon_{c}$. The left-hand-side dependent variable $\triangle_{99,05} \& \triangle_{07,09} Ln(HPI_{c})$ is the stacked growth rate of the house price index at county $c$ 99-05 and 07-09 in either non-recourse state or recourse state. The key variable of interest $\triangle_{99,05} Ln(PLNJM_{c})$ is the growth rate of the dollar amount of private-label mortgages at the county $c$ 99-05. $Dum_{NRC}$ is the dummy variable for counties in non-recourse states. $Controls_{c}$ indicates control variables at county $c$ in the start year 1999. In either boom or bust period, we have two endogenous variables here: $\triangle_{99,05} Ln(PLNJM_{c})$ is instrumented by $\triangle_{99,05}\text{givNetExp}_{m}$ and $\triangle_{99,05} Ln(PLNJM_{c}) \times Dum_{APL}$ is instrumented by $\triangle_{99,05}\text{givNetExp}_{m} \times Dum_{APL}$. For each of the first-stage F-tests of two endogenous variables, we report Sanderson-Windmeijer (2016) robust (clustered) statistics. To evaluate the overall strength of instruments, we report the p-value of robust (clustered) Kleibergen-Paap test statistics calculated by Windmeijer (2021). Each regression is weighted by the natural logarithm of housing units in 1999. Standard errors are clustered at the CBSA level. ***, **, and * indicate significance at the 1\%, 5\%, and 10\% levels, respectively.
} % end of small font size
} % end of caption
\label{table_Robust.NRCvsRC.HPI.D99t05.D07t09}

\resizebox{\columnwidth}{!}{%

\begin{tabular}{l*{4}{c}}
\toprule
\textbf{TSLS Estimates}              &\multicolumn{4}{c}{House Price Growth (07USD, 99-05 or 07-09, An)} \\
            \cmidrule{2-5} 
            &\multicolumn{1}{c}{(1)}&\multicolumn{1}{c}{(2)}&\multicolumn{1}{c}{(3)}&\multicolumn{1}{c}{(4)}\\
            
\midrule

PLNJM Dollar Growth (07USD, 99-05, An) x Dum99t05&    0.507***&    0.446***&    0.445***&    0.478***\\
               &  (0.115)   &  (0.124)   &  (0.142)   &  (0.160)   \\
\addlinespace
PLNJM Dollar Growth (07USD, 99-05, An) x Dum07t09&   -0.888***&   -0.824***&   -0.854***&   -0.882***\\
               &  (0.212)   &  (0.220)   &  (0.259)   &  (0.259)   \\
\addlinespace
PLNJM Dollar Growth (99-05) x Dum99t05 x DumNRC&    0.096   &    0.089   &    0.029   &    0.021   \\
               &  (0.060)   &  (0.057)   &  (0.059)   &  (0.062)   \\
\addlinespace
PLNJM Dollar Growth (99-05) x Dum07t09 x DumNRC&   -0.226** &   -0.233** &   -0.224** &   -0.190*  \\
               &  (0.104)   &  (0.103)   &  (0.106)   &  (0.101)   \\
\addlinespace
\midrule
DumPeriod  &    Y        &  Y   &   Y    & Y       \\
Basic Controls x DumPeriod &            &  Y   &   Y    & Y        \\
Housing Controls x DumPeriod &           &      & Y       & Y        \\
Demographic Controls x DumPeriod &            &      &        &  Y        \\

\midrule
Obs            &     1576   &     1576   &     1418   &     1418   \\
Cluster SE     &     CBSA   &     CBSA   &     CBSA   &     CBSA   \\
Weight         & Ln(HU99)   & Ln(HU99)   & Ln(HU99)   & Ln(HU99)   \\
SW F-Stat: PLNJM (99 to 05) 1st-Stage &    24.19   &    21.68   &    17.28   &    14.18   \\
SW F-Stat: PLNJMxDumNRC (99 to 05) 1st-Stage &    34.61   &    35.66   &    32.51   &    33.14   \\
KP Robust (99 to 05) UnderID P-Value &   0.0007   &   0.0007   &   0.0024   &   0.0048   \\
CoefEqual\_Chi2 &   20.803   &   15.475   &   11.278   &   11.215   \\
CoefEqual\_PValue&    0.000   &    0.000   &    0.001   &    0.001   \\

\bottomrule

\end{tabular}

} % end of resize box

\end{table}

\clearpage 
%---------------------------------------------------------------

%%%%%%%%%%%%%%%%%%%%%%%%%%%%%%%%%%%%%%%%%%%%%%%%
% table_Robust.NJDvsJD.HPI.D99t05.D07t09
%%%%%%%%%%%%%%%%%%%%%%%%%%%%%%%%%%%%%%%%%%%%%%%%

%---------------------------------------------------------------

%%%%%%%%%%%%%%%%%%%%%%%%%%%%%%%%%%%%%%%%%%%%%%%%
% table_Robust.NJDvsJD.HPI.D99t05.D07t09
%%%%%%%%%%%%%%%%%%%%%%%%%%%%%%%%%%%%%%%%%%%%%%%%

\noindent 

\begin{table}[h!]
\centering
\caption{
\textbf{Robustness Test for Non-Judicial States vs. Judicial States.} \\
Stacked 2SLS Regressions of House Price Growth in Boom (99-05) and Bust (07-09) Periods on PLNJM Growth (99-05)  \smallskip \newline
{\scriptsize
This table reports stacked 2SLS regression $\triangle_{99,05} \& \triangle_{07,09} Ln(HPI_{c}) = \beta_{Boom} * \triangle_{99,05} Ln(PLNJM_{c}) \times Dum_{99,05} + \beta_{Bust} * \triangle_{99,05} Ln(PLNJM_{c}) \times Dum_{07,09}  + \beta_{NJD, Boom} * \triangle_{99,05} Ln(PLNJM_{c}) \times Dum_{99,05} \times Dum_{NJD} + \beta_{NJD, Bust} * \triangle_{99,05} Ln(PLNJM_{c}) \times Dum_{07,09} \times Dum_{NJD} + \gamma_{Boom} * \bm{Controls_{c}} \times Dum_{99,05} + \gamma_{Bust} * \bm{Controls_{c}} \times Dum_{07,09} + \epsilon_{c}$. The left-hand-side dependent variable $\triangle_{99,05} \& \triangle_{07,09} Ln(HPI_{c})$ is the stacked growth rate of the house price index at county $c$ 99-05 and 07-09 in either judicial state or non-judicial state. The key variable of interest $\triangle_{99,05} Ln(PLNJM_{c})$ is the growth rate of the dollar amount of private-label mortgages at the county $c$ 99-05. $Dum_{NJD}$ is the dummy variable for counties in states where foreclosure of a delinquent property needs judicial judgment. $Controls_{c}$ indicates control variables at county $c$ in the start year 1999. In either boom or bust period, we have two endogenous variables here: $\triangle_{99,05} Ln(PLNJM_{c})$ is instrumented by $\triangle_{99,05}\text{givNetExp}_{m}$ and $\triangle_{99,05} Ln(PLNJM_{c}) \times Dum_{APL}$ is instrumented by $\triangle_{99,05}\text{givNetExp}_{m} \times Dum_{APL}$. For each of the first-stage F-tests of two endogenous variables, we report Sanderson-Windmeijer (2016) robust (clustered) statistics. To evaluate the overall strength of instruments, we report the p-value of robust (clustered) Kleibergen-Paap test statistics calculated by Windmeijer (2021). Each regression is weighted by the natural logarithm of housing units in 1999. Standard errors are clustered at the CBSA level. ***, **, and * indicate significance at the 1\%, 5\%, and 10\% levels, respectively.
} % end of small font size
} % end of caption
\label{table_Robust.NJDvsJD.HPI.D99t05.D07t09}

\resizebox{\columnwidth}{!}{%

\begin{tabular}{l*{4}{c}}
\toprule
\textbf{TSLS Estimates}              &\multicolumn{4}{c}{House Price Growth (07USD, 99-05 or 07-09, An)} \\
            \cmidrule{2-5} 
            &\multicolumn{1}{c}{(1)}&\multicolumn{1}{c}{(2)}&\multicolumn{1}{c}{(3)}&\multicolumn{1}{c}{(4)}\\
            
\midrule
PLNJM Dollar Growth (99-05, 07USD, An) x Dum99t05&    0.498***&    0.421***&    0.421***&    0.460***\\
               &  (0.106)   &  (0.113)   &  (0.139)   &  (0.155)   \\
\addlinespace
PLNJM Dollar Growth (99-05, 07USD, An) x Dum07t09&   -0.848***&   -0.753***&   -0.767***&   -0.809***\\
               &  (0.210)   &  (0.219)   &  (0.260)   &  (0.245)   \\
\addlinespace
PLNJM Dollar Growth (99-05) x Dum99t05 x DumNJD&    0.042   &    0.065*  &    0.049   &    0.036   \\
               &  (0.044)   &  (0.034)   &  (0.030)   &  (0.031)   \\
\addlinespace
PLNJM Dollar Growth (99-05) x Dum07t09 x DumNJD&   -0.140** &   -0.183***&   -0.177***&   -0.152***\\
               &  (0.060)   &  (0.052)   &  (0.055)   &  (0.054)   \\
\addlinespace
\midrule
DumPeriod  &    Y        &  Y   &   Y    & Y     \\
Basic Controls x DumPeriod &            &  Y   &   Y    & Y    \\
Housing Controls x DumPeriod &           &      & Y       & Y     \\
Demographic Controls x DumPeriod &            &      &        &  Y    \\
\midrule
Obs            &     1576   &     1576   &     1418   &     1418   \\
Cluster SE     &     CBSA   &     CBSA   &     CBSA   &     CBSA   \\
Weight         & Ln(HU99)   & Ln(HU99)   & Ln(HU99)   & Ln(HU99)   \\
SW F-Stat: PLNJM (99 to 05) 1st-Stage &    21.28   &    19.13   &    16.37   &    13.10   \\
SW F-Stat: PLNJMxDumNJD (99 to 05) 1st-Stage &    45.87   &    51.44   &    55.14   &    53.92   \\
KP Robust (99 to 05) UnderID P-Value &   0.0011   &   0.0014   &   0.0042   &   0.0082   \\
CoefEqual\_Chi2 &   21.058   &   14.372   &    9.719   &   10.961   \\
CoefEqual\_PValue&    0.000   &    0.000   &    0.002   &    0.001   \\

\bottomrule

\end{tabular}

} % end of resize box

\end{table}

\clearpage 
%---------------------------------------------------------------

%%%%%%%%%%%%%%%%%%%%%%%%%%%%%%%%%%%%%%%%%%%%%%%%
% table_Robust.SandvsNone.HPI.D99t05.D07t09
%%%%%%%%%%%%%%%%%%%%%%%%%%%%%%%%%%%%%%%%%%%%%%%%

%---------------------------------------------------------------

%%%%%%%%%%%%%%%%%%%%%%%%%%%%%%%%%%%%%%%%%%%%%%%%
% table_Robust.SandvsNone.HPI.D99t05.D07t09
%%%%%%%%%%%%%%%%%%%%%%%%%%%%%%%%%%%%%%%%%%%%%%%%

\noindent 

\begin{table}[h!]
\centering
\caption{
\textbf{Robustness Test for Sand States vs. Non-Sand States.} \\
Stacked 2SLS Regressions of House Price Growth in Boom (99-05) and Bust (07-09) Periods on PLNJM Growth (99-05)  \smallskip \newline
{\scriptsize
This table reports stacked 2SLS regression $\triangle_{99,05} \& \triangle_{07,09} Ln(HPI_{c}) = \beta_{Boom} * \triangle_{99,05} Ln(PLNJM_{c}) \times Dum_{99,05} + \beta_{Bust} * \triangle_{99,05} Ln(PLNJM_{c}) \times Dum_{07,09}  + \beta_{Sand, Boom} \times Dum_{99,05} \times Dum_{Sand} + \beta_{Sand, Bust} * \times Dum_{07,09} \times Dum_{Sand} + \gamma_{Boom} * \bm{Controls_{c}} \times Dum_{99,05} + \gamma_{Bust} * \bm{Controls_{c}} \times Dum_{07,09} + \epsilon_{c}$. The left-hand-side dependent variable $\triangle_{99,05} \& \triangle_{07,09} Ln(HPI_{c})$ is the stacked growth rate of the house price index at county $c$ 99-05 and 07-09 in either sand state or non-sand state. The key variable of interest $\triangle_{99,05} Ln(PLNJM_{c})$ is the growth rate of the dollar amount of private-label mortgages at the county $c$ 99-05. $Dum_{Sand}$ is the dummy variable for counties in four sand states.  $Controls_{c}$ indicates control variables at county $c$ in the start year 1999. We use the gravity model-based instrumental variable ($\triangle_{99,05}\text{givNetExp}_{m}$) as IV for $\triangle_{99,05}Ln(PLNJM_{c})$. For the first-stage F-test of the non-stacked sample (99-05), we report kleibergen-Paap (2006) robust (clustered) statistics and Montiel Olea-Pflueger (2013) efficient statistics. Each regression is weighted by the natural logarithm of housing units in 1999. Standard errors are clustered at the CBSA level. ***, **, and * indicate significance at the 1\%, 5\%, and 10\% levels, respectively.
} % end of small font size
} % end of caption
\label{table_Robust.SandvsNone.HPI.D99t05.D07t09}

\resizebox{\columnwidth}{!}{%

\begin{tabular}{l*{6}{c}}
\toprule
\textbf{TSLS Estimates}              &\multicolumn{5}{c}{House Price Growth (07USD, 99-05 or 07-09, An)} \\
            \cmidrule{2-6} 
            &\multicolumn{1}{c}{(1)}&\multicolumn{1}{c}{(2)}&\multicolumn{1}{c}{(3)}&\multicolumn{1}{c}{(4)}\\
            
\midrule

PLNJM Dollar Growth (99-05, 07USD, An) x Dum99t05&    0.408***&    0.364***&    0.382***&    0.421** \\
               &  (0.125)   &  (0.130)   &  (0.147)   &  (0.172)   \\
\addlinespace
PLNJM Dollar Growth (99-05, 07USD, An) x Dum07t09&   -0.621***&   -0.575***&   -0.615***&   -0.648***\\
               &  (0.180)   &  (0.175)   &  (0.209)   &  (0.217)   \\
\addlinespace
Dum\_Sand\_xD99t05&    0.042***&    0.036** &    0.028** &    0.023   \\
               &  (0.015)   &  (0.015)   &  (0.014)   &  (0.015)   \\
\addlinespace
Dum\_Sand\_xD07t09 &   -0.113***&   -0.109***&   -0.106***&   -0.095***\\
               &  (0.018)   &  (0.018)   &  (0.020)   &  (0.020)   \\
\addlinespace
\midrule
DumPeriod  &    Y        &  Y   &   Y    & Y      \\
Basic Controls x DumPeriod &            &  Y   &   Y    & Y      \\
Housing Controls x DumPeriod &           &      & Y       & Y     \\
Demographic Controls x DumPeriod &            &      &        &  Y      \\
\midrule
Obs            &     1576   &     1576   &     1418   &     1418   \\
Cluster SE     &     CBSA   &     CBSA   &     CBSA   &     CBSA   \\
Weight         & Ln(HU99)   & Ln(HU99)   & Ln(HU99)   & Ln(HU99)   \\
KP F-Stat (99 to 05, non-stacked sample) &    15.44   &    13.97   &    11.66   &    9.242   \\
MOP F-Stat (99 to 05, non-stacked sample)    &    15.44   &    13.97   &    11.66   &    9.242   \\
CoefEqual\_Chi2 &   13.973   &   11.865   &    9.047   &    8.545   \\
CoefEqual\_PValue&    0.000   &    0.001   &    0.003   &    0.003   \\

\bottomrule

\end{tabular}

} % end of resize box

\end{table}

\clearpage 
%---------------------------------------------------------------

%%%%%%%%%%%%%%%%%%%%%%%%%%%%%%%%%%%%%%%%%%%%%%%%
% table_Robust.StCapGainTax.HPI.D99t05.D07t09
%%%%%%%%%%%%%%%%%%%%%%%%%%%%%%%%%%%%%%%%%%%%%%%%

%---------------------------------------------------------------

%%%%%%%%%%%%%%%%%%%%%%%%%%%%%%%%%%%%%%%%%%%%%%%%
% table_Robust.StCapGainTax.HPI.D99t05.D07t09
%%%%%%%%%%%%%%%%%%%%%%%%%%%%%%%%%%%%%%%%%%%%%%%%

\noindent 

\begin{table}[h!]
\centering
\caption{
\textbf{Robustness Test for State Capital Gain Tax Rates.} \\
Stacked 2SLS Regressions of House Price Growth in Boom (99-05) and Bust (07-09) Periods on PLNJM Growth (99-05)  \smallskip \newline
{\scriptsize
This table reports stacked 2SLS regression $\triangle_{99,05} \& \triangle_{07,09} Ln(HPI_{c}) = \beta_{99,05} * \triangle_{99,05} Ln(PLNJM_{c}) \times Dum_{99,05} + \beta_{07,09} * \triangle_{99,05} Ln(PLNJM_{c}) \times Dum_{07,09} + \beta_{Tax, Boom} \times StateCapGainTax_{s} \times Dum_{00,06}  + \beta_{Tax, Bust} \times StateCapGainTax_{s} * \times Dum_{07,10} + \gamma_{99,05}* \bm{Controls_{c}} \times Dum_{99,05} + \gamma_{07,09}* \bm{Controls_{c}} \times Dum_{07,09} + \epsilon_{period, c}$. The left-hand-side dependent variable $\triangle_{99,05} \& \triangle_{07,09} Ln(HPI_{c})$ is the stacked growth rate of the house price index at county $c$ 99-05 and 07-09. The key variable of interest $\triangle_{99,05} Ln(PLNJM_{c})$ is the growth rate of the dollar amount of private-label mortgages at the county $c$ 99-05. $StateCapGainTax_{s}$ is the 2005 state-level capital gain tax rate. $Controls_{c}$ indicates control variables at county $c$ in the start year 1999. We add state-level capital gain tax rate interacted with period dummies. We use the gravity model-based instrumental variable ($\triangle_{99,05}\text{givNetExp}_{m}$) as IV for $\triangle_{99,05}Ln(PLNJM_{c})$. For the first-stage F-test of the non-stacked sample (99-05), we report kleibergen-Paap (2006) robust (clustered) statistics and Montiel Olea-Pflueger (2013) efficient statistics. Each regression is weighted by the natural logarithm of housing units in 1999. Standard errors are clustered at the CBSA level. ***, **, and * indicate significance at the 1\%, 5\%, and 10\% levels, respectively.
} % end of small font size
} % end of caption
\label{table_Robust.StCapGainTax.HPI.D99t05.D07t09}

\resizebox{\columnwidth}{!}{%

\begin{tabular}{l*{4}{c}}
\toprule
\textbf{TSLS Estimates}              &\multicolumn{4}{c}{House Price Growth (07USD, 99-05 or 07-09, An)} \\
            \cmidrule{2-5} 
            &\multicolumn{1}{c}{(1)}&\multicolumn{1}{c}{(2)}&\multicolumn{1}{c}{(3)}&\multicolumn{1}{c}{(4)}\\
            
\midrule

PLNJM Dollar Growth (99-05, 07USD, An) x Dum99t05 &    0.547***&    0.495***&    0.476***&    0.514***\\
               &  (0.116)   &  (0.123)   &  (0.147)   &  (0.168)   \\
\addlinespace
PLNJM Dollar Growth (99-05, 07USD, An) x Dum07t09 &   -0.943***&   -0.895***&   -0.902***&   -0.945***\\
               &  (0.224)   &  (0.227)   &  (0.266)   &  (0.260)   \\
\addlinespace
State Capital Gain Tax x Dum99t05&    0.246** &    0.309***&    0.251** &    0.270** \\
               &  (0.115)   &  (0.112)   &  (0.097)   &  (0.111)   \\
\addlinespace
State Capital Gain Tax x Dum07t09&   -0.287   &   -0.422** &   -0.395** &   -0.442** \\
               &  (0.181)   &  (0.173)   &  (0.174)   &  (0.193)   \\
\addlinespace
\midrule
DumPeriod  &    Y        &  Y   &   Y    & Y      \\
Basic Controls x DumPeriod &            &  Y   &   Y    & Y      \\
Housing Controls x DumPeriod &           &      & Y       & Y     \\
Demographic Controls x DumPeriod &            &      &        &  Y       \\
\midrule
Obs            &     1576   &     1576   &     1418   &     1418   \\
Cluster SE     &     CBSA   &     CBSA   &     CBSA   &     CBSA   \\
Weight         & Ln(HU99)   & Ln(HU99)   & Ln(HU99)   & Ln(HU99)   \\
KP F-Stat (99 to 05, non-stacked sample)   &    19.69   &    17.75   &    15.06   &    11.87   \\
MOP F-Stat (99 to 05, non-stacked sample)   &    19.69   &    17.75   &    15.06   &    11.87   \\
CoefEqual\_Chi2 &   22.017   &   18.233   &   12.244   &   12.601   \\
CoefEqual\_PValue&    0.000   &    0.000   &    0.000   &    0.000   \\

\bottomrule

\end{tabular}

} % end of resize box

\end{table}

\pagebreak 
%---------------------------------------------------------------
%---------------------------------------------------------------
% Empirical: Verification Tests
%---------------------------------------------------------------
%---------------------------------------------------------------

%---------------------------------------------------------------

%%%%%%%%%%%%%%%%%%%%%%%%%%%%%%%%%%%%%%%%%%%%%%%%
% table_BEATotExConstEmp.D91t07.4Reg
%%%%%%%%%%%%%%%%%%%%%%%%%%%%%%%%%%%%%%%%%%%%%%%%

%---------------------------------------------------------------

%%%%%%%%%%%%%%%%%%%%%%%%%%%%%%%%%%%%%%%%%%%%%%%%
% table_BEATotExConstEmp.D91t07.4Reg
%%%%%%%%%%%%%%%%%%%%%%%%%%%%%%%%%%%%%%%%%%%%%%%%

\noindent 

\begin{table}[h!]
\centering
\caption{
\textbf{Four Regressions of BEA Total (excluding construction) Employment Share Growth (91-07) on Net Export Growth (91-07)} \smallskip \newline
{\footnotesize 
This table reports OLS, reduced-form, first stage, and second stage results of 2SLS regression $\triangle_{91,07} \text{TotExConstEmpShr}_{c} = \beta * \triangle_{91,07} \text{NetExp}_{m} + \gamma* \bm{Controls_{c}} + \alpha + \epsilon_{c}$. The left-hand-side dependent variable $\triangle_{91,07} \text{TotExConstEmpShr}_{c}$ is the increase of BEA total (excluding construction) employment share per working-age population at county $c$ 91-07 and the key variable of interest $\triangle_{91,07} \text{NetExp}_{m}$ is the growth rate of net export at the metropolitan area (CBSA03 code) $m$ 91-07. $Controls_{c}$ indicates control variables at county $c$ in 1991. We use the gravity model-based instrumental variable ($\triangle_{91,07}\text{givNetExp}_{m}$) as IV for $\triangle_{91,07}\text{NetExp}_{m}$. For the first-stage F-test, we report robust (clustered) Kleibergen-Paap (2006) F-statistics and Montiel Olea-Pflueger (2013) efficient statistics. Each regression is weighted by the natural logarithm of the county-level population in 1991. Standard errors are clustered at the CBSA level. ***, **, and * indicate significance at the 1\%, 5\%, and 10\% levels, respectively.
} % end of small font size
} % end of caption
\label{table_BEATotExConstEmp.D91t07.4Reg}

\resizebox{\columnwidth}{!}{%

\begin{tabular}{l*{3}{c}}
\toprule
Dep Var (Panel A, B, and C)                     &\multicolumn{3}{c}{\small Total (ex const) Emp Share Growth (1991-2007, annualized)} \\
            \cmidrule{2-4} 
            &\multicolumn{1}{c}{(1)}&\multicolumn{1}{c}{(2)}&\multicolumn{1}{c}{(3)}\\
            
\midrule
\multicolumn{4}{l}{\textbf{Panel A. OLS estimates}} \\
\addlinespace
Net Export Growth (91-07, An)&    0.392***&    0.214** &    0.175** \\
               &  (0.097)   &  (0.105)   &  (0.087)   \\
\addlinespace
R2-adj         &  0.00975   &   0.0844   &    0.150   \\
\addlinespace

\midrule
\multicolumn{4}{l}{\textbf{Panel B. Reduced-form estimates}} \\
\addlinespace
GIV Net Export Growth (91-07, An)&    0.531***&    0.397***&    0.343***\\
               &  (0.137)   &  (0.115)   &  (0.103)   \\
\addlinespace
R2-adj         &  0.00880   &   0.0867   &    0.152   \\
\addlinespace

\midrule
\multicolumn{4}{l}{\textbf{Panel C . 2SLS estimates}} \\
\addlinespace
Net Export Growth (91-07, An)&    0.493***&    0.376***&    0.326***\\
               &  (0.102)   &  (0.101)   &  (0.091)   \\

\addlinespace
\addlinespace

Dep Var (Panel D): &\multicolumn{3}{c}{Net Export Growth (91-07, An)} \\ 
\midrule 
\multicolumn{4}{l}{\textbf{Panel D . First-stage estimates}} \\
\addlinespace
GIV Net Export Growth (91-07, An)&    1.077***&    1.057***&    1.051***\\
               &  (0.117)   &  (0.110)   &  (0.108)   \\
\addlinespace
KP F-Stat      &    85.18   &    92.68   &    94.55   \\
MOP F-Stat     &    85.18   &    92.68   &    94.55   \\

\addlinespace
\midrule
\multicolumn{4}{l}{\textbf{Controls (for all Panels)}} \\
Basic Controls          &       &  Y   &   Y    \\
Demographic Controls  &         &      &  Y      \\
\midrule              
Obs            &     1041   &     1041   &     1041   \\
Cluster SE     &     CBSA   &     CBSA   &     CBSA    \\
Weight         &{\footnotesize Ln(TotPop91)}   &{\footnotesize Ln(TotPop91)}   &{\footnotesize Ln(TotPop91)}   \\
\bottomrule

\end{tabular}

} % end of resize box

\end{table}

\pagebreak 
%---------------------------------------------------------------

%%%%%%%%%%%%%%%%%%%%%%%%%%%%%%%%%%%%%%%%%%%%%%%%
% table_IRSHHIncome.D91t07.4Reg
%%%%%%%%%%%%%%%%%%%%%%%%%%%%%%%%%%%%%%%%%%%%%%%%

%---------------------------------------------------------------

%%%%%%%%%%%%%%%%%%%%%%%%%%%%%%%%%%%%%%%%%%%%%%%%
% table_IRSHHIncome.D91t07.4Reg
%%%%%%%%%%%%%%%%%%%%%%%%%%%%%%%%%%%%%%%%%%%%%%%%

\noindent 

\begin{table}[h!]
\centering
\caption{
\textbf{Four Regressions of IRS Household Income Growth (91-07) on Net Export Growth (91-07)} \smallskip \newline
{\footnotesize 
This table reports OLS, reduced-form, first stage, and second stage results of 2SLS regression $\triangle_{91,07} Ln(HHIncome_{c}) = \beta * \triangle_{91,07} \text{NetExp}_{m} + \gamma* \bm{Controls_{c}} + \alpha + \epsilon_{c}$. The left-hand-side dependent variable $\triangle_{91,07} Ln(HHIncome_{c})$ is the growth rate of the average household income (deflated to 2007 USD) at county $c$ 91-07, and the key variable of interest $\triangle_{91,07} \text{NetExp}_{m}$ is the growth rate of net export at the metropolitan area (CBSA03 code) $m$ 91-07. $Controls_{c}$ indicates control variables at county $c$ in 1991. We use the gravity model-based instrumental variable ($\triangle_{91,07}\text{givNetExp}_{m}$) as IV for $\triangle_{91,07}\text{NetExp}_{m}$. For the first-stage F-test, we report Kleibergen-Paap (2006) robust (clustered) statistics and Montiel Olea-Pflueger (2013) efficient statistics. Each regression is weighted by the natural logarithm of the number of households in 1991. Standard errors are clustered at the CBSA level. ***, **, and * indicate significance at the 1\%, 5\%, and 10\% levels, respectively.
} % end of small font size
} % end of caption
\label{table_IRSHHIncome.D91t07.4Reg}

\resizebox{\columnwidth}{!}{%

\begin{tabular}{l*{3}{c}}
\toprule
Dep Var (Panel A, B, and C)                     &\multicolumn{3}{c}{IRS Household Income Growth (1991-2007, An, 07USD)} \\
            \cmidrule{2-4} 
            &\multicolumn{1}{c}{(1)}&\multicolumn{1}{c}{(2)}&\multicolumn{1}{c}{(3)}\\
            
\midrule
\multicolumn{4}{l}{\textbf{Panel A. OLS estimates}} \\
\addlinespace
Net Export Growth (91-07, An)&    1.230***&    1.049***&    0.993***\\
               &  (0.148)   &  (0.123)   &  (0.115)   \\
\addlinespace
R2-adj         &   0.0742   &    0.240   &    0.330   \\
\addlinespace

\midrule
\multicolumn{4}{l}{\textbf{Panel B. Reduced-form estimates}} \\
\addlinespace
GIV Net Export Growth (91-07, An)&    1.246***&    1.081***&    1.098***\\
               &  (0.237)   &  (0.192)   &  (0.177)   \\
\addlinespace
R2-adj         &   0.0368   &    0.216   &    0.313   \\
\addlinespace

\midrule
\multicolumn{4}{l}{\textbf{Panel C . 2SLS estimates}} \\
\addlinespace
Net Export Growth (91-07, An)&    1.149***&    1.016***&    1.036***\\
               &  (0.180)   &  (0.155)   &  (0.141)   \\
\addlinespace
\addlinespace

Dep Var (Panel D): &\multicolumn{3}{c}{Net Export Growth (91-07, An)} \\ 
\midrule 
\multicolumn{4}{l}{\textbf{Panel D . First-stage estimates}} \\
\addlinespace
GIV Net Export Growth (91-07, An)&    1.085***&    1.064***&    1.059***\\
               &  (0.117)   &  (0.111)   &  (0.109)   \\
\addlinespace
KP F-Stat     &    86.22   &    91.76   &    93.73   \\
MOP F-Stat  &    86.22   &    91.76   &    93.73   \\

\addlinespace
\midrule
\multicolumn{4}{l}{\textbf{Controls (for all Panels)}} \\
Basic Controls &            &  Y   &   Y         \\
Demographic Controls  &         &     &  Y      \\
\midrule              
Obs            &     1074   &     1074   &     1074   \\
Cluster SE     &     CBSA   &     CBSA   &     CBSA     \\
Weight         &{\footnotesize Ln(NumHH91)}   &{\footnotesize Ln(NumHH91)}   &{\footnotesize Ln(NumHH91)}     \\
\bottomrule

\end{tabular}

} % end of resize box

\end{table}

\pagebreak 
%---------------------------------------------------------------

%%%%%%%%%%%%%%%%%%%%%%%%%%%%%%%%%%%%%%%%%%%%%%%%
% table_WorkAgePop.D91t07.4Reg
%%%%%%%%%%%%%%%%%%%%%%%%%%%%%%%%%%%%%%%%%%%%%%%%

%---------------------------------------------------------------

%%%%%%%%%%%%%%%%%%%%%%%%%%%%%%%%%%%%%%%%%%%%%%%%
% table_WorkAgePop.D91t07.4Reg
%%%%%%%%%%%%%%%%%%%%%%%%%%%%%%%%%%%%%%%%%%%%%%%%

\noindent 

\begin{table}[h!]
\centering
\caption{
\textbf{Four Regressions of Total and Working-Age Population Growth (91-07) on Net Export Growth (91-07)} \smallskip \newline
{\footnotesize 
This table reports OLS, reduced-form, first stage, and second stage results of 2SLS regression $\triangle_{91,07} Ln(Pop_{c}) = \beta * \triangle_{91,07} \text{NetExp}_{m} + \gamma* \bm{Controls_{c}} + \alpha + \epsilon_{c}$. The left-hand-side dependent variable $\triangle_{91,07} Ln(Pop_{c})$ is the growth rate of the total population growth (in column 1-3) or working-age population (age 15 - 64) growth (in column 4-6) at county $c$ 91-07, and the key variable of interest $\triangle_{91,07} \text{NetExp}_{m}$ is the growth rate of net export at the metropolitan area (CBSA03 code) $m$ 91-07. $Controls_{c}$ indicates control variables at county $c$ in 1991. We use the gravity model-based instrumental variable ($\triangle_{91,07}\text{givNetExp}_{m}$) as IV for $\triangle_{91,07}\text{NetExp}_{m}$. For the first-stage F-test, we report Kleibergen-Paap (2006) robust (clustered) statistics and Montiel Olea-Pflueger (2013) efficient statistics. Each regression is weighted by the natural logarithm of the number of households in 1991. Standard errors are clustered at the CBSA level. ***, **, and * indicate significance at the 1\%, 5\%, and 10\% levels, respectively.
} % end of small font size
} % end of caption
\label{table_WorkAgePop.D91t07.4Reg}

\resizebox{\columnwidth}{!}{%

\begin{tabular}{l*{6}{c}}
\toprule
Dep Var (Panel A, B, and C)                     &\multicolumn{3}{c}{Total Population Growth (91-07, An)}  &\multicolumn{3}{c}{Working-Age Population Growth (91-07, An)} \\
            \cmidrule{2-7} 
            &\multicolumn{1}{c}{(1)}&\multicolumn{1}{c}{(2)}&\multicolumn{1}{c}{(3)}&\multicolumn{1}{c}{(4)}&\multicolumn{1}{c}{(5)}&\multicolumn{1}{c}{(6)}\\
            
\midrule
\multicolumn{7}{l}{\textbf{Panel A. OLS estimates}} \\
\addlinespace
Net Export Growth (91-07, An)&    0.729***&    0.842***&    0.761***&    0.979***&    1.163***&    1.093***\\
               &  (0.259)   &  (0.250)   &  (0.236)   &  (0.249)   &  (0.237)   &  (0.228)   \\
\addlinespace
R2-adj         &  0.00739   &    0.132   &    0.168   &   0.0144   &    0.122   &    0.146   \\
\addlinespace

\midrule
\multicolumn{7}{l}{\textbf{Panel B. Reduced-form estimates}} \\
\addlinespace
GIV Net Export Growth (91-07, An)&    0.759** &    0.859***&    0.795** &    1.033***&    1.185***&    1.137***\\
               &  (0.328)   &  (0.329)   &  (0.318)   &  (0.331)   &  (0.334)   &  (0.319)   \\
\addlinespace
R2-adj         &  0.00348   &    0.126   &    0.165   &  0.00740   &    0.112   &    0.137   \\
\addlinespace

\midrule
\multicolumn{7}{l}{\textbf{Panel C . 2SLS estimates}} \\
\addlinespace
Net Export Growth (91-07, An)&    0.700** &    0.807***&    0.750** &    0.952***&    1.113***&    1.074***\\
               &  (0.308)   &  (0.305)   &  (0.302)   &  (0.304)   &  (0.296)   &  (0.291)   \\
               
\addlinespace
\addlinespace

Dep Var (Panel D): &\multicolumn{6}{c}{Net Export Growth (91-07, An)} \\ 
\midrule 
\multicolumn{7}{l}{\textbf{Panel D . First-stage estimates}} \\
\addlinespace
GIV Net Export Growth (91-07, An)&    1.085***&    1.064***&    1.059***&    1.085***&    1.064***&    1.059***\\
               &  (0.117)   &  (0.111)   &  (0.109)   &  (0.117)   &  (0.111)   &  (0.109)   \\
\addlinespace
KP F-Stat      &    86.20   &    91.72   &    93.75   &    86.20   &    91.72   &    93.75   \\
MOP F-Stat     &    86.20   &    91.72   &    93.75   &    86.20   &    91.72   &    93.75   \\

\addlinespace
\midrule
\multicolumn{7}{l}{\textbf{Controls (for all Panels)}} \\
Basic Controls &            &  Y   &   Y   &            &  Y   &   Y        \\
Demographic Controls  &         &      & Y       &      &         & Y     \\
\midrule              
Obs            &     1075   &     1075   &     1075   &     1075   &     1075   &     1075   \\
Cluster SE     &     CBSA   &     CBSA   &     CBSA   &     CBSA   &     CBSA   &     CBSA   \\
Weight         &Ln(NumHH91)   &Ln(NumHH91)   &Ln(NumHH91)   &Ln(NumHH91)   &Ln(NumHH91)   &Ln(NumHH91)   \\
\bottomrule

\end{tabular}

} % end of resize box

\end{table}

%\pagebreak 
%---------------------------------------------------------------

%%%%%%%%%%%%%%%%%%%%%%%%%%%%%%%%%%%%%%%%%%%%%%%%
% table_IRSNumHH.D92t06.4Reg
% this table might go to business cycle paper
%%%%%%%%%%%%%%%%%%%%%%%%%%%%%%%%%%%%%%%%%%%%%%%%

%\input{Table/table_IRSNumHH.D92t06.4Reg}

%------------------------------------------------------------
%----------------------------------------------------------------------

%----------------------------------------------------------------------
% end of document
%----------------------------------------------------------------------

\end{document}